\documentclass[10pt,letterpaper]{article}
\usepackage[top=0.85in,left=2.75in,footskip=0.75in]{geometry}

\usepackage{amsmath,amssymb}

\usepackage{changepage}

\usepackage{textcomp,marvosym}

\usepackage{cite}

\usepackage{nameref,hyperref}

\usepackage[right]{lineno}

\usepackage[table]{xcolor}

\usepackage{array}

\newcolumntype{+}{!{\vrule width 2pt}}

\newlength\savedwidth

\raggedright
\usepackage[aboveskip=1pt,labelfont=bf,labelsep=period,justification=raggedright,singlelinecheck=off]{caption}

\makeatletter
\renewcommand{\@biblabel}[1]{\quad#1.}
\makeatother

\usepackage{lastpage,fancyhdr,graphicx}
\usepackage{epstopdf}
\fancyheadoffset[L]{2.25in}
\fancyfootoffset[L]{2.25in}
\usepackage[utf8]{inputenc}
\usepackage[T1]{fontenc}
\usepackage{bm}
\usepackage{setspace}
\usepackage{epsfig}
\usepackage{amsmath}
\usepackage{lscape}
\usepackage[justification=justified]{caption}
\usepackage{url}
\usepackage{xcolor}

\usepackage{amsmath,amsthm,amssymb,mathtools,bm}
\usepackage{algorithm,algorithmic}
\usepackage{enumerate}
\graphicspath{{./figs/}}

\usepackage{graphicx}
\usepackage{multirow}
\usepackage{xparse}
\usepackage{array}
\usepackage{tabularx}%
\usepackage{booktabs}

\usepackage{subcaption}

\newcommand{\dtoprule}{\specialrule{1pt}{0pt}{0.4pt}%
            \specialrule{0.3pt}{0pt}{\belowrulesep}%
            }
\newcommand{\dbottomrule}{\specialrule{0.3pt}{0pt}{0.4pt}%
            \specialrule{1pt}{0pt}{\belowrulesep}%
            }

\newcommand\norm[1]{\left\Vert#1\right\Vert}
\newcommand{\bfA}{\bf A}
\newcommand{\cG}{\mathcal{G}}
\newcommand{\bfX}{\bf X}

\newcommand{\bfD}{\bf D}

\newcommand{\real}{\mathbb{R}}
\newcommand{\Ex}{\mathbb{E}}
\newcommand{\ind}[1]{[#1]}

\DeclareMathOperator*{\argmin}{arg\,min}
\DeclareMathOperator{\trace}{Tr}

\NewDocumentCommand{\boxDictZero}{m O{0.6in} O{0.6in}}{\parbox[c]{#2}{\includegraphics[width=#3]{online_cvxNDL_figures/dictionary_plots/#1/#1_dictionary_0.png}}}
\NewDocumentCommand{\boxDictOne}{m O{0.6in} O{0.6in}}{\parbox[c]{#2}{\includegraphics[width=#3]{online_cvxNDL_figures/dictionary_plots/#1/#1_dictionary_1.png}}}
\NewDocumentCommand{\boxDictTwo}{m O{0.6in} O{0.6in}}{\parbox[c]{#2}{\includegraphics[width=#3]{online_cvxNDL_figures/dictionary_plots/#1/#1_dictionary_2.png}}}
\NewDocumentCommand{\boxDictThree}{m O{0.6in} O{0.6in}}{\parbox[c]{#2}{\includegraphics[width=#3]{online_cvxNDL_figures/dictionary_plots/#1/#1_dictionary_3.png}}}
\NewDocumentCommand{\boxDictFour}{m O{0.6in} O{0.6in}}{\parbox[c]{#2}{\includegraphics[width=#3]{online_cvxNDL_figures/dictionary_plots/#1/#1_dictionary_4.png}}}
\NewDocumentCommand{\boxDictFive}{m O{0.6in} O{0.6in}}{\parbox[c]{#2}{\includegraphics[width=#3]{online_cvxNDL_figures/dictionary_plots/#1/#1_dictionary_5.png}}}
\NewDocumentCommand{\boxDictSix}{m O{0.6in} O{0.6in}}{\parbox[c]{#2}{\includegraphics[width=#3]{online_cvxNDL_figures/dictionary_plots/#1/#1_dictionary_6.png}}}
\NewDocumentCommand{\boxDictSeven}{m O{0.6in} O{0.6in}}{\parbox[c]{#2}{\includegraphics[width=#3]{online_cvxNDL_figures/dictionary_plots/#1/#1_dictionary_7.png}}}
\NewDocumentCommand{\boxDictEight}{m O{0.6in} O{0.6in}}{\parbox[c]{#2}{\includegraphics[width=#3]{online_cvxNDL_figures/dictionary_plots/#1/#1_dictionary_8.png}}}
\NewDocumentCommand{\boxDictNine}{m O{0.6in} O{0.6in}}{\parbox[c]{#2}{\includegraphics[width=#3]{online_cvxNDL_figures/dictionary_plots/#1/#1_dictionary_9.png}}}
\NewDocumentCommand{\boxDictOneZero}{m O{0.6in} O{0.6in}}{\parbox[c]{#2}{\includegraphics[width=#3]{online_cvxNDL_figures/dictionary_plots/#1/#1_dictionary_10.png}}}
\NewDocumentCommand{\boxDictOneOne}{m O{0.6in} O{0.6in}}{\parbox[c]{#2}{\includegraphics[width=#3]{online_cvxNDL_figures/dictionary_plots/#1/#1_dictionary_11.png}}}
\NewDocumentCommand{\boxDictOneTwo}{m O{0.6in} O{0.6in}}{\parbox[c]{#2}{\includegraphics[width=#3]{online_cvxNDL_figures/dictionary_plots/#1/#1_dictionary_12.png}}}
\NewDocumentCommand{\boxDictOneThree}{m O{0.6in} O{0.6in}}{\parbox[c]{#2}{\includegraphics[width=#3]{online_cvxNDL_figures/dictionary_plots/#1/#1_dictionary_13.png}}}
\NewDocumentCommand{\boxDictOneFour}{m O{0.6in} O{0.6in}}{\parbox[c]{#2}{\includegraphics[width=#3]{online_cvxNDL_figures/dictionary_plots/#1/#1_dictionary_14.png}}}
\NewDocumentCommand{\boxDictOneFive}{m O{0.6in} O{0.6in}}{\parbox[c]{#2}{\includegraphics[width=#3]{online_cvxNDL_figures/dictionary_plots/#1/#1_dictionary_15.png}}}
\NewDocumentCommand{\boxDictOneSix}{m O{0.6in} O{0.6in}}{\parbox[c]{#2}{\includegraphics[width=#3]{online_cvxNDL_figures/dictionary_plots/#1/#1_dictionary_16.png}}}
\NewDocumentCommand{\boxDictOneSeven}{m O{0.6in} O{0.6in}}{\parbox[c]{#2}{\includegraphics[width=#3]{online_cvxNDL_figures/dictionary_plots/#1/#1_dictionary_17.png}}}
\NewDocumentCommand{\boxDictOneEight}{m O{0.6in} O{0.6in}}{\parbox[c]{#2}{\includegraphics[width=#3]{online_cvxNDL_figures/dictionary_plots/#1/#1_dictionary_18.png}}}
\NewDocumentCommand{\boxDictOneNine}{m O{0.6in} O{0.6in}}{\parbox[c]{#2}{\includegraphics[width=#3]{online_cvxNDL_figures/dictionary_plots/#1/#1_dictionary_19.png}}}
\NewDocumentCommand{\boxDictTwoZero}{m O{0.6in} O{0.6in}}{\parbox[c]{#2}{\includegraphics[width=#3]{online_cvxNDL_figures/dictionary_plots/#1/#1_dictionary_20.png}}}
\NewDocumentCommand{\boxDictTwoOne}{m O{0.6in} O{0.6in}}{\parbox[c]{#2}{\includegraphics[width=#3]{online_cvxNDL_figures/dictionary_plots/#1/#1_dictionary_21.png}}}
\NewDocumentCommand{\boxDictTwoTwo}{m O{0.6in} O{0.6in}}{\parbox[c]{#2}{\includegraphics[width=#3]{online_cvxNDL_figures/dictionary_plots/#1/#1_dictionary_22.png}}}
\NewDocumentCommand{\boxDictTwoThree}{m O{0.6in} O{0.6in}}{\parbox[c]{#2}{\includegraphics[width=#3]{online_cvxNDL_figures/dictionary_plots/#1/#1_dictionary_23.png}}}
\NewDocumentCommand{\boxDictTwoFour}{m O{0.6in} O{0.6in}}{\parbox[c]{#2}{\includegraphics[width=#3]{online_cvxNDL_figures/dictionary_plots/#1/#1_dictionary_24.png}}}

\NewDocumentCommand{\dictZero}{m O{0.55in}}{\includegraphics[width=#2]{online_cvxNDL_figures/dictionary_plots/#1/#1_dictionary_0.png}}
\NewDocumentCommand{\dictOne}{m O{0.55in}}{\includegraphics[width=#2]{online_cvxNDL_figures/dictionary_plots/#1/#1_dictionary_1.png}}
\NewDocumentCommand{\dictTwo}{m O{0.55in}}{\includegraphics[width=#2]{online_cvxNDL_figures/dictionary_plots/#1/#1_dictionary_2.png}}
\NewDocumentCommand{\dictThree}{m O{0.55in}}{\includegraphics[width=#2]{online_cvxNDL_figures/dictionary_plots/#1/#1_dictionary_3.png}}
\NewDocumentCommand{\dictFour}{m O{0.55in}}{\includegraphics[width=#2]{online_cvxNDL_figures/dictionary_plots/#1/#1_dictionary_4.png}}
\NewDocumentCommand{\dictFive}{m O{0.55in}}{\includegraphics[width=#2]{online_cvxNDL_figures/dictionary_plots/#1/#1_dictionary_5.png}}
\NewDocumentCommand{\dictSix}{m O{0.55in}}{\includegraphics[width=#2]{online_cvxNDL_figures/dictionary_plots/#1/#1_dictionary_6.png}}
\NewDocumentCommand{\dictSeven}{m O{0.55in}}{\includegraphics[width=#2]{online_cvxNDL_figures/dictionary_plots/#1/#1_dictionary_7.png}}
\NewDocumentCommand{\dictEight}{m O{0.55in}}{\includegraphics[width=#2]{online_cvxNDL_figures/dictionary_plots/#1/#1_dictionary_8.png}}
\NewDocumentCommand{\dictNine}{m O{0.55in}}{\includegraphics[width=#2]{online_cvxNDL_figures/dictionary_plots/#1/#1_dictionary_9.png}}
\NewDocumentCommand{\dictOneZero}{m O{0.55in}}{\includegraphics[width=#2]{online_cvxNDL_figures/dictionary_plots/#1/#1_dictionary_10.png}}
\NewDocumentCommand{\dictOneOne}{m O{0.55in}}{\includegraphics[width=#2]{online_cvxNDL_figures/dictionary_plots/#1/#1_dictionary_11.png}}
\NewDocumentCommand{\dictOneTwo}{m O{0.55in}}{\includegraphics[width=#2]{online_cvxNDL_figures/dictionary_plots/#1/#1_dictionary_12.png}}
\NewDocumentCommand{\dictOneThree}{m O{0.55in}}{\includegraphics[width=#2]{online_cvxNDL_figures/dictionary_plots/#1/#1_dictionary_13.png}}
\NewDocumentCommand{\dictOneFour}{m O{0.55in}}{\includegraphics[width=#2]{online_cvxNDL_figures/dictionary_plots/#1/#1_dictionary_14.png}}
\NewDocumentCommand{\dictOneFive}{m O{0.55in}}{\includegraphics[width=#2]{online_cvxNDL_figures/dictionary_plots/#1/#1_dictionary_15.png}}
\NewDocumentCommand{\dictOneSix}{m O{0.55in}}{\includegraphics[width=#2]{online_cvxNDL_figures/dictionary_plots/#1/#1_dictionary_16.png}}
\NewDocumentCommand{\dictOneSeven}{m O{0.55in}}{\includegraphics[width=#2]{online_cvxNDL_figures/dictionary_plots/#1/#1_dictionary_17.png}}
\NewDocumentCommand{\dictOneEight}{m O{0.55in}}{\includegraphics[width=#2]{online_cvxNDL_figures/dictionary_plots/#1/#1_dictionary_18.png}}
\NewDocumentCommand{\dictOneNine}{m O{0.55in}}{\includegraphics[width=#2]{online_cvxNDL_figures/dictionary_plots/#1/#1_dictionary_19.png}}
\NewDocumentCommand{\dictTwoZero}{m O{0.55in}}{\includegraphics[width=#2]{online_cvxNDL_figures/dictionary_plots/#1/#1_dictionary_20.png}}
\NewDocumentCommand{\dictTwoOne}{m O{0.55in}}{\includegraphics[width=#2]{online_cvxNDL_figures/dictionary_plots/#1/#1_dictionary_21.png}}
\NewDocumentCommand{\dictTwoTwo}{m O{0.55in}}{\includegraphics[width=#2]{online_cvxNDL_figures/dictionary_plots/#1/#1_dictionary_22.png}}
\NewDocumentCommand{\dictTwoThree}{m O{0.55in}}{\includegraphics[width=#2]{online_cvxNDL_figures/dictionary_plots/#1/#1_dictionary_23.png}}
\NewDocumentCommand{\dictTwoFour}{m O{0.55in}}{\includegraphics[width=#2]{online_cvxNDL_figures/dictionary_plots/#1/#1_dictionary_24.png}}

\newcommand{\threeDict}[3]{\parbox[l][0.57in]{6in}{#1\hspace{0.12in}#2\hspace{0.12in}#3}}

\newcolumntype{G}{>{\hsize=0.7\hsize\linewidth=\hsize}X} 
\newcolumntype{D}{>{\hsize=0.3\hsize\linewidth=\hsize\centering\arraybackslash}X}

\newcolumntype{M}{>{\hsize=0.25\hsize\linewidth=\hsize}X} 
\newcolumntype{L}{>{\hsize=0.4\hsize\linewidth=\hsize\centering\arraybackslash}X}
\newcolumntype{J}{>{\hsize=0.8\hsize\linewidth=\hsize}X}
\newcolumntype{S}{>{\hsize=0.2\hsize\linewidth=\hsize\centering\arraybackslash}X}

\begin{document}

\vspace*{0.2in}

\begin{flushleft}
{\Large
\textbf\newline{Interpretable Online Network Dictionary Learning for Inferring Long-Range Chromatin Interactions} 
}
\newline
\\
Vishal Rana\textsuperscript{1},
Jianhao Peng\textsuperscript{1},
Chao Pan\textsuperscript{1},
Hanbaek Lyu\textsuperscript{2},
Albert Cheng\textsuperscript{3},
Minji Kim\textsuperscript{4},
Olgica Milenkovic\textsuperscript{1*}
\\
\bigskip
\textbf{1} Department of Electrical and Computer Engineering, University of Illinois, Urbana-Champaign.
\\
\textbf{2} Department of Mathematics, University of Wisconsin - Madison.
\\
\textbf{3} School of Biological and Health Systems Engineering, Arizona State University, Phoenix.
\\
\textbf{4} Department of Computational Medicine and Bioinformatics, University of Michigan, Ann Arbor.
\\
\bigskip

%
%





*Corresponding author: milenkov@illinois.edu

\end{flushleft}
\section*{Abstract}
 
Dictionary learning (DL), implemented via matrix factorization (MF), is commonly used in computational biology to tackle ubiquitous clustering problems. The method is favored due to its conceptual simplicity and relatively low computational complexity. However, DL algorithms produce results that lack interpretability in terms of real biological data. Additionally, they are not optimized for graph-structured data and hence often fail to handle them in a scalable manner. 

In order to address these limitations, we propose a novel DL algorithm called \emph{online convex network dictionary learning} (online cvxNDL). Unlike classical DL algorithms, online cvxNDL is implemented via MF and designed to handle extremely large datasets by virtue of its online nature. Importantly, it enables the interpretation of dictionary elements, which serve as cluster representatives, through convex combinations of real measurements. Moreover, the algorithm can be applied to data with a network structure by incorporating specialized subnetwork sampling techniques.

To demonstrate the utility of our approach, we apply cvxNDL on 3D-genome RNAPII ChIA-Drop data with the goal of identifying important long-range interaction patterns (long-range dictionary elements). ChIA-Drop probes higher-order interactions, and produces data in the form of hypergraphs whose nodes represent genomic fragments. The hyperedges represent observed physical contacts. Our hypergraph model analysis has the objective of creating an interpretable dictionary of long-range interaction patterns that accurately represent global chromatin physical contact maps. Through the use of dictionary information, one can also associate the contact maps with RNA transcripts and infer cellular functions.

To accomplish the task at hand, we focus on RNAPII-enriched ChIA-Drop data from \emph{Drosophila Melanogaster} S2 cell lines. Our results offer two key insights. First, we demonstrate that online cvxNDL retains the accuracy of classical DL (MF) methods while simultaneously ensuring unique interpretability and scalability. Second, we identify distinct collections of proximal and distal interaction patterns involving chromatin elements shared by related processes across different chromosomes, as well as patterns unique to specific chromosomes. To associate the dictionary elements with biological properties of the corresponding chromatin regions, we employ Gene Ontology (GO) enrichment analysis and perform multiple RNA coexpression studies.

\textbf{Availability and Implementation:} The code and test datasets are available at: \url{https://github.com/rana95vishal/chromatin_DL/} \\
\section*{Author summary}

We introduce a novel method for dictionary learning termed \emph{online convex Network Dictionary Learning} (online cvxNDL). The method operates in an online manner and utilizes representative subnetworks of a network dataset as dictionary elements. A key feature of online cvxNDL is its ability to work with graph-structured data and generate dictionary elements that represent convex combinations of real data points, thus ensuring interpretability.

Online cvxNDL is used to investigate long-range chromatin interactions in S2 cell lines of \emph{Drosophila Melanogaster} obtained through RNAPII ChIA-Drop measurements represented as hypergraphs. The results show that dictionary elements can accurately and efficiently reconstruct the original interactions present in the data, even when subjected to convexity constraints. To shed light on the biological relevance of the identified dictionaries, we perform Gene Ontology enrichment and RNA-seq coexpression analyses. These studies uncover multiple long-range interaction patterns that are chromosome-specific. Furthermore, the findings affirm the significance of convex dictionaries in representing TADs cross-validated by imaging methods (such as $3$-color FISH (fluorescence in situ hybridization)).


\section*{Introduction}
\label{sec:intro}

Dictionary learning (DL) is a widely used method in learning and computational biology for approximating a matrix through sparse linear combinations of dictionary elements. DL has been used in various applications such as clustering, denoising, data compression, and extracting low-dimensional patterns~\cite{elad2006image, mairal2007sparse, cichocki2008non, ye2014multitask,lu2020community,zhu2017detecting,shao2017robust,zhang2019silico}.  For example, DL is used to cluster data points since dictionary elements essentially represent centroids of clusters. DL can perform denoising by combining only the highest-score dictionary elements to reconstruct the input; in this case, the low-score dictionary elements reflect the distortion in the data due to noise. DL can also perform efficient data compression by storing only the dictionary elements and associated weights needed for reconstruction. In addition, DL can be used to extract low-dimensional patterns from complex high-dimensional inputs. 

However, standard DL methods~\cite{paatero1994positive, paatero1997least} suffer from interpretability and scalability issues and are primarily applied to \emph{unstructured} data. To address interpretability issues for unstructured data, convex matrix factorization was introduced in~\cite{ding2010convex}. Convex matrix factorization requires that the dictionary elements be convex combinations of real data points, thereby introducing a constraint that adds to the computational complexity of the method. At the same time, to improve scalability, DL and convex DL algorithms can be adapted to online settings~\cite{mairal2010online, peng2019online}. 
Network DL (NDL), introduced in~\cite{lyu2023sampling}, operates on graph-structured data and samples subnetworks via Markov Chain Monte Carlo 
 (MCMC) methods~\cite{lyu2023sampling,lyu2020online, lyu2021learning} to efficiently and accurately identify a small number of subnetwork dictionary elements that best explain subgraph-level interactions of the entire global network. These dictionary elements learned by the original NDL algorithm only provide `latent' subgraph structures that are not necessarily associated with specific subgraphs in the network. When applied to gene interaction networks, such latent subnetworks cannot be associated with specific genomic regions or viewed as physical interactions between genomic loci, making the method biologically uninterpretable.

To address the shortcoming of online NDL, we propose online cvxNDL, a novel NDL method that combines the MCMC sampling technique from~\cite{lyu2023sampling} with convexity constraints on the matrix representation of sampled subnetworks. These constraints are handled through the concept of ``dictionary element representatives,'' which are essentially adjacency matrices of real subnetworks of the input network. The representatives are used as building blocks of actual dictionary elements. More precisely, dictionary elements are convex combinations of small subsets of representatives. This allows us to map the dictionary element entries to actual genomic regions and view them as real physical interactions. The online learning component is handled via sequential updates of the best choice of representative elements, complementing the approach proposed in~\cite{peng2019online} for unstructured data. This formulation ensures interpretability of the results and allows for scaling to large datasets.

The utility of online cvxNDL is demonstrated by performing an extensive analysis of 3D chromatin interaction data generated by the RNAPII ChIA-Drop~\cite{zheng2019multiplex} technique.  Chromatin 3D structures play a crucial role in gene regulation~\cite{li2012extensive,tang2015ctcf} and have traditionally been measured using ``bulk'' sequencing methods, such as Hi-C~\cite{lieberman2009comprehensive} and ChIA-PET~\cite{li2010chia, fullwood2009oestrogen}. However, due to the proximity ligation step, these methods can only capture pairwise contacts and fail to extract potential multiway interactions that exist in the cell. Further, these methods operate on a population of millions of molecules and therefore only provide information about population averages. ChIA-Drop, by contrast, mitigates these issues by employing droplet-based barcode-linked sequencing to capture multiway chromatin interactions at the single-molecule level, enabling the detection of short- and long-range interactions involving multiple genomic loci. Note that, more specifically, RNAPII ChIA-Drop data elucidates interactions among regulatory elements such as enhancers and promoters, which warrants contrasting/combining it with RNA-seq data.

The cvxNDL method is first tested on synthetic data, and, subsequently, on real-world RNAPII ChIA-Drop data pertaining to chromosomes of \emph{Drosophila Melanogaster} Schneider 2 (S2) phagocytic cell lines\footnote{Due to the limited number of complete ChIA-Drop datasets, we only report findings for cell-lines also studied in~\cite{zheng2019multiplex}.}. For simplicity, we will henceforth refer to the latter as ChIA-Drop data. Our findings are multi-fold. 

First, we provide dictionary elements that can be used to represent chromatin interactions in a succinct and highly accurate manner.

Second, we discover significant differences between the long-range interactions captured by dictionary elements of different chromosomes. These differences can also be summarized via the average distance between interacting genomic loci and the densities of interactions. 

Third, we perform Gene Ontology (GO) enrichment analysis to gain insights into the collective functionality of the genomic regions represented by the dictionary elements of different chromosomes. As an example, for chromosomes 2L and 2R, our GO enrichment analysis reveals significant enrichment in several important terms related to reproduction, oocyte differentiation, and embryonic development. Likewise, chromosomes 3L and 3R are enriched in key GO terms associated with blood circulation and response to heat and cold.

Fourth, to further validate the utility of the dictionary elements, we perform an RNA-Seq coexpression analysis using data from independent experiments conducted on \emph{Drosophila Melanogaster} S2 cell lines, available through the NCBI Sequence Read Archive~\cite{ziemann2019digital}. We show that genes associated with a given dictionary element exhibit high levels of coexpression, as validated on TAD interactions T1-T4 and R1-R4~\cite{zheng2019multiplex}. Notably, a small subset of our dictionary elements is able to accurately represent these TAD regions and their multiway interactions, confirming the capability of our method to effectively capture complex patterns of both short- and long-range interactions. In addition, we map our dictionary elements onto interaction networks, including the STRING protein-protein interaction network~\cite{szklarczyk2019string}, as well as large gene expression repositories like FlyMine. We observe closely coordinated coexpression among the identified genes, further supporting the biological relevance of the identified dictionary elements.

With its unique features, our new interpretable method for dictionary learning adds to the growing literature on machine learning approaches that aim to elucidate properties of chromatin interactions~\cite{wang2022dloopcaller, xie2020characterizing, zhang2022clnn, tian2022mcibox}.

\section*{Results and Discussion}
\label{sec:results}

\begin{figure}[p]
\begin{subfigure}{\textwidth}
  \centering
  \includegraphics[width=.88\linewidth]
  {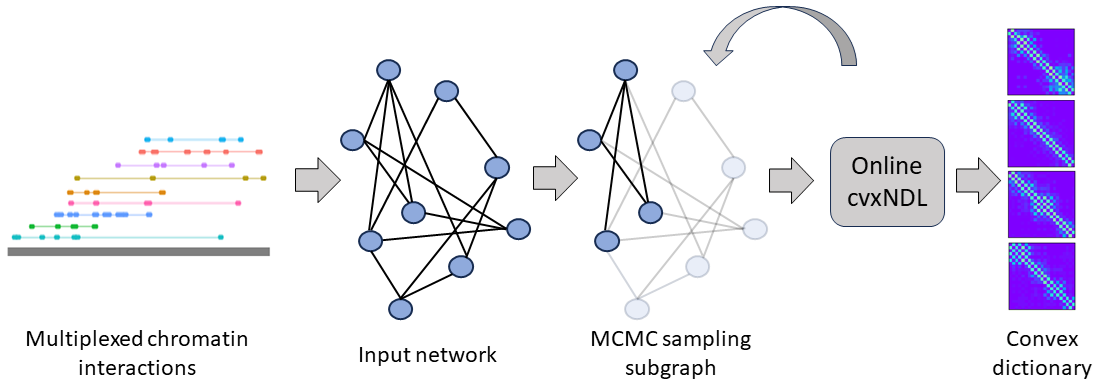}
  \caption{}
  \label{fig:cvxNDL_workflow}
\end{subfigure}
\hfill
\begin{subfigure}{0.5\textwidth}
  \centering
  \includegraphics[width=.9\linewidth]{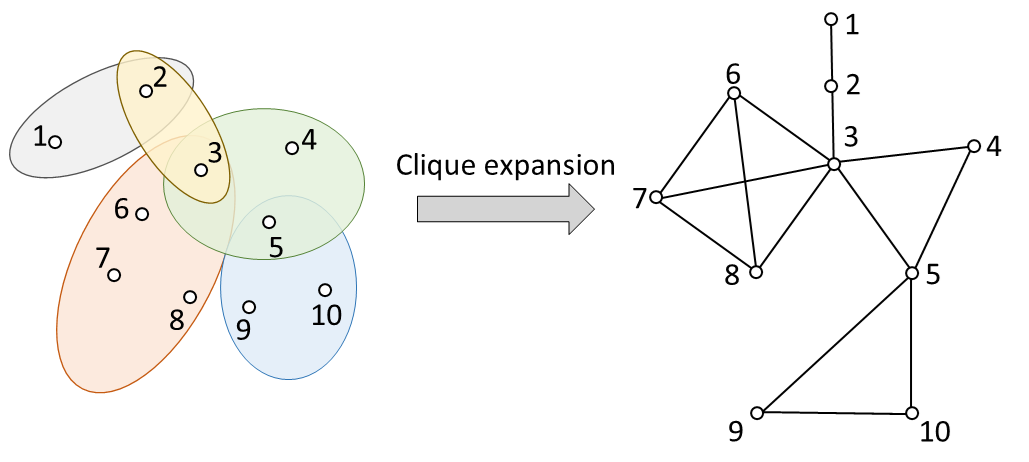}
  \caption{}
  \label{fig:clique_expansion_a}
\end{subfigure}
\begin{subfigure}{.5\textwidth}
  \centering
  \includegraphics[width=.9\linewidth]{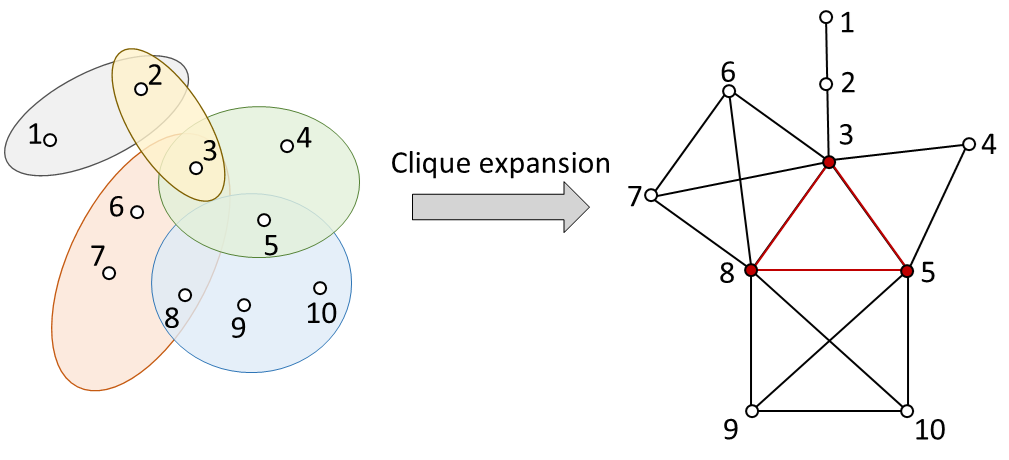}
  \caption{}
  \label{fig:clique_expansion_b}
\end{subfigure}
\hfill
\begin{subfigure}{.5\textwidth}
  \centering
  \includegraphics[width=.98\linewidth]{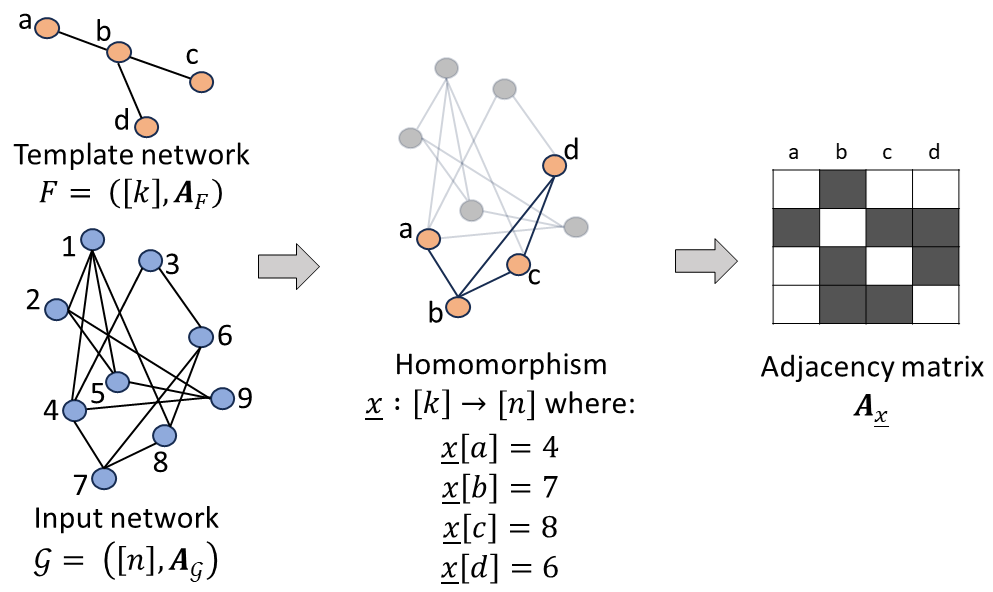}
  \caption{}
  \label{fig:motif_sample_img}
\end{subfigure}
\begin{subfigure}{.5\textwidth}
  \centering
  \includegraphics[width=.94\linewidth]{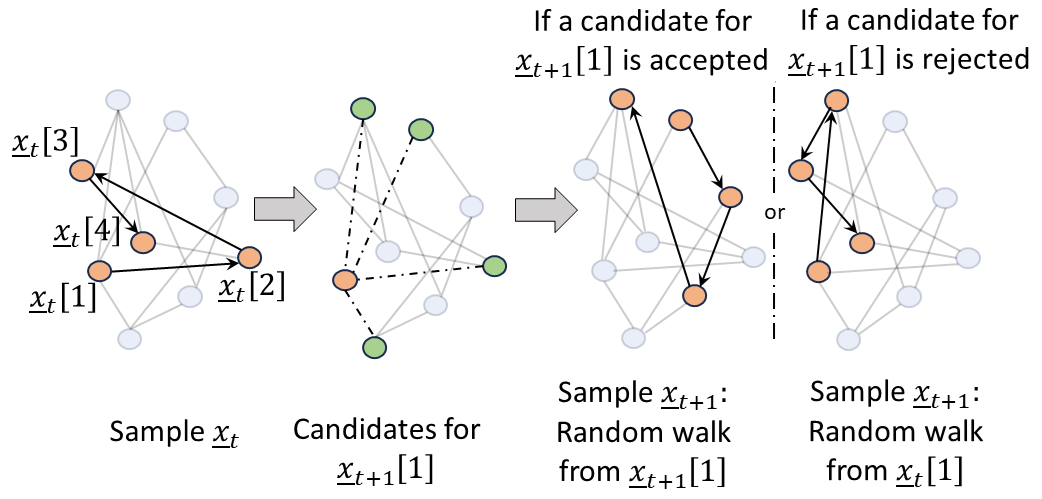}
  \caption{}
  \label{fig:mcmc_img}
\end{subfigure}
\vspace{0.1in}
\caption{(a) Workflow of the dictionary learning method. Multiway (multiplexed) chromatin interactions represented as hyperedges are \emph{clique expanded} into standard networks and combined to create input networks for the algorithm. MCMC subnetwork sampling is then used to generate samples for initialization and online updates during iterative optimization of the objective function, resulting in convex dictionary elements. (b) Illustration of the clique expansion process. Hyperedges are subsets of indexed nodes shaded with the same color. (c) Illustration of clique expansion distortion. There is no hyperedge including nodes $3,$ $5$, and $8$ (colored red), and this $3$-clique only exists due to shared nodes/edges of ``real'' hyperedges. Such distortion is negligible when the number of large hyperedges is limited. (d) Subnetwork sampling and the notion of a \emph{motif homomorphism}. These correspond to subnetworks of the input network induced by a fixed number of nodes that contain a template motif topology. The set of homomorphisms $\mathtt{Hom}(F, \cG)$ for a network $\cG$ and the template network $F$ are defined in the Methods Section (Equation~\ref{eq:homomorphism}). Also depicted are an example homomorphism $\underline{x}\in \mathtt{Hom}(F, \cG)$ and its induced adjacency matrix $\mathbf{A}_{\underline{x}}$ for an input network $\mathcal{G}$ with $9$ nodes. The template $F$ is a star network on $4$ nodes. In the adjacency matrix, a black field indicates $1$, while a white field indicates $0$. (e) Workflow of the MCMC sampling algorithm for path homomorphisms. Given a sample $\underline{x}_t$ at time $t,$ obtained via a directed random walk from an initial state in the input network, $\underline{x}_t[1]$, we generate a sample $\underline{x}_{t + 1}$ at time $t+1$ by choosing uniformly at random a node $v$ from the neighborhood of $\underline{x}_t[1]$ (marked in green) and calculating a  probability of acceptance $\beta$. If node $v$ is accepted, we initiate a new directed random walk from $v$, otherwise, we restart a directed random walk from $\underline{x}_t[1]$ to generate a new sample.}
\label{fig:pipeline}
\end{figure}

We first provide an intuitive, high-level overview of the steps of the interpretable dictionary learning method, as illustrated in Figure~\ref{fig:pipeline}. The figure describes the most important global ideas behind our novel online cvxNDL pipeline. A rigorous mathematical formulation of the problem and relevant analyses are delegated to the Methods Section, while detailed algorithmic methods are available in the Supplement Section $2$.

Chromatin interactions are commonly represented as contact maps. A contact map can be viewed as a hypergraph, where nodes represent genomic loci and two or more such nodes are connected through hyperedges to represent experimentally observed multiway chromatin interactions. Since it is challenging to work with hypergraphs directly, the first step is to transform a hypergraph into an ordinary network (graph), which we tacitly assume is connected. For this purpose, we employ \emph{clique expansion}~\cite{agarwal2005beyond,zhou2006learning}, as shown in Figure~\ref{fig:clique_expansion_a}. Clique expansion converts a hyperedge into a clique (a fully connected network) and therefore preserves all interactions encapsulated by the hyperedge. However, large hyperedges covering roughly $10$ or more nodes in the network can introduce distortion by creating new cliques that do not correspond to any multiway interaction, as shown in Figure~\ref{fig:clique_expansion_b}~\cite{li2017inhomogeneous}. The frequency of such large hyperedges and the total number of hyperedges in chromatin interaction data is limited (i.e., the hypergraph is sparse, see Supplement Table~$1$). This renders the distortion due to the hypergraph-to-network conversion process negligible.

To generate an online sample from the clique-expanded input network, we use a subnetwork sampling procedure shown in Figure~\ref{fig:motif_sample_img}. We consider a small template network consisting of a fixed number of nodes and search for induced subnetworks in the input that contain the template network topology. These induced subnetworks can be rigorously characterized via \emph{homomorphisms} and are discussed in detail in the Methods Section. An example of a homomorphism is shown in Figure~\ref{fig:motif_sample_img}. Throughout our analysis, we will \emph{exclusively focus on path homomorphisms} because they are most suitable for the biological problem investigated. To generate a sequence of online samples from the input network, we employ MCMC sampling. Given a path sample at discrete time $t$, the next sample at time $t+1$ is generated by selecting a new node uniformly at random from the neighborhood of the sample at time $t$ and calculating its probability of acceptance $\beta$, explained in the Methods Section. If this new node is accepted, we perform a \emph{directed} random walk starting at the selected node, otherwise, we restart the random walk from the first node of the sample at time $t$. Note that the input network is undirected while only the sampling method requires a directed walk as the order of the labeled nodes matters. (see Figure~\ref{fig:mcmc_img}).

\begin{figure}[p]
\begin{subfigure}{\textwidth}
  \centering
  \includegraphics[width=\linewidth]{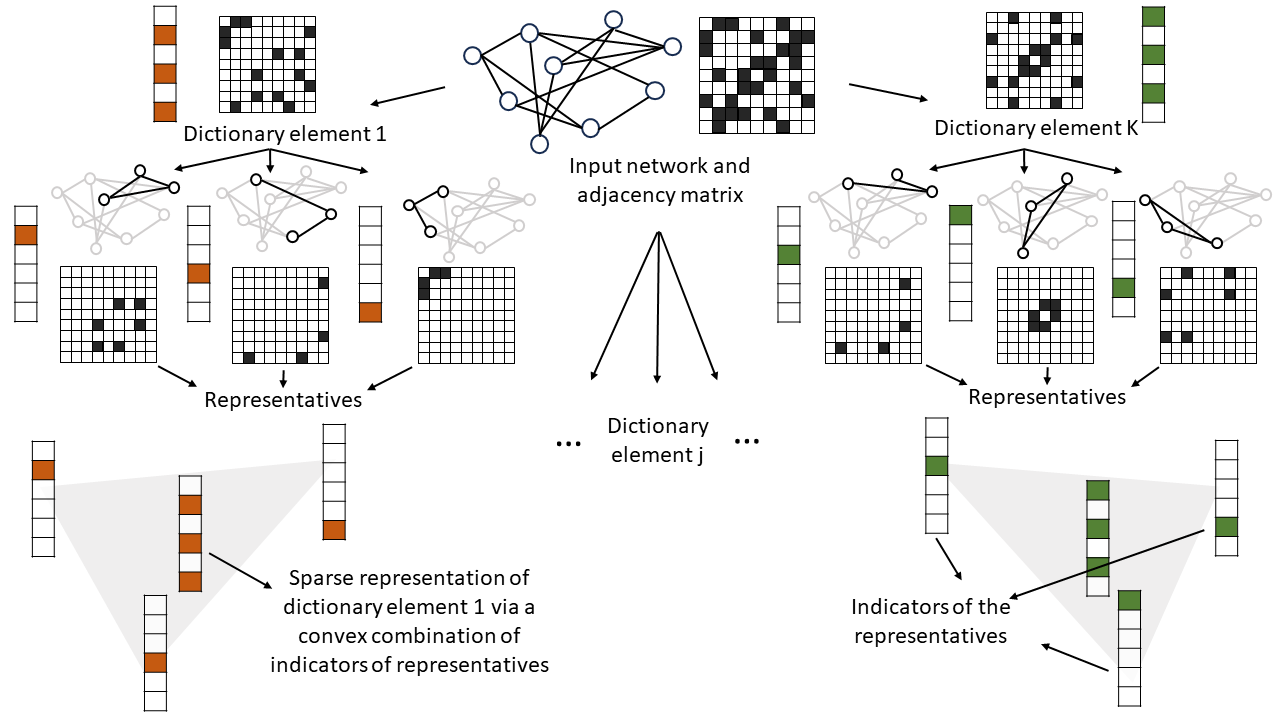}
  \caption{}
  \vspace{0.3cm}
  \label{fig:dictionary_organization}
\end{subfigure}
\begin{subfigure}{\textwidth}
  \centering
  \includegraphics[width=\linewidth]{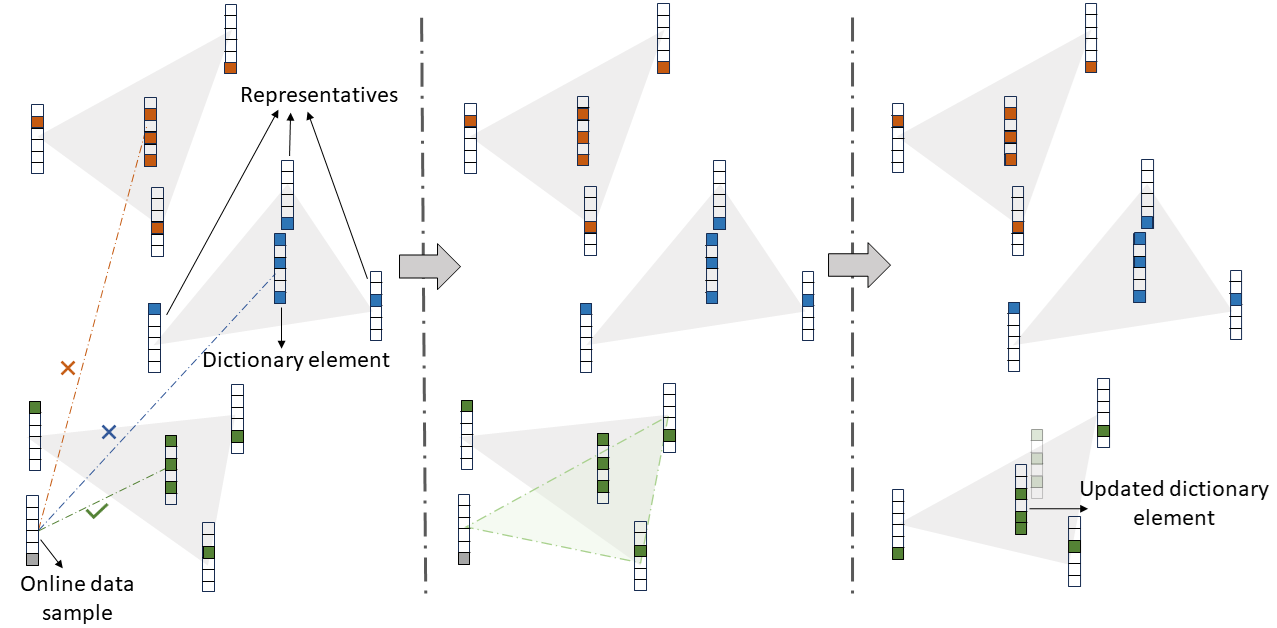}
  \caption{}
  \vspace{0.3cm}
  \label{fig:cvxNDL_alg}
\end{subfigure}
\caption{(a) Organization of a dictionary comprising $K$ dictionary elements that are convex combinations of real representative subnetworks. Each dictionary element itself is a sparse \emph{convex combination} of a set of representatives which are small subnetworks of the input real-world network. In the example, there are $6$ options for the representatives, and inclusion of a representative into a dictionary element is indicated by a colored entry in a $6$-dimensional indicator column-vector. Each of the $6$ representatives corresponds to a subnetwork of the input network with a  fixed number of nodes ($3$ for our example). The dictionary element is generated by a convex combination of the corresponding adjacency matrices of its corresponding representative subnetworks. For the example, the resulting dictionary elements are $9 \times 9$ matrices. (b) Illustration of the representative region update. When an online data sample is observed, the distance of the sample to each of the current dictionary elements is computed and the sample is assigned to the representative region of the nearest dictionary element. From this expanded set of representatives, one representative is carefully selected for removal to improve the objective. The new dictionary element is then obtained as an optimized convex combination of the updated set of representatives.} 
\label{fig:dict_org}
\end{figure}

MCMC sampling is used to generate a sequence of samples to initialize a dictionary with $K$ \emph{dictionary elements,} where $K$ is chosen based on the properties of the dataset. Each of the dictionary elements is represented as a convex combination of a \emph{small} (sparse) set of \emph{representatives} that are real biological observations. The convex hull of these representatives is termed the \emph{representative region} of the dictionary element. As a result, the vertices of the representative regions comprise a collection of MCMC-generated real-world samples. Figure~\ref{fig:dictionary_organization} shows the organization of a dictionary as a collection of dictionary elements, representatives, and representative regions.

After initialization, we perform iterative optimization of the DL objective function using online samples, again generated via the MCMC method. More precisely, at each iteration, we compute the distance between the new sample and every current estimate of dictionary elements. Subsequently, we assign the sample to the representative region of the nearest dictionary element, which leads to an increase in the size of the set of representatives associated with the dictionary element. From this expanded set of representatives, we carefully select one representative for removal, maximizing the improvement in the quality of our dictionary element and the objective function. It is possible that the removed representative is the newly added data sample assigned to the representative region. In this case, the dictionary element remains unchanged. Otherwise, it is obtained as a convex combination of the updated set of representatives. After observing sufficiently many online samples, the algorithm converges to an accurate set of dictionary elements or the procedure terminates without convergence (in which case we declare a failure and restart the learning process). In our experiments, we never terminated with failure, but due to the lack of provable convergence guarantees for real-world datasets, such scenarios cannot be precluded. The update procedure is shown in Figure~\ref{fig:cvxNDL_alg}.

\begin{figure}[htb]
\begin{subfigure}{\textwidth}
  \centering
  \includegraphics[width=.95\linewidth]{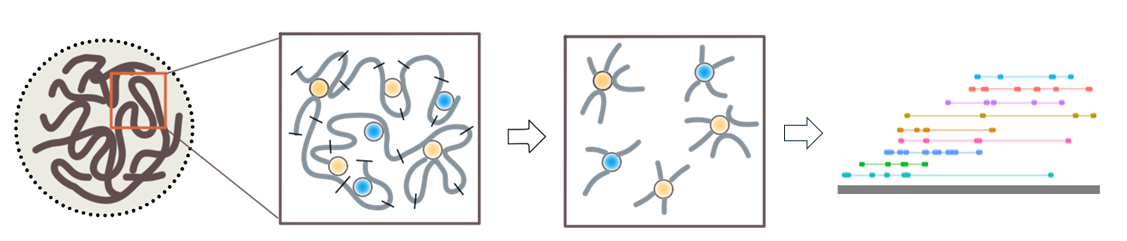}
  \caption{}
  \label{fig:sfig1}
\end{subfigure}
\caption{Generation of ChIA-Drop data. ChIA-Drop~\cite{zheng2019multiplex} adopts a droplet-based barcode-linked technique to reveal multiway chromatin interactions at a single molecule level. Chromatin samples are crosslinked and fragmented without a proximity ligation step. The samples are enriched for informative fragments through antibody pull-down.}
\label{fig:chia_drop}
\end{figure}

We applied the method outlined above to RNAPII-enriched ChIA-Drop data from \emph{Drosophila Melanogaster} S2 cells, using a dm3 reference genome~\cite{zheng2019multiplex}, to learn dictionaries of chromatin interactions. 
Figure~\ref{fig:chia_drop} provides an illustration of the ChIA-Drop pipeline. 

We preprocessed the RNAPII ChIA-Drop data to remove fragments mapped to the repetitive regions in the genome and performed an MIA-Sig enrichment test with FDR $0.1$~\cite{kim2019mia}. Only the hyperedges that passed this test were used in our subsequent analysis. To facilitate the analysis, we binned chromosomal genetic sequences into $500$ bp regions and used the midpoint of each fragment for mapping. These bins of $500$ consecutive bases form the nodes of the hypergraph for each chromosome, while the set of filtered multiway interactions form the hyperedges. The dataset hence includes $45,938$, $42,292$, $49,072$, and $55,795$ nodes and $36,140$, $28,387$, $53,006$, $45,530$ hyperedges for chromosome chr2L, chr2R, chr3L and chr3R  respectively. The distribution of the hyperedge sizes is given in Supplement Table~$1$. To create networks from hypergraphs, we converted the multiway interactions into cliques. The clique-expanded input network has $113,606$, $85,316$, $161,590$, and $143,370$ edges respectively. Although the ChIA-Drop data comprises interactions from six chromosomes {chr2L, chr2R, chr3L, chr3R, chr4 and chrX}, since chr4 and chrX are relatively short regions and most of the functional genes are located on chr2L, chr2R, chr3L, and chr3R, we focus our experiments only on the latter.

In the experiments, we set the number of dictionary elements to $K = 25$. The number of dictionary elements $K$ is selected to achieve the best trade-off between accuracy and complexity of the learned dictionary representations. Small values of $K$ do not fully capture the diversity of multiway interactions present in the data, while very large values result in unnecessarily redundant representations. The latter can also obscure important interactions by capturing the inherent noise in ChIA-Drop data, and contribute to representation distortion~\cite{li2017inhomogeneous}. After testing our method for multiple different values of $K,$ we settled for $K=25$. Clearly, other datasets may benefit from a different choice of the parameter $K$, which has to be fine-tuned. Also, as template subnetworks, we use \emph{paths}, since paths are the simplest and most common network motifs, especially in chromatin interaction data (most contact measurements are proximal due to the linear chromosome order). Once again, by optimizing via trial-and-error, we select paths including $21$ nodes (i.e., $21 \times 500$ bases). Both the choice of the subnetwork (motif) and its number of constituent nodes is data dependent.

MCMC sampling for initialization, as well as for subsequent online optimization steps, was performed before running the online optimization process to improve the efficiency of our implementation. We sampled $20,000$ subnetworks from each of the four chromosomes to ensure sufficient coverage of the input network. From this pool of subnetworks, we randomly selected $500$ subnetworks to initialize our dictionaries, ensuring that each dictionary element had at least $10$ representatives (which suffice to get quality initializations for the dictionary elements themselves). Each online step involved sampling an additional subnetwork and we iterated this procedure up to $1$ million times, as needed for convergence (see Figure~\ref{fig:cvxNDL_workflow}).

\begin{figure}[htb]
\begin{subfigure}{\textwidth}
    \centering
    \includegraphics[width = 1\linewidth]{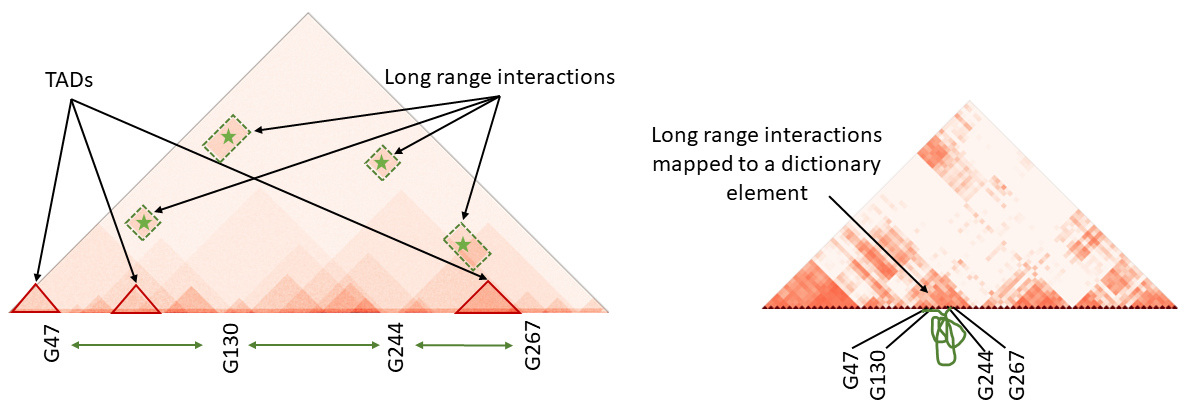}
\end{subfigure}
\begin{subfigure}{\textwidth}
    \centering
    \includegraphics[width = 1\linewidth]{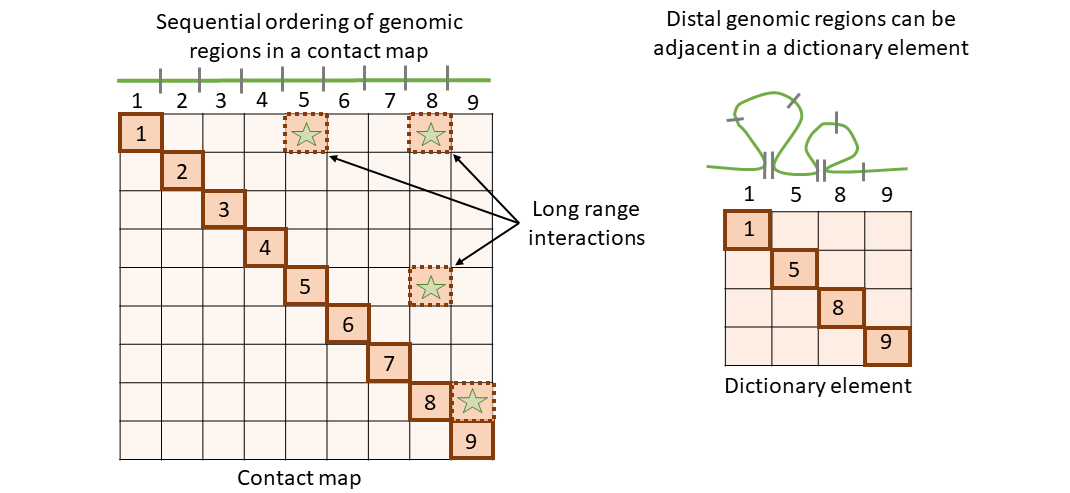}
\end{subfigure}
\caption{A dictionary element, represented as a matrix, consists of both proximal and distal interacting genomic regions. The elements on the diagonal are not necessarily indexed by adjacent (consecutive) genomic fragments, as explained by the example in the second row. There, off-diagonal long-range interactions in the $9 \times 9$ matrix are included in a $3 \times 3$ dictionary element whose diagonal elements are not in consecutive order.}
    \label{fig:long_range}
\end{figure}

At this point, it is crucial to observe that the dictionary elements learned by online cvxNDL effectively capture \emph{long-range interactions} because each dictionary element may include distal genomic regions that are not adjacent in the genomic order. In other words, the diagonal entries of our dictionary elements \emph{do not exclusively represent consecutive genomic regions} as in standard chromatin contact maps; instead, they may include \emph{both} nonconsecutive (long-range)  and consecutive (short-range, adjacent) interactions. This point is explained in detail in Figure~\ref{fig:long_range}. Another relevant remark is that without the convexity constraint, dictionary element entries could not have been meaningfully mapped back (associated) to genomic regions and viewed as \emph{real physical interactions between genomic loci}.

\begin{figure}[p]
\begin{subfigure}{.5\textwidth}
  \centering
  \includegraphics[width=.8\linewidth]{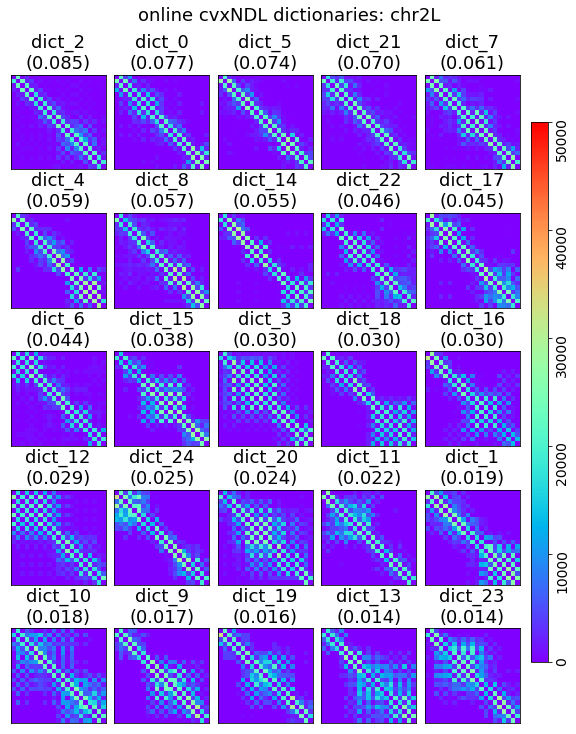}
  \caption{chr2L}
  \label{fig:chr2L_dict_all}
\end{subfigure}%
\begin{subfigure}{.5\textwidth}
  \centering
  \includegraphics[width=.8\linewidth]{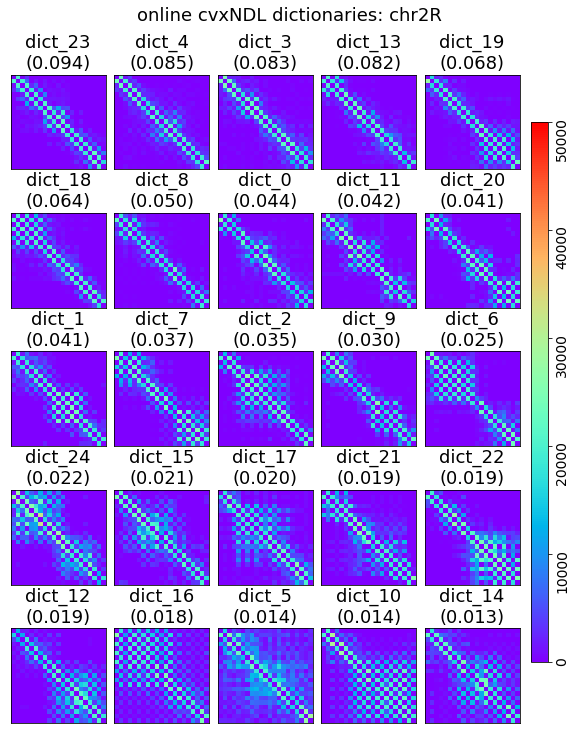}
  \caption{chr2R}
  \label{fig:chr2R_dict_all}
\end{subfigure}
\hfill
\begin{subfigure}{.5\textwidth}
  \centering
  \includegraphics[width=.8\linewidth]{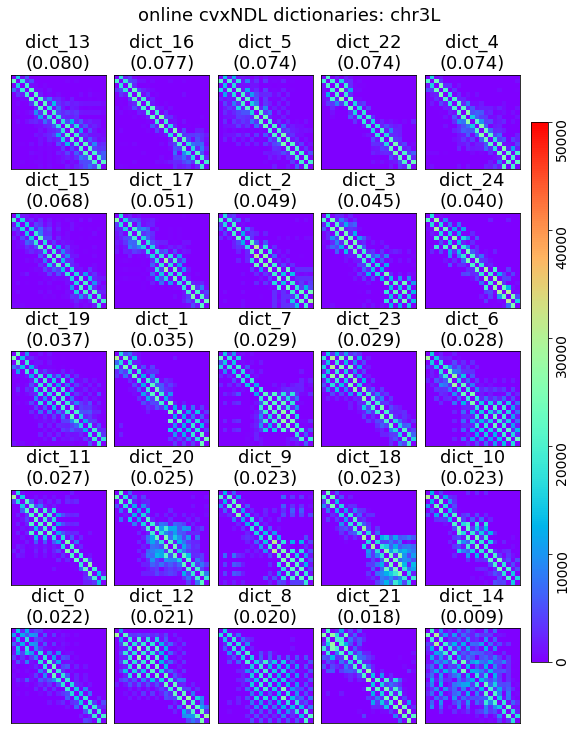}
  \caption{chr3L}
  \label{fig:chr3L_dict_all}
\end{subfigure}%
\begin{subfigure}{.5\textwidth}
  \centering
  \includegraphics[width=.8\linewidth]{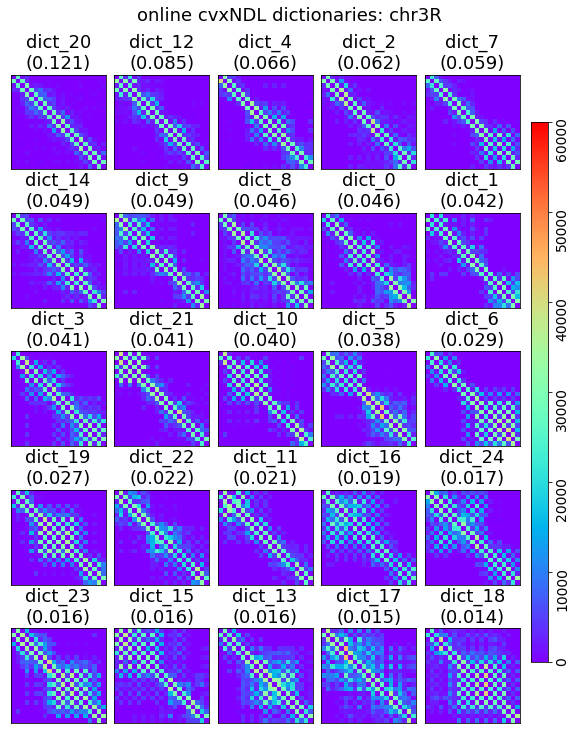}
  \caption{chr3R}
  \label{fig:chr3R_dict_all}
\end{subfigure}
\caption{Dictionary elements for \emph{Drosophila} chromosomes 2L, 2R, 3L and 3R obtained using online cvxNDL. Each subplot contains $25$ dictionary elements for the corresponding chromosome and each block in the subplots corresponds to one dictionary element. The elements are ordered by their importance score. Note that the ``diagonals'' in the dictionary elements do not exclusively represent localized topologically associated domains (TADs) as in standard chromatin contact maps; instead, they can also capture long-range interactions. This is due to the fact that the indices of the dictionary element matrices represent genomic regions that may be far apart in the genome. In contrast, standard contact maps have indices that correspond to continuously ordered genomic regions, so that the diagonals truly represent TADs (see Figure~\ref{fig:long_range}). The color-code captures the actual locations of the genomic regions involved in the representatives and their dictionary elements. The most interesting dictionary elements are those that contain both dark blue and light blue/green and red spectrum colors (since they involve long-range interactions). This is especially the case for chr3L and chr3R.}
\label{fig:online_cvx_dictionaries}
\end{figure}

The dictionary elements generated from the \emph{Drosophila} ChIA-Drop data for chr2L, chr2R, chr3L, and chr3R using the online cvxNDL method are shown in Figure~\ref{fig:online_cvx_dictionaries}. Each subplot corresponds to one chromosome and has $25$ dictionary elements ordered with respect to their \emph{importance scores}, capturing the relevance and frequency of use of the dictionary element, and formally defined in the Methods Section. Each element is color-coded based on the genomic location of the genes covered by their representatives. Hence, dictionary elements represent combinations of experimentally observed interaction patterns, uniquely capturing the significance of the genomic locations involved in the corresponding interactions. We also report the density and median distance between all consecutive pairs of interacting loci (connected nodes) of all dictionary elements in Supplement Tables~$2$ and $3$. 

\begin{figure}[p]
\begin{subfigure}{.5\textwidth}
  \centering
  \includegraphics[width=.8\linewidth]{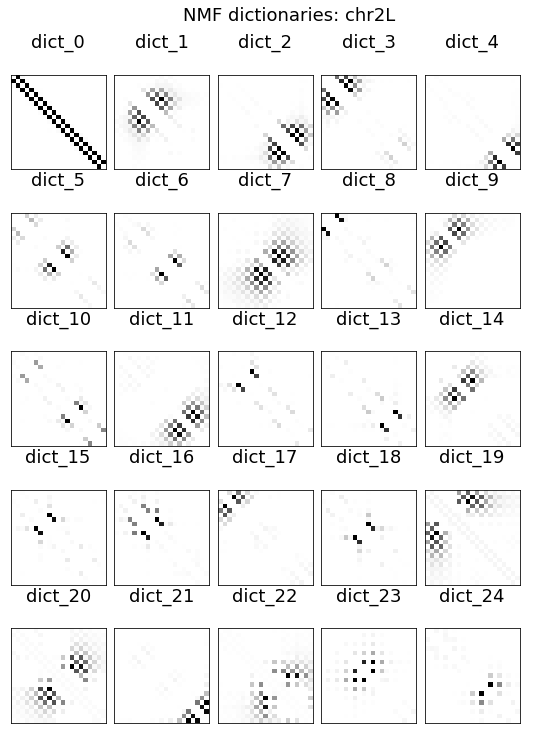}
  \caption{NMF}
  \label{fig:cmp_nmf_chr2L_dict}
\end{subfigure}%
\begin{subfigure}{.5\textwidth}
  \centering
  \includegraphics[width=.8\linewidth]{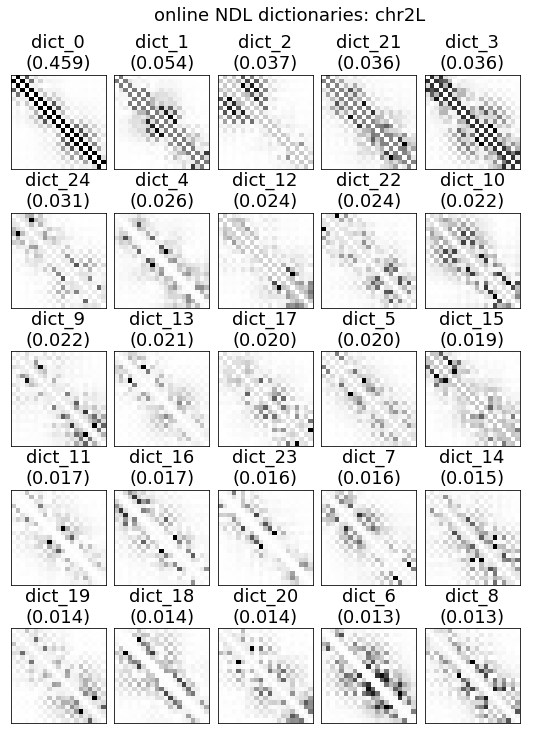}
  \caption{Online NDL}
  \label{fig:cmp_omf_chr2L_dict}
\end{subfigure}
\begin{subfigure}{.5\textwidth}
  \centering
  \includegraphics[width=.8\linewidth]{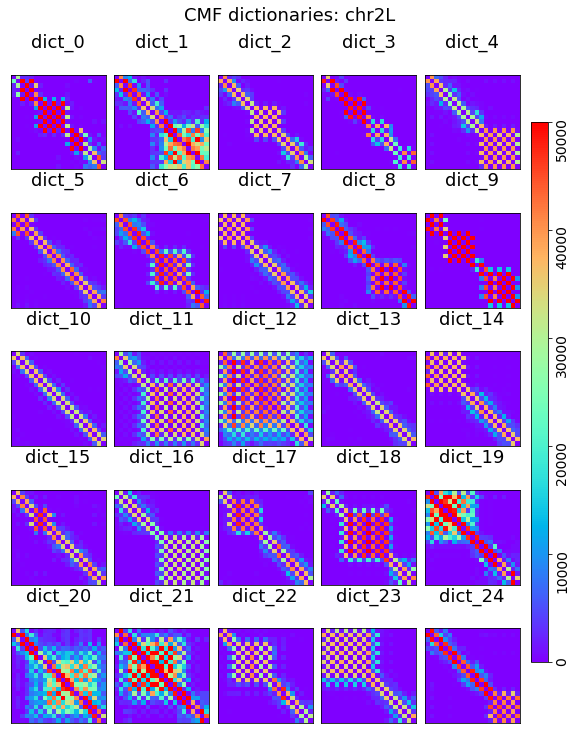}
  \caption{CMF}
  \label{fig:cmp_cmf_chr2L_dict}
\end{subfigure}%
\begin{subfigure}{.5\textwidth}
  \centering
  \includegraphics[width=.8\linewidth]{figures_main/chr2L_dictionary_all.png}
  \caption{Online cvxNDL}
  \label{fig:cmp_cvxNDL_chr2L_dict}
\end{subfigure}
\caption{Dictionary elements for \emph{Drosophila} chromosome chr2L generated by NMF~(\ref{fig:cmp_nmf_chr2L_dict}), online NDL~(\ref{fig:cmp_omf_chr2L_dict}), CMF~(\ref{fig:cmp_cmf_chr2L_dict}) and online cvxNDL~(\ref{fig:cmp_cvxNDL_chr2L_dict}). NMF and CMF are learned off-line, using a total of $20,000$ samples. Note that these algorithms do not scale and cannot work with larger number of samples such as those used in online cvxNDL. The color-coding is performed in the same manner as for the accompanying online cvxNDL results. Columns of the dictionary elements in the second row are color-coded based on the genome locations of the representatives. As biologically meaningful locations can be determined only via convex methods, the top row corresponding to NMF and online NDL results is black-and-white.}
    \label{fig:dictionary_comparison_among_methods}
\end{figure}

Note that our algorithm is the first method for online learning of convex (interpretable) network dictionaries. We can therefore only compare its \emph{representation accuracy} to that of nonnegative matrix factorization (NMF), convex matrix factorization (CMF), and online network dictionary learning (online NDL). A visual comparison of the dictionaries formed through online cvxNDL and the aforementioned methods for chr2L is provided in Figure~\ref{fig:dictionary_comparison_among_methods}.

Classical NMF does not allow the mapping of results back to real interacting genomic regions. While the dictionary elements obtained via CMF are interpretable, they tend to mostly comprise widely spread genomic regions since they do not use the network information. The dictionary elements generated by online cvxNDL have smaller yet relevant spreads that are more likely to capture meaningful long-range interactions. In contrast to online cvxNDL, both NMF and CMF are not scalable to large datasets, rendering them unsuitable for handling current and future high-resolution datasets such as those generated by ChIA-Drop. Compared to online NDL, online cvxNDL also has a more balanced distribution of importance scores. For example, in Figure~\ref{fig:dictionary_comparison_among_methods}(b), dict\_0 has score $0.459$, while the scores in Figure~\ref{fig:dictionary_comparison_among_methods}(d) are all $\leq 0.085$. Moreover, akin to standard NMF, NDL fails to provide interpretable results since the dictionary elements cannot be mapped back to real interacting genomic loci. 

Results for other chromosomes are reported in the Supplement Section $4$. Recall that both online cvxNDL and online NDL use a $k$-path as the template. 

\begin{figure}[htb]
        \centering
        \begin{subfigure}[t]{0.3\textwidth}
                \includegraphics[width=\textwidth]{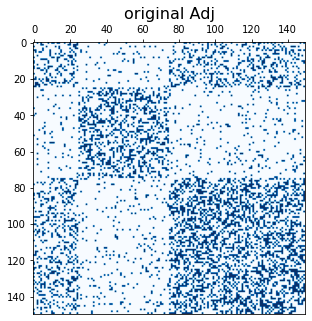}
                \caption{Original adjacency}
        \end{subfigure}
        \quad
        \begin{subfigure}[t]{0.3\textwidth}
                \includegraphics[width=\textwidth]{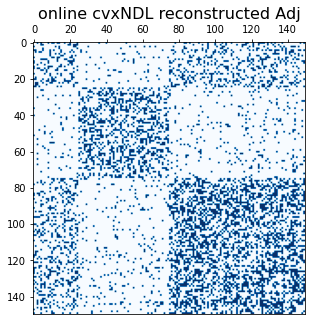}
                \caption{Online cvxNDL}
        \end{subfigure}
        \quad
        \begin{subfigure}[t]{0.3\textwidth}
                \includegraphics[width=\textwidth]{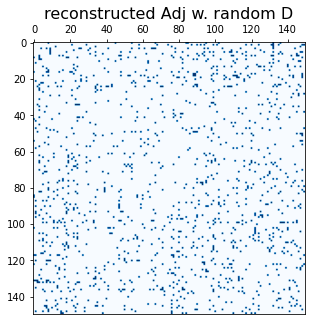}
                \caption{Random dictionaries}
        \end{subfigure}
        
        \begin{subfigure}[t]{0.3\textwidth}
                \includegraphics[width=\textwidth]{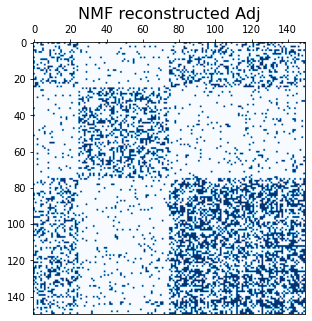}
                \caption{NMF}
        \end{subfigure}
        \quad
        \begin{subfigure}[t]{0.3\textwidth}
                \includegraphics[width=\textwidth]{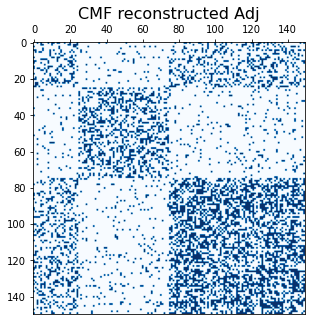}
                \caption{CMF}
        \end{subfigure}
        \quad
        \begin{subfigure}[t]{0.3\textwidth}
                \includegraphics[width=\textwidth]{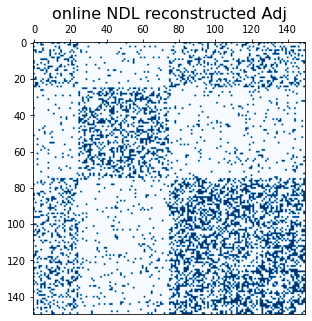}
                \caption{Online NDL}
        \end{subfigure}
        \caption{Original adjacency matrix and reconstructed adjacency matrices based on different DL methods, including randomly selected dictionaries. The figure illustrates the fact that the additional convexity constraint does not compromise the quality of interaction representation/reconstruction in a visual manner. For more rigorous analytical accuracy comparisons Table~\ref{tab:online_cxvNDL_avg_precision}.}\label{fig:synth_recon_adj}
\end{figure}

\noindent\textbf{Reconstruction Accuracy.} Once a dictionary is constructed, one can use the network reconstruction algorithm from~\cite{lyu2020online} to recover a subnetwork or the whole network by locally approximating subnetworks via dictionary elements. The accuracy of approximation in this case measures the ``expressibility'' of the dictionary with respect to the network. All methods, excluding randomly generated dictionaries used for illustrative purposes only, can accurately reconstruct the input network. For a quantitative assessment, the average precision-recall score for all methods is plotted in Table~\ref{tab:online_cxvNDL_avg_precision}. As expected, random dictionaries have the lowest scores across all chromosomes, while all other methods are of comparable quality. This means that interpretable methods, such as our online cvxNDL, do not introduce representation distortions (CMF also learns interpretable dictionaries; however, it is substantially more expensive computationally when compared to our method but does not ensure that network topology is respected). A zoomed-in sample-based reconstruction result for chr2L is shown in Supplement Figure~$6$, while the reconstruction results for the entire contact maps of chr2L, chr2R, chr3L, and chr3R are available in Supplement Figures~$7$-$10$. Additionally, for synthetic data, Figure~\ref{fig:synth_recon_adj} shows the reconstructed adjacency matrices for various dictionary learning methods, further confirming the validity of findings for the chromatin data. More detailed results for synthetic data are available in Supplement Section~$3$. 

\begin{table}[!h]
\centering
\caption{Average Precision Recall for different DL methods, for all chromosomes as well as synthetic datasets. Methods that return interpretable dictionaries are indicated by the superscript $i$ while methods that are scalable to large datasets are indicated by the superscript $s$. Online cvxNDL is both interpretable and scalable while maintaining performance on par with other noninterpretable and nonscalable methods.} 
\label{tab:online_cxvNDL_avg_precision}
\begin{tabular}{lccccc}
\dtoprule
 & chr2L & chr2R & chr3L & chr3R & Synthetic \\ 
 \cmidrule(lr){2-6}  
Online cvxNDL$^{i,s}$ & 0.995{4} & 0.998{6} & 0.983{0} & 0.98{76} & 0.97{47}\\
Online NDL$^{s}$ & 0.995{5} & 0.998{6} & 0.983{4} & 0.98{80} & 0.9728\\
NMF & 0.9952 & 0.9985 & 0.9829 & 0.9873 & 0.97{74}\\
CMF$^{i}$ & 0.9951 & 0.9985 & 0.9824 & 0.9870 & 0.9731\\
Random Dict. & 0.0007 & 0.2547 & 0.5276 & 0.0796 & 0.1922\\
\dbottomrule
\end{tabular}
\end{table}

\noindent\textbf{Gene Ontology Enrichment Analysis.} As each dictionary element is associated with a set of representatives that correspond to real observed subnetworks, their nodes can be mapped back to actual genomic loci. This allows one to create lists of genes covered by at least one node included in the representatives. 

To gain insights into the functional annotations of the genes associated with the dictionary elements, we conducted a Gene Ontology (GO) enrichment analysis using the annotation category ``Biological Process'' from \url{http://geneontology.org}, with the reference list \emph{Drosophila Melanogaster}. This analysis was performed for each dictionary element. Our candidate set for enriched GO terms was selected with a false discovery rate (FDR) threshold of $< 0.05$. Note that the background genes used for comparison are all genes from all chromosomes (the default option). We also utilized the hierarchical structure of GO terms~\cite{musen2015protege}, where terms are represented as nodes in a directed acyclic graph, and their relationships are described via arcs in the digraph (i.e., each ``child'' GO term is more specific than its ``parent'' term and where one child may have multiple parents). 

We further refined our results by running additional processing steps. For each GO term, we identified all the paths between the term and the root node and then removed any intermediate parent GO term from the enriched GO terms set. By iteratively performing this filtering process for each dictionary element, we created a list of the most specific GO terms associated with each element. More details about the procedure are available in the Supplement Section~$6$.

We report the most frequently enriched GO terms for each chromosome, along with the corresponding dictionary elements exhibiting enrichment for chr3R in Table~\ref{tab:chr3R_GO_top_bottom}. The results for other chromosomes are available in the Supplement Tables~$4$-$6$. Notably, the most frequent GO terms are related to regulatory functions, reflecting the significance of RNA Polymerase II. We also observe that dictionary elements for chr2L and chr2R are enriched in GO terms associated with reproduction and embryonic development. Similarly, chr3L and 3R are enriched in GO terms for blood circulation and responses to heat and cold.

We report the number of GO terms associated with each dictionary element, along with their importance scores in Supplement Tables~$10$-$13$. Dictionary elements with higher importance scores tend to exhibit a larger number of enriched GO terms while dictionary elements with $0$ enriched GO terms generally have small importance scores.

\begin{table}[htb]
\scriptsize
\centering
\caption{The $5$ most enriched GO terms for genes covered by dictionary elements from chr3R. Column `\#' indicates the number of dictionary elements that show enrichment for the given GO term. Also reported are up to $3$ dictionary elements with the largest importance score in the dictionary, along with the ``density'' $\rho$ of interactions in the dictionary element (defined in the Methods section) and median distance $d_{\text{med}}$ of all adjacent pairs of nodes in its representatives.}
\label{tab:chr3R_GO_top_bottom}
\begin{tabularx}{\textwidth}{L|c|L}    
\dtoprule
Most frequent GO term & \# & Top 3 dictionaries\\ 
\midrule
(GO:0001819)   Positive regulation of cytokine production & 7 & \threeDict{\dictTwoZero{chr3R}}{\dictSeven{chr3R}}{\dictNine{chr3R}} $\scriptstyle \rho = 0.126, 0.146, 0.157$ $\scriptstyle d_{\text{med}} = 12791, 12830, 11930$
 \vspace{0.07cm}
	\\ 

	\hline
(GO:0008015) Blood   circulation & 7 & \threeDict{\dictTwoZero{chr3R}}{\dictOneTwo{chr3R}}{\dictFour{chr3R}} $\scriptstyle \rho = 0.126, 0.142, 0.138$ $\scriptstyle d_{\text{med}} = 12791, 13455, 13674$
 \vspace{0.07cm}
	\\ 
	\hline
(GO:0045948) Positive regulation of   translational initiation & 5 & \threeDict{\dictTwoZero{chr3R}}{\dictFour{chr3R}}{\dictOneFour{chr3R}} $\scriptstyle \rho = 0.126, 0.138, 0.162$ $\scriptstyle d_{\text{med}} = 12791 , 13674 , 12572 $
 \vspace{0.07cm}
	\\ 
	\hline
(GO:0042177) Negative regulation of protein catabolic process & 5 & \threeDict{\dictTwoZero{chr3R}}{\dictOneTwo{chr3R}}{\dictFour{chr3R}} $\scriptstyle \rho = 0.126, 0.142, 0.138$ $\scriptstyle d_{\text{med}} = 12791, 13455, 13674$
 \vspace{0.07cm}
	\\ 
	\hline
(GO:0043065) Positive   regulation of apoptotic process & 4 & \threeDict{\dictTwoZero{chr3R}}{\dictSeven{chr3R}}{\dictThree{chr3R}} $\scriptstyle \rho = 0.126, 0.146, 0.179$ $\scriptstyle d_{\text{med}} = 12791, 12830, 11748 $
 \vspace{0.07cm}
	\\
\dbottomrule
\end{tabularx}
\end{table}

\textbf{RNA-Seq Coexpression Analysis.} The ChIA-Drop dataset~\cite{zheng2019multiplex} used in our analysis was accompanied by a single noisy RNA-Seq replicate. To address this issue, we retrieved $20$ collections of RNA-Seq data corresponding to untreated S2 cell lines of \emph{Drosophila Melanogaster} from the Digital Expression Explorer (DEE2) repository. DEE2 provides uniformly processed RNA-Seq data sourced from the publicly available NCBI Sequence Read Archive (SRA)~\cite{ziemann2019digital}. The list of sample IDs is available in Supplement Table~$14$. 

To ensure consistent normalization across all samples, we used the trimmed mean of M values (TMM) method~\cite{robinson2010scaling}, available through the edgeR package~\cite{robinson2010edger}. This is of crucial importance when jointly analyzing samples from multiple sources. We selected the most relevant genes by filtering the list of covered genes and retaining only those with more than $95\%$ overlap with the gene promoter regions, as defined in the \emph{Ensembl} genome browser. Subsequently, for each dictionary element, we collected all genes covered by it and then calculated the pairwise Pearson correlation coefficient of expressions of pairs of genes in the set. To visualize the underlying coexpression clusters within the genes, we performed  hierarchical clustering, the results of which are shown in Supplement Section~$7$ and Figure~\ref{fig:RTAD_pearson}. The latter corresponds to the R1-R4 and T1-T4 genomic regions of chr2L to be discussed in what follows.

Additionally, we conducted control experiments by constructing dictionary elements through random sampling of genes from the list of all genes on each of the chromosomes. For these randomly constructed dictionaries, we carried out a coexpression analysis as described above. We observed that the mean of coexpressions of all pairs of genes in a randomly constructed dictionary element is significantly lower compared to the mean of the online cvxNDL dictionary elements. Specifically, for dictionary elements generated using online cvxNDL, the mean coexpression values for all pairs of genes covered by the $25$ dictionary elements, and for each of the four chromosomes, 2L, 2R, 3L, and 3R, were found to be $0.419, 0.383, 0.411,$ and $0.407,$ respectively. The corresponding values for randomly constructed dictionaries were found to be $0.333, 0.329, 0.323,$ and $0.337,$ respectively. To determine if these differences are statistically significant, we employed the two-sample Kolmogorov-Smirnov test~\cite{massey1951kolmogorov}, comparing the empirical cumulative distribution functions (ECDFs) of pairwise coexpression values of the learned and randomly constructed dictionaries. The null hypothesis used was ``the two sets of dictionary elements are drawn from the same underlying distribution.'' The null hypotheses for all four chromosomes were rejected, with p-values equal to $3.6 \times 10^{-9}$, $8.5 \times 10^{-6}$, $3.6 \times 10^{-9}$, and $2.5 \times 10^{-7}$ for chr2L, chr2R, chr3L, and chr3R, respectively. This indicates that the learned dictionary elements indeed capture meaningful biological patterns of chromatin interactions.

\begin{figure}[htb]
     \centering
     \subfloat[]{\includegraphics[width=0.47\textwidth]{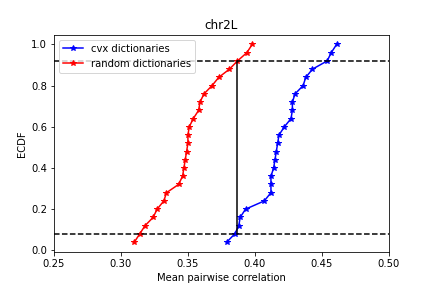}}
     \hfill
     \subfloat[]{\includegraphics[width=0.47\textwidth]{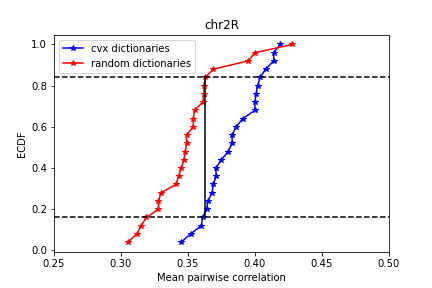}}
     \hfill
     \subfloat[]{\includegraphics[width=0.47\textwidth]{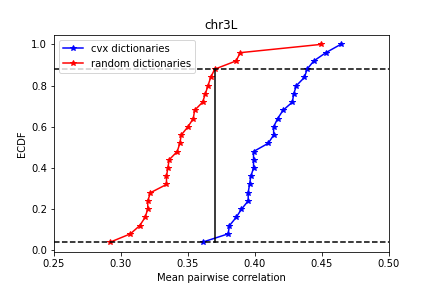}}
      \hfill
      \subfloat[]{\includegraphics[width=0.47\textwidth]{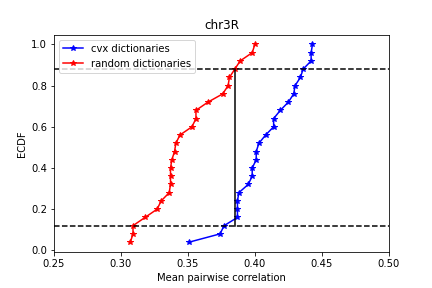}}
        \caption{Empirical cumulative distribution functions (ECDF) of mean pairwise coexpressions of genes covered by random and online cvxNDL dictionary elements ((a) for chr2L, (b) for chr2R, (c) for chr3L and (d) for chr3R). The results are based on the two-sample Kolmogorov-Smirnov test, and the null hypothesis described in the main text.}
        \label{fig:dee2_KS}
\end{figure}

To further evaluate our results, we also examined the well-documented R1-R4 and T1-T4 TAD interactions on chr2L, reported in~\cite{zheng2019multiplex}. The results of the coexpression analysis for these genomic regions are reported in Figure~\ref{fig:RTAD_pearson}. The mean pairwise correlation between genes belonging to the R1-R4 genomic regions equals $0.422,$ which is comparable to the mean value $0.419$ of the results obtained via online cvxNDL. We also calculated the intersection of the set of genes within the R1-R4 genomic regions and the set of genes covered by online cvxNDL dictionary elements identified for chr2L. We observed that the top $5$ online cvxNDL dictionary elements cover $38$ out of $85$ genes in the R1-R4 genomic regions. This is to be contrasted with the results for random dictionary elements, which cover only $7$ genes. Table~\ref{tab:R_region_cover} describes these and related findings in more detail.

\begin{figure}[!htbp]
     \centering
     \subfloat[]{\includegraphics[width=0.4\textwidth]{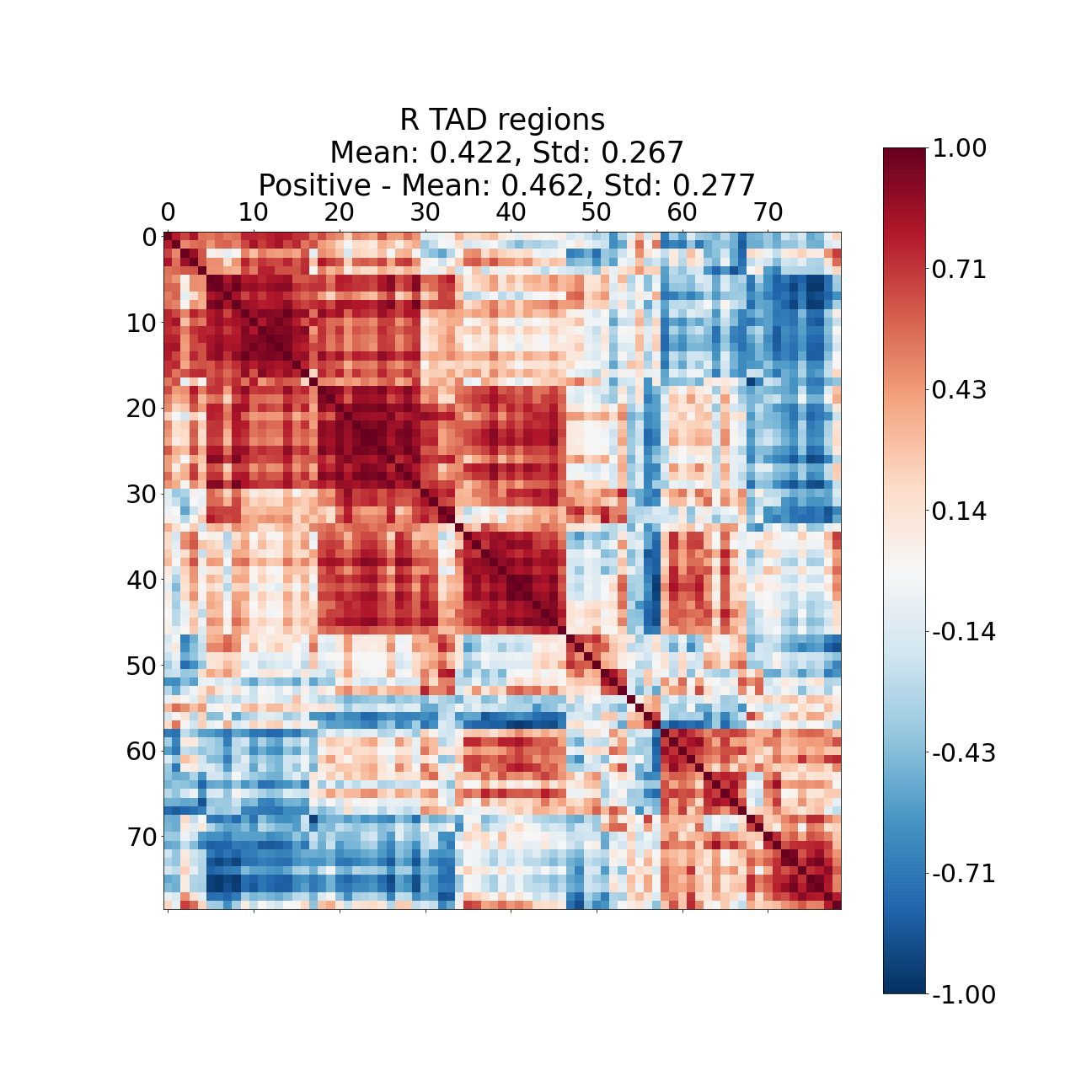}}
     \hfill
     \subfloat[]{\includegraphics[width=0.4\textwidth]{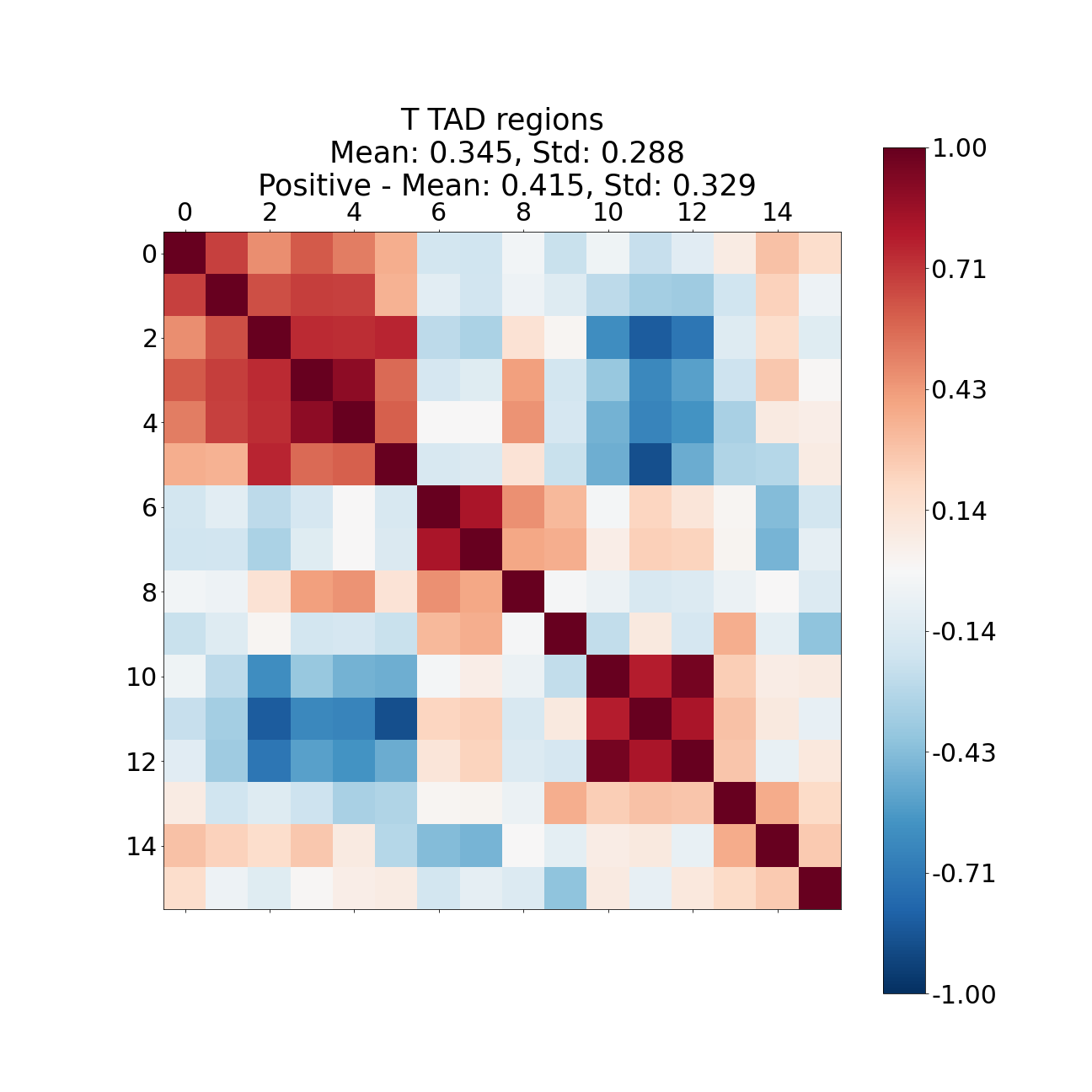}}
     \hfill
     \subfloat[]{\includegraphics[width=0.4\textwidth]{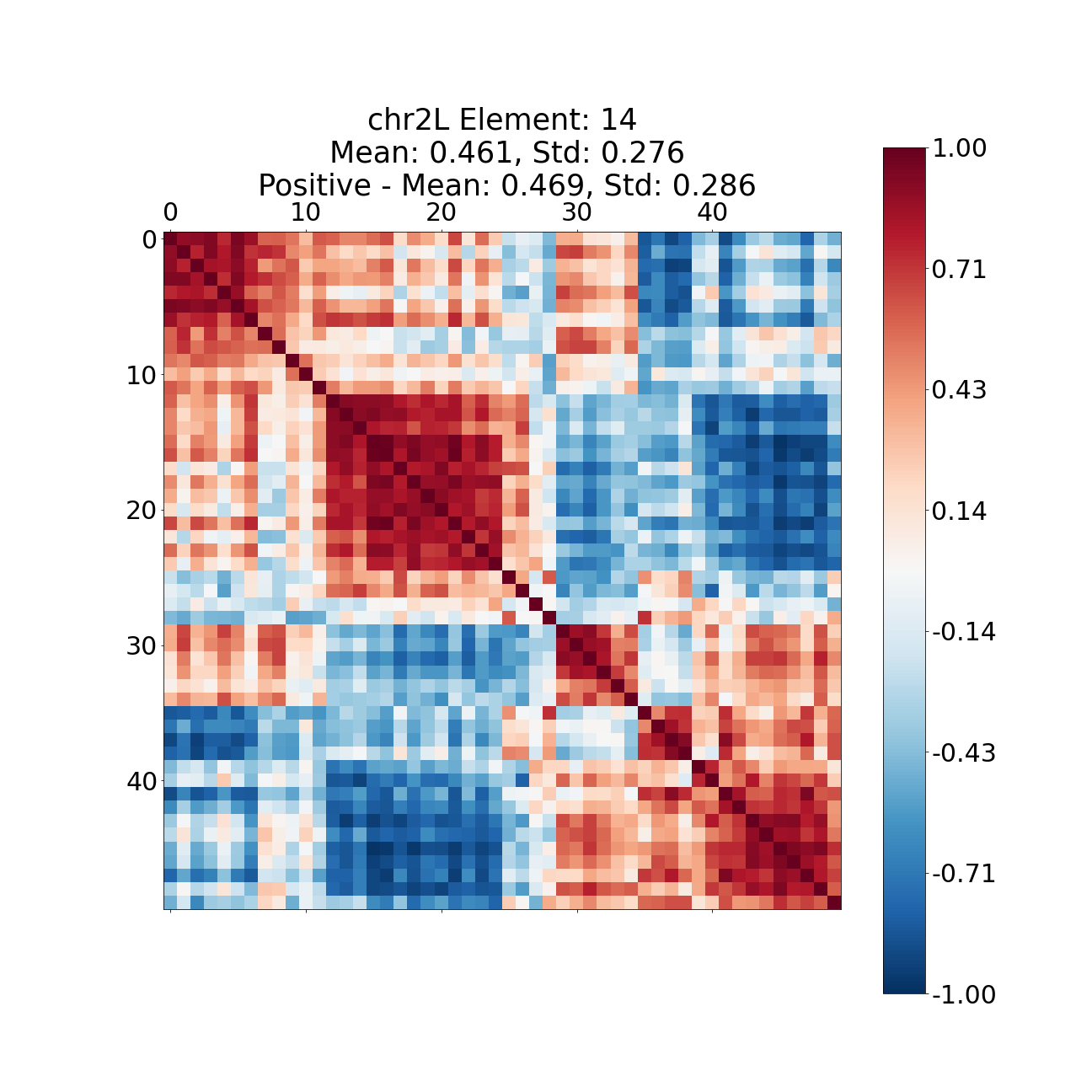}}
      \hfill
      \subfloat[]{\includegraphics[width=0.4\textwidth]{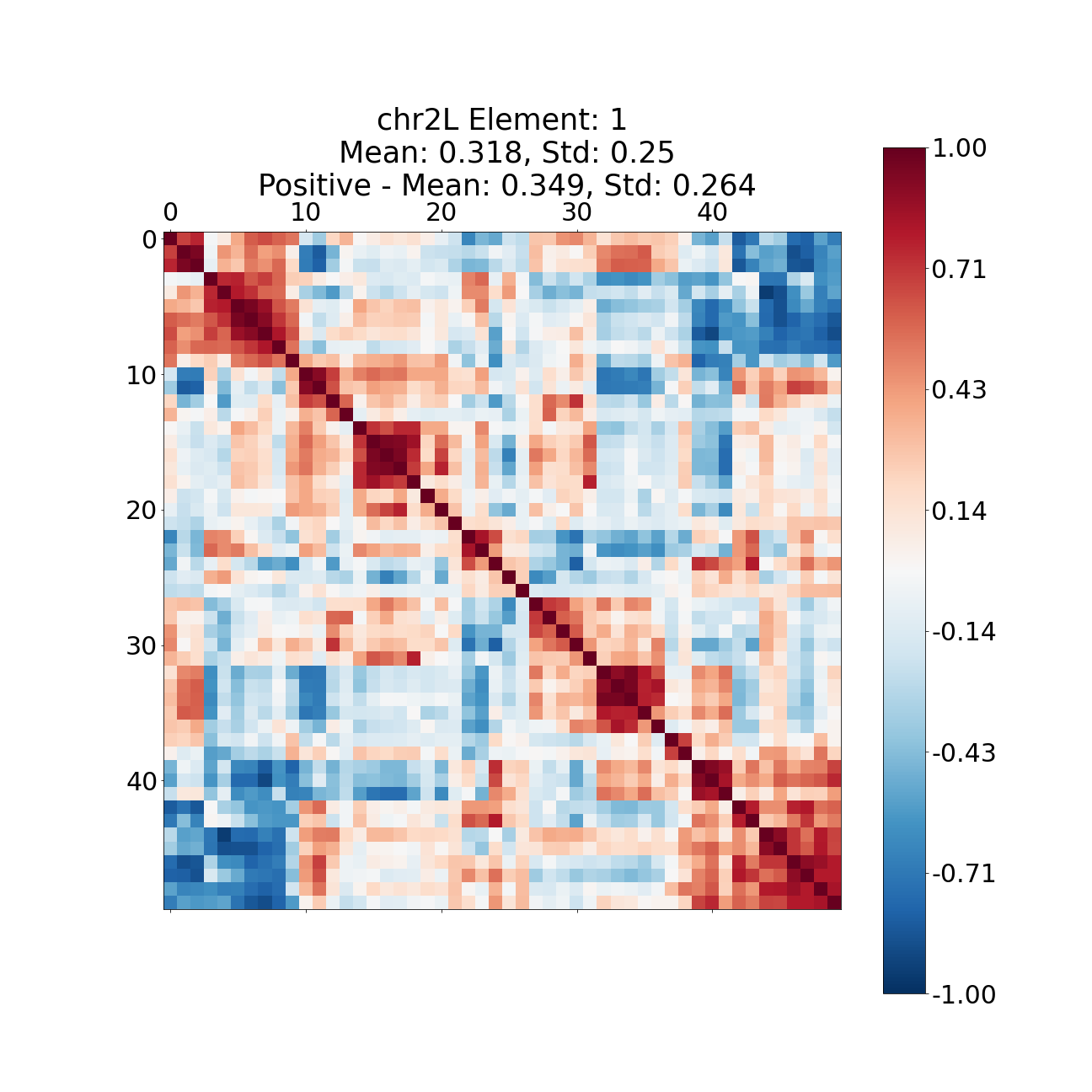}}
        \caption{Pairwise coexpression of genes covered by (a) the R1-R4 genomic regions, (b) the T1-T4 genomic regions, (c) an online cvxNDL dictionary element, and (d) a randomly constructed dictionary element. We calculated the mean and standard deviation of absolute pairwise coexpression values, and the mean and standard deviation of coexpression values specifically for all positively correlated gene pairs. The mean coexpression values within TADs and dictionary elements are similar to each other and generally higher than those of randomly constructed dictionary elements. Note that the plot (b) is of coarser resolution due to the small number of genes covered  when compared to the cases (a), (c), (d).}
        \label{fig:RTAD_pearson}
\end{figure}

\begin{table}[htb]
\centering
\begin{tabular}{c||c  c  c ||c  c  c}
\dtoprule
\multicolumn{1}{c}{} &
\multicolumn{3}{c}{Online cvxNDL}    &
\multicolumn{3}{c}{Random}    \\ 
\cmidrule(r){1-7}
&Dictionary & & & Dictionary &  &  \\
& element id & Intersection & Cumulative &element id & Intersection & Cumulative \\
\cmidrule(r){1-5} \cmidrule(lr){5-7}
1 & 1   & 15 & 15   & 20  &	3 &	3\\
2 & 11 & 12 & 24 & 0   & 1  &4\\
3 & 12 & 12 & 30 & 1   & 1  &5\\
4 & 7   & 11 & 35 & 21 & 1  & 6\\
5 & 21 & 10 & 38 & 17 & 1  &7\\
\dbottomrule
\end{tabular}
\vspace{0.1cm}
\caption{Intersection between the set of genes within the R1-R4 genomic regions and the sets of genes covered by online cvxNDL dictionary elements for chr2L. We determined the sizes of the intersections of the set of genes covered by each dictionary element and the genes in the R1-R4 genomic region and arranged them in decreasing order. The top $5$ dictionary elements in this order cumulatively contain $38$ out of the $85$ genes within the R1-R4 genomic regions. This is in sharp contrast with randomly generated dictionary elements, where the top $5$ elements with maximum intersection cover only $7$ genes.}
\label{tab:R_region_cover}
\end{table}

Finally, we mapped genes covered by our dictionary elements onto nodes of the STRING protein-protein interaction network~\cite{szklarczyk2019string}. These mappings allow us to determine the confidence of pairwise gene interactions. These, and related results based on  FlyMine~\cite{lyne2007flymine} data, a large gene expression repository for \emph{Drosophila Melanogaster}, are available in Supplement Section~$8$.


\section*{Methods}
\label{sec:methods}

\textbf{Notation.} Sets of consecutive integers are denoted by $[l]=\{1,\ldots,l\}$. The symbol $\mathbb{N}$ is reserved for the natural numbers. Capital letters are reserved for matrices (bold font) and random variables (RVs) (regular font). Vectors are denoted by lower-case underlined letters. For a matrix of dimension $d \times n$ over the reals, ${\bf A} \in \real^{d \times n}$, ${\bf A}[i,:]$ is used to denote the $i^{\text{th}}$ row and ${\bf A}[:,i]$ the $i^{\text{th}}$ column of ${\bf A}$. The entry in row $i$, column $j$ is denoted by ${\bf A}[i, j]$. Similarly, $\underline{x}[l]$ is used to denote the $l^{\text{th}}$ coordinate of a deterministic vector $\underline{x} \in \mathbb{R}^{d}$. Furthermore, we use the standard notation for the $\ell_1$ and Frobenius norm of matrices, $\norm{{\bf A}}_1 = \sum_{i,j}|{\bf A}[i,j]|$ and $\norm{{\bf A}}_F^2 = \sum_{i,j}{\bf A}[i,j]^2$, respectively. 

A network $\cG = ([n], \bfA)$ is an ordered pair of sets, the node set $[n]$, and the set of edges represented by their adjacency matrix $\bfA$. Our underlying assumption is that the network is connected, which means that every node can be reached from every other node. Also, $\mathbf{A}[i,j] = \mathbf{A}[j, i] \in \{0, 1\}$, indicating the presence or absence of an undirected edge between nodes $i,j$. In addition, $\text{Col}({\bfA})$ stands for the set of columns of $\bfA$, while $\text{cvx}(\bfA)$ stands for the convex hull of $\text{Col}({\bfA})$.

\noindent\textbf{Online DL.} We first formulate the online DL problem. Assume that $N$ input data samples are generated by a random process and organized in matrices $(\mathbf{X}_t)_{t\in \mathbb{N}} \in \real^{d \times N}$ indexed by time $t$. For $N=1$, $\mathbf{X}_t$ reduces to a column vector that encodes a $d$-dimensional signal. Given an online, sequentially observed data stream $(\mathbf{X}_t)_{t\in \mathbb{N}}$, the goal is to find a sequence of dictionary matrices $(\mathbf{D}_t)_{t \in \mathbb{N}}, \mathbf{D}_t \in \real^{d \times K}$, 
and codes $( {\bm{\Lambda}}_t)_{t \in \mathbb{N}},  {\bm{\Lambda}}_t \in \real^{K \times N}$, such that when $t \rightarrow \infty$ almost surely we have
\begin{equation}
  \norm{\mathbf{X}_t - \mathbf{D}_{t}  {\bm{\Lambda}}_t}_F^2 \rightarrow \min_{\bfD,  {\bm{\Lambda}}} \Ex_{\bfX}{\norm{\bfX - \bfD  {\bm{\Lambda}}}_F^2}.
  \label{eq:NDL_loss}
\end{equation}
The expected loss in Equation~\ref{eq:NDL_loss} can be minimized by iteratively updating $ {\bm{\Lambda}}_t$ and $\mathbf{D}_t$ every time a new data sample $\mathbf{X}_t$ is observed. The approximation error of $\bfD$ for a single data sample $\bfX$ is chosen as 
\begin{equation}
  l(\mathbf{X}, \mathbf{D}) = \min_{ {\bm{\Lambda}} \in \real^{K \times N}} \norm{\bfX - \bfD  {\bm{\Lambda}}}_F^2 + \lambda \norm{ {\bm{\Lambda}}}_1.
  \label{eq:NDL_single_loss}
\end{equation}
The second term represents a sparsity-enforcing regularizer. Furthermore, the empirical $f_t$ and surrogate loss $\hat f_t$ for $\bfD$ are defined as
\begin{align}
  f_t (\bfD) &= (1 - w_t)f_{t - 1}(\mathbf{D}) + w_t l(\mathbf{X}_t, \mathbf{D}), t \geq 1,
  \label{eq:NDL_impirical_loss_W}\\
  \hat f_t(\bfD) &= (1 - w_t)\hat f_{t - 1}(\mathbf{D}) + w_t ({\norm{\mathbf{X}_t - \mathbf{D}  {\bm{\Lambda}}}_F^2 + \lambda \norm{ {\bm{\Lambda}}}_1}),
  \label{eq:NDL_surrogate_loss_W}
\end{align}
where the weight $w_t$ determines the sensitivity of the algorithm to the newly observed data. The online DL algorithm first updates the code matrix ${\bm{\Lambda}}_t$ by solving Equation~(\ref{eq:NDL_single_loss}) with $l(\mathbf{X}_t, \mathbf{D}_{t - 1})$, then updates the dictionary matrix $\mathbf{D}_t$ by minimizing~(\ref{eq:NDL_surrogate_loss_W}) via
\begin{equation}
  \mathbf{D}_t = \argmin_{\mathbf{D} \in \real^{d \times r}}({\trace({\mathbf{D} \mathbf{A}_t \mathbf{D}^T}) - 2\trace({\mathbf{D} \mathbf{B}_t})}),
  \label{eq:NDL_update_W}
\end{equation}
where $\mathbf{A}_t = (1 - w_t) \mathbf{A}_{t - 1} + w_t {\bm{\Lambda}}_t {\bm{\Lambda}}_t^T$ and $\mathbf{B}_t = (1 - w_t) \mathbf{B}_{t - 1} + w_t {\bm{\Lambda}}_t \mathbf{X}_t^T$ are the aggregated history of the input data and their codes, respectively. For simplicity, we set $w_t = \frac{1}{t}$.

To add convexity constraints, we introduce for each dictionary element a \emph{representative set (region)} $\hat \bfX_t^{(i)} \in \real^{d \times N_i}, i\in [K],$ where $N_i$ is the size of the representative set for dictionary element $\mathbf{D}_t[:, i]$, and $N=\sum_{i=1}^{K}N_i$. The representative set for a dictionary element is a small subcollection of real data samples observed up to time $t$ that best explain the dictionary element they are assigned to. The set of representatives is updated after observing a sample, the inclusion of which provides a better estimate of the dictionary element compared to the previous set. Since the representative set is bounded in size, if a new sample is included, an already existing sample has to be removed (see Figure~\ref{fig:cvxNDL_alg}). Formally, the optimization objective is of the form
\begin{equation}
    \min_{\mathbf{D} \in \text{cvx}(\hat\bfX), \hat\bfX} \hat f_t(\mathbf{D}) = \min_{\bfD \in \text{cvx}(\hat\bfX), \hat\bfX} \left(1 - \frac{1}{t}\right)\hat f_{t - 1}(\mathbf{D}) +
    \frac{1}{t} \left(\norm{\mathbf{X}_t - \mathbf{D} {\bm{\Lambda}}_t}_F^2 + \lambda \norm{{\bm{\Lambda}}_t}_1\right).
    \label{eq:online_cvxNDL_obj}
\end{equation}

\noindent\textbf{MCMC sampling of subnetworks (sample generation).} For NDL, it is natural to let the columns of $\mathbf{X}_t$ be vectorized adjacency matrices of $N$ subnetworks. Hence one needs to efficiently sample meaningful subnetworks from a (large) network. In image DL problems, samples can be generated directly from the image using adjacent rows and columns. However, such a sampling technique cannot be applied to arbitrary network data. Selecting nodes along with their one-hop neighbors at random may produce subnetworks of vastly different sizes and the results do not capture meaningful long-range interactions. It is also difficult to trim such subnetworks to uniform sizes. Furthermore, sampling a fixed number of nodes uniformly at random from sparse networks produces disconnected subnetworks with high probability and is not an acceptable approach either. 

To address these problems, we consider ``subnetwork sampling'' introduced in~\cite{lyu2023sampling,lyu2020online} where we fix a template network $F=([k], \mathbf{A}_F)$ of $k$ nodes and seek subnetworks induced by $k$ nodes in the input network $\cG$, with the constraint that the subnetwork \emph{contains} (but does not necessarily equals) the template $F$ topology. Given an input network $\cG = ([n], \bfA)$ and a template network $F = ([k], \mathbf{A}_F)$, we define a set of homomorphisms as a vector of the form
\begin{equation}
  \mathtt{Hom}(F, \cG) = \left\{\underline{x} : [k] \rightarrow [n] \left\vert \prod_{1\leq i,j \leq k} \mathbf{A}[\underline{x}[i], \underline{x}[j]]^{\mathbf{A}_F[i, j]} = 1\right. \right\},
  \label{eq:homomorphism}
\end{equation}
where we by default assume that $0^0=1$.
For each homomorphism $\underline{x} \in \mathtt{Hom}(F, \cG)$, denote its induced adjacency matrix by $\mathbf{A}_{\underline{x}}$, where $\mathbf{A}_{\underline{x}}[a, b]=\mathbf{A}[\underline{x}[a], \underline{x}[b]]$, $1 \leq a, b \leq k$. The adjacency matrix $\mathbf{A}_{\underline{x}}$ represents one sample from the input network $\cG$. An example homomorphism is shown in Figure~\ref{fig:motif_sample_img}, where the input network $\cG$ contains $n=9$ nodes and the template network $F$ is a star network that contains $k=4$ nodes. One proper homomorphism in this case is $\underline{x}[a] = 9,$ $\underline{x}[b] = 6,$ $\underline{x}[c] = 4,$ $\underline{x}[d] = 7$, which gives rise to an adjacency matrix $\mathbf{A}_{\underline{x}}$ as depicted. A homomorphism can be sampled using the rejection sampling algorithm presented in the Supplement Section~$2$,  Algorithm~$1$. Our choice of template network, as already mentioned, is a \textit{k-path}, i.e., a path joining $k$ nodes. Paths are a simple and natural choice for networks with long average path lengths, such as chromatin interaction networks. It is also the same choice of template used in standard NDL. As a final remark, we note that a $k$-path homomorphism leads to a sample of dimension $d=k^2$, as we will flatten its $k \times k$ adjacency matrix into a single vector.

Although rejection sampling can be used repeatedly to generate several homomorphisms, it is highly inefficient. To efficiently generate a sequence of sample adjacency matrices $\mathbf{A}_{\underline{x}_t}$ from $\cG$, the MCMC sampling algorithm is used instead, while rejection sampling is only used to initialize the MCMC algorithm. 

Next, for a homomorphism $\underline{x}_t$, let $\mathcal{N}[\underline{x}_t\ind{1}]$ ($\mathcal{N}$ for short) denote the set of neighbors of $\underline{x}_t\ind{1}.$ We first choose a node $v\in \mathcal{N}$ from the neighborhood of $\underline{x}_t\ind{1}$ uniformly at random, i.e. with probability $P(v) = \frac{1}{|\mathcal{N}|}.$ We also calculate the probability of acceptance $\beta$ for the selected node $v.$ For a $k-$path template used in our approach, the value of $\beta$ is given by
\begin{equation}
    \beta = \min{\left\{\frac{\sum_{c \in [n]} A^{k-1}[v,c]}{\sum_{c \in [n]}A^{k-1}[\underline{x}_t[1],c]}\frac{\sum_{c \in [n]} A[\underline{x}_t[1],c]}{\sum_{c \in [n]}A[v,c]},1\right\}},
\end{equation}
following the guidelines from~\cite{lyu2023sampling,lyu2020online}.

Next, we draw a value $u \in [0, 1]$ uniformly at random. If $u < \beta$, we accept $\underline{x}_{(t+1)}\ind{1} = v$, otherwise we reject $v$ and reset $\underline{x}_{(t + 1)}\ind{1} = \underline{x}_t\ind{1}$. We then perform a directed random walk from $\underline{x}_{t+1}[1]$ of length equal to $k-1$ to obtain $\underline{x}_{(t + 1)}\ind{2}, \ldots,\underline{x}_{(t + 1)}\ind{k}$. An illustration of the sampling procedure is shown in Figure~\ref{fig:mcmc_img}, while the detailed algorithm is presented in the Supplement Section~$2$, Algorithm~$2$.

\noindent\textbf{Online convex NDL (online cvxNDL).} We start by initializing the dictionary $\mathbf{D}_0$ and representative sets $\{\hat \bfX_0^{(i)}\}, i\in [K],$ for each dictionary element. The algorithm for initialization is presented in the Supplement Section~$2$ Algorithm~$3$. After initialization, we perform iterative optimization to generate $\mathbf{D}_t$ and $\{\hat \bfX_t^{(i)}\}, i\in [K],$ to reduce the loss at round $t$. At each iteration, we use MCMC sampling to obtain a $k$-node random subnetwork as sample $\mathbf{X}_t$, and then update the codes $\Lambda_t$ based on the dictionary $\mathbf{D}_{t-1}$ by solving the optimization problem in Equation~(\ref{eq:NDL_single_loss}). Then we assign the current sample to a representative set of the closest dictionary element, say $\mathbf{D}_{t-1}[:,j]$, and jointly update its representative set $\hat \bfX_t^{(j)}$ and all dictionaries $\mathbf{D}_t$ as shown in Figure~\ref{fig:cvxNDL_alg}. The iterative update algorithm for online cvxNDL is presented in the Supplement Section~$2$ Algorithm~$4$.

The output of the algorithm is a dictionary matrix $\mathbf{D}_T\in\mathbb{R}^{k^2\times K}$, where each column is a flattened vector of a dictionary element of size $k \times k$, and the representative sets $\{\hat \bfX_T^{(i)}\}, i\in [K],$ for each dictionary element. Each representative set $\hat\bfX_T^{(i)} \in \real^{k^2 \times N_i}$ contains $N_i$ history-sampled subnetworks from the input network as its columns which are called the representatives of the dictionary element. The convex hull of all representatives of a dictionary element forms the representative region of the dictionary element. We can easily convert both the dictionary elements and representatives back to $k\times k$ adjacency matrices. Due to the added convexity constraint, each dictionary element $\mathbf{D}_T[:, j]$ at the final step $T$ has the \emph{interpretable} form:
\begin{align}
    \mathbf{D}_T[:, j] = \sum_{i \in [N_j]} w_{j, i} \hat \bfX_T^{(j)}[:, i],\;\; \mbox{ s.t. } \sum_{i\in[N_j]} w_{j, i} = 1,
    w_{j, i} & \geq 0, i\in[N_j], j\in[K].
\end{align}
The weight $w_{j,i}, i\in[N_j],$ is the \textit{convex coefficient} of the $i^{\text{th}}$ representative of dictionary element $\mathbf{D}_T[:, j]$. Dictionary elements learned from the data stream can be used to reconstruct the input network by multiplying it with the dictionary element weights from Equation~(\ref{eq:NDL_single_loss}). The $j^{\text{th}}$ index of the weight vector corresponds to the contribution of dictionary element $\mathbf{D}_{T-1}[:,j]$ to the reconstruction. Similarly to what was done in~\cite{lyu2020online}, we can also define the \textit{importance score} for each dictionary element as
\begin{equation}
    \gamma(i) = \frac{\mathbf{A}_t[i, i]^2 }{\sum_{j \in [K]} \mathbf{A}_t[j, j]^2}. \label{eq:imp_score}
\end{equation}
We use the importance scores, as described in the previous sections, to determine the most frequently used interactions in the dictionary construction, as well as the most typical and important long-range interactions. 

To conclude, we point out that the \emph{density} $\rho$ of interactions in a dictionary element is defined as
$$ \rho=\frac{1}{k^2}\, \sum_{i,j=1}^{k} \mathbf{D}_T[i,j].$$

\clearpage

\section*{Funding and Acknowledgement}
The work was supported by the National Science Foundation grants $\#1956384$ and $\#2206296$ and grant CZI DAF$2022-249217$. The authors gratefully
acknowledge many useful discussions with Dr. Yijun Ruan.

\section*{Supporting information}

Supplemental material, including figures and tables, is available in the Supplement file. The online cvxNDL code and test datasets are available at: \url{https://github.com/rana95vishal/chromatin_DL/}


\nolinenumbers

%
%
%

\bibliography{main.bib}

\end{document}


\vspace*{0.2in}

\begin{flushleft}
{\Large
\textbf\newline{Supplement - Interpretable Online Network Dictionary Learning for Inferring Long-Range Chromatin Interactions} 
}
\newline
\\
Vishal Rana,
Jianhao Peng,
Chao Pan,
Hanbaek Lyu,
Albert Cheng,
Minji Kim,
Olgica Milenkovic
\end{flushleft}

\section{Motivation}
\begin{figure}[h]
    \centering
    \subfloat[]{\includegraphics[width = 0.31\linewidth]{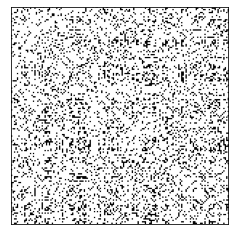}}
    \hfill
    \subfloat[]{\includegraphics[width = 0.37\linewidth]{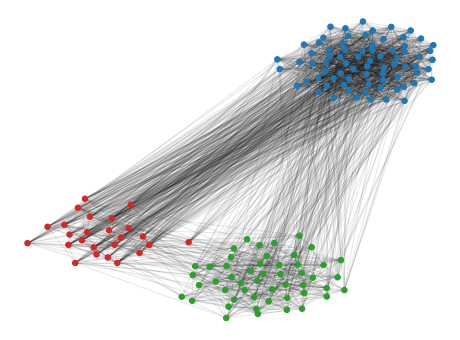}}
    \hfill
    \subfloat[]{\includegraphics[width = 0.31\linewidth]{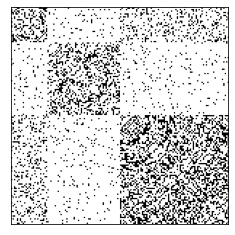}}
    \caption{\textbf{(a)} Observed adjacency matrix of a three-cluster model, where points are arranged in linear order with dense interactions existing both at short- and long-range. \textbf{(b)} The underlying cluster structure. \textbf{(c)} The reordered adjacency matrix that reveals all interaction classes.}
    \hfill
    \label{fig:sbm_cluster_example}
\end{figure}

Dictionary learning (DL), a form of nonnegative matrix factorization
(MF), has been widely used in the analysis of biological data. However, \emph{efficient}, and \emph{biologically interpretable} computational methods for analyzing long-distance multiplexed chromatin interactions at a single-cell level are still lacking. This gap exists primarily because classical DL methods are not tailored for network data analysis. Furthermore, these interactions cannot be easily visualized or predicted via classical clustering approaches. This issue is best illustrated by Figure~\ref{fig:sbm_cluster_example}, where a part of the contact map contains three hidden clusters, colored red, green, and blue~\cite{holland1983stochastic}. When using a linear chromatin order, the particular structure of the clusters is not observable. By rearranging the rows/columns, the cluster structure becomes apparent within the adjacency matrix.
To mitigate this issue, we propose a novel online convex network dictionary learning algorithm (online cvxNDL) that imposes ``convexity'' constraints on the sampled subgraph patterns to address the issue of interpretability. Furthermore, due to its online nature, it scales to large graph-structured datasets. The detailed algorithmic implementations are described in the next section.

\section{Algorithmic Details}
\label{supp_sec:alg}
The algorithms presented in this section describe the detailed steps of implementation outlined in the Methods Section.
\subsection{MCMC Sampling of Subnetworks}
We use the MCMC sampling in conjunction with subnetwork sampling to generate online samples. We seek samples in the form of subnetworks induced by $k$ nodes in the original input network $\cG$ such that these subnetworks contain the template $F$ topology. Given an input network $\cG = (V, \mathbf{A})$ and a template network $F = ([k], \mathbf{A}_F)$, we define a set of homomorphisms as a vector of the form (with the assumption that $0^0=1$):
\begin{equation*}
  \mathtt{Hom}(F, \cG) = \left\{\underline{x} : [k] \rightarrow [n] \left\vert \prod_{1\leq i,j \leq k} \mathbf{A}[\underline{x}[i], \underline{x}[j]]^{\mathbf{A}_F[i, j]} = 1\right. \right\}.
\end{equation*}
Algorithm~\ref{alg:rej_sample} outlines how to use rejection sampling to obtain one homomorphism $\underline{x}$ (an illustrative example is presented in Figure $1$(d) in the main text).  In this work, we use a $k$-path as the template network, where a $k$-path represents a directed path from node $1$ to $k$. Paths serve as a simple and natural choice for networks containing inherent long paths, such as chromatin interaction networks, where most contact measurements are due to proximity in the linear chromosome order.
\begin{algorithm}
   \caption{Rejection Sampling of Homomorphisms}
   \label{alg:rej_sample}
   \begin{algorithmic}[1]
    \STATE \textbf{input:} Network $\cG = ([n], \bfA)$, template $F = ([k], \mathbf{A}_F)$ (under the assumption that there exists at least one homomorphism $F \rightarrow \cG$).
    \WHILE{true} 
      \STATE Sample $\underline{x} = (\underline{x}\ind{1}, \underline{x}\ind{2}, \dots, \underline{x}\ind{k}) \in [n]^{k}$ so that $\underline{x}\ind{i}$'s are i.i.d.
      \IF{$\prod_{1\leq i,j\leq k}\mathbf{A}[\underline{x}\ind{i}, \underline{x}\ind{j}]^{\mathbf{A}_F[i, j]} > 0$}
        \STATE \textbf{break}
      \ENDIF
    \ENDWHILE 
    \RETURN A homomorphism $\underline{x}: F \rightarrow \cG$.
  \end{algorithmic}
\end{algorithm}

While we can find different homomorphisms from the input $\cG$ by iteratively executing Algorithm~\ref{alg:rej_sample}, this method is computationally expensive. To efficiently generate a sequence of sample adjacency matrices $\mathbf{A}_{\underline{x}_t}$ from $\cG$, the MCMC sampling algorithm gradually changes the sampled subnetwork based on previous samples as described in Algorithm~\ref{alg:MCMC_motif}. An illustrative example is shown in Figure $1$(e) in the main text. This sampling algorithm was introduced in~\cite{lyu2023sampling,lyu2020online}.
\begin{algorithm}
  \caption{The MCMC Sampling Algorithm}
  \label{alg:MCMC_motif}
  \begin{algorithmic}[1]
    \STATE \textbf{input:} Network $\cG = ([n], \bfA)$, template $F = ([k], \mathbf{A}_F)$, and one homomorphism $\underline{x}: F \rightarrow \cG$.
    \STATE Sample $v \in \mbox{Neighbor}(\underline{x}\ind{1})$ with probability
      $P(v) = \frac{1}{\mathcal{N}[\underline{x}[1]]}$.
    \STATE Compute the acceptance probability\\
      $\beta = \min{\left\{\frac{\sum_{c \in [n]} A^{k-1}[v,c]}{\sum_{c \in [n]}A^{k-1}[\underline{x}[1],c]}\frac{\sum_{c \in [n]} A[\underline{x}[1],c]}{\sum_{c \in [n]}A[v,c]},1\right\}}$.
    \STATE Sample $u$ uniformly at random from $[0, 1]$.
    \IF{$u < \beta$}
      \STATE $\underline{x}'\ind{1} = v$
    \ELSE
      \STATE $\underline{x}'\ind{1} = \underline{x}\ind{1}$
    \ENDIF
    \FOR{$s = 2, 3, \dots, k$}
      \STATE Sample $w \in [n]$ with probability
        $P_s(w) = \frac{\mathbf{A}[\underline{x}'\ind{s - 1}, w]}{\sum_{c \in V} \mathbf{A}[\underline{x}'\ind{s - 1}, c]}$.
      \STATE $\underline{x}'\ind{s} = w$
    \ENDFOR
    \RETURN New homomorphism $\underline{x}': F \rightarrow \cG$.
  \end{algorithmic}
\end{algorithm}

\subsection{Online Convex NDL (online cvxNDL)}
Our online cvxNDL algorithm consists of two parts: initialization and iterative optimization. For initialization, we compute an initial choice for the dictionary elements $\mathbf{D}_0$ and initialize the representative regions $\hat \bfX_0^{(j)},~\forall j\in [K]$ using i.i.d. sampling of homomorphisms (Algorithm~\ref{alg:ndl_initialize}). Note that we use i.i.d. sampling of homomorphisms only during the initialization step, and MCMC sampling afterwards. Upon initialization, we iteratively optimize the dictionary and the representative regions in the next phase (Algorithm~\ref{alg:online_cvxNDL}). The output of the latter algorithm is the final dictionary $\mathbf{D}_T$ and the corresponding representative regions for all dictionary elements $\hat \bfX_T^{(j)},~\forall j\in [K]$. Due to the added convexity constraint, each dictionary element $\mathbf{D}_T[:, j]$ at the final step $T$ has the following interpretable form:
\begin{align*}
    \mathbf{D}_T[:, j] = \sum_{i \in [N_j]} w_{j, i} \hat \bfX_T^{(j)}[:, i],\;\mbox{s.t. } \sum_{i\in[N_j]} w_{j, i} = 1,
    w_{j, i} \geq 0, i\in[N_j], j\in[K].
\end{align*}
The weight $w_{j,i}, i\in[N_j]$ is the convex coefficient of the $i^{\text{th}}$ representative of dictionary element $\mathbf{D}_T[:, j]$.
\begin{algorithm}[hbtp]
   \caption{Initialization}
   \label{alg:ndl_initialize}
   \begin{algorithmic}[1]
      \STATE \textbf{input:} Use rejection sampling in Algorithm~\ref{alg:rej_sample} to sample i.i.d homomorphisms $\underline{x}_1, \underline{x}_2, \dots, \underline{x}_N$.
      \STATE For each homomorphism, define an adjacency matrix such that: $\mathbf{A}_{\underline{x}_i}[a, b] = \mathbf{A}[\underline{x}_i\ind{a}, \underline{x}_i\ind{b}]$. Flatten the adjacency matrices into vectors: $\underline{x}_1, \underline{x}_2, \dots,\underline{x}_{N}$, $\underline{x}_i \in \real^d, d = k^2$ and collect them in $\hat \bfX \in \real^{d\times N}$.
      \STATE Run $K$-means on $\hat \bfX$ to generate the cluster indicator matrix $\mathbf{H} \in \{0, 1\}^{N \times K}$ and determine the initial cluster sizes (subsequent representative set sizes) $N_i, i\in [K]$.
      \STATE Compute $\mathbf{D}_0$ and $\xhati{0} \in \real^{d\times N_i},~\forall i\in [K],$ according to:
      \[\mathbf{D}_0 = \hat\bfX\;\mathbf{H}\;{ \text{diag}(1/N_1, \ldots, 1/N_K)}\]
      and summarize the initial representative sets of the clusters into matrices $\xhati{0}, \, i=[K]$.
   \RETURN $\mathbf{D}_0$, $\{\xhati{0}\}_{i\in [K]}$.
\end{algorithmic}
\end{algorithm}
\begin{algorithm}
   \caption{Online cvxNDL}
   \label{alg:online_cvxNDL}
   \begin{algorithmic}[1]
      \STATE \textbf{input:} Network $\cG = ([n], \bfA)$, template $F = ([k], \mathbf{A}_F)$, a parameter $\lambda \in \real$, max number of iterations $T$, and number of dictionary elements $K$.
      \STATE \textbf{initialization:} Compute $\mathbf{D}_0$, $\{\xhati{0}\}_{i\in[K]}$ using Algorithm~\ref{alg:ndl_initialize}. Set $\mathbf{A}_0=\mathbf{0}$, $\mathbf{B}_0=\mathbf{0}$.
      \label{initialize}
      \FOR {$t= 1$ to $T$}
	  \STATE MCMC sample a homomorphism $\underline{x}_t$ (Algorithm~\ref{alg:MCMC_motif}). Find its adjacency matrix $\mathbf{A}_{\underline{x}_t}[a, b] = \mathbf{A}[\underline{x}_t\ind{a}, \underline{x}_t\ind{b}]$ and flatten it to $\underline{x}_t$. 
          \STATE Update $\mathbf{\Lambda}_t$\label{update_alpha} according to:
        \begin{gather}
          \mathbf{\Lambda_t} =  \argmin_{\mathbf{\Lambda} \in \real^{K\times 1}}  \frac{1}{2} \norm{\underline{x}_t-\mathbf{D}_{t-1} \mathbf{\Lambda}}_2^2  + \lambda \norm{\mathbf{\Lambda}}_1. \label{eq:classify_data}
        \end{gather}
        \STATE Set $\mathbf{A}_t = \frac1t({({t-1}) \mathbf{A}_{t-1} + \mathbf{\Lambda}_t \mathbf{\Lambda}_t^T})$ \label{update_At}~~and~~ $\mathbf{B}_t = \frac1t({({t-1}) \mathbf{B}_{t-1} + \underline{x}_t~\mathbf{\Lambda}_t^T})$.\label{update_Bt}

    \STATE Choose the index of the basis $i_t$ to be updated according to 
    $i_t =\argmax_{j \in [k]} {\mathbf{\Lambda}_t}\ind j$
    \label{algoclassif}
    \STATE Generate the augmented representative regions $\{\hat{\mathbf{Y}}^{l}_t\}_{l\in [N_{i_t}]\cup \{0\}}$: \label{update_xhat}
    	\begin{equation}\label{eq:candidate_rep_region}
	\begin{gathered}
	    	\hat{\mathbf{Y}}^{0}_t = \hat{\mathbf{X}}^{i_t}_{t-1}	\\
		\{\hat{\mathbf{Y}}^{l}_t\}_{l \in [N_{i_t}]} :~\hat{\mathbf{Y}}^{l}_t\ind j =
		\begin{cases}
	 		\hat{\mathbf{X}}^{i_t}_{t-1}\ind j, &\text{ if } j\in [N_i]\setminus l\\
	 	      	\underline{x}_t, &\text{ if } j= l.\\
	 	\end{cases}
	\end{gathered}
   	\end{equation}
      \STATE Update $\{\hat{\mathbf{X}}^ t_{i}\}_{i\in[K]}$ and $\mathbf{D}_t$ \label{dic_update} by executing the following two steps \label{step:jointly_update}
      \begin{itemize}
      \item Compute $l^\star, \hat{\mathbf{D}}^{\star}$ by solving the optimization problems:
    \[
	      l^\star, \hat{\mathbf{D}}^{\star} =
         \displaystyle \argmin_{\substack{l,~\mathbf{D} \; \text{ s.t. }\\ \mathbf{D}\ind j \in
         \text{cvx}\{\hat{\mathbf{X}}^{j}_{t-1}\} ~j\ne i_t,\\ \mathbf{D}\ind{i_t}\in \text{cvx}\{\hat{\mathbf{Y}}^{l}_t\}}}
         \frac12 \trace(\mathbf{D}^T\mathbf{D}\mathbf{A}_t) - \trace(\mathbf{D}^T \mathbf{B}_t).
        \label{eq:dic_update}
    \]
      \item Set
      \[\begin{aligned}	
      	      	      \hat{\mathbf{X}}^{i}_{t}  &= \begin{cases}
      	      	      	\hat{\mathbf{Y}}^ {l^\star}_t, &\text{ if } i=i_t\\
      	      	        \hat{\mathbf{X}}^{i}_{t-1}, &\text{ if }  i\in [K]\setminus i_t,
      	      	      \end{cases} \\
      	      	      \mathbf{D}_t &= \hat{\mathbf{D}}^{\star}.
 	      \end{aligned}\]
      \end{itemize}
   \ENDFOR
  \RETURN $\mathbf{D}_T$, $\hat{ \mathbf{X}}_T^{(i)},~\forall i\in [K]$.
\end{algorithmic}
\end{algorithm}
\clearpage
\section{Synthetic Data Analysis}
\label{supp_sec:synthetic}
We tested our online cvxNDL method on a network (graph) generated by Stochastic Block Model (SBM)~\cite{holland1983stochastic}, containing $150$ nodes with $3$ clusters of size $25, 50, 75$. Due to the small size of the synthetic set, we fixed the number of dictionary elements to $K = 6$ and used a path of length $11$ as our template. In the initialization step, we sampled $30$ subgraphs from the input synthetic data network, with each dictionary element represented by at least $3$ representatives. The maximum number of iterations of the online method was set to $1,000$. 

\begin{figure}[htb]
  \centering
  \subfloat[NMF]{\includegraphics[width = 0.45\linewidth]{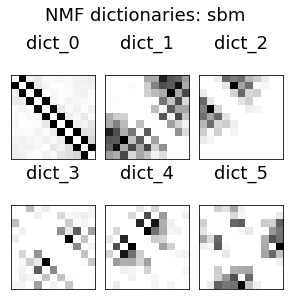}}
  \subfloat[Online NDL]{\includegraphics[width = 0.45\linewidth]{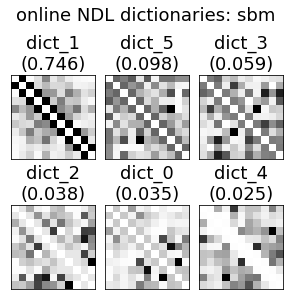}}
  \hfill
  \subfloat[CMF]{\includegraphics[width = 0.45\linewidth]{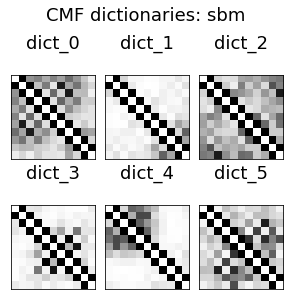}}
  \subfloat[Online cvxNDL]{\includegraphics[width = 0.45\linewidth]{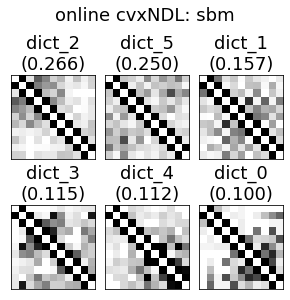}}
  \hfill
  \caption{Dictionary elements generated by different methods on an SBM synthetic dataset. Numbers in parenthesis are the importance scores for online NDL and online cvxNDL.}
  \label{fig:synth_dict_all}
\end{figure}

We compared online cvxNDL with various baseline methods, including NMF, CMF, and online NDL. The learned dictionary elements for different methods are shown in Figure~\ref{fig:synth_dict_all}. The dictionary elements in online NDL and online cvxNDL are ordered by their importance score defined as $\gamma(i) = \frac{\mathbf{A}_t[i, i]^2 }{\sum_{j \in [K]} \mathbf{A}_t[j, j]^2}$. Each square block in the subplots indicates one dictionary element in the form of an adjacency matrix. The color-shade reflects the values in the adjacency matrix, with black corresponding to $1$ (the largest value) and white corresponding to $0$ (the smallest value).

From the results, we can see that dictionaries generated using NMF only contain partial interaction structures and are hard to interpret. The two convex methods, CMF and online cvxNDL, contain the template structure in all learned dictionary elements and show stronger off-diagonal connectivity, which is expected as the input data has slightly stronger connections between the first and last cluster than other pairs (See Figure~\ref{fig:sbm_cluster_example}). Online NDL dictionary elements represent ``a middle ground'' between NMF and online cvxNDL. Dictionary elements $2$, $0$, and $4$ resemble those generated by NMF, while dictionary elements $1$, $5$, and $3$ are similar to the ones generated by online cvxNDL, although with weaker connectivity. Also, the importance score distributions of online NDL and online cvxNDL differ substantially. In online NDL, dictionary element $1$ in Figure~\ref{fig:synth_dict_all} is the dominant component in representations, whereas, in online cxvNDL, the top two dictionary elements (dictionary elements $2$ and $5$ in~\ref{fig:synth_dict_all}) share similar scores and the dictionary elements, in general, have a more balanced distribution of importance scores. From the original adjacency, we can see that there are indeed two different connectivity patterns in the network captured by online cvxNDL.

\clearpage
\textbf{Reconstruction accuracy:} To validate the reliability of our learned dictionaries for representing the global interactions, we reconstructed the whole graph by aggregating the regenerated subgraphs: $\hat{\mathbf{x}}_i = \mathbf{D}_T\alphab_i$ from the same MCMC sampling stream. For each method we selected the top-$m$ edges after aggregation to reconstruct the original adjacency matrix, where $m$ is the number of edges in the original adjacency matrix. The original and the reconstructed adjacency matrices are shown in Figure~$7$ in the main text. For comparison, we also added the reconstructed adjacency achieved when using random dictionary elements. From the results, we can see that all baseline methods, as well as online cvxNDL, almost perfectly reconstruct the original network, while, clearly random dictionaries do not capture any meaningful information. We also report the average precision recall score for each method, both for synthetic and real datasets as listed in Table~$1$ in the main text.

\section{ChIA-Drop Dataset}

The preprocessed and binned RNAPII ChIA-Drop data includes $45,938$, $42,292$, $49,072$, and $55,795$ nodes and $36,140$, $28,387$, $53,006$, $45,530$ hyperedges for chromosome chr2L, chr2R, chr3L and chr3R  respectively. The size distribution of hyperedges is given in Table~\ref{tab:hyperedge_size}. The clique-expanded input network has $113,606$, $85,316$, $161,590$, and $143,370$ edges respectively. 

\begin{table}[!h]
\centering
\scriptsize
\caption{Number of hyperedges of various sizes observed in the ChIA-Drop data for various chromosomes.}
\label{tab:hyperedge_size}
\begin{tabular}{l||c|c|c|c}
\hline
\hline
\begin{tabular}[c]{@{}l@{}}hyperedge \\ sizes\end{tabular} & chr2L & chr2R & chr3L & chr3R \\ \hline
2&  28373&	22951&	42175&	35585\\
3&	5723&	4018&	8103&	7379\\
4&	1307&	936&	1804&	1700\\
5&	424&	275&	533&	479\\
6&	136&	94&		196&	187\\
7&	60&		41&		82&		69\\
8&	48&		29&		38&		31\\
9&	21&		15&		28&		22\\
10&	8&		5&		16&		7\\
11&	7&		6&		9&		8\\
12&	11&		2&		7&		9\\
13&	5&		2&		5&		7\\
14&	7&		2&		2&		5\\
15&	4&		2&		1&		4\\
16&	3&		2&		1&		4\\
17&	1&		2&		2&		0\\
18&	2&		1&		1&		1\\
19&	0&		1&		0&		0\\
$\geq$20&		1&		4&		4&		7\\
\hline\hline
\end{tabular} 
\end{table}

The dictionary elements for each of the $4$ chromosomes are presented in Figure~$5$ in the main text. The density or complexity of dictionary elements, defined as $ \rho=\frac{1}{k^2}\, \sum_{i,j=1}^{k} \mathbf{D}_T[i,j],$ is reported in Table~\ref{tab:full_density} while the median distance of pairwise interacting nodes in all representatives of a dictionary element is reported in Table~\ref{tab:full_median_distance}.

\begin{table}[!h]
\centering
\scriptsize
\caption{Density of dictionary elements, reported for all chromosomes.}
\label{tab:full_density}
\begin{tabular}{l||c|c|c|c}
\hline\hline
\begin{tabular}[c]{@{}l@{}}Dictionary \\ element\end{tabular} & chr2L & chr2R & chr3L & chr3R \\ \hline
1 & 0.146 & 0.158 & 0.168 & 0.161 \\ 
2 & 0.188 & 0.165 & 0.156 & 0.157 \\ 
3 & 0.134 & 0.185 & 0.141 & 0.140 \\ 
4 & 0.220 & 0.147 & 0.159 & 0.179 \\ 
5 & 0.145 & 0.146 & 0.142 & 0.139 \\ 
6 & 0.132 & 0.297 & 0.148 & 0.173 \\ 
7 & 0.162 & 0.189 & 0.191 & 0.184 \\ 
8 & 0.158 & 0.184 & 0.164 & 0.147 \\ 
9 & 0.148 & 0.136 & 0.210 & 0.183 \\ 
10 & 0.177 & 0.166 & 0.168 & 0.157 \\ 
11 & 0.220 & 0.261 & 0.163 & 0.161 \\ 
12 & 0.168 & 0.162 & 0.145 & 0.157 \\ 
13 & 0.204 & 0.203 & 0.186 & 0.142 \\ 
14 & 0.225 & 0.142 & 0.148 & 0.205 \\ 
15 & 0.142 & 0.229 & 0.262 & 0.163 \\ 
16 & 0.173 & 0.184 & 0.143 & 0.205 \\ 
17 & 0.189 & 0.263 & 0.127 & 0.224 \\ 
18 & 0.161 & 0.219 & 0.152 & 0.251 \\ 
19 & 0.182 & 0.159 & 0.183 & 0.242 \\ 
20 & 0.187 & 0.156 & 0.170 & 0.193 \\ 
21 & 0.231 & 0.157 & 0.199 & 0.126 \\ 
22 & 0.143 & 0.195 & 0.165 & 0.150 \\ 
23 & 0.162 & 0.201 & 0.134 & 0.175 \\ 
24 & 0.223 & 0.141 & 0.167 & 0.212 \\ 
25 & 0.167 & 0.212 & 0.140 & 0.208 \\ 
\hline\hline
\end{tabular} 
\end{table}

\begin{table}[!h]
\centering
\scriptsize
\caption{Median distance of pairwise interacting nodes within each dictionary element and for each chromosome.}
\label{tab:full_median_distance}
\begin{tabular}{l||c|c|c|c}
\hline\hline
\begin{tabular}[c]{@{}l@{}}dictionary \\ element\end{tabular} & chr2L & chr2R & chr3L & chr3R \\ \hline
1 & 10758 & 6738 & 7328 & 14753 \\ 
2 & 8523 & 7688 & 12934 & 14760 \\ 
3 & 9906 & 8759 & 9539 & 12666 \\ 
4 & 8354 & 7158 & 12690 & 11748 \\ 
5 & 9847 & 7651 & 10412 & 13674 \\ 
6 & 8547 & 6953 & 10608 & 15598 \\ 
7 & 10024 & 9383 & 11994 & 13498 \\ 
8 & 8870 & 9226 & 10399 & 12830 \\ 
9 & 10692 & 7085 & 14414 & 12493 \\ 
10 & 11220 & 6414 & 9466 & 11930 \\ 
11 & 10455 & 10711 & 10130 & 11421 \\
12 & 8488 & 7656 & 11694 & 9398 \\ 
13 & 9979 & 7706 & 14206 & 13455 \\ 
14 & 10591 & 8251 & 8689 & 12540 \\ 
15 & 10928 & 7284 & 10532 & 12572 \\
16 & 10268 & 7143 & 8849 & 13842 \\ 
17 & 8545 & 9681 & 9978 & 15184 \\ 
18 & 8675 & 6859 & 8558 & 11974 \\ 
19 & 9854 & 7882 & 8501 & 18233 \\ 
20 & 9314 & 8199 & 10532 & 11592 \\
21 & 9343 & 8872 & 9728 & 12791 \\ 
22 & 8105 & 6418 & 10214 & 13301 \\
23 & 8870 & 7418 & 11012 & 14239 \\
24 & 9527 & 8764 & 10010 & 12692 \\
25 & 11072 & 9711 & 13471 & 11316 \\ 
\hline\hline
\end{tabular} 
\end{table}
\clearpage
\subsection{Results for Baseline Methods Applied to ChIA-Drop Datasets}
\label{supp_sec:online_cvxNDL_fig}
\begin{figure}[!h]
    \vspace{-0.2in}
    \centering
    \subfloat[chr2L]{\includegraphics[width=0.4\linewidth]{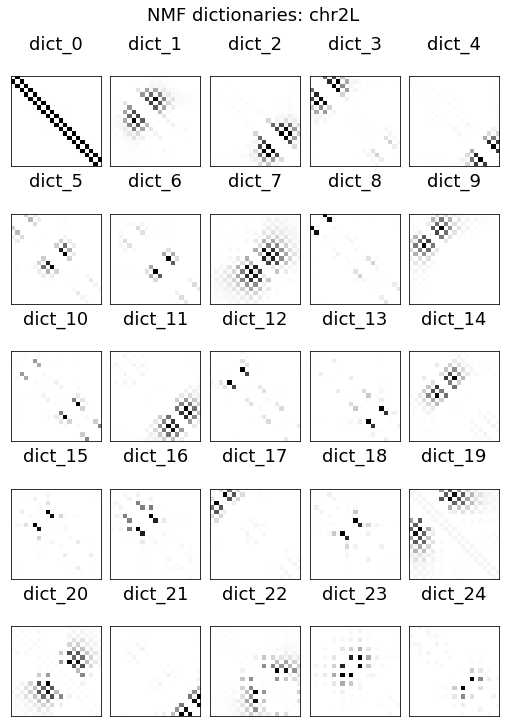}}
    \hfill
    \subfloat[chr2R]{\includegraphics[width=0.4\linewidth]{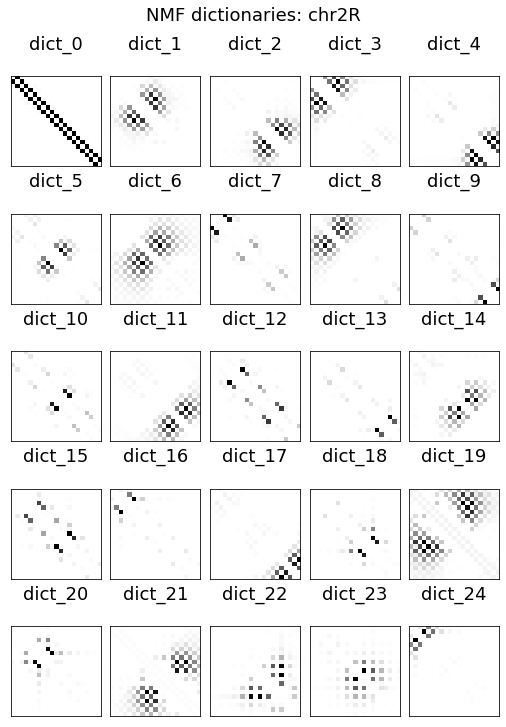}}
    \break
    \subfloat[chr3L]{\includegraphics[width=0.4\linewidth]{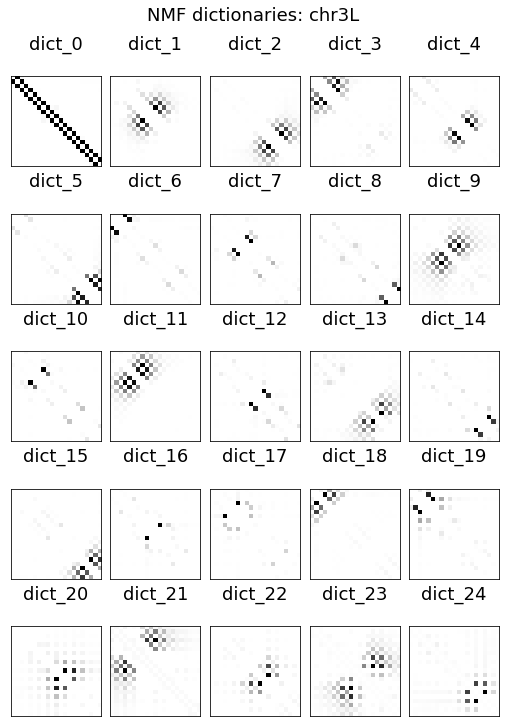}}
    \hfill
    \subfloat[chr3R]{\includegraphics[width=0.4\linewidth]{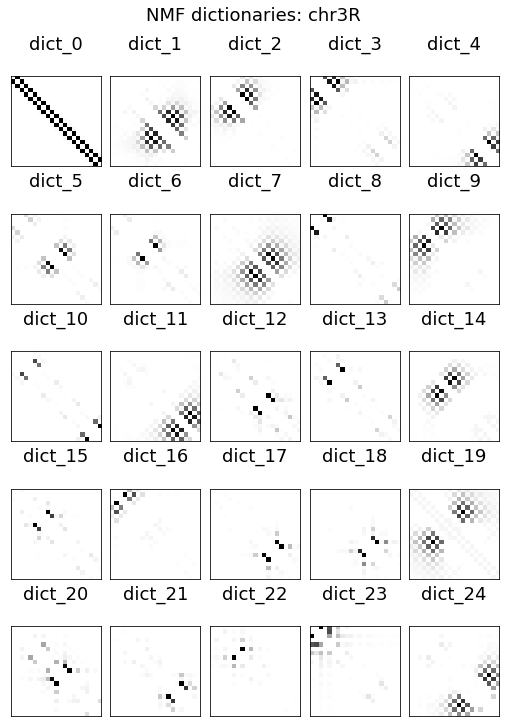}}
    \caption{Dictionaries learned by NMF for chr2L, 2R, 3L and 3R.}
    \label{supp_fig:nmf_all_dicts}
\end{figure}
\newpage
\begin{figure}[!h]
    \centering
    \subfloat[chr2L]{\includegraphics[width=0.45\linewidth]{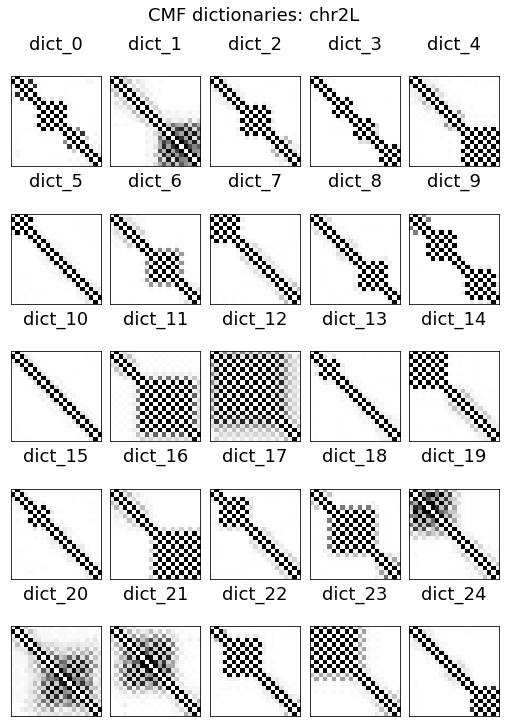}}
    \hfill
    \subfloat[chr2R]{\includegraphics[width=0.45\linewidth]{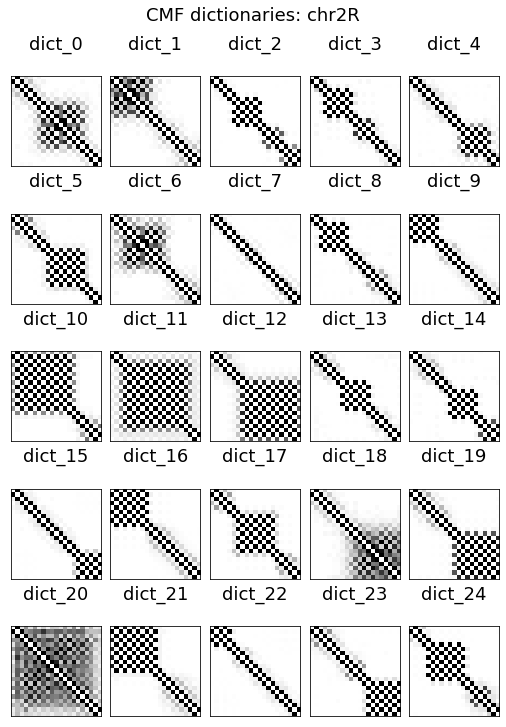}}
    \break
    \subfloat[chr3L]{\includegraphics[width=0.45\linewidth]{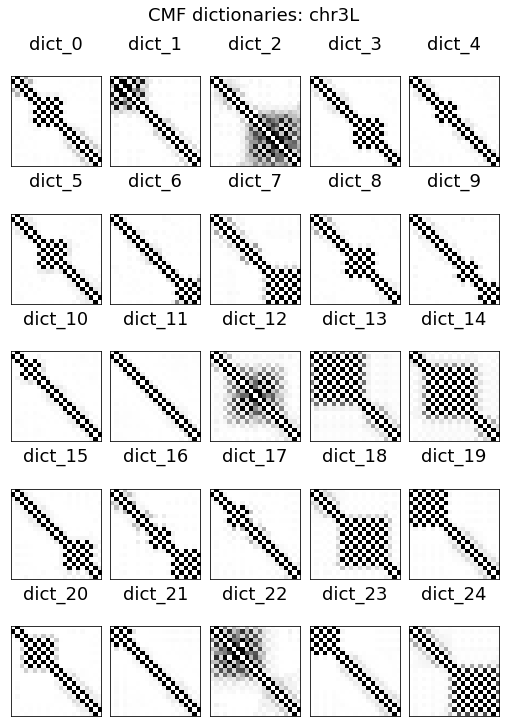}}
    \hfill
    \subfloat[chr3R]{\includegraphics[width=0.45\linewidth]{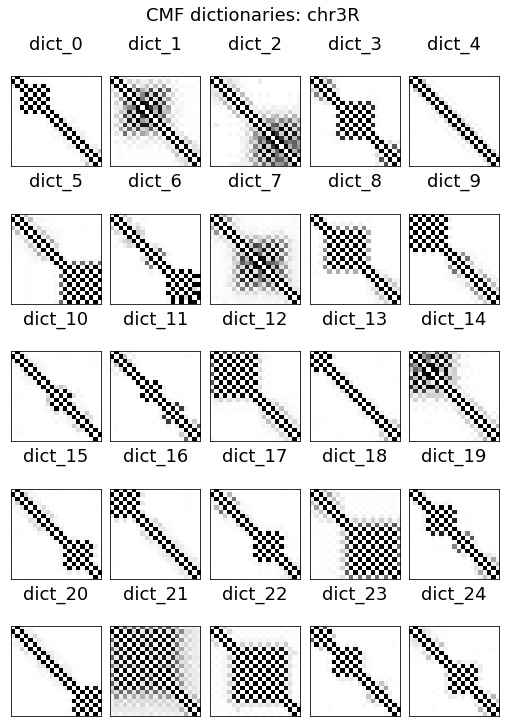}}
    \caption{Dictionaries learned by CMF for chr2L, 2R, 3L and 3R.}
    \label{supp_fig:cmf_all_dicts}
\end{figure}
\newpage
\begin{figure}[!h]
    \centering
    \subfloat[chr2L]{\includegraphics[width=0.45\linewidth]{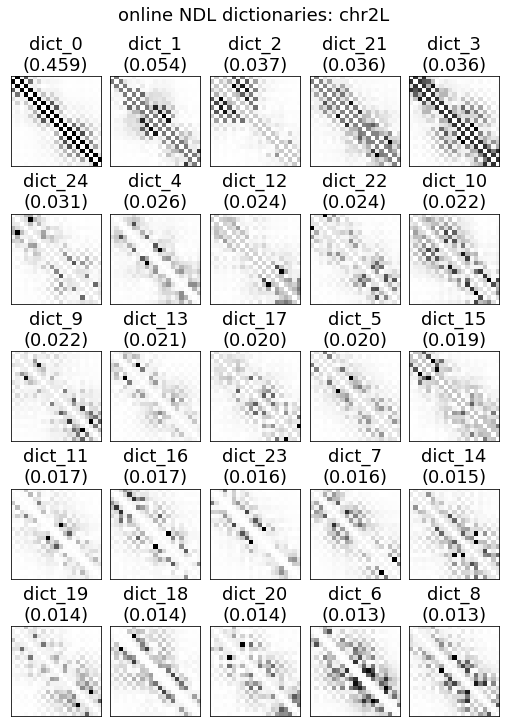}}
    \hfill
    \subfloat[chr2R]{\includegraphics[width=0.45\linewidth]{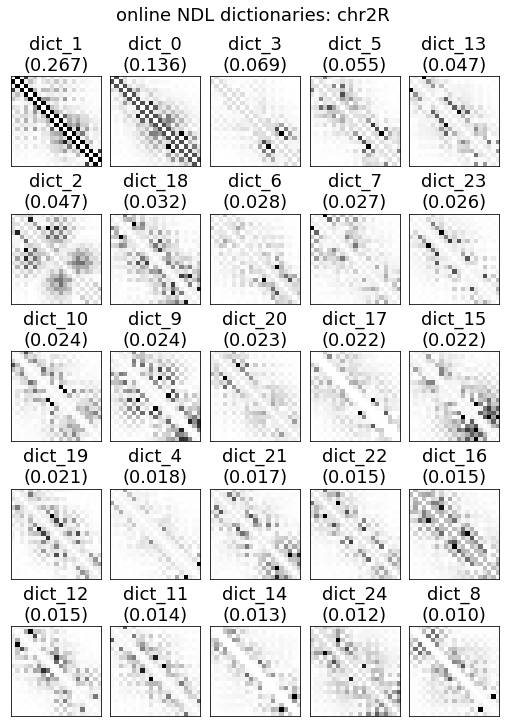}}
    \break
    \subfloat[chr3L]{\includegraphics[width=0.45\linewidth]{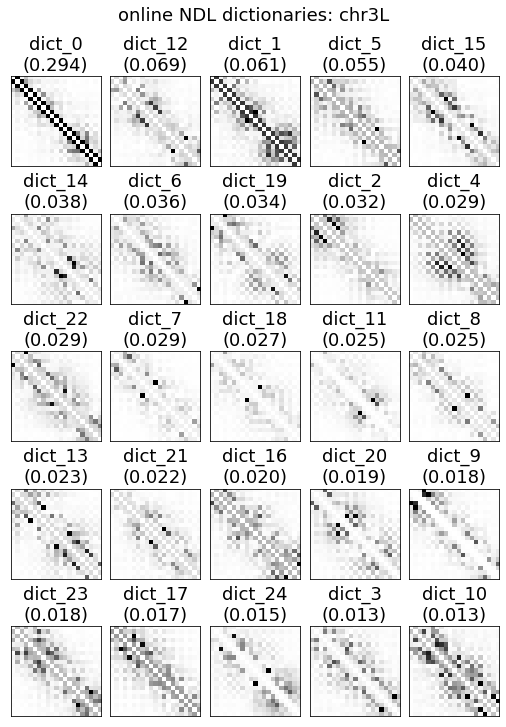}}
    \hfill
    \subfloat[chr3R]{\includegraphics[width=0.45\linewidth]{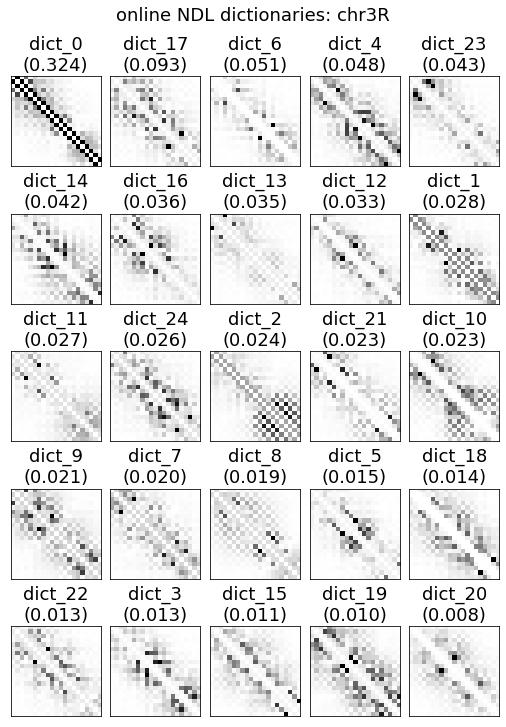}}
    \caption{Dictionaries learned by online NDL for chr2L, 2R, 3L and 3R.}
    \label{supp_fig:omf_all_dicts}
\end{figure}
\clearpage
\section{Reconstruction of ChIA-Drop Contact Maps}
\label{supp_sec:adj_recon}

\begin{figure}[!h]
    \centering
    \subfloat[Reconstruction of sample $\#15657$]{\includegraphics[width = 0.48\linewidth]{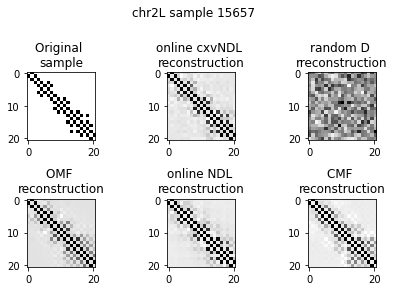}}
    \subfloat[Reconstruction of sample $\#8814$]{\includegraphics[width = 0.48\linewidth]{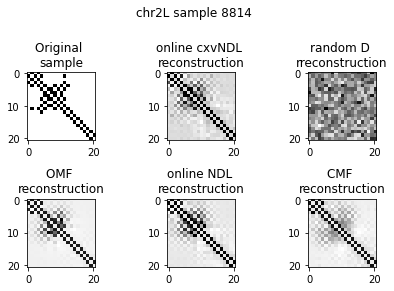}}
    \hfill
    \subfloat[Reconstruction of sample $\#2019$]{\includegraphics[width = 0.48\linewidth]{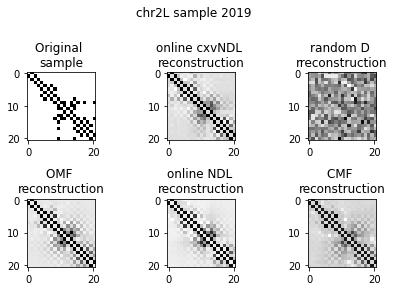}}
    \subfloat[Reconstruction of sample $\#9632$]{\includegraphics[width = 0.48\linewidth]{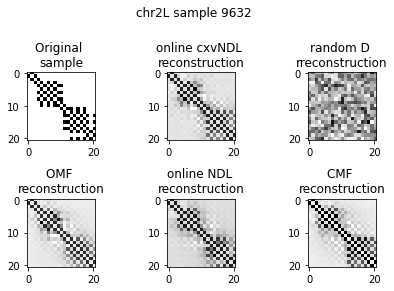}}
    \caption{Reconstructed adjacency matrices for chr2L obtained using different methods and random dictionaries. OMF stands for Ordinary (Standard) MF or NMF.}
    \label{fig:sample_recon_chr2L}
\end{figure}

The reconstructions for $4$ randomly selected subnetwork samples are shown in Figure~\ref{fig:sample_recon_chr2L}, providing a means to visually assess the accuracy of reconstructed small-scale interactions.

\newpage

\begin{figure}[!h]
    \centering
    \subfloat[Original adjacency matrix]{\includegraphics[width = 0.33\linewidth]{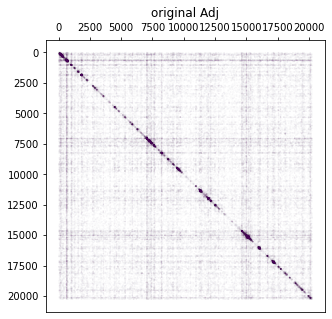}}
    \hfill
    \subfloat[Online cvxNDL]{\includegraphics[width = 0.33\linewidth]{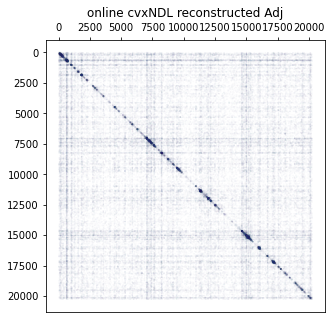}}
    \hfill
    \subfloat[Random dictionaries]{\includegraphics[width = 0.33\linewidth]{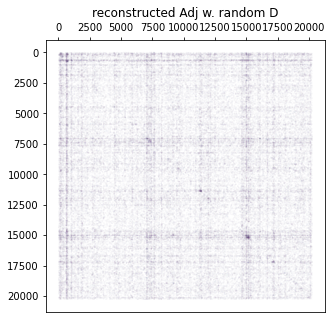}}
    \hfill
    \subfloat[NMF]{\includegraphics[width = 0.33\linewidth]{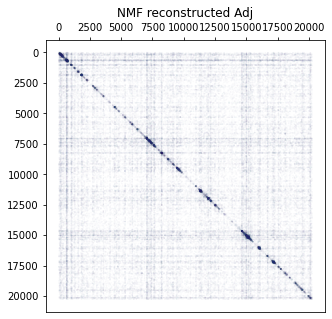}}
    \hfill
    \subfloat[CMF]{\includegraphics[width = 0.33\linewidth]{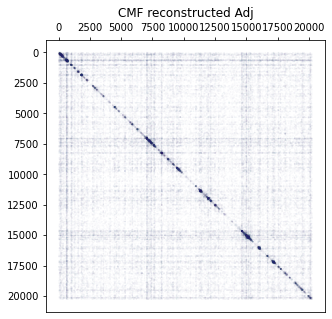}}
    \hfill
    \subfloat[Online NDL]{\includegraphics[width = 0.33\linewidth]{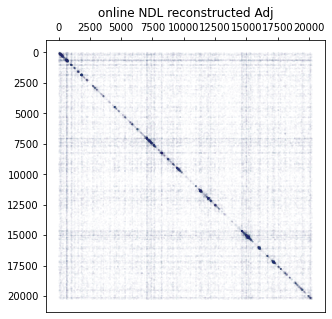}}
    \caption{Comparison of network reconstructions obtained using different baseline methods and random dictionaries for \emph{Drosophila} chromosome 2L. (a): The original adjacency matrix; (b, c, d, e, f): Reconstructed network adjacency matrices with online cxvNDL, random dictionary elements, NMF, CMF and online NDL, respectively.}
    \label{fig:adj_recon_chr2L}
\end{figure}

\newpage
\begin{figure}[!h]
    \centering
    \subfloat[Original adjacency matrix]{\includegraphics[width = 0.33\linewidth]{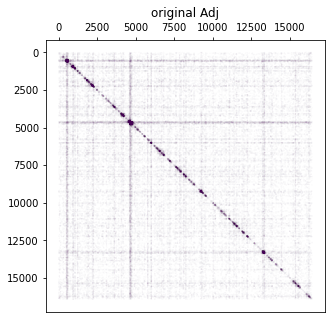}}
    \hfill
    \subfloat[Online cvxNDL]{\includegraphics[width = 0.33\linewidth]{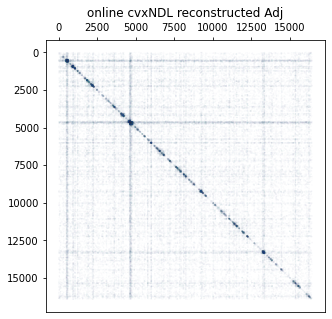}}
    \hfill
    \subfloat[Random dictionaries]{\includegraphics[width = 0.33\linewidth]{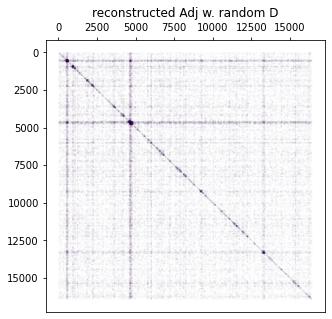}}
    \hfill
    \subfloat[NMF]{\includegraphics[width = 0.33\linewidth]{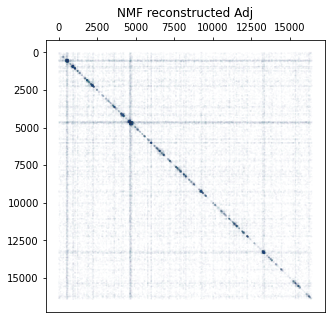}}
    \hfill
    \subfloat[CMF]{\includegraphics[width = 0.33\linewidth]{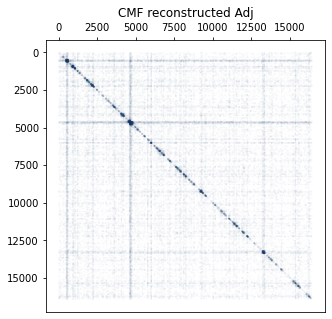}}
    \hfill
    \subfloat[Online NDL]{\includegraphics[width = 0.33\linewidth]{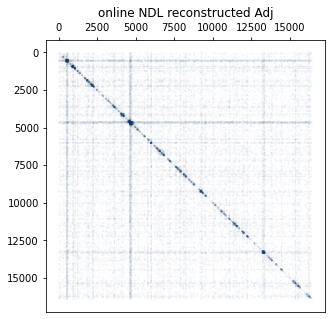}}
    \hfill
    \caption{Reconstructed network comparisons based on different baseline methods and random dictionaries, applied on \emph{Drosophila} chromosome 2R. (a): The original adjacency matrix. (b, c, d, e, f): Reconstructed network adjacency matrices with online cxvNDL, random dictionary elements, NMF, CMF and online NDL.}
    \label{fig:adj_recon_chr2R}
\end{figure}
\newpage
\begin{figure}[!h]
    \centering
    \subfloat[Original adjacency matrix]{\includegraphics[width = 0.33\linewidth]{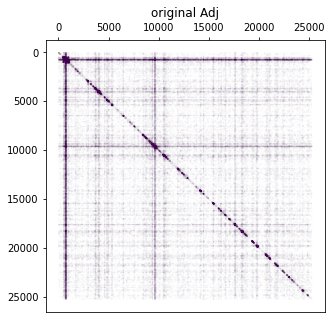}}
    \hfill
    \subfloat[Online cvxNDL]{\includegraphics[width = 0.33\linewidth]{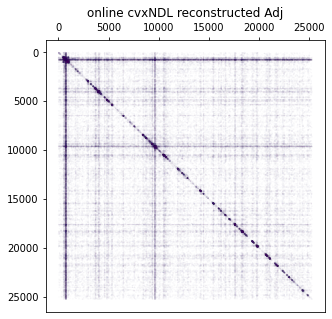}}
    \hfill
    \subfloat[Random dictionaries]{\includegraphics[width = 0.33\linewidth]{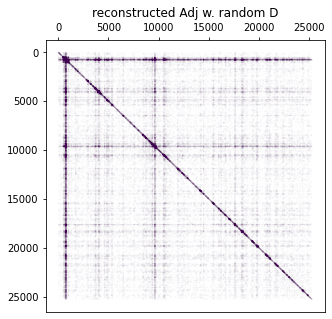}}
    \hfill
    \subfloat[NMF]{\includegraphics[width = 0.33\linewidth]{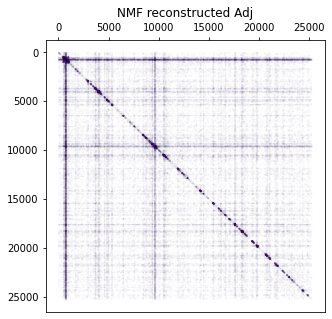}}
    \hfill
    \subfloat[CMF]{\includegraphics[width = 0.33\linewidth]{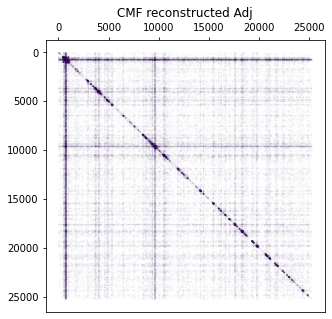}}
    \hfill
    \subfloat[Online NDL]{\includegraphics[width = 0.33\linewidth]{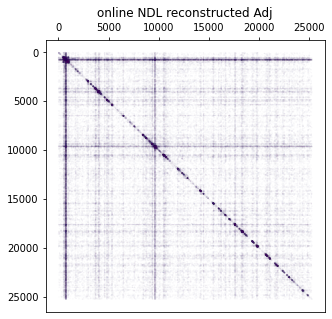}}
    \caption{Reconstructed network comparisons based on different baseline methods and random dictionaries, applied on \emph{Drosophila} chromosome 3L. (a): The original adjacency matrix. (b, c, d, e, f): Reconstructed network adjacency matrices with online cxvNDL, random dictionary elements, NMF, CMF and online NDL.}
    \label{fig:adj_recon_chr3L}
\end{figure}
\newpage
\begin{figure}[!h]
    \centering
    \subfloat[Original adjacency matrix]{\includegraphics[width = 0.33\linewidth]{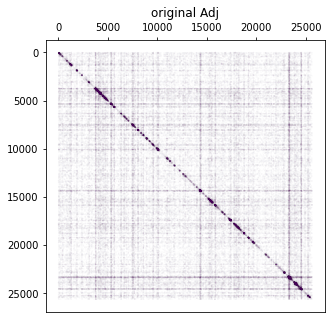}}
    \hfill
    \subfloat[Online cvxNDL]{\includegraphics[width = 0.33\linewidth]{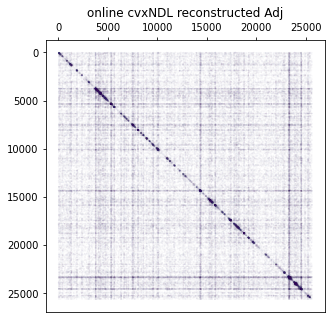}}
    \hfill
    \subfloat[Random dictionaries]{\includegraphics[width = 0.33\linewidth]{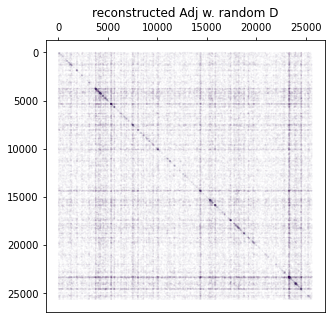}}
    \hfill
    \subfloat[NMF]{\includegraphics[width = 0.33\linewidth]{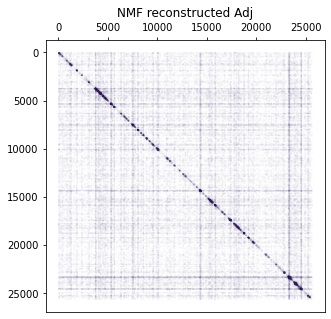}}
    \hfill
    \subfloat[CMF]{\includegraphics[width = 0.33\linewidth]{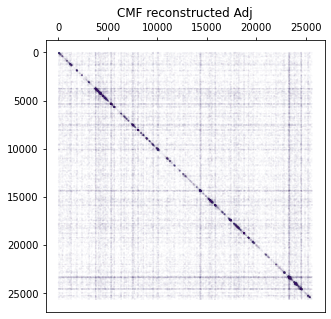}}
    \hfill
    \subfloat[Online NDL]{\includegraphics[width = 0.33\linewidth]{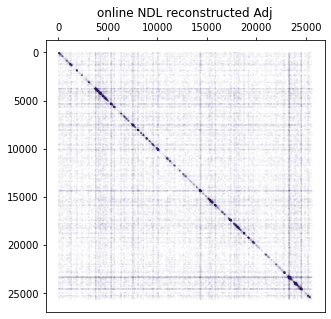}}
    \caption{Reconstructed network comparisons based on different baseline methods and random dictionaries, applied on \emph{Drosophila} chromosome 3R. (a): The original adjacency matrix. (b, c, d, e, f): Reconstructed network adjacency with online cxvNDL, random dictionary elements, NMF, CMF and online NDL.}
    \label{fig:adj_recon_chr3R}
\end{figure}

\clearpage
\section{Gene Ontology Enrichment Analysis}
\label{supp_sec:online_cvxNDL_tab}
\begin{figure}[!h]
  \centering
  \includegraphics[width=\linewidth]{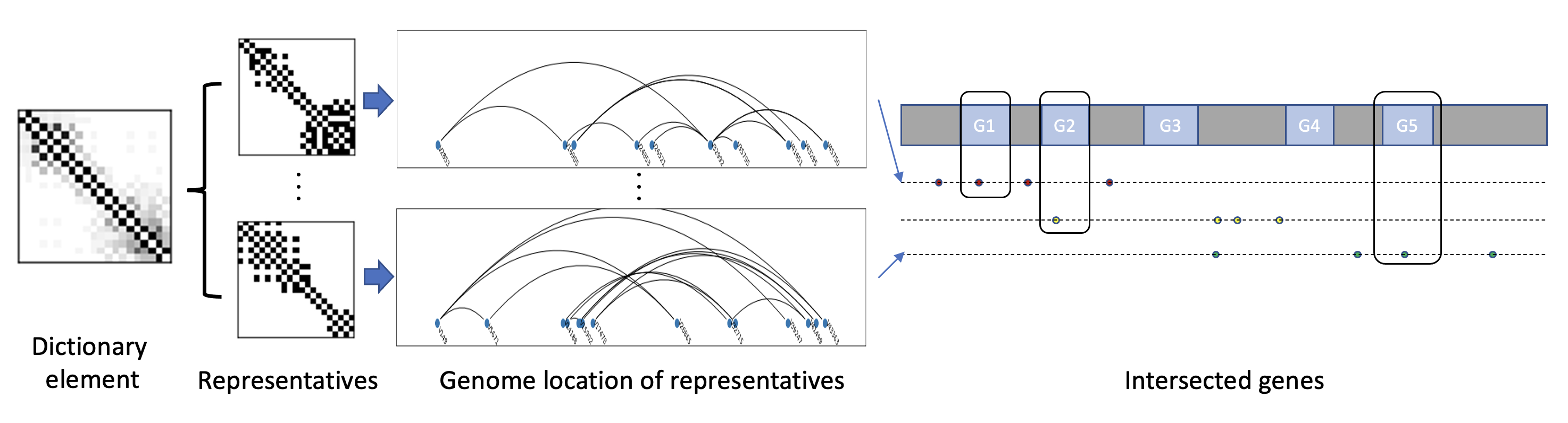}
  \caption{GO enrichment analysis workflow. Each dictionary element is associated with a collection of real subnetwork representatives. These comprise nodes that can be mapped to the genome to identify their locations. A gene is said to cover the node if the genomic fragment corresponding to the node is fully contained within the gene.}
  \label{fig:GO_enrichment}
\end{figure}

To associate a biological function with each dictionary element, we performed a gene ontology (GO) enrichment analysis for each element and the corresponding chromosome. Recall that as a results of the convexity constraint, every dictionary element has its corresponding set of representatives that capture real observed subgraphs which can be mapped back to actual genomic locations. Of most interest is the set of genes that covers at least one vertex in at least one of the representatives, as described in Figure~\ref{fig:GO_enrichment}.

Using the set of representative genes, we run the GO enrichment analysis using the annotation category ``Biological Process'' from \url{http://geneontology.org}, with the reference list \emph{Drosophila Melanogaster} for each dictionary element. For further analysis, we only selected results with false discovery rate (FDR) $< 0.05$ and hence obtained candidate sets of enriched GO terms. Note that there may be inherently enriched GO terms for each dictionary element due to the sampling bias. To remove this bias, we ran another GO enrichment analysis with all genes on each chromosome and used those results to filter out the background GO terms for each dictionary element. 

Furthermore, we utilized the hierarchical structure of GO terms~\cite{musen2015protege}, where terms are represented as nodes in a directed acyclic graph, and their relationships are described via arcs in the digraph. A child GO term is considered more specific than a parent GO term. Since the GO graph is not a strict hierarchy (a child node may have multiple parent nodes), to further improve the results, we performed the following processing. For each GO term: i) we first find all the paths between the term and the root node (which is ``Biological process'' in our setting), and ii) we remove all intermediate parent GO terms from its enriched GO terms set. By iteratively repeating this filtering process for each dictionary element, we derived a set of the most specific GO terms for each dictionary element.

\subsection{Dictionary Elements Associated with GO Terms}
\label{supp_sec:go_big_tables}
We investigated the most frequently enriched GO terms as well as the least frequently enriched GO terms for each chromosome and identified the corresponding dictionary elements where they were found to be enriched. The results are shown in Tables~\ref{tab:chr2L_GO_top_bottom} to~\ref{tab:chr3R_GO_top_bottom_sup}. For each dictionary element, we computed its density (complexity) $\rho$ via $\rho = \frac{1}{k^2}\sum_{i,j}\mathbf{D}_{i, j}$ and the median genomic distance between all consecutive pairs of nodes, denoted by $d_{\text{med}}$. The full set of results for the densities and median distances for all dictionary elements and all chromosomes is provided in Tables~\ref{tab:full_density} and~\ref{tab:full_median_distance}.

Note that the \emph{Drosophila} S2 cells are embryonic cells, and most GO terms found are related to cellular reproductive process or developmental process, as expected. From the tables, one can also see that different dictionary elements reflect different biological processes and for the same GO term, the dictionary elements share similar patterns. For example, in Table~\ref{tab:chr2L_GO_top_bottom}, we can see that dictionary elements $19$ and $12$ share very similar structural patterns, and both of them are enriched in biosynthetic processes of antibacterial peptides. On the other hand, dictionary elements $13$ and $8$ have a pattern that differs from that of $19$ and $12$, and they are enriched in dorsal/ventral lineage restriction processes. We also found that dictionary elements with GO term \textit{peripheral nervous system development}, \textit{celluar response to organic substance}, and \textit{neuroblast fate determination} have relatively lower density and smaller median node distances than the top $2$ enriched GO terms, \textit{regulation of reproductive process} and \textit{muscle cell cellular homeostasis}. The difference in density and median distance is also reflected by the significantly different dictionary patterns observed, such as dictionary element $12$ and dictionary element $5$; the former element has a much higher density and median distance than the latter.

There are also a few shared GO terms that are enriched in both chr2L and chr2R ($11$ shared terms in total) and in both chr3L and chr3R ($3$ shared terms in total). The results are reported in Table~\ref{tab:shared_GO_chr2L2R} and~\ref{tab:shared_GO_chr3L3R}. We found that there are very few shared terms between the two chromosomes when compared to the roughly one hundred uniquely enriched GO terms for each chromosome. Most of the shared terms also have ``similar'' patterns (which can be seen visually or through a simple computation of the $\ell_2$ distance between their flattened adjacency matrices) of their corresponding dictionary elements.

\begin{table}
\scriptsize
\centering
\caption{The $5$ most and least enriched GO terms within the span of dictionary elements for chr2L. Column `\#' indicates the number of dictionary elements that show enrichment for the given GO term. Also reported are up to $3$ dictionary elements with the largest importance score in the dictionary, along with the ``density'' $\rho$ of interactions in the dictionary element and median distance $d_{\text{med}}$ of all adjacent pairs of nodes in its representatives.}
\label{tab:chr2L_GO_top_bottom}
\begin{tabularx}{\textwidth}{M|c|L||M|c|S}
\toprule
most frequent GO term & \# & top 3 dictionaries & least frequent GO term & \# & dictionary \\ 
\midrule
(GO:2000241) regulation of reproductive process & 5 & \cthreeDict{\dictTwo{chr2L}[0.55in]}{\dictTwoOne{chr2L}[0.55in]}{\dictSix{chr2L}[0.55in]}  $\scriptstyle{\rho = 0.134, 0.142, 0.161}$ $\scriptstyle d_{\text{med}} = 9906, 8105, 10024$
	& (GO:0007485) imaginal disc-derived male genitalia development & 1 & \coneDict{\dictTwoOne{chr2L}} $\scriptstyle \rho = 0.142$ $\scriptstyle d_{\text{med}} = 8105 $ \\ 
	\hline
(GO:0046716) muscle cell cellular homeostasis & 4 & \cthreeDict{\dictOneFour{chr2L}[0.55in]}{\dictSix{chr2L}[0.55in]}{\dictOneTwo{chr2L}[0.55in]} $\scriptstyle \rho = 0.141, 0.161, 0.203$ $\scriptstyle d_{\text{med}} = 10928, 10024, 9979 $ 
	&  (GO:0008347) glial cell migration  & 1 & \coneDict{\dictFive{chr2L}} $\scriptstyle \rho =0.132$ $\scriptstyle d_{\text{med}} = 8547 $ \\ 
	\hline
(GO:0007422) peripheral nervous system development & 3 & \cthreeDict{\dictFive{chr2L}[0.55in]}{\dictSeven{chr2L}[0.55in]}{\dictEight{chr2L}[0.55in]} $\scriptstyle \rho = 0.132, 0.158, 0.147$ $\scriptstyle d_{\text{med}} = 8547, 8870, 10692$
 	& (GO:0002920)  regulation of humoral immune   response & 1 & \coneDict{\dictTwoOne{chr2L}} $\scriptstyle \rho = 0.142$ $\scriptstyle d_{\text{med}} = 8105 $  \\  
 	\hline
(GO:0071310) cellular response to organic substance & 3 & \cthreeDict{\dictTwo{chr2L}[0.55in]}{\dictTwoOne{chr2L}[0.55in]}{\dictSeven{chr2L}[0.55in]} $\scriptstyle \rho = 0.134, 0.142, 0.158$ $\scriptstyle d_{\text{med}} = 9906, 8105, 8870$
	& (GO:0016075)  rRNA catabolic process  & 1 & \coneDict{\dictEight{chr2L}} $\scriptstyle \rho = 0.147$ $\scriptstyle d_{\text{med}} = 10692$ \\ 
	\hline
(GO:0007400) neuroblast fate determination & 3 & \cthreeDict{\dictFive{chr2L}[0.55in]}{\dictTwoOne{chr2L}[0.55in]}{\dictEight{chr2L}[0.55in]} $\scriptstyle \rho = 0.132, 0.142, 0.147$   $\scriptstyle d_{\text{med}} = 8547, 8105, 10692 $
	& (GO:0008258)  head involution & 1 & \coneDict{\dictEight{chr2L}} $\scriptstyle \rho = 0.147$ $\scriptstyle d_{\text{med}} = 10692 $ \\ 
\bottomrule
\end{tabularx}
\end{table}
\begin{table}
\scriptsize
\centering
\caption{The $5$ most and least enriched GO terms within the span of dictionary elements for chr2R. Column `\#' indicates the number of dictionary elements that show enrichment for the given GO term. Also reported are up to $3$ dictionary elements with the largest importance score in the dictionary, along with the ``density'' $\rho$ of interactions in the dictionary element and median distance $d_{\text{med}}$ of all adjacent pairs of nodes in its representatives.}
\label{tab:chr2R_GO_top_bottom}
\begin{tabularx}{\textwidth}{M|c|L||M|c|S}
\toprule
most frequent GO term & \# & top 3 dictionaries & least frequent GO term & \# & dictionary \\ 
\midrule
(GO:0030706) germarium-derived   oocyte differentiation  & 6 & \threeDict{\dictTwoThree{chr2R}}{\dictFour{chr2R}}{\dictThree{chr2R}} $\scriptstyle \rho = 0.140, 0.145, 0.146$ $\scriptstyle d_{\text{med}} = 8764,  7651, 7158$
	& (GO:0050803) regulation of synapse   structure or activity  & 1 & \oneDict{\dictTwoThree{chr2R}} $\scriptstyle \rho = 0.140 $ $\scriptstyle d_{\text{med}} = 8764 $ \\ 
	\hline
(GO:0001700) embryonic   development via the syncytial blastoderm  & 5 & \threeDict{\dictFour{chr2R}}{\dictOneThree{chr2R}}{\dictEight{chr2R}} $\scriptstyle \rho = 0.145, 0.141, 0.136$ $\scriptstyle d_{\text{med}} = 7651, 8251, 7085$
	&(GO:0007498) mesoderm development   & 1 & \oneDict{\dictOneFive{chr2R}} $\scriptstyle \rho = 0.183$ $\scriptstyle d_{\text{med}} = 7143 $ \\ 
	\hline
(GO:0007451) dorsal/ventral   lineage restriction, imaginal disc  & 4 & \threeDict{\dictTwoThree{chr2R}}{\dictEight{chr2R}}{\dictZero{chr2R}} $\scriptstyle \rho = 0.140, 0.136, 0.157$ $\scriptstyle d_{\text{med}} = 8764, 7085, 6738$
	& (GO:0010638) positive regulation of   organelle organization & 1 & \oneDict{\dictFour{chr2R}} $\scriptstyle \rho = 0.145$ $\scriptstyle d_{\text{med}} = 7651 $ \\ 
	\hline
(GO:0006964) positive   regulation of biosynthetic process of antibacterial peptides active against   Gram-negative bacteria  & 3 & \threeDict{\dictFour{chr2R}}{\dictOneNine{chr2R}}{\dictOneTwo{chr2R}} $\scriptstyle \rho = 0.145, 0.156, 0.202 $ $\scriptstyle d_{\text{med}} = 7651, 8199, 7706$
	& (GO:0043277) apoptotic cell clearance    & 1 & \oneDict{\dictEight{chr2R}} $\scriptstyle \rho = 0.136$ $\scriptstyle d_{\text{med}} = 7085 $ \\ 
	\hline
(GO:0045476) nurse   cell apoptotic process  & 3 & \threeDict{\dictOneThree{chr2R}}{\dictOneEight{chr2R}}{\dictEight{chr2R}} $\scriptstyle \rho = 0.141, 0.159, 0.136$ $\scriptstyle d_{\text{med}} = 8251, 7882, 7085 $
	& (GO:0001707) mesoderm formation    & 1 & \oneDict{\dictOneFive{chr2R}} $\scriptstyle \rho = 0.183$ $\scriptstyle d_{\text{med}} = 7143 $ \\
\bottomrule
\end{tabularx}
\end{table}
\begin{table}
\scriptsize
\centering
\caption{The $5$ most and least enriched GO terms within the span of dictionary elements for chr3L. Column `\#' indicates the number of dictionary elements that show enrichment for the given GO term. Also reported are up to $3$ dictionary elements with the largest importance score in the dictionary, along with the ``density'' $\rho$ of interactions in the dictionary element and median distance $d_{\text{med}}$ of all adjacent pairs of nodes in its representatives.}
\label{tab:chr3L_GO_top_bottom}
\begin{tabularx}{\textwidth}{M|c|L||M|c|S}
\toprule
most frequent GO term & \# & top 3 dictionaries & least frequent GO term & \# & dictionary \\ 
\midrule
(GO:0009631) cold   acclimation  & 2 & \ctwoDict{\dictFive{chr3L}}{\dictOneSeven{chr3L}} $\scriptstyle \rho = 0.148, 0.152$ \break $\scriptstyle d_{\text{med}} = 10608, 8558 $
	& (GO:0035070) salivary gland histolysis    & 1 & \coneDict{\dictOneFive{chr3L}} $\scriptstyle \rho = 0.143 $ $\scriptstyle d_{\text{med}} = 8849 $ \\ 
	\hline
(GO:0009408) response   to heat & 2 & \ctwoDict{\dictOneThree{chr3L}}{\dictOneSeven{chr3L}} $\scriptstyle \rho = 0.147, 0.152$ \break $\scriptstyle d_{\text{med}} = 8689, 8558 $
	& (GO:0046843) dorsal appendage formation   & 1 & \coneDict{\dictOneThree{chr3L}} $\scriptstyle \rho = 0.147$ $\scriptstyle d_{\text{med}} = 8689 $ \\
	\hline
(GO:0007616) long-term   memory & 2 & \ctwoDict{\dictOneThree{chr3L}}{\dictOneSix{chr3L}} $\scriptstyle \rho = 0.147, 0.126$ \break $\scriptstyle d_{\text{med}} = 8689 , 9978 $
	& (GO:0007097) nuclear migration & 1 & \coneDict{\dictTwoTwo{chr3L}} $\scriptstyle \rho = 0.134$ $\scriptstyle d_{\text{med}} = 11012 $ \\
	\hline
(GO:0061077) chaperone-mediated   protein folding & 2 & \ctwoDict{\dictFive{chr3L}}{\dictOneSeven{chr3L}} $\scriptstyle \rho = 0.148, 0.152$ \break $\scriptstyle d_{\text{med}} = 10608 , 8558 $
	& (GO:0035071) salivary gland cell autophagic   cell death & 1 & \coneDict{\dictOneFive{chr3L}} $\scriptstyle \rho = 0.143$ $\scriptstyle d_{\text{med}} = 8849 $ \\
	\hline
(GO:0008587) imaginal   disc-derived wing margin morphogenesis & 2 & \ctwoDict{\dictOneSix{chr3L}}{\dictOneSeven{chr3L}} $\scriptstyle \rho = 0.126, 0.152$ \break $\scriptstyle d_{\text{med}} = 9978,  8558$
	& (GO:0007528) neuromuscular junction   development & 1 & \coneDict{\dictOneThree{chr3L}} $\scriptstyle \rho = 0.147$ $\scriptstyle d_{\text{med}} = 8689 $ \\
\bottomrule
\end{tabularx}
\end{table}
\begin{table}
\scriptsize
\centering
\caption{The $5$ most and least enriched GO terms within the span of dictionary elements for chr3R. Column `\#' indicates the number of dictionary elements that show enrichment for the given GO term. Also reported are up to $3$ dictionary elements with the largest importance score in the dictionary, along with the ``density'' $\rho$ of interactions in the dictionary element and median distance $d_{\text{med}}$ of all adjacent pairs of nodes in its representatives.}
\label{tab:chr3R_GO_top_bottom_sup}
\begin{tabularx}{\textwidth}{M|c|L||M|c|S}    
\toprule
most frequent GO term & \# & top 3 dictionaries & least frequent GO term & \# & dictionary \\ 
\midrule
(GO:0001819)   positive regulation of cytokine production & 7 & \threeDict{\dictTwoZero{chr3R}}{\dictSeven{chr3R}}{\dictNine{chr3R}} $\scriptstyle \rho = 0.126, 0.146, 0.157$ $\scriptstyle d_{\text{med}} = 12791, 12830, 11930$
	& (GO:0061448) connective tissue   development & 1 & \oneDict{\dictOneTwo{chr3R}} $\scriptstyle \rho = 0.142$ $\scriptstyle d_{\text{med}} = 13455 $ \\ 
	\hline
(GO:0008015) blood   circulation & 7 & \threeDict{\dictTwoZero{chr3R}}{\dictOneTwo{chr3R}}{\dictFour{chr3R}} $\scriptstyle \rho = 0.126, 0.142, 0.138$ $\scriptstyle d_{\text{med}} = 12791, 13455, 13674$
	& (GO:0051282) regulation of sequestering of calcium ion & 1 & \oneDict{\dictTwoZero{chr3R}} $\scriptstyle \rho = 0.126$ $\scriptstyle d_{\text{med}} = 12791 $ \\ 
	\hline
(GO:0045948) positive regulation of   translational initiation & 5 & \threeDict{\dictTwoZero{chr3R}}{\dictFour{chr3R}}{\dictOneFour{chr3R}} $\scriptstyle \rho = 0.126, 0.138, 0.162$ $\scriptstyle d_{\text{med}} = 12791 , 13674 , 12572 $
	& (GO:0043123) positive regulation of I-kappaB kinase/NF-kappaB   signaling & 1 & \oneDict{\dictOneThree{chr3R}} $\scriptstyle \rho = 0.204$ $\scriptstyle d_{\text{med}} = 12540 $ \\ 
	\hline
(GO:0042177) negative regulation of protein   catabolic process & 5 & \threeDict{\dictTwoZero{chr3R}}{\dictOneTwo{chr3R}}{\dictFour{chr3R}} $\scriptstyle \rho = 0.126, 0.142, 0.138$ $\scriptstyle d_{\text{med}} = 12791, 13455, 13674$
	& (GO:0007435) salivary gland morphogenesis & 1 & \oneDict{\dictOneThree{chr3R}} $\scriptstyle \rho = 0.204$ $\scriptstyle d_{\text{med}} = 12540 $ \\ 
	\hline
(GO:0043065) positive   regulation of apoptotic process & 4 & \threeDict{\dictTwoZero{chr3R}}{\dictSeven{chr3R}}{\dictThree{chr3R}} $\scriptstyle \rho = 0.126, 0.146, 0.179$ $\scriptstyle d_{\text{med}} = 12791, 12830, 11748 $
	& (GO:0045738) negative regulation of DNA repair & 1 & \oneDict{\dictEight{chr3R}} $\scriptstyle \rho = 0.183$ $\scriptstyle d_{\text{med}} = 12493 $ \\
\bottomrule
\end{tabularx}
\end{table}
\begin{table}
\scriptsize
\centering
\caption{GO terms shared between chr2L and chr2R.}
\label{tab:shared_GO_chr2L2R}
\begin{tabularx}{\textwidth}{M|M||M}    
\toprule
GO\_term & chr2L dictionaries & chr2R dictionaries \\ \midrule
(GO:0016325) oocyte   microtubule cytoskeleton organization & \threeDict{\dictFive{chr2L}}{\dictSeven{chr2L}}{\dictSix{chr2L}} & \oneDict{\dictOneFour{chr2R}} \\ \hline 
(GO:1901701) cellular response   to oxygen-containing compound & \twoDict{\dictTwo{chr2L}}{\dictSeven{chr2L}} & \oneDict{\dictEight{chr2R}} \\ \hline
(GO:0007298) border follicle   cell migration & \twoDict{\dictTwo{chr2L}}{\dictTwoOne{chr2L}} & \threeDict{\dictFour{chr2R}}{\dictThree{chr2R}}{\dictOneEight{chr2R}} \\\hline
(GO:0043410) positive   regulation of MAPK cascade & \twoDict{\dictTwo{chr2L}}{\dictEight{chr2L}} & \twoDict{\dictFour{chr2R}}{\dictEight{chr2R}} \\\hline
(GO:0016049) cell growth & \oneDict{\dictTwoOne{chr2L}} & \oneDict{\dictEight{chr2R}} \\\hline
(GO:0035331) negative   regulation of hippo signaling & \oneDict{\dictEight{chr2L}} & \oneDict{\dictFour{chr2R}} \\\hline
(GO:0051962) positive   regulation of nervous system development & \oneDict{\dictSeven{chr2L}} & \oneDict{\dictOneFive{chr2R}} \\\hline
(GO:0060322) head   development & \oneDict{\dictEight{chr2L}} & \oneDict{\dictFour{chr2R}} \\\hline
(GO:0007293) germarium-derived   egg chamber formation & \oneDict{\dictEight{chr2L}} & \fourDict{\dictTwoThree{chr2R}}{\dictFour{chr2R}}{\dictOneThree{chr2R}}{\dictOneFive{chr2R}} \\\hline
(GO:0002164) larval   development & \oneDict{\dictSix{chr2L}} & \oneDict{\dictOneFive{chr2R}} \\\hline
(GO:0007420) brain   development & \oneDict{\dictSix{chr2L}} & \twoDict{\dictFour{chr2R}}{\dictOneEight{chr2R}} \\
\bottomrule
\end{tabularx}
\end{table}

\begin{table}
\scriptsize
\centering
\caption{GO terms shared between chr3L and chr3R.}
\label{tab:shared_GO_chr3L3R}
\begin{tabularx}{\textwidth}{M|L||M}    
\toprule
GO\_term & chr3L dictionaries & chr3R dictionaries \\ \midrule
(GO:0070373)   negative regulation of ERK1 and ERK2 cascade & \fourDict{\dictOneThree{chr3L}}{\dictTwoTwo{chr3L}}{\dictThree{chr3L}}{\dictOne{chr3L}} & \oneDict{\dictEight{chr3R}} \\ \hline
(GO:0007140) male meiotic   nuclear division & \oneDict{\dictTwoThree{chr3L}} & \oneDict{\dictTwoFour{chr3R}} \\ \hline
(GO:0046777) protein   autophosphorylation & \oneDict{\dictTwoTwo{chr3L}} & \oneDict{\dictEight{chr3R}} \\
\bottomrule
\end{tabularx}
\end{table}

\clearpage
\subsection{Additional Results}
\label{supp_sec:density_detail}
Here we report more detailed results for each dictionary element, including its number of enriched GO terms and importance scores (Tables~\ref{tab:go2l},~\ref{tab:go2r},~\ref{tab:go3l},~\ref{tab:go3r}).
\begin{table}[!h]
\centering
\scriptsize
\caption{Number of enriched GO terms for each dictionary element identified for chr2L.}
\begin{tabular}{l|l|l|l|l} 
\toprule
\multicolumn{1}{c|}{\# GO terms} & \multicolumn{1}{c|}{\# GO terms} & \multicolumn{1}{c|}{\# GO terms} & \multicolumn{1}{c|}{\# GO terms} & \multicolumn{1}{c}{\# GO terms} \\ 
\midrule
\boxDictZero{chr2L} 2 & \boxDictFive{chr2L} 15 & \boxDictOneZero{chr2L} 0 & \boxDictOneFive{chr2L} 0 & \boxDictTwoZero{chr2L} 0 \\
\boxDictOne{chr2L} 0 & \boxDictSix{chr2L} 19 & \boxDictOneOne{chr2L} 2 & \boxDictOneSix{chr2L} 2 & \boxDictTwoOne{chr2L} 27 \\
\boxDictTwo{chr2L} 20 & \boxDictSeven{chr2L} 24 & \boxDictOneTwo{chr2L} 1 & \boxDictOneSeven{chr2L} 0 & \boxDictTwoTwo{chr2L} 1 \\
\boxDictThree{chr2L} 0 & \boxDictEight{chr2L} 31 & \boxDictOneThree{chr2L} 0 & \boxDictOneEight{chr2L} 0 & \boxDictTwoThree{chr2L} 0 \\
\boxDictFour{chr2L} 0 & \boxDictNine{chr2L} 0 & \boxDictOneFour{chr2L} 6 & \boxDictOneNine{chr2L} 0 & \boxDictTwoFour{chr2L} 0 \\
\bottomrule
\end{tabular} \label{tab:go2l}
\end{table}

\begin{table}[!h]
\centering
\scriptsize
\caption{Number of enriched GO terms for each dictionary element identified for chr2R.}
\begin{tabular}{l|l|l|l|l} 
\toprule
\multicolumn{1}{c|}{\# GO terms} & \multicolumn{1}{c|}{\# GO terms} & \multicolumn{1}{c|}{\# GO terms} & \multicolumn{1}{c|}{\# GO terms} & \multicolumn{1}{c}{\# GO terms} \\ 
\midrule
\boxDictZero{chr2R} 4 & \boxDictFive{chr2R} 0 & \boxDictOneZero{chr2R} 0 & \boxDictOneFive{chr2R} 23 & \boxDictTwoZero{chr2R} 6 \\
\boxDictOne{chr2R} 1 & \boxDictSix{chr2R} 0 & \boxDictOneOne{chr2R} 1 & \boxDictOneSix{chr2R} 0 & \boxDictTwoOne{chr2R} 0 \\
\boxDictTwo{chr2R} 0 & \boxDictSeven{chr2R} 1 & \boxDictOneTwo{chr2R} 2 & \boxDictOneSeven{chr2R} 0 & \boxDictTwoTwo{chr2R} 8 \\
\boxDictThree{chr2R} 12 & \boxDictEight{chr2R} 17 & \boxDictOneThree{chr2R} 9 & \boxDictOneEight{chr2R} 8 & \boxDictTwoThree{chr2R} 10 \\
\boxDictFour{chr2R} 40 & \boxDictNine{chr2R} 0 & \boxDictOneFour{chr2R} 5 & \boxDictOneNine{chr2R} 7 & \boxDictTwoFour{chr2R} 2 \\
\bottomrule
\end{tabular} \label{tab:go2r}
\end{table}

\begin{table}[!h]
\centering
\scriptsize
\caption{Number of enriched GO terms for each dictionary element identified for chr3L.}
\begin{tabular}{l|l|l|l|l} 
\toprule
\multicolumn{1}{c|}{\# GO terms} & \multicolumn{1}{c|}{\# GO terms} & \multicolumn{1}{c|}{\# GO terms} & \multicolumn{1}{c|}{\# GO terms} & \multicolumn{1}{c}{\# GO terms} \\ 
\midrule
\boxDictZero{chr3L} 0 & \boxDictFive{chr3L} 6 & \boxDictOneZero{chr3L} 2 & \boxDictOneFive{chr3L} 10 & \boxDictTwoZero{chr3L} 0 \\
\boxDictOne{chr3L} 3 & \boxDictSix{chr3L} 1 & \boxDictOneOne{chr3L} 0 & \boxDictOneSix{chr3L} 14 & \boxDictTwoOne{chr3L} 0 \\
\boxDictTwo{chr3L} 0 & \boxDictSeven{chr3L} 1 & \boxDictOneTwo{chr3L} 1 & \boxDictOneSeven{chr3L} 9 & \boxDictTwoTwo{chr3L} 4 \\
\boxDictThree{chr3L} 3 & \boxDictEight{chr3L} 0 & \boxDictOneThree{chr3L} 16 & \boxDictOneEight{chr3L} 4 & \boxDictTwoThree{chr3L} 3 \\
\boxDictFour{chr3L} 3 & \boxDictNine{chr3L} 0 & \boxDictOneFour{chr3L} 0 & \boxDictOneNine{chr3L} 0 & \boxDictTwoFour{chr3L} 0 \\
\bottomrule
\end{tabular}  \label{tab:go3l}
\end{table}

\begin{table}[!h]
\centering
\scriptsize
\caption{Number of enriched GO terms for each dictionary element identified for chr3R.}
\begin{tabular}{l|l|l|l|l} 
\toprule
\multicolumn{1}{c|}{\# GO terms} & \multicolumn{1}{c|}{\# GO terms} & \multicolumn{1}{c|}{\# GO terms} & \multicolumn{1}{c|}{\# GO terms} & \multicolumn{1}{c}{\# GO terms} \\ 
\midrule
\boxDictZero{chr3R} 15 & \boxDictFive{chr3R} 2 & \boxDictOneZero{chr3R} 5 & \boxDictOneFive{chr3R} 8 & \boxDictTwoZero{chr3R} 124 \\
\boxDictOne{chr3R} 9 & \boxDictSix{chr3R} 2 & \boxDictOneOne{chr3R} 0 & \boxDictOneSix{chr3R} 0 & \boxDictTwoOne{chr3R} 10 \\
\boxDictTwo{chr3R} 13 & \boxDictSeven{chr3R} 14 & \boxDictOneTwo{chr3R} 16 & \boxDictOneSeven{chr3R} 0 & \boxDictTwoTwo{chr3R} 4 \\
\boxDictThree{chr3R} 7 & \boxDictEight{chr3R} 25 & \boxDictOneThree{chr3R} 57 & \boxDictOneEight{chr3R} 0 & \boxDictTwoThree{chr3R} 0 \\
\boxDictFour{chr3R} 20 & \boxDictNine{chr3R} 1 & \boxDictOneFour{chr3R} 6 & \boxDictOneNine{chr3R} 0 & \boxDictTwoFour{chr3R} 4 \\
\bottomrule
\end{tabular} \label{tab:go3r}
\end{table}
\newpage

\clearpage
\section{RNA-Seq Coexpression Analysis}
\label{supp:rna_seq_validation}
The ChIA-Drop dataset~\cite{zheng2019multiplex} used for learning dictionaries of chromatin interactions lacks RNA-Seq replicates, posing a challenge when trying to validate our results through coexpression analysis. To address this limitation, we retrieved RNA-Seq data corresponding to untreated S2 cell lines of \emph{Drosophila Melanogaster} from the Digital Expression Explorer (DEE2) repository. DEE2 provides uniformly processed RNA-Seq data sourced from the publicly available NCBI Sequence Read Archive (SRA)~\cite{ziemann2019digital}. In total, we retrieved $20$ samples from untreated S2 cell lines with their IDs reported in Table~\ref{tab:srr_id}.

\begin{table}[!h]
\centering
\scriptsize
\caption{Sample IDs retrieved from NCBI Sequence Read Archive for RNA-Seq coexpression analysis.}
\label{tab:srr_id}
\begin{tabular}{c c c c c}
\hline
SRR12191916& SRR12191917& SRR12191918& SRR12191920& SRR12191921 \\ 
[0.05cm]
SRR12191923& SRR12191927& SRR2442878& SRR2442879& SRR3065067 \\
[0.05cm]
SRR5340065& SRR5340066& SRR5340069& SRR5340070& SRR5340071 \\
[0.05cm]
SRR5340072& SRR6930637& SRR8108628& SRR8108629& SRR8108630\\
[0.05cm]
\hline
\end{tabular} 
\end{table}

To ensure consistent normalization across all samples, we use the trimmed mean of M values (TMM) method~\cite{robinson2010scaling}, available through the edgeR package~\cite{robinson2010edger}. This is of crucial importance when jointly analyzing samples from multiple sources. We selected the most relevant genes by filtering the list of covered genes and retaining only those with more than $95\%$ overlap with the gene promoter regions, as defined in the \emph{Ensmbl} browser. Subsequently, for each dictionary element, we collected all genes covered by it and calculated the pairwise Pearson correlation coefficient of expressions of pairs of genes in the set. For a pair of random variables $X_1$ and $X_2$, the correlation coefficient is defined as
\begin{align*}
	\rho_{X_1X_2} = \frac{\text{Covariance}(X_1,X_2)}{\text{Var}(X_1)\text{Var}(X_2)}\\
\end{align*}
For two genes $G_1$ and $G_2$, let $X_1$ and $X_2$ be vectors of normalized read counts. The Pearson correlation coefficient can be written as 
\begin{align*}
	\rho_{G_1G_2} = \frac{\sum_{i=1}^n(x_{1i} - \bar{x}_1)(x_{2i} - \bar{x}_2)}{\sqrt{\sum_{i=1}^n(x_{1i}-\bar{x}_1)^2}\sqrt{\sum_{i=1}^n(x_{2i}-\bar{x}_2)^2}}
\end{align*}
where
\begin{align*}
& n \text{ is the number of samples,}\\
& \bar{x}_1 = \frac{\sum_{i=1}^n x_{1i}}{n} \text{ and } \bar{x}_2 = \frac{\sum_{i=1}^n x_{2i}}{n} \text{ are sample means.}\\
\end{align*}
To visualize the underlying coexpression clusters within the genes, we performed hierarchical clustering. We report the mean correlation statistics as well as mean statistics for positively correlated genes for each dictionary element. Correlation plots for all dictionary elements are shown in Figures~\ref{fig:dee2_pearson2L},~\ref{fig:dee2_pearson2R},~\ref{fig:dee2_pearson3L} and~\ref{fig:dee2_pearson3R}.

\begin{figure}[h]
     \centering
     \begin{subfigure}[b]{0.32\textwidth}
         \centering
         \includegraphics[width=\textwidth]{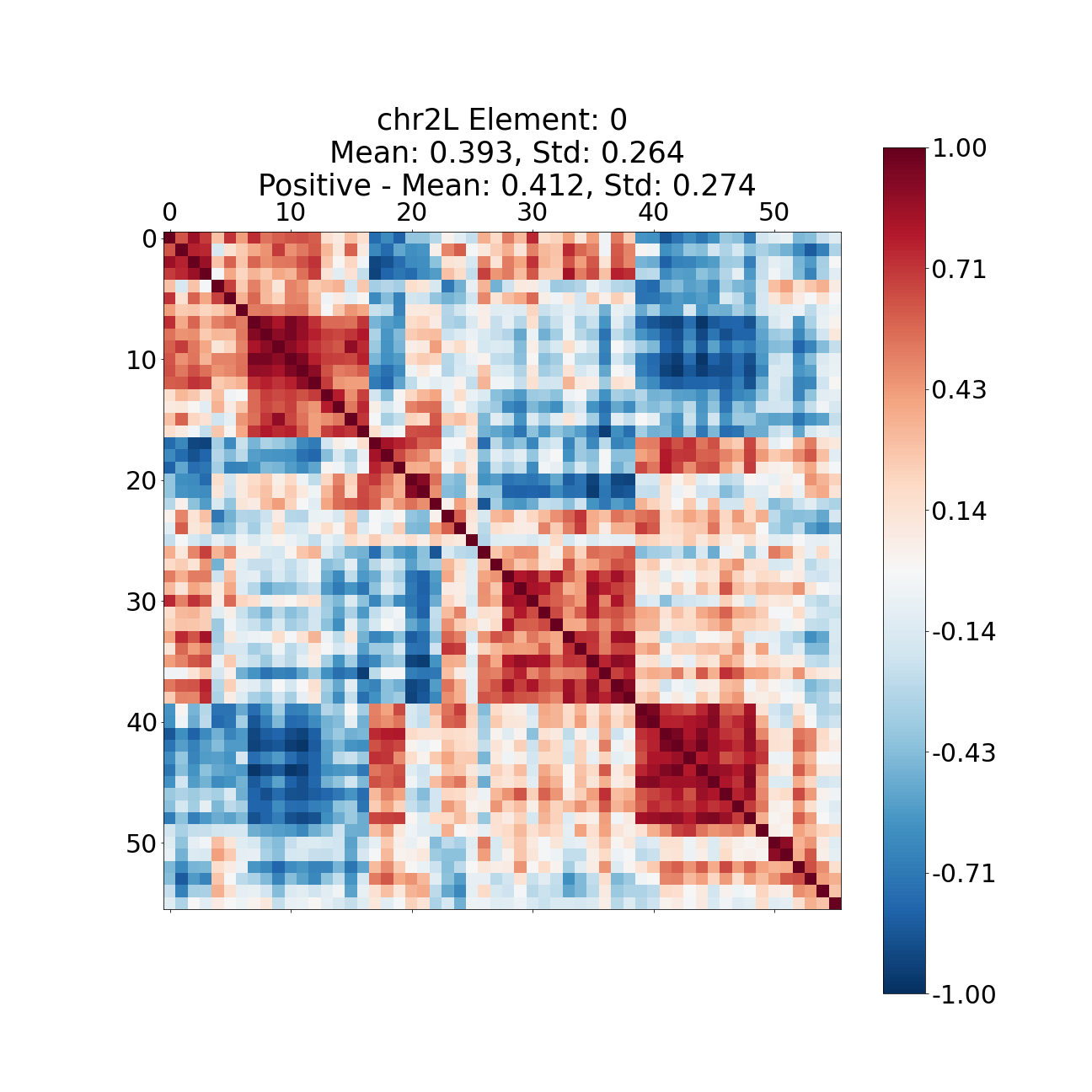}
     \end{subfigure}
     \hfill
     \begin{subfigure}[b]{0.32\textwidth}
         \centering
         \includegraphics[width=\textwidth]{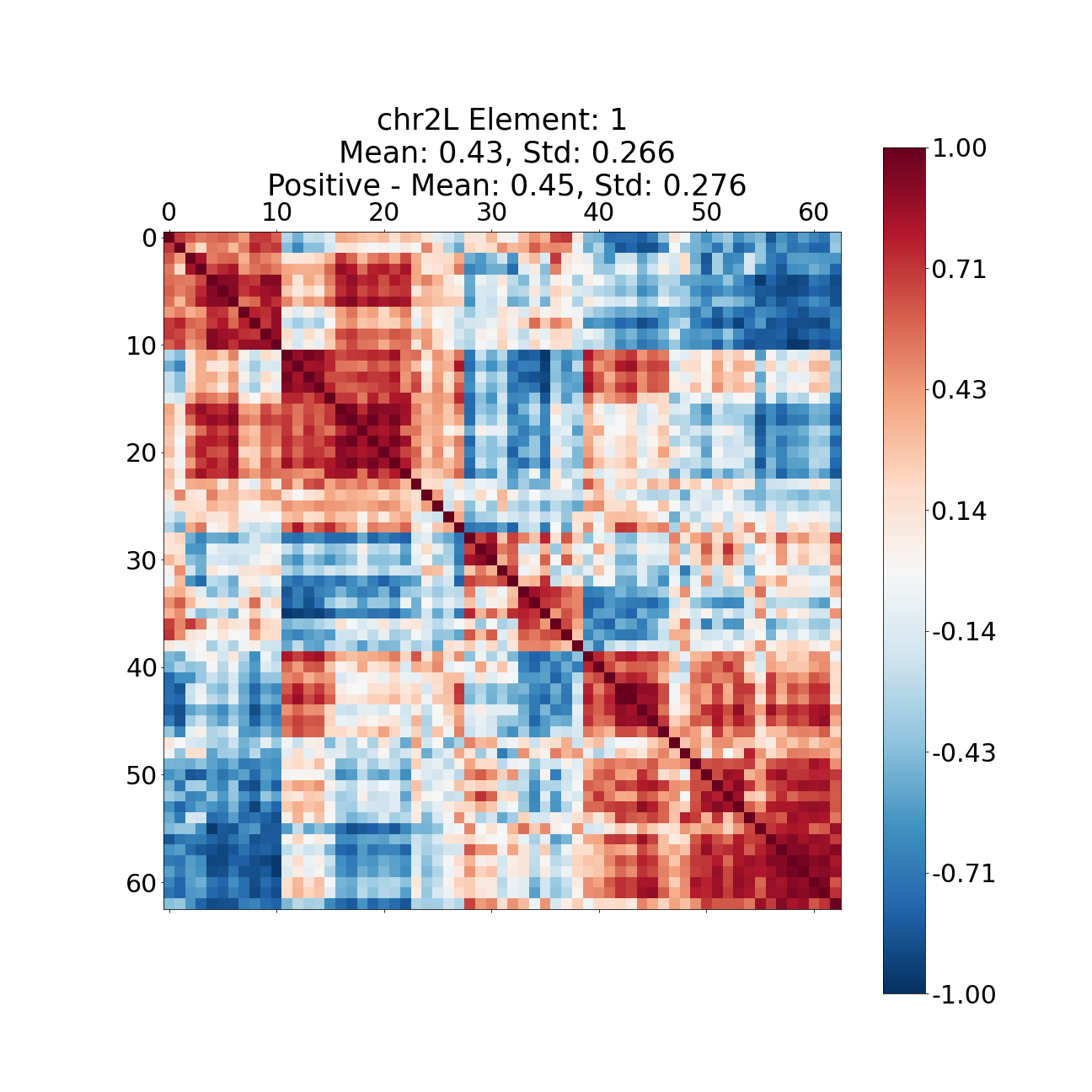}
     \end{subfigure}
     \hfill
     \begin{subfigure}[b]{0.32\textwidth}
         \centering
         \includegraphics[width=\textwidth]{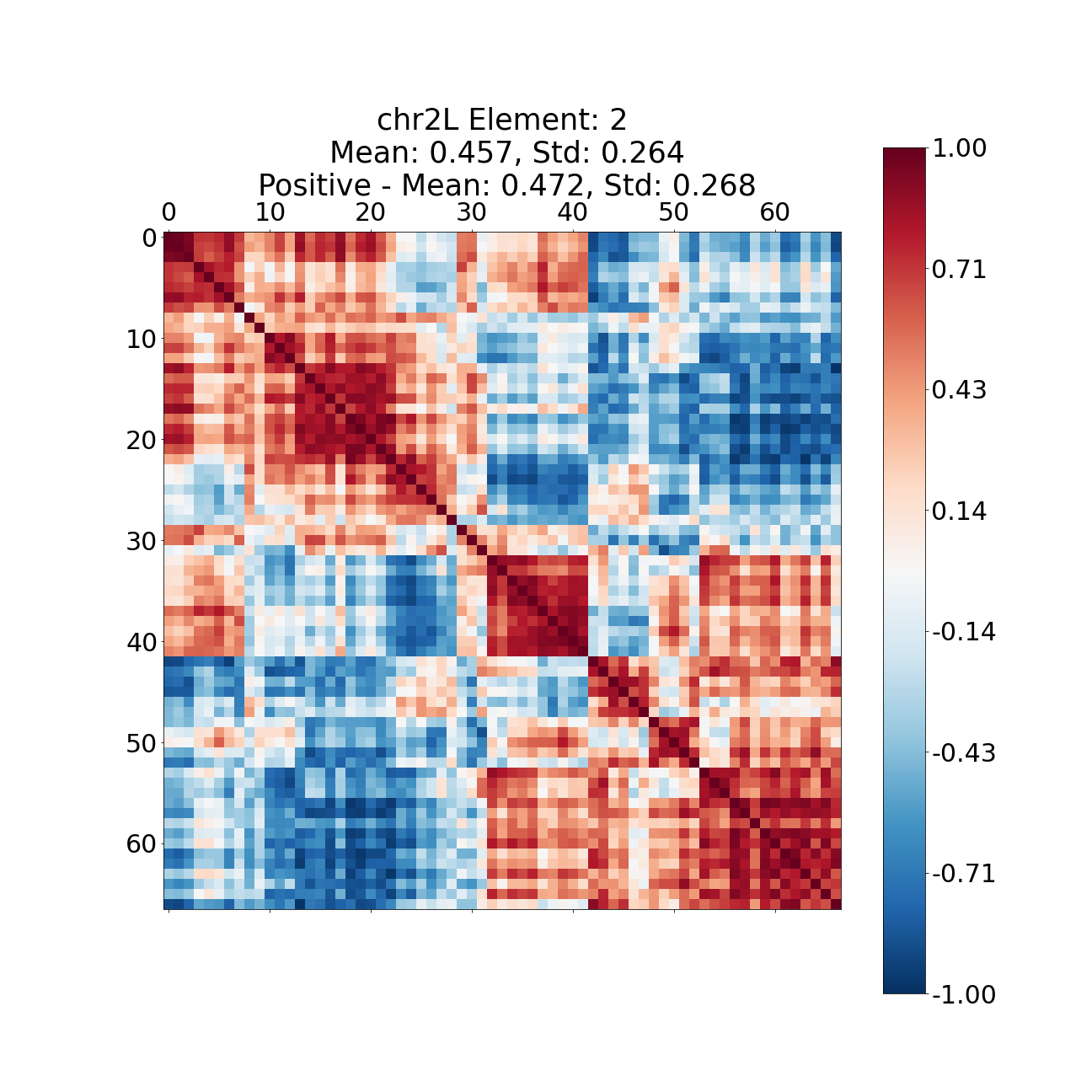}
     \end{subfigure}
      \hfill
     \begin{subfigure}[b]{0.32\textwidth}
         \centering
         \includegraphics[width=\textwidth]{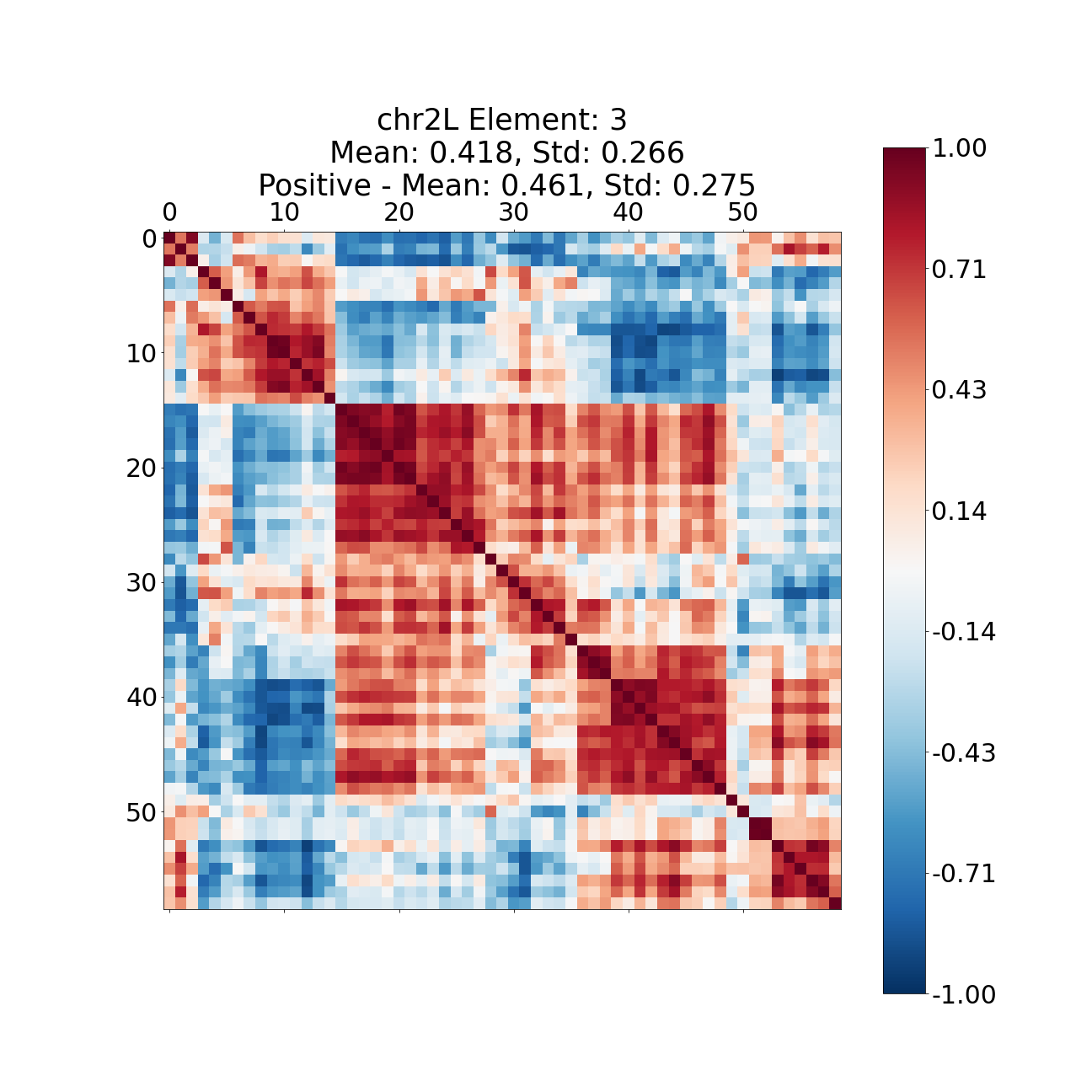}
     \end{subfigure}
     \hfill
     \begin{subfigure}[b]{0.32\textwidth}
         \centering
         \includegraphics[width=\textwidth]{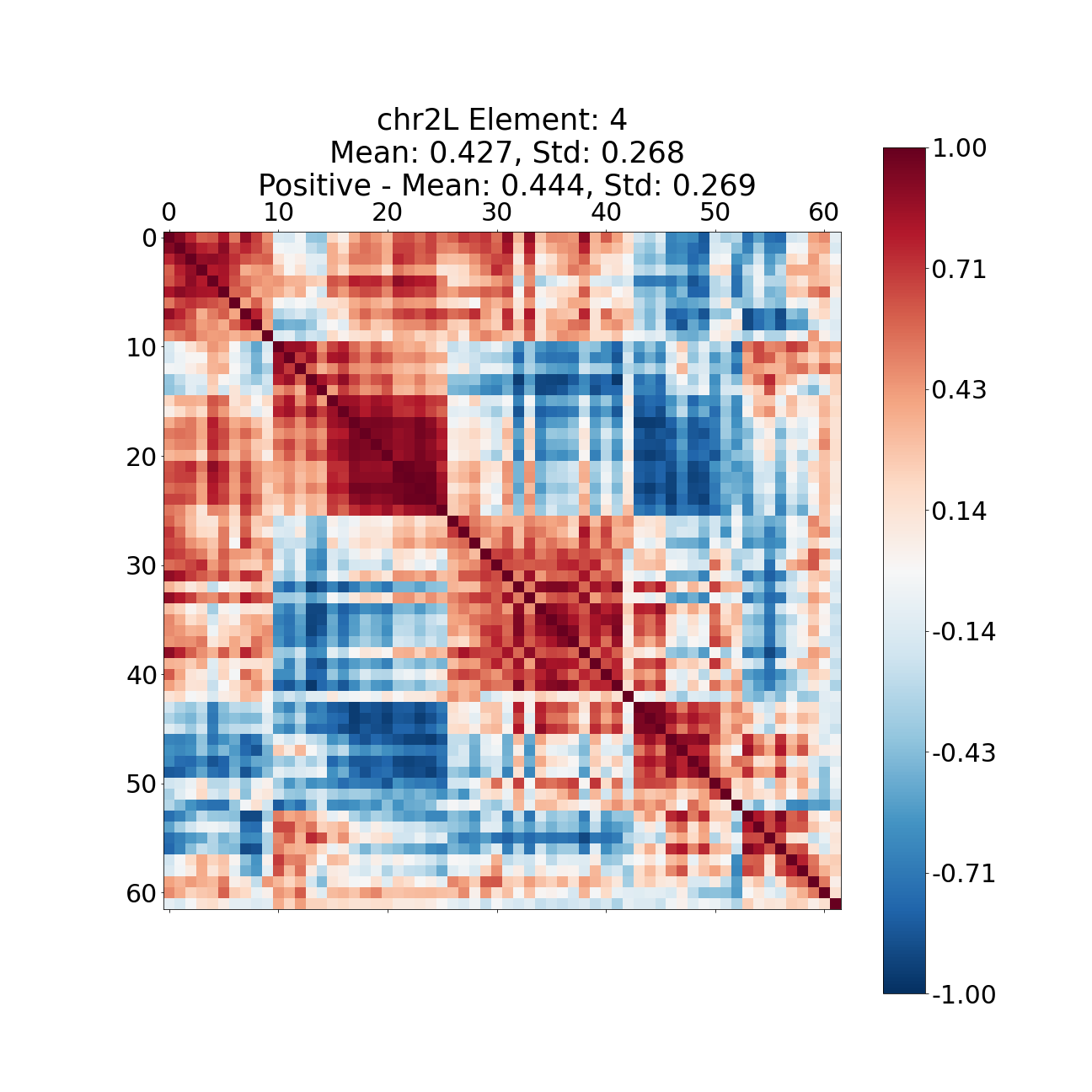}
     \end{subfigure}
     \hfill
     \begin{subfigure}[b]{0.32\textwidth}
         \centering
         \includegraphics[width=\textwidth]{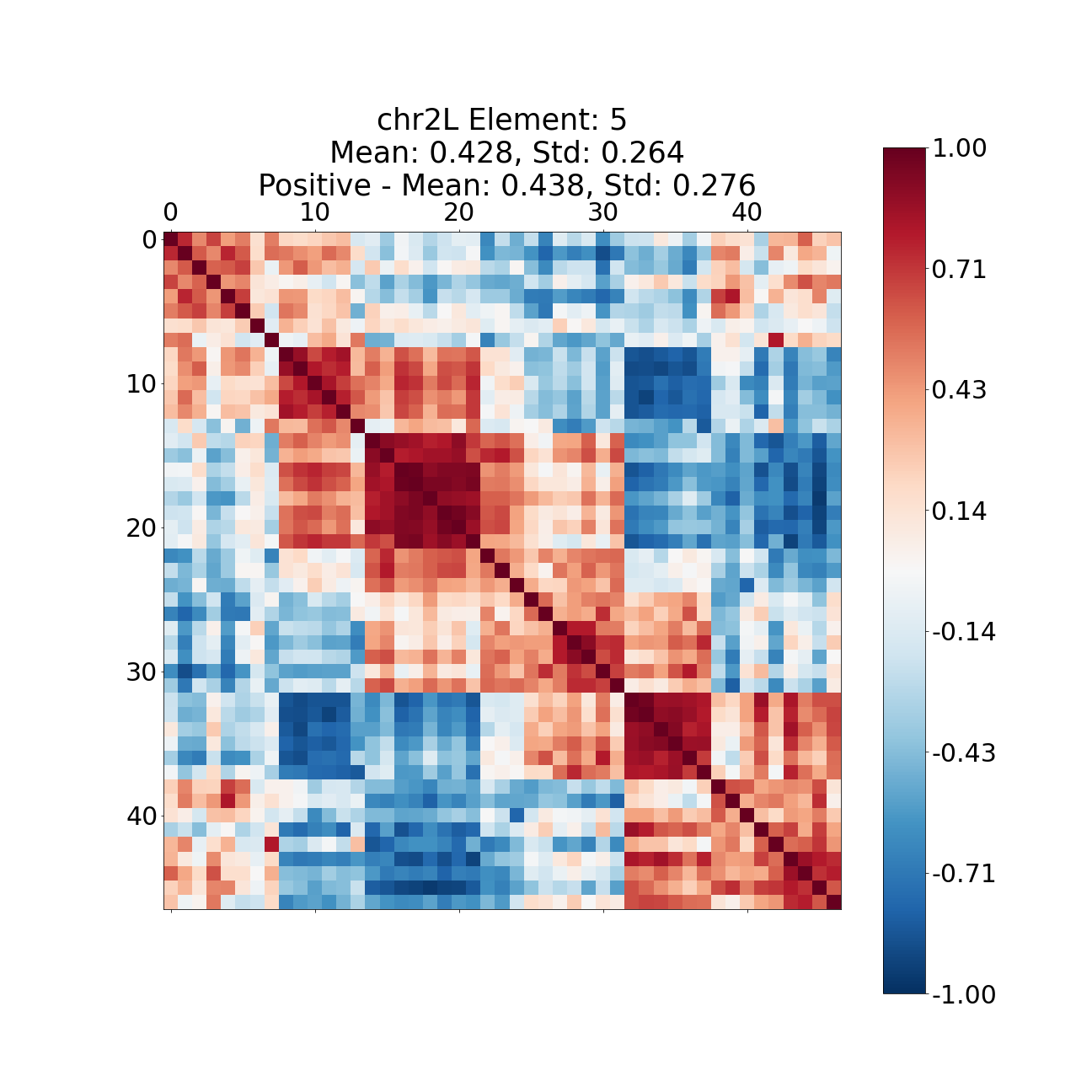}
     \end{subfigure}
     \hfill
     \begin{subfigure}[b]{0.32\textwidth}
         \centering
         \includegraphics[width=\textwidth]{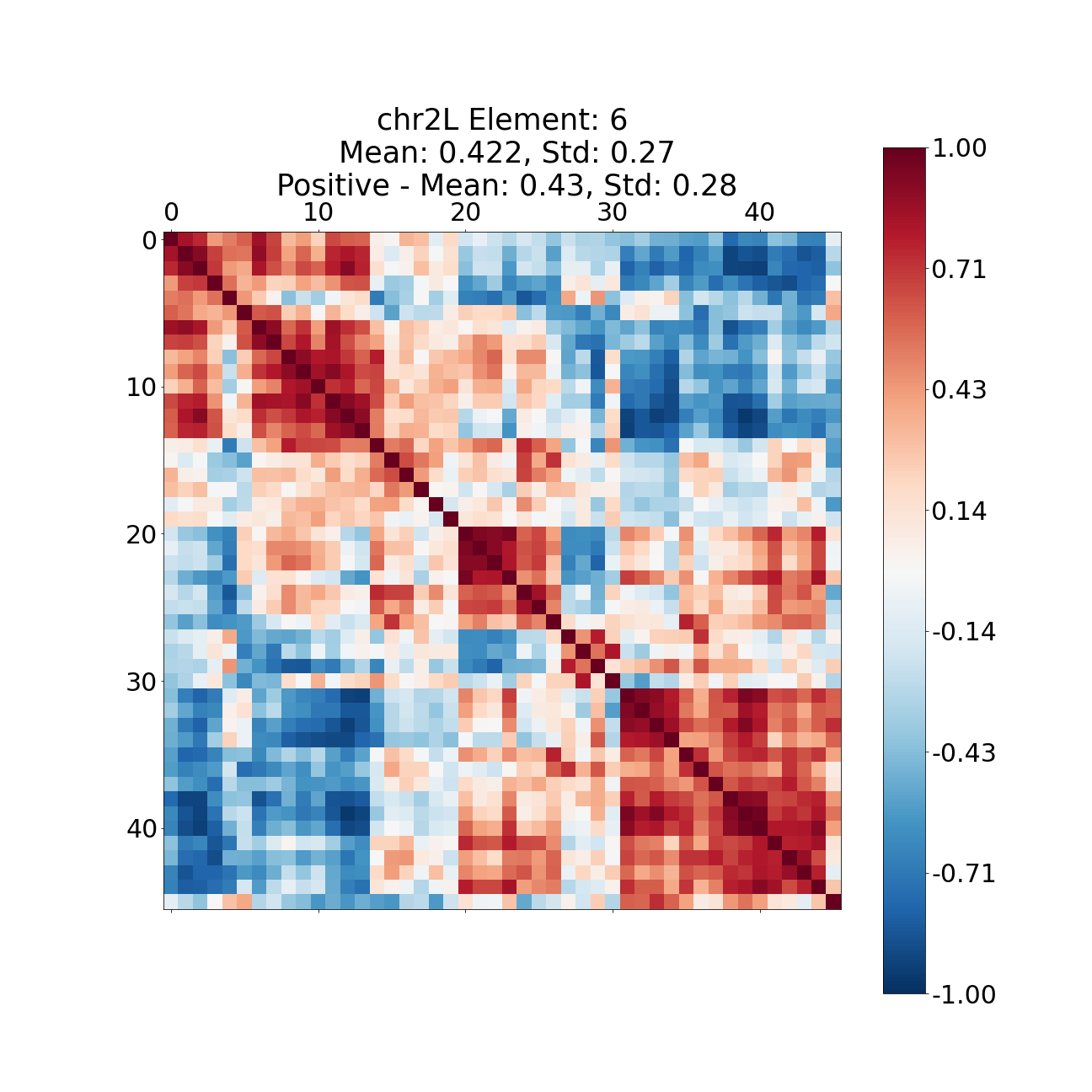}
     \end{subfigure}
      \hfill
     \begin{subfigure}[b]{0.32\textwidth}
         \centering
         \includegraphics[width=\textwidth]{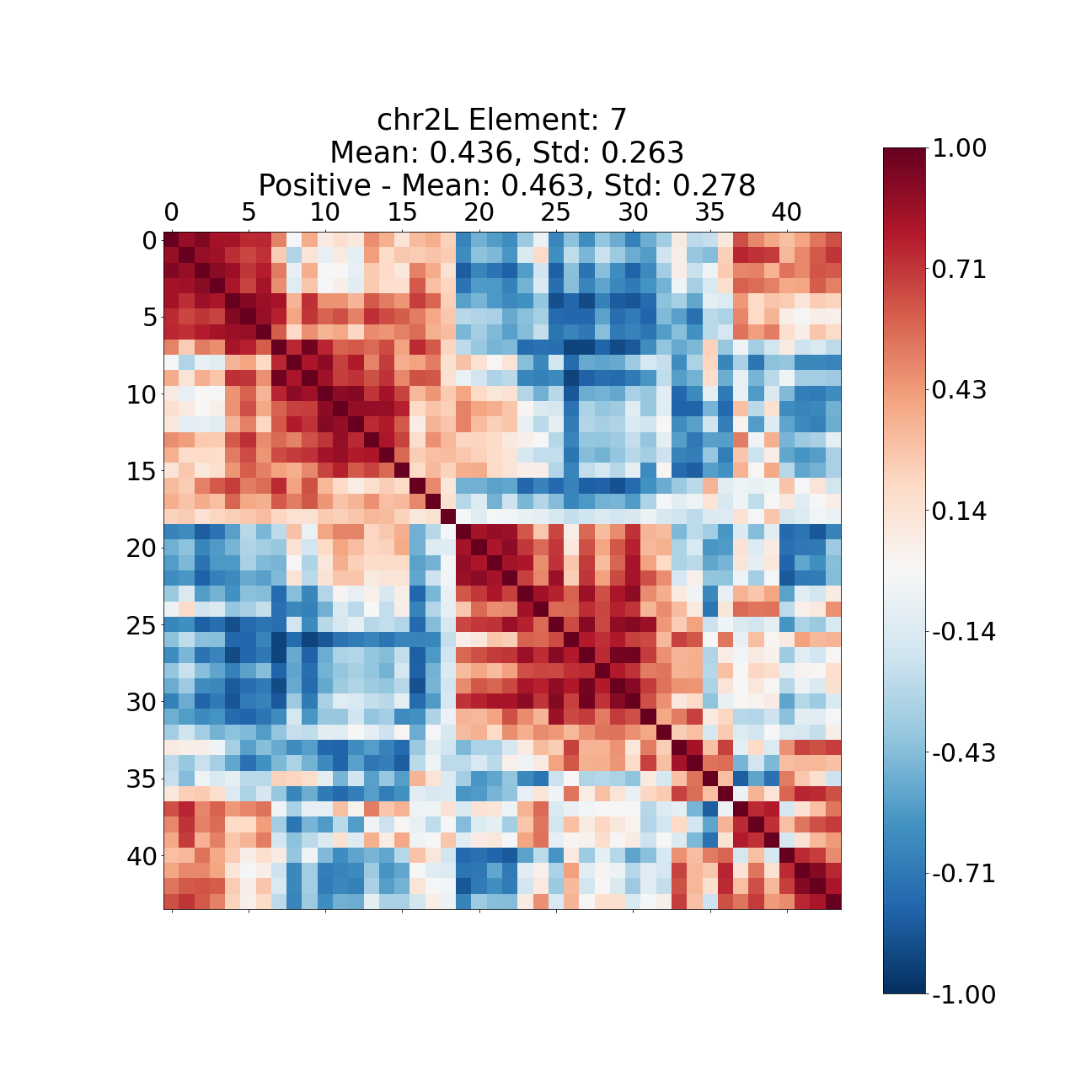}
     \end{subfigure}
     \hfill
     \begin{subfigure}[b]{0.32\textwidth}
         \centering
         \includegraphics[width=\textwidth]{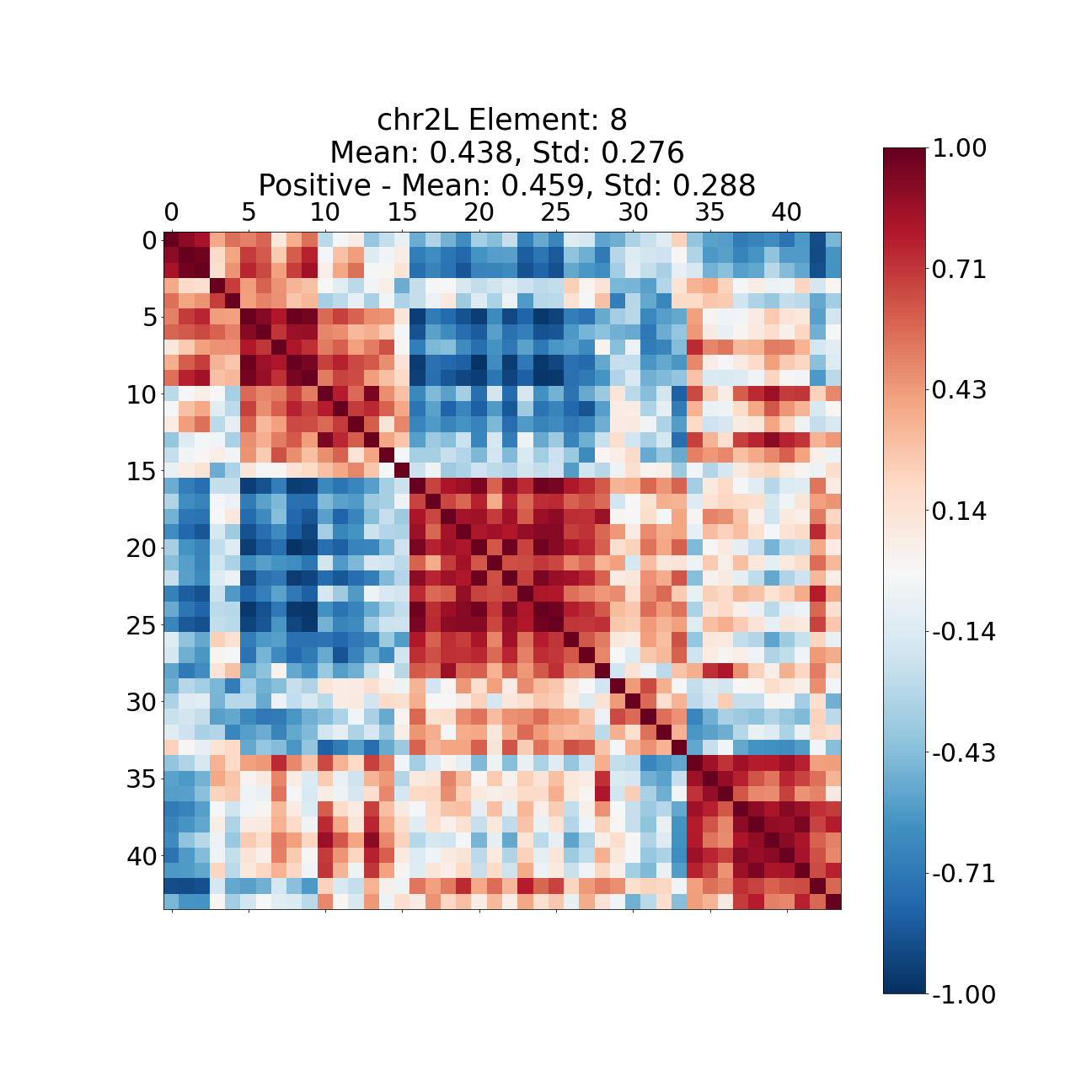}
     \end{subfigure}
     \hfill
     \begin{subfigure}[b]{0.32\textwidth}
         \centering
         \includegraphics[width=\textwidth]{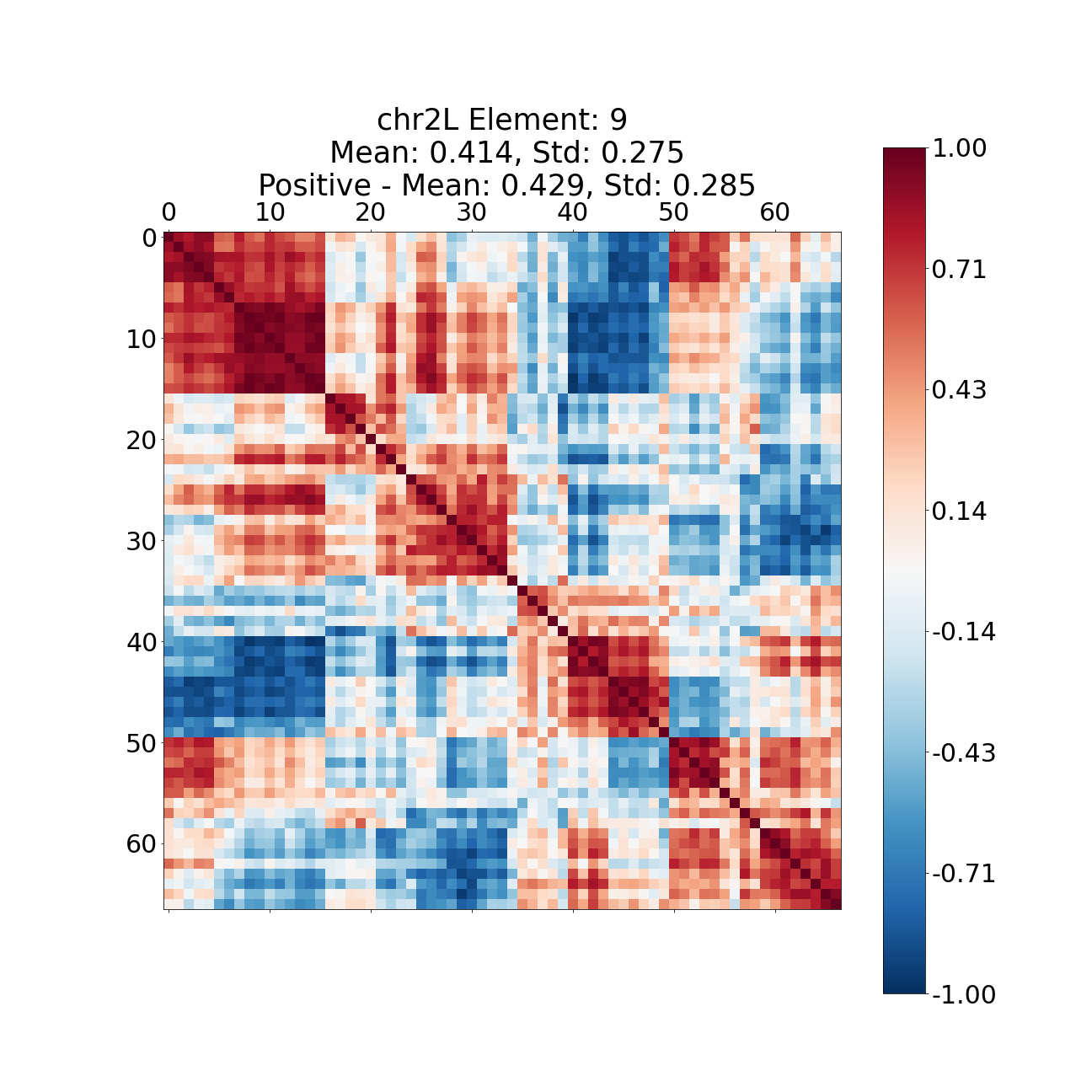}
     \end{subfigure}
     \hfill
     \begin{subfigure}[b]{0.32\textwidth}
         \centering
         \includegraphics[width=\textwidth]{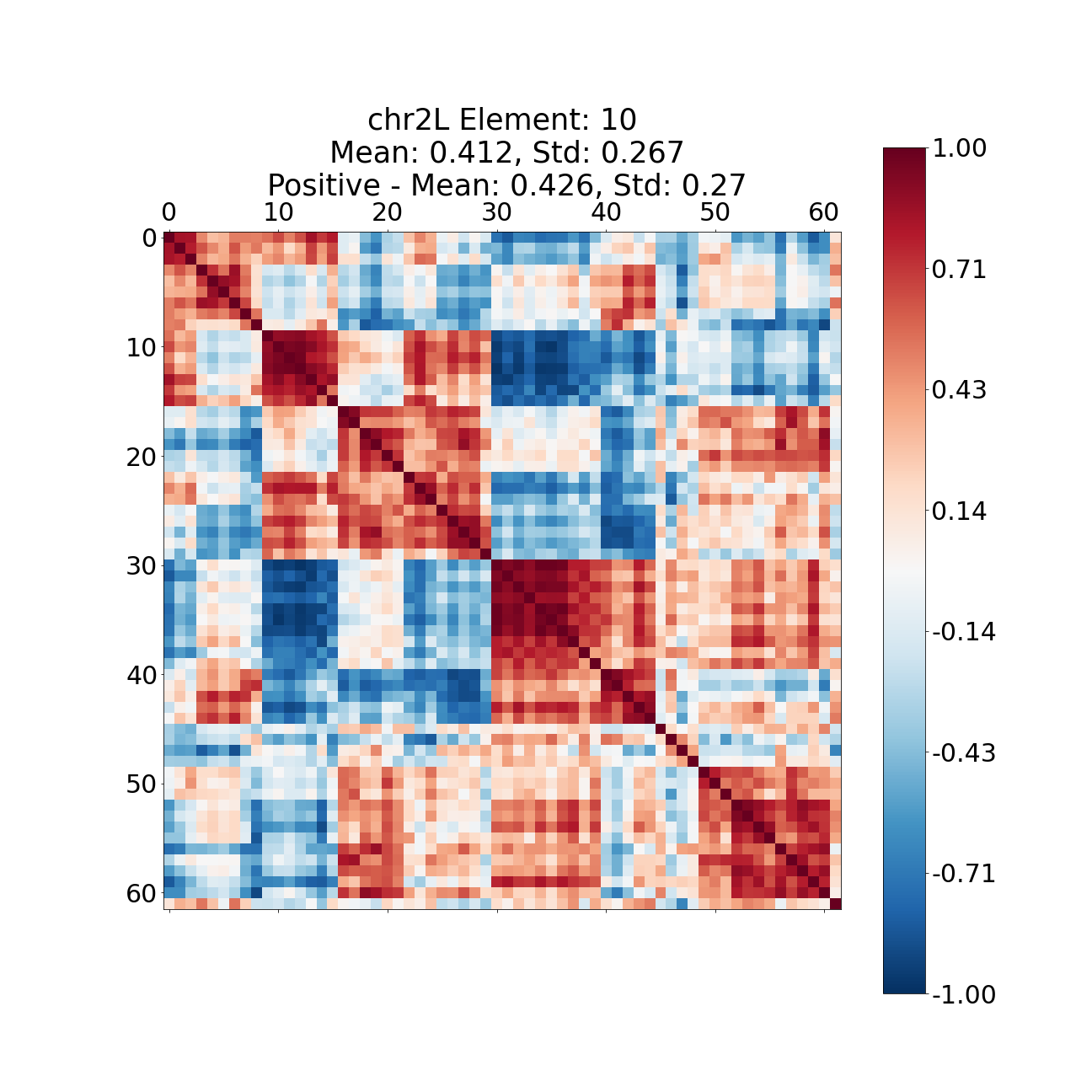}
     \end{subfigure}
      \hfill
     \begin{subfigure}[b]{0.32\textwidth}
         \centering
         \includegraphics[width=\textwidth]{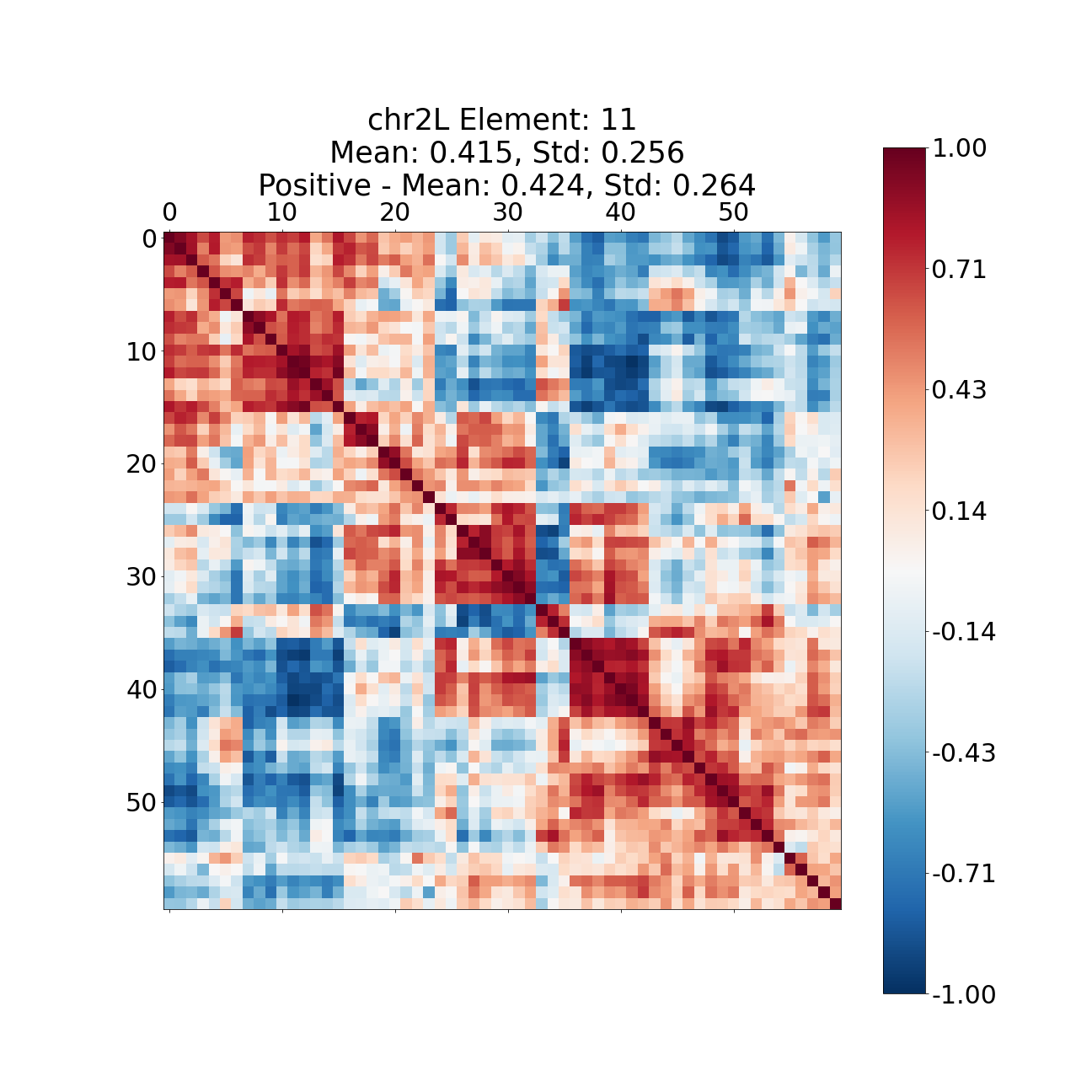}
     \end{subfigure}
        \caption{Pairwise coexpression of genes covered by various dictionary elements for chr 2L obtained through online cvxNDL. We calculated the mean and standard deviation of absolute pairwise coexpression values, along with the mean and standard deviation of coexpression values specifically for all positively correlated gene pairs.}
        \label{fig:dee2_pearson2L}
\end{figure}

\begin{figure}[h]
\ContinuedFloat
     \centering
     \begin{subfigure}[b]{0.32\textwidth}
         \centering
         \includegraphics[width=\textwidth]{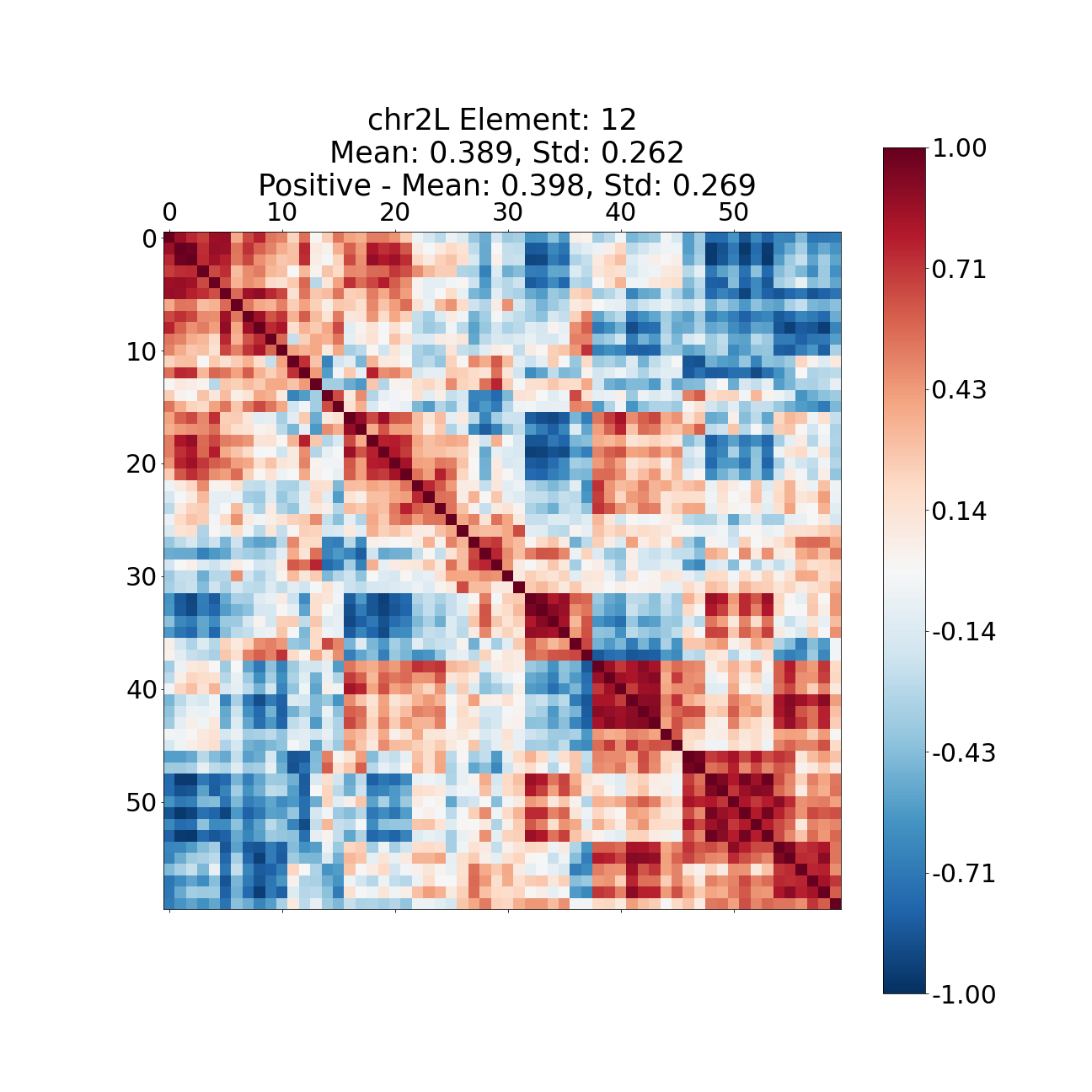}
     \end{subfigure}
     \hfill
     \begin{subfigure}[b]{0.32\textwidth}
         \centering
         \includegraphics[width=\textwidth]{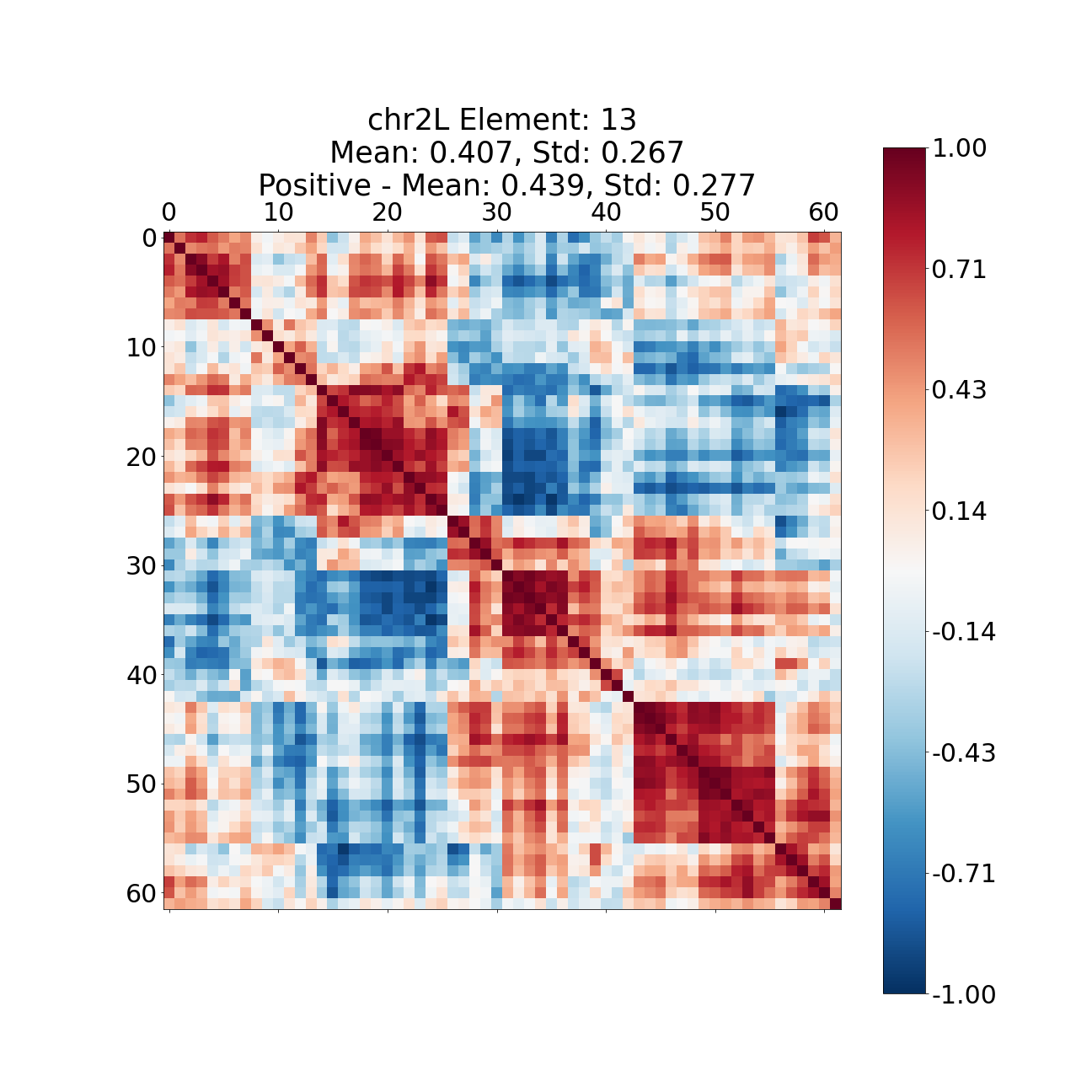}
     \end{subfigure}
     \hfill
     \begin{subfigure}[b]{0.32\textwidth}
         \centering
         \includegraphics[width=\textwidth]{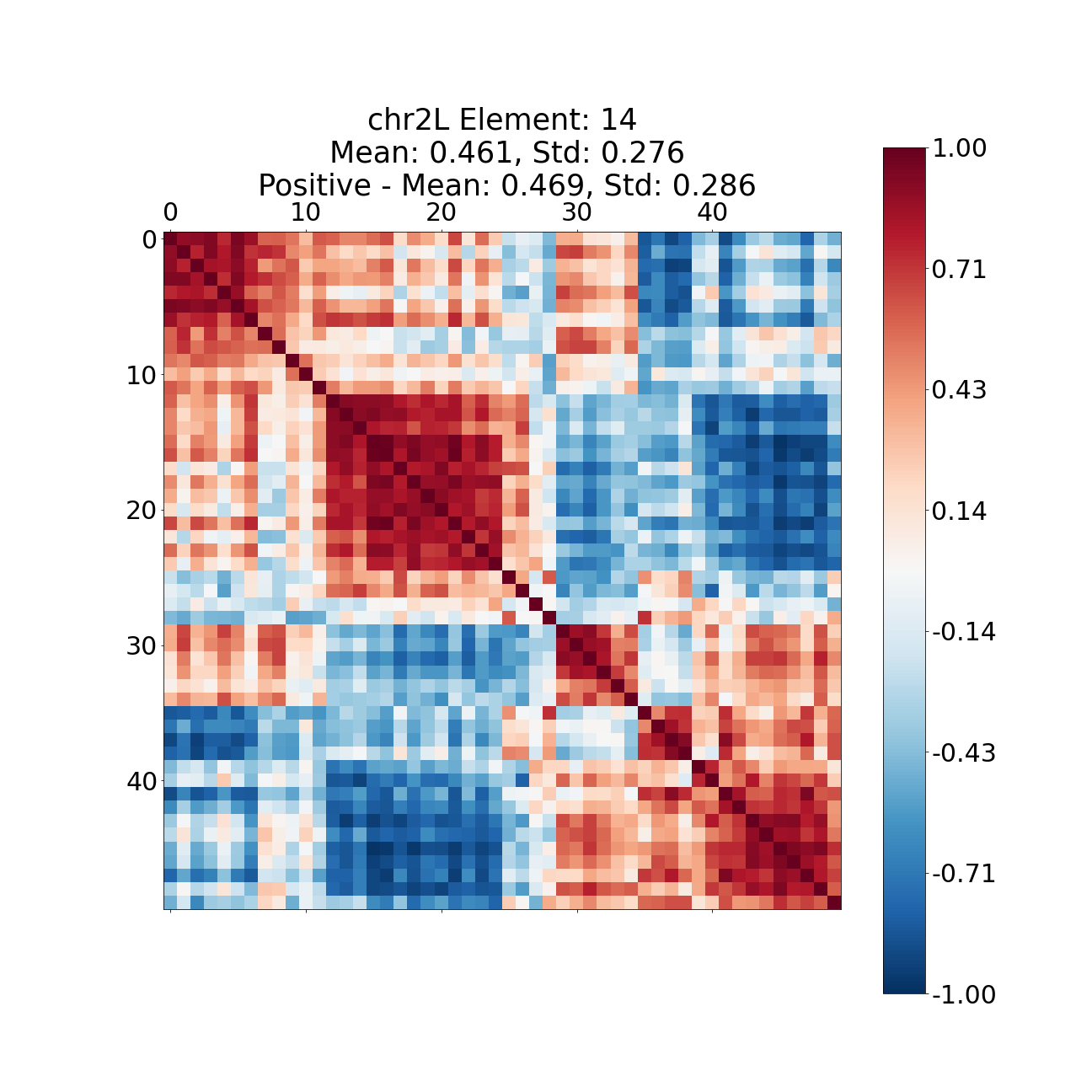}
     \end{subfigure}
      \hfill
     \begin{subfigure}[b]{0.32\textwidth}
         \centering
         \includegraphics[width=\textwidth]{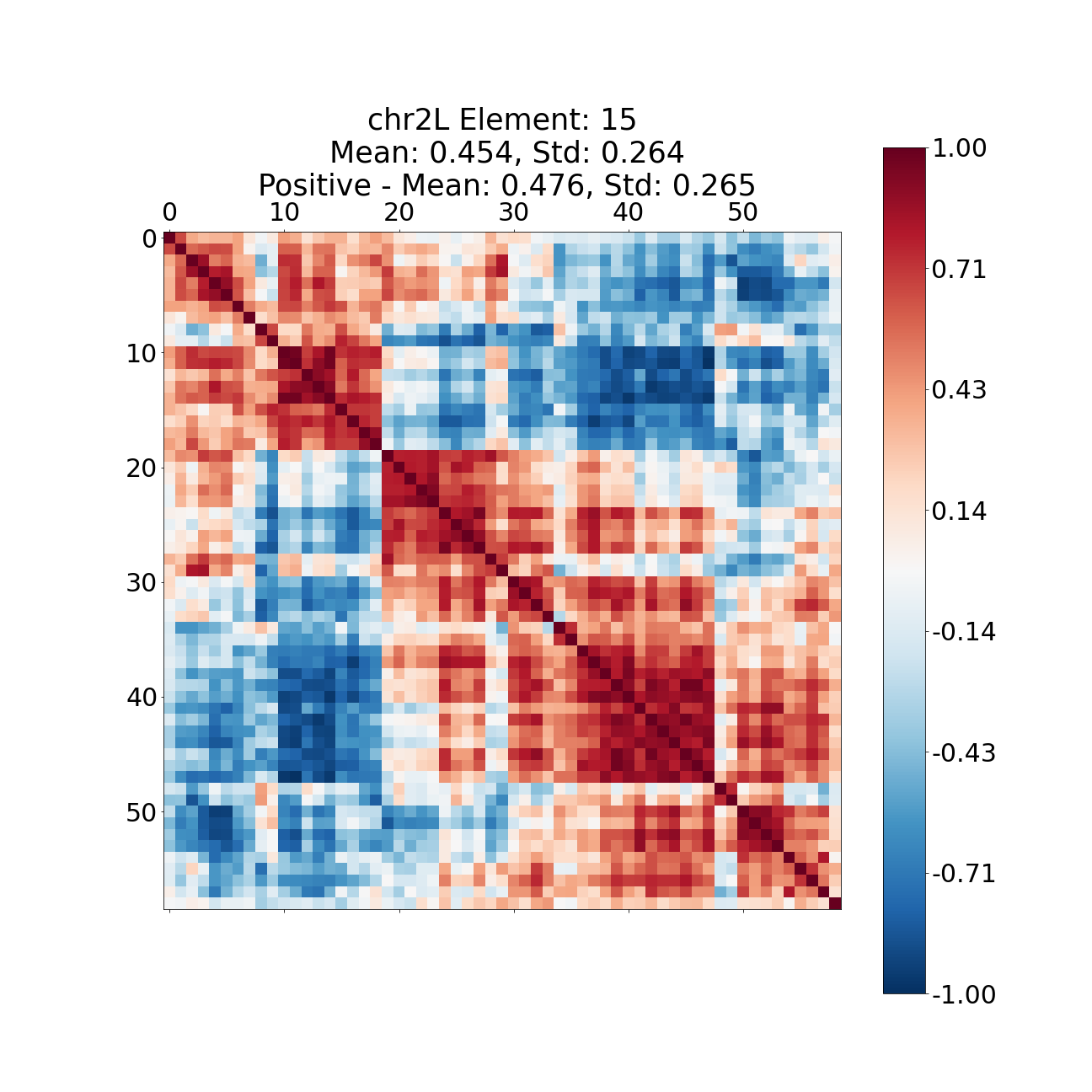}
     \end{subfigure}
     \hfill
     \begin{subfigure}[b]{0.32\textwidth}
         \centering
         \includegraphics[width=\textwidth]{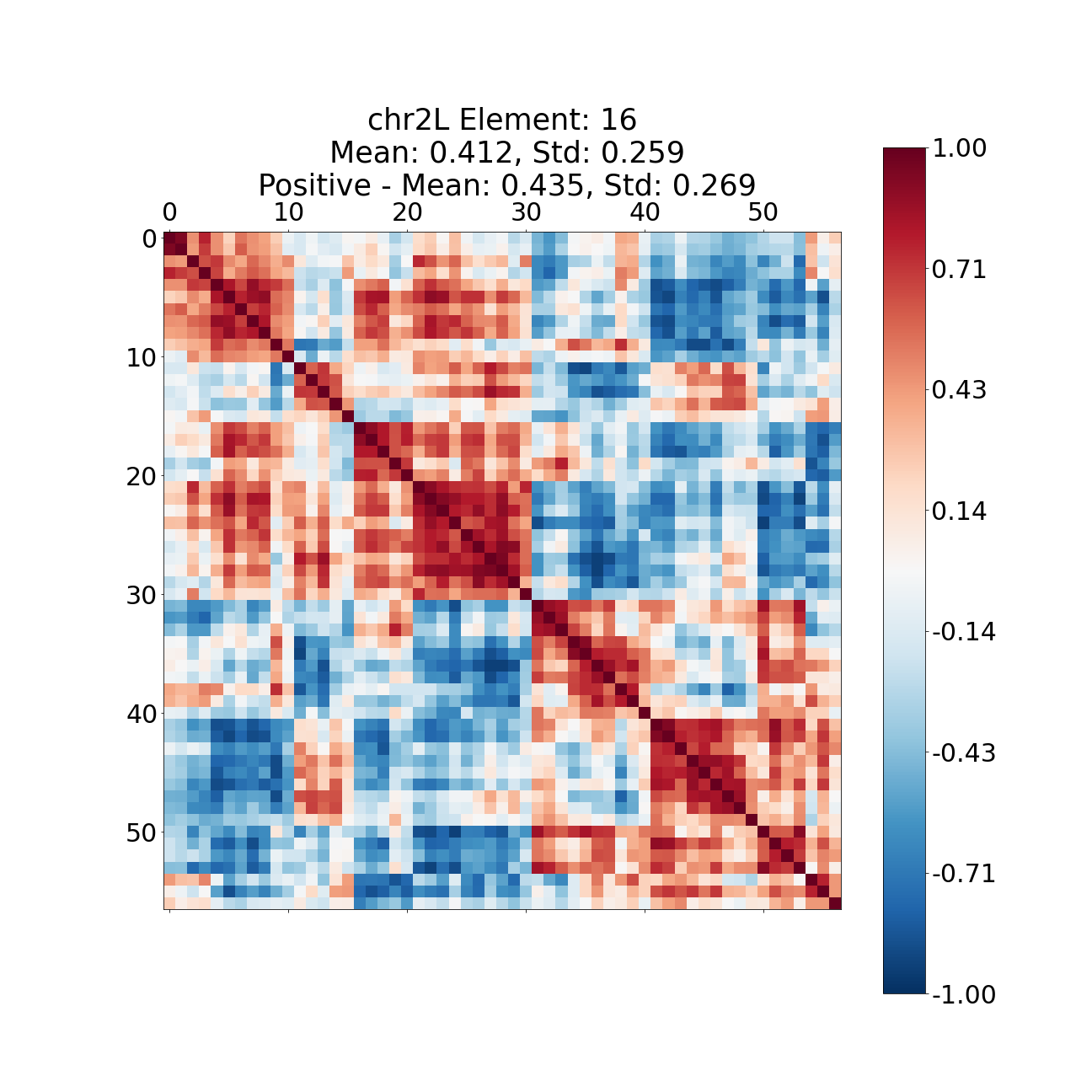}
     \end{subfigure}
     \hfill
     \begin{subfigure}[b]{0.32\textwidth}
         \centering
         \includegraphics[width=\textwidth]{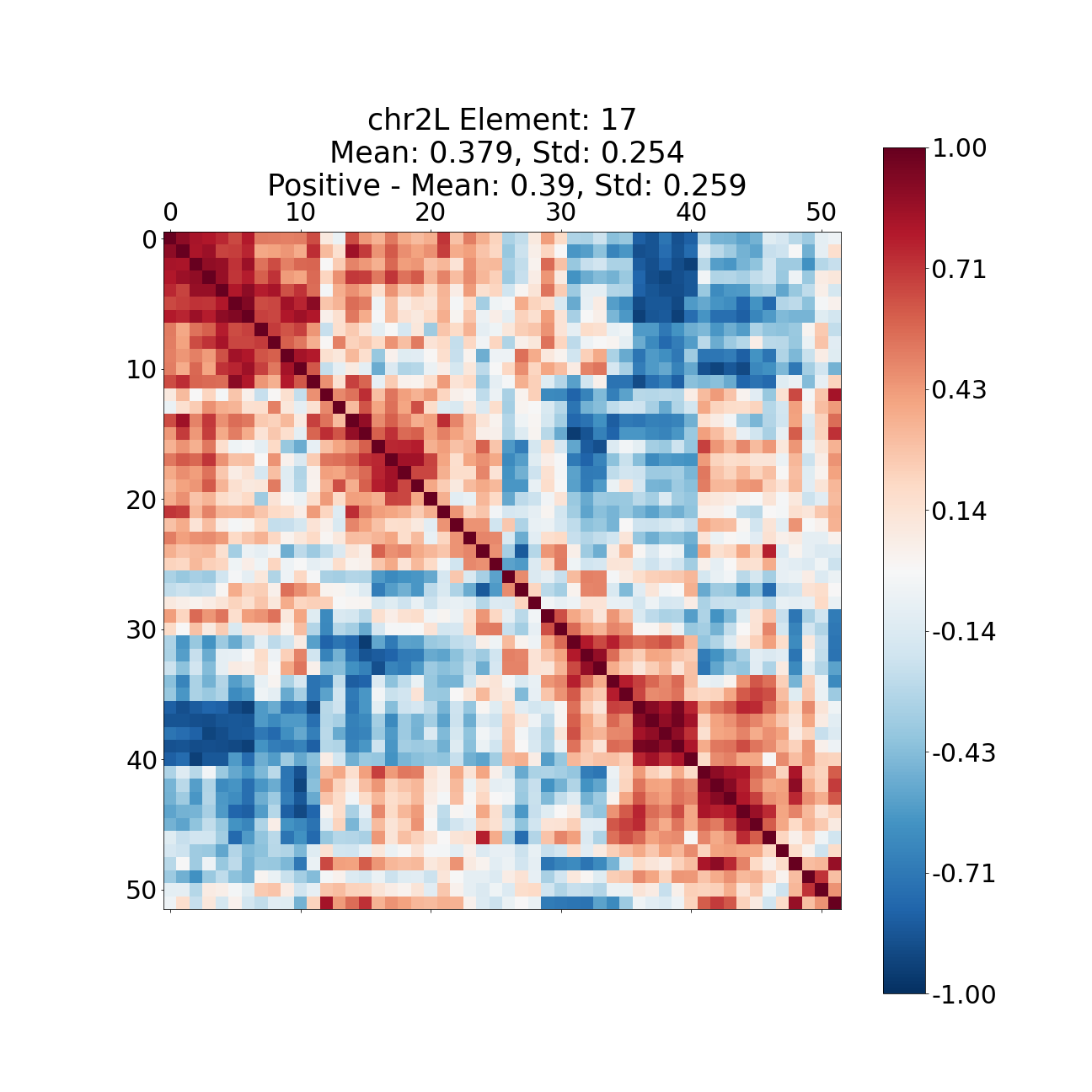}
     \end{subfigure}
     \hfill
     \begin{subfigure}[b]{0.32\textwidth}
         \centering
         \includegraphics[width=\textwidth]{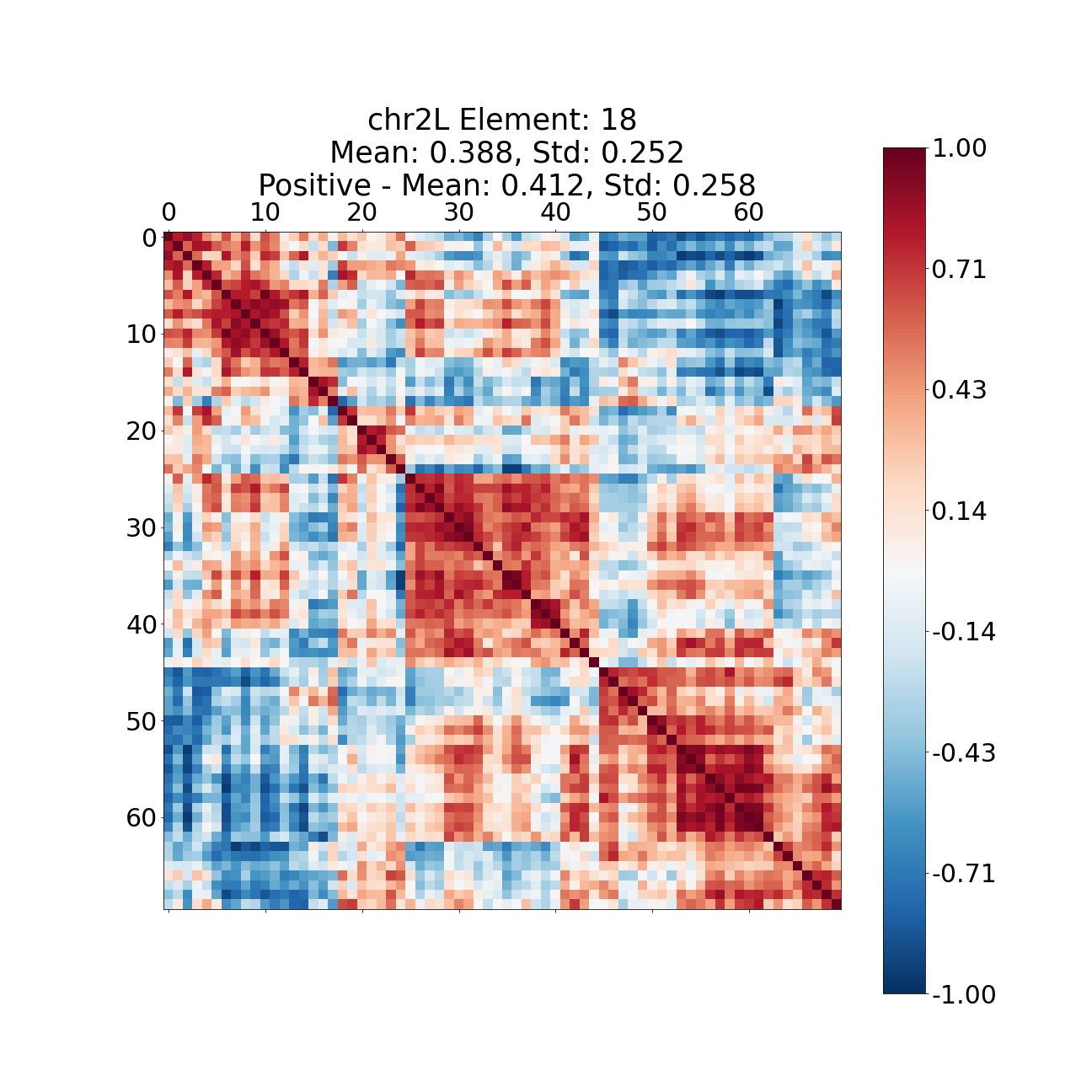}
     \end{subfigure}
      \hfill
     \begin{subfigure}[b]{0.32\textwidth}
         \centering
         \includegraphics[width=\textwidth]{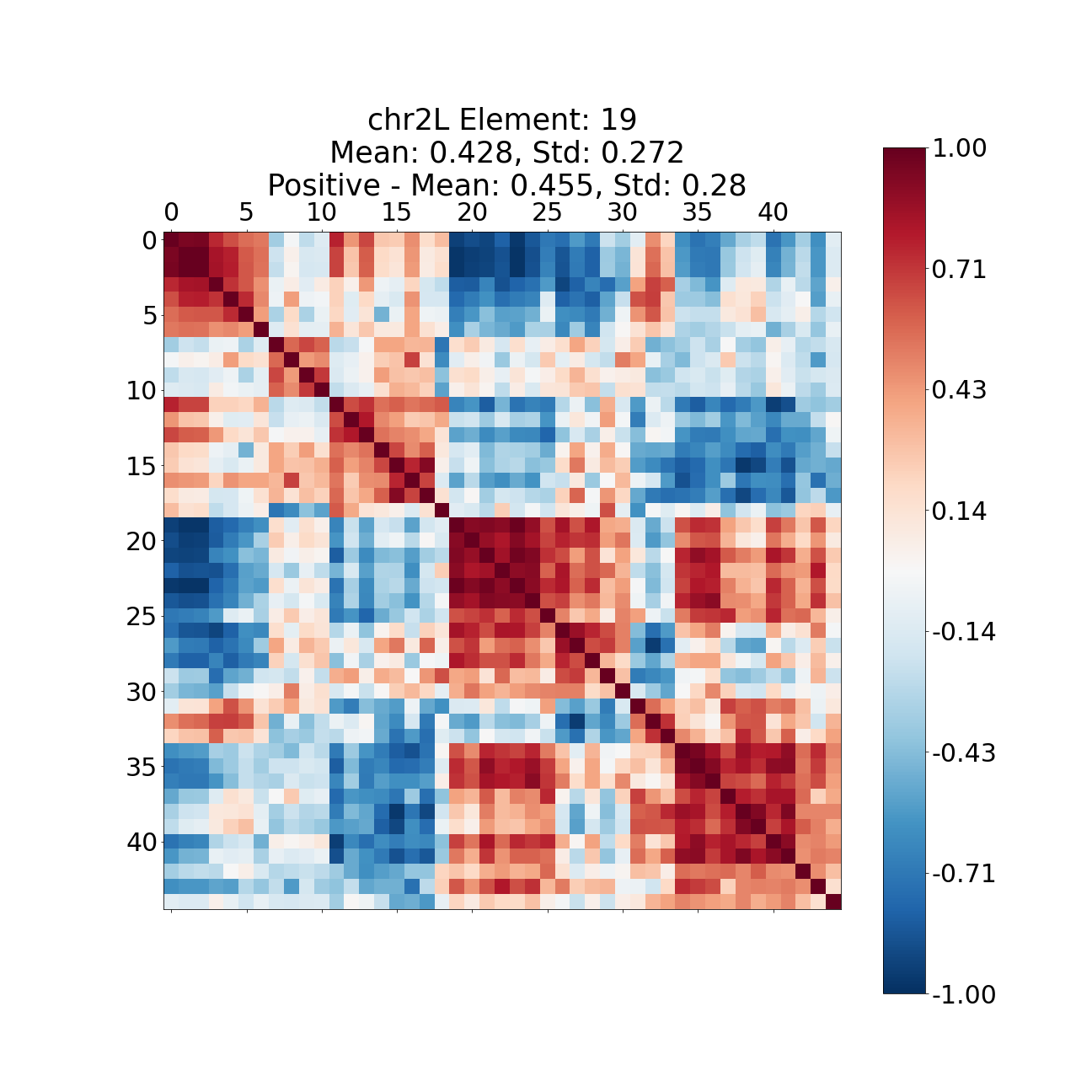}
     \end{subfigure}
     \hfill
     \begin{subfigure}[b]{0.32\textwidth}
         \centering
         \includegraphics[width=\textwidth]{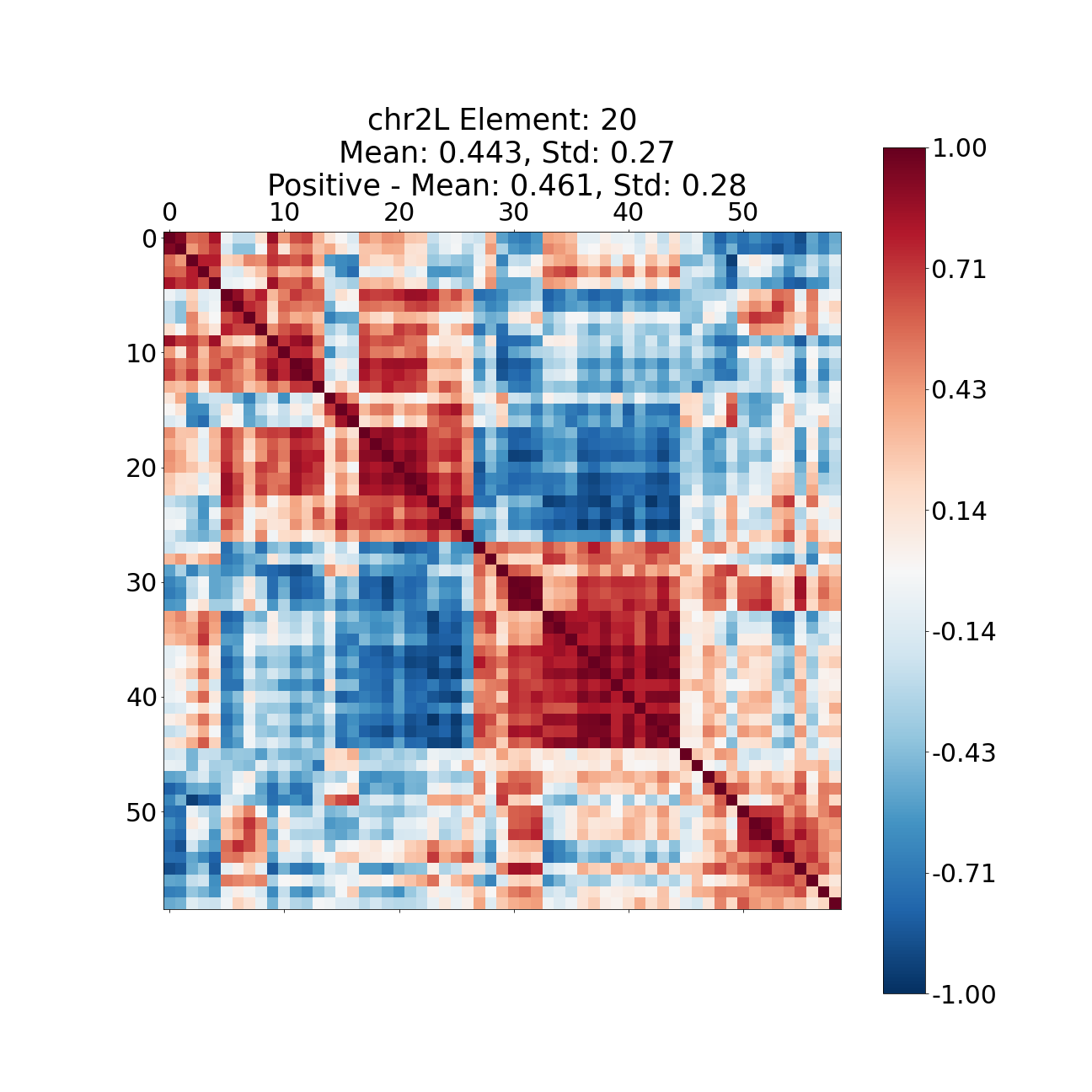}
     \end{subfigure}
     \hfill
     \begin{subfigure}[b]{0.32\textwidth}
         \centering
         \includegraphics[width=\textwidth]{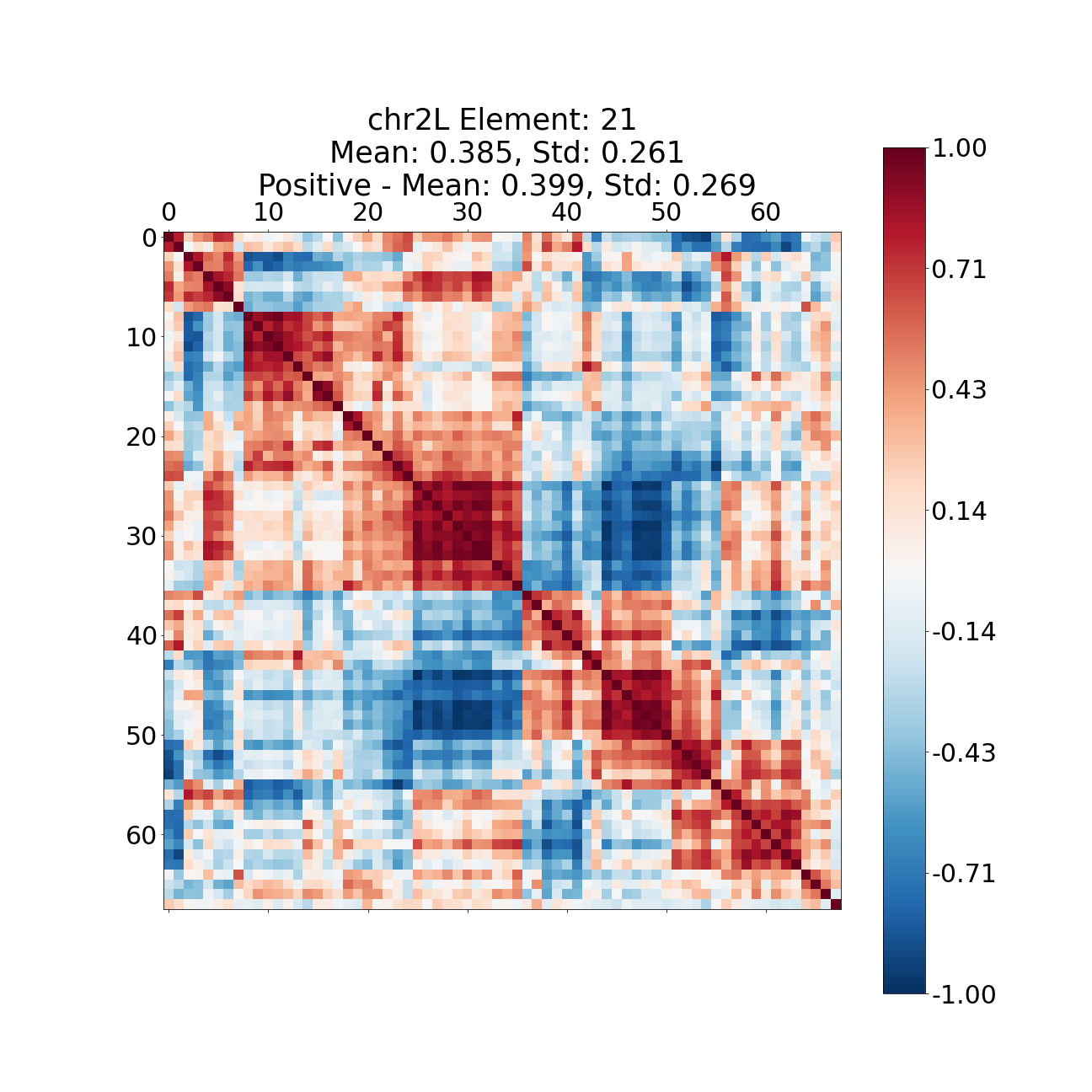}
     \end{subfigure}
     \hfill
     \begin{subfigure}[b]{0.32\textwidth}
         \centering
         \includegraphics[width=\textwidth]{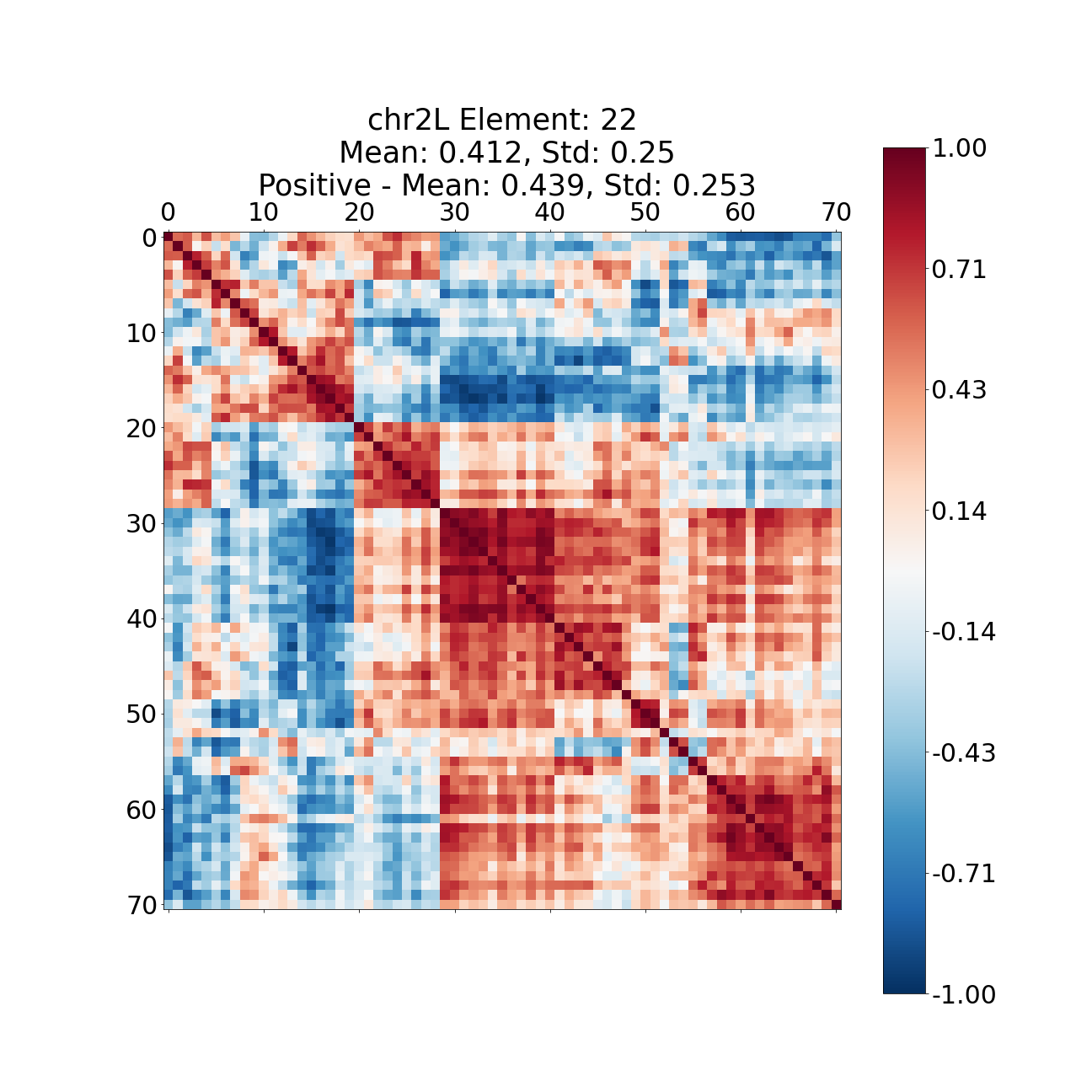}
     \end{subfigure}
      \hfill
     \begin{subfigure}[b]{0.32\textwidth}
         \centering
         \includegraphics[width=\textwidth]{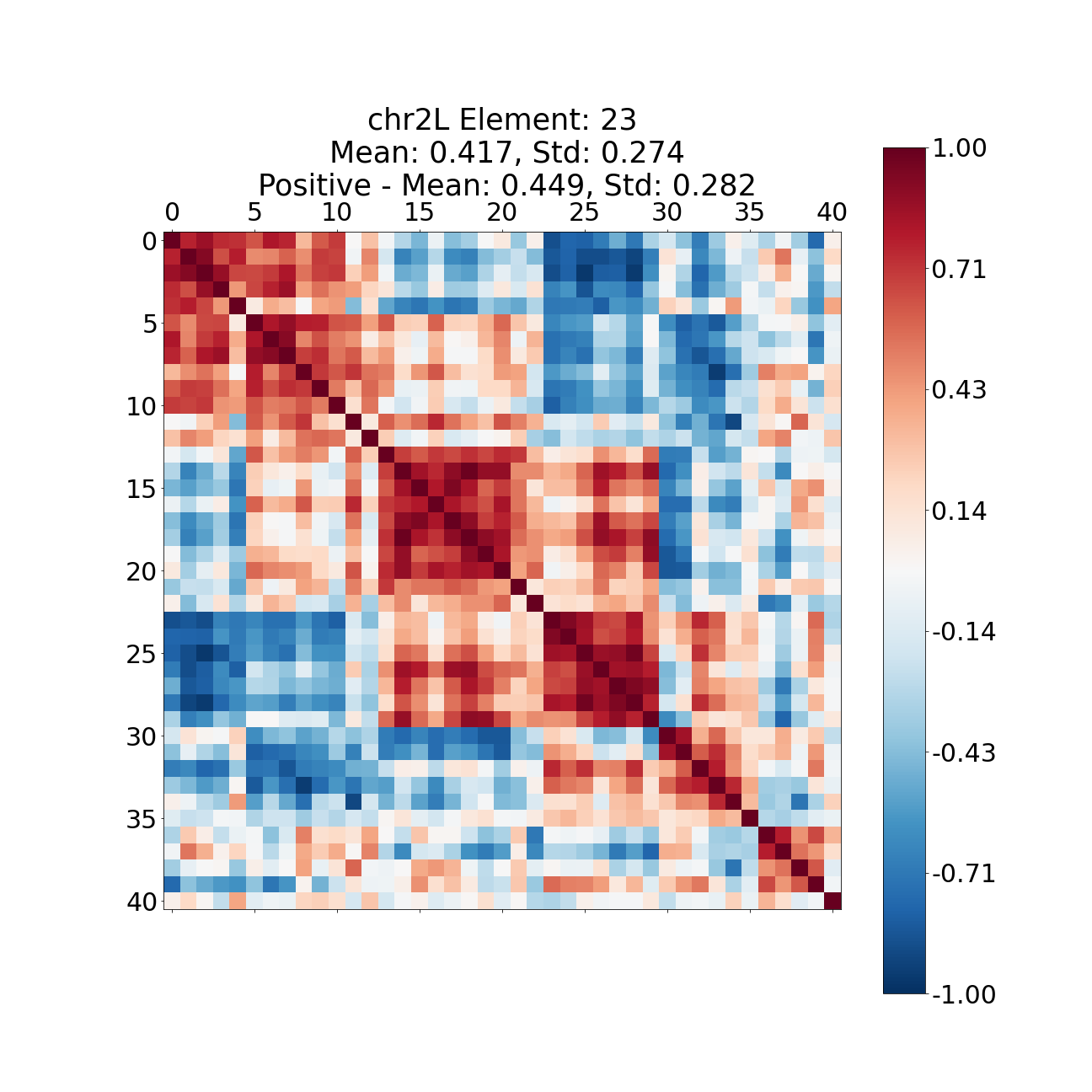}
     \end{subfigure}
        \caption{Pairwise coexpression of genes covered by various dictionary elements for chr 2L obtained through online cvxNDL. We calculated the mean and standard deviation of absolute pairwise coexpression values, along with the mean and standard deviation of coexpression values specifically for all positively correlated gene pairs.}
\end{figure}

\begin{figure}[h]
\ContinuedFloat
     \centering
     \begin{subfigure}[b]{0.32\textwidth}
         \centering
         \includegraphics[width=\textwidth]{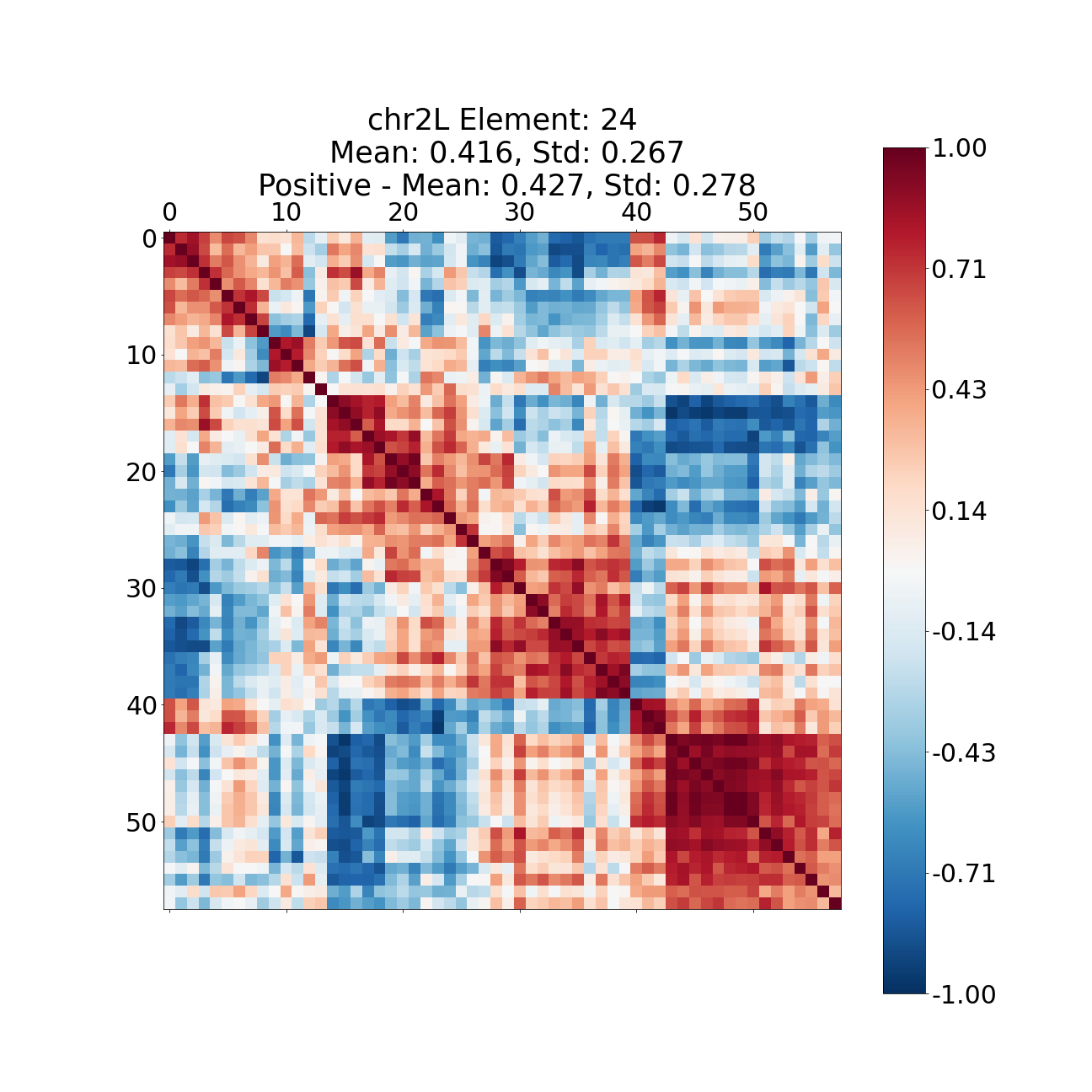}
     \end{subfigure}
        \caption{Pairwise coexpression of genes covered by various dictionary elements for chr 2L obtained through online cvxNDL. We calculated the mean and standard deviation of absolute pairwise coexpression values, along with the mean and standard deviation of coexpression values specifically for all positively correlated gene pairs.}
\end{figure}


\begin{figure}[h]
     \centering
     \begin{subfigure}[b]{0.32\textwidth}
         \centering
         \includegraphics[width=\textwidth]{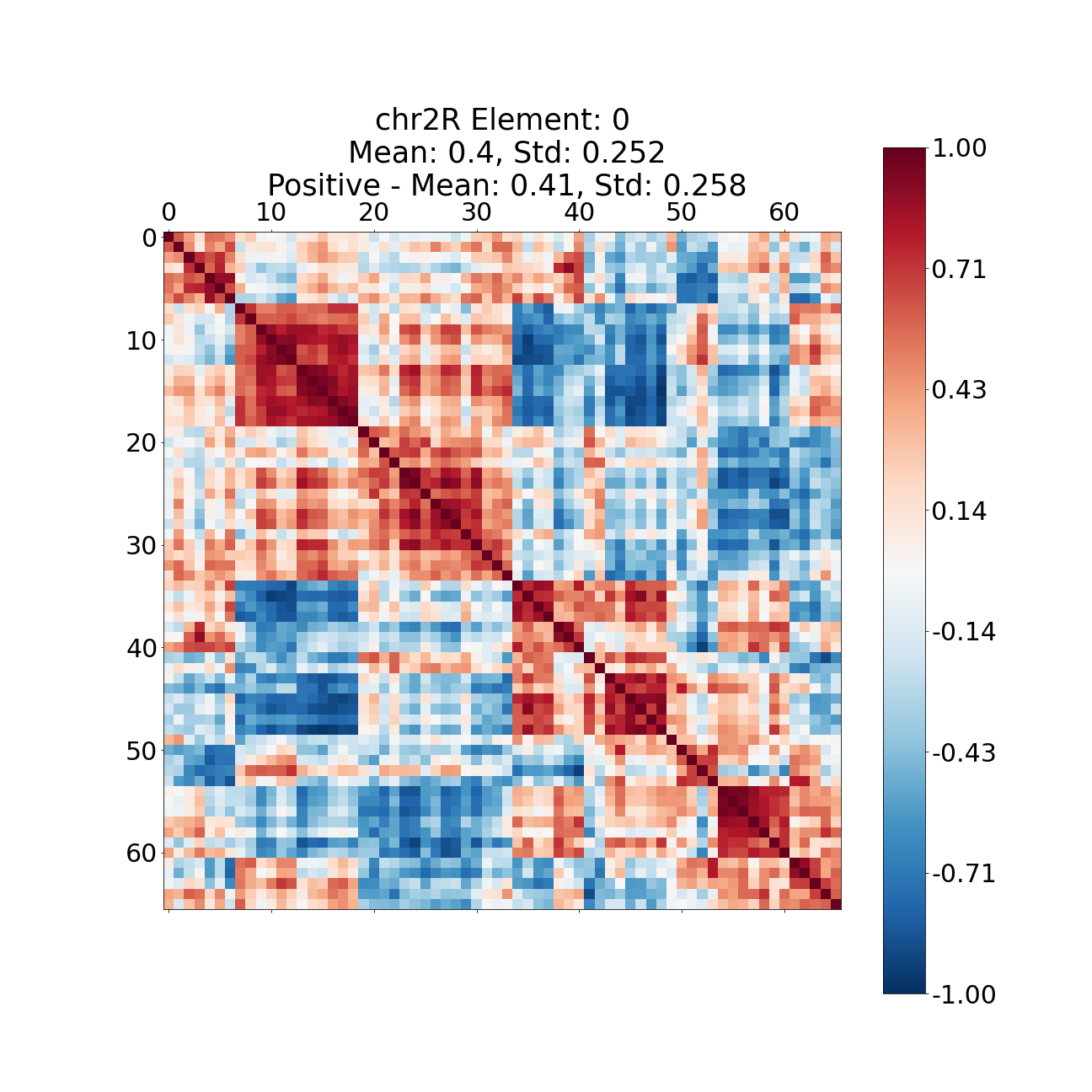}
     \end{subfigure}
     \hfill
     \begin{subfigure}[b]{0.32\textwidth}
         \centering
         \includegraphics[width=\textwidth]{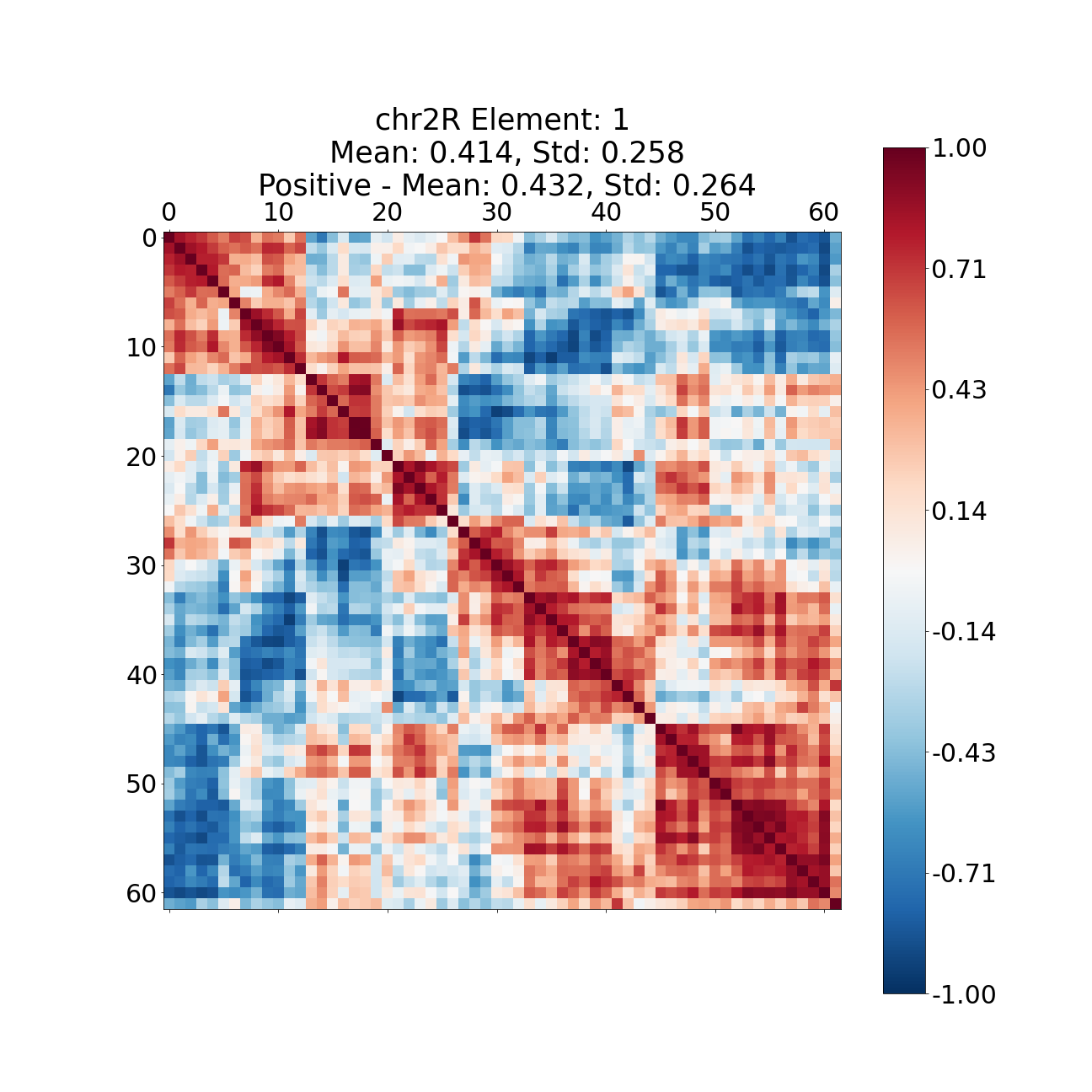}
     \end{subfigure}
     \hfill
     \begin{subfigure}[b]{0.32\textwidth}
         \centering
         \includegraphics[width=\textwidth]{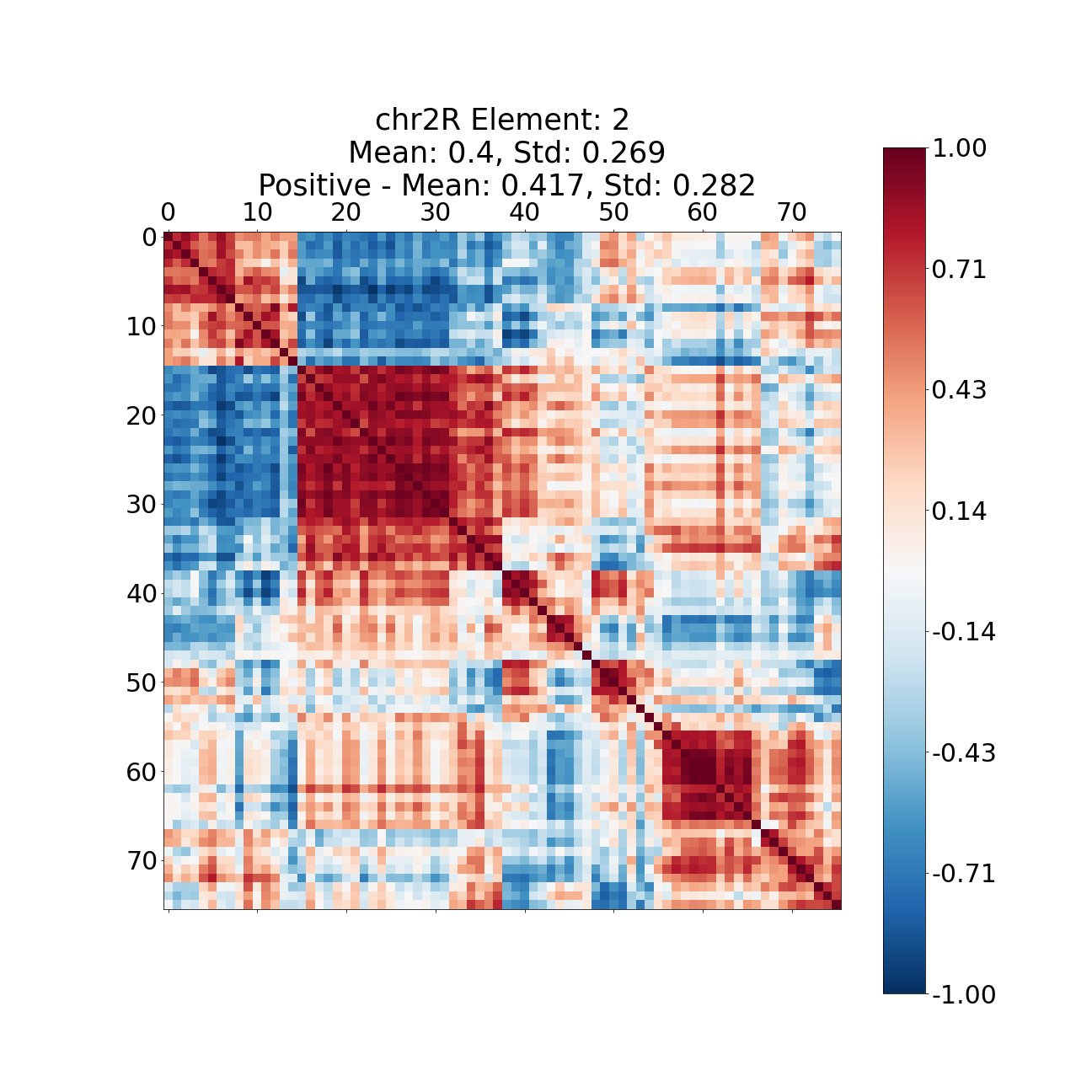}
     \end{subfigure}
      \hfill
     \begin{subfigure}[b]{0.32\textwidth}
         \centering
         \includegraphics[width=\textwidth]{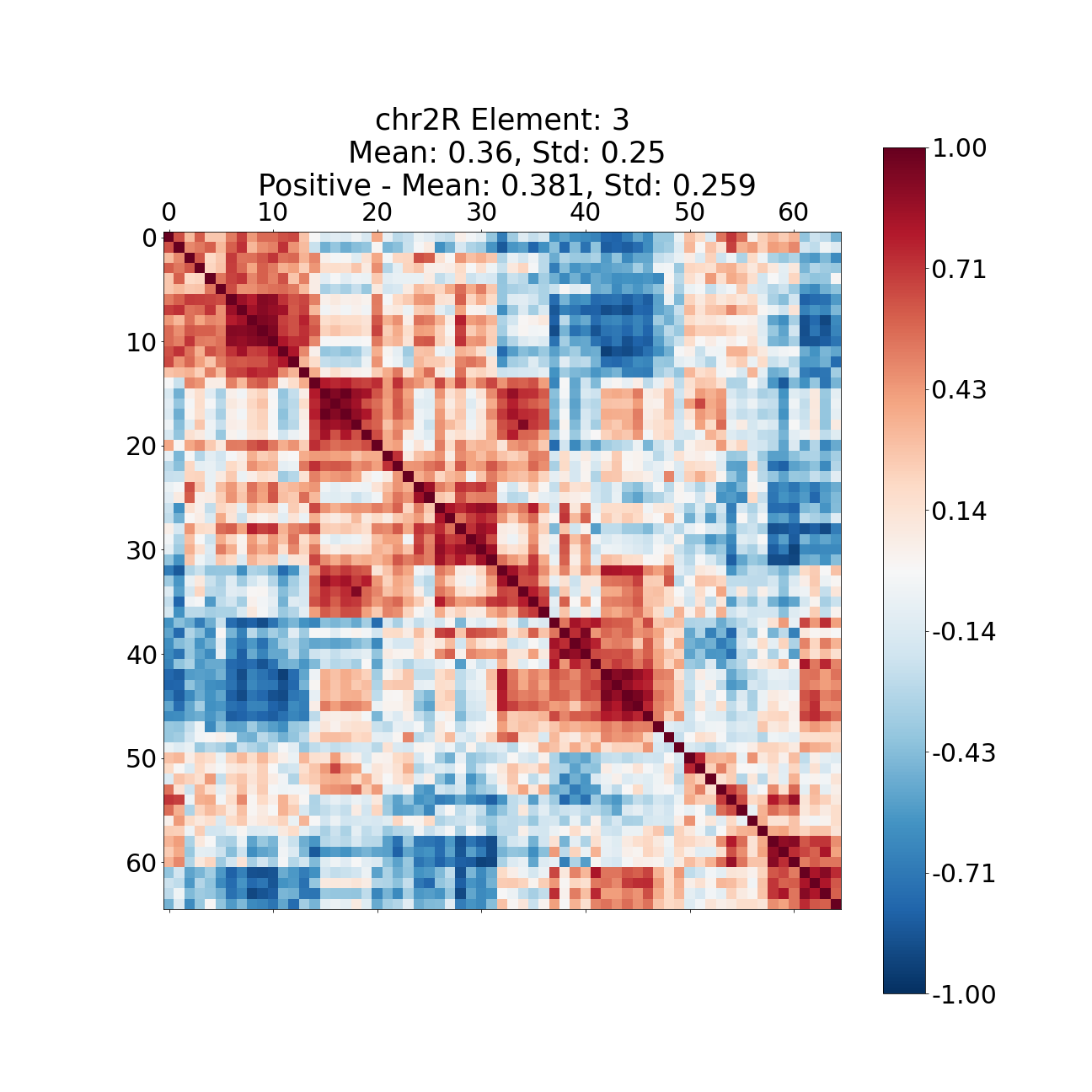}
     \end{subfigure}
     \hfill
     \begin{subfigure}[b]{0.32\textwidth}
         \centering
         \includegraphics[width=\textwidth]{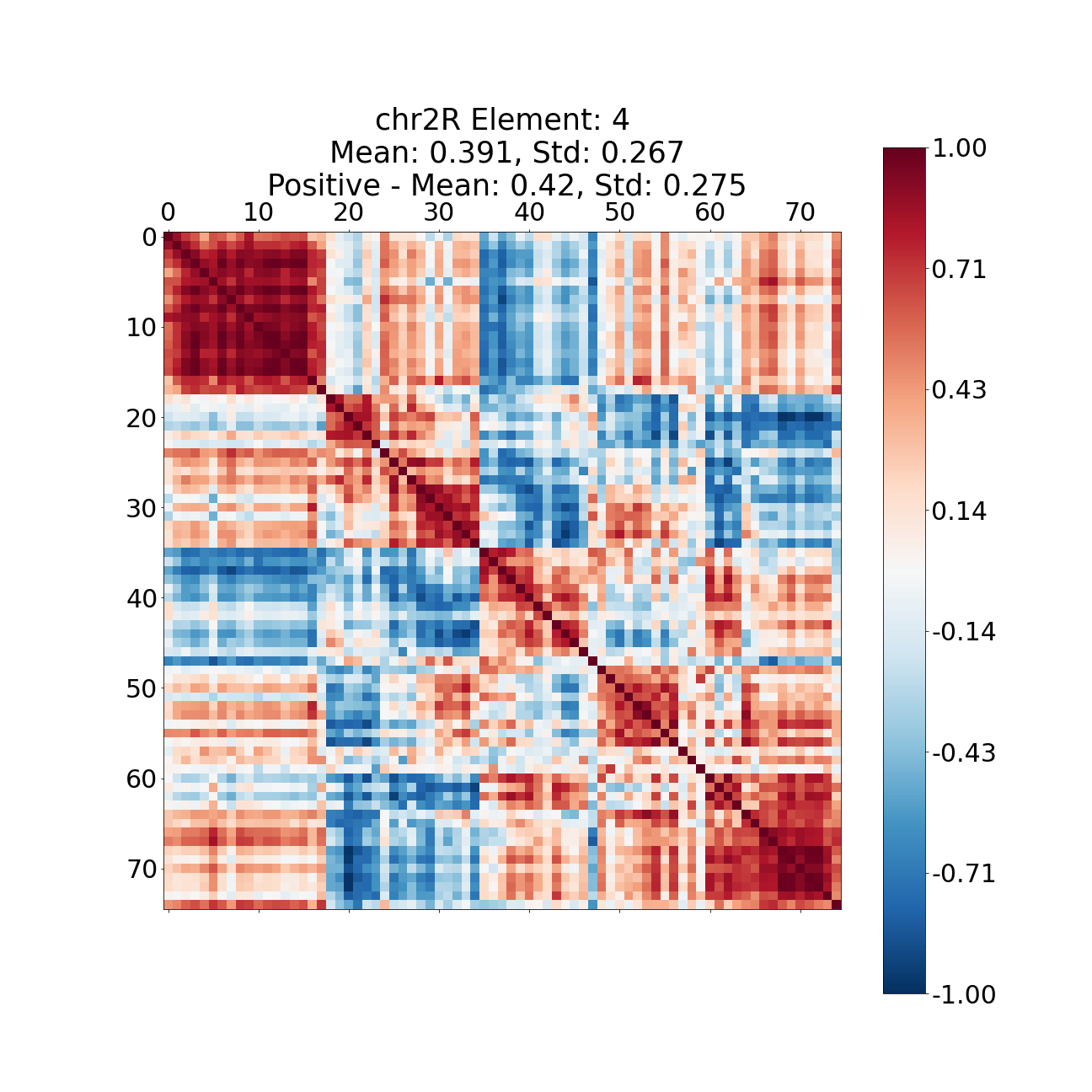}
     \end{subfigure}
     \hfill
     \begin{subfigure}[b]{0.32\textwidth}
         \centering
         \includegraphics[width=\textwidth]{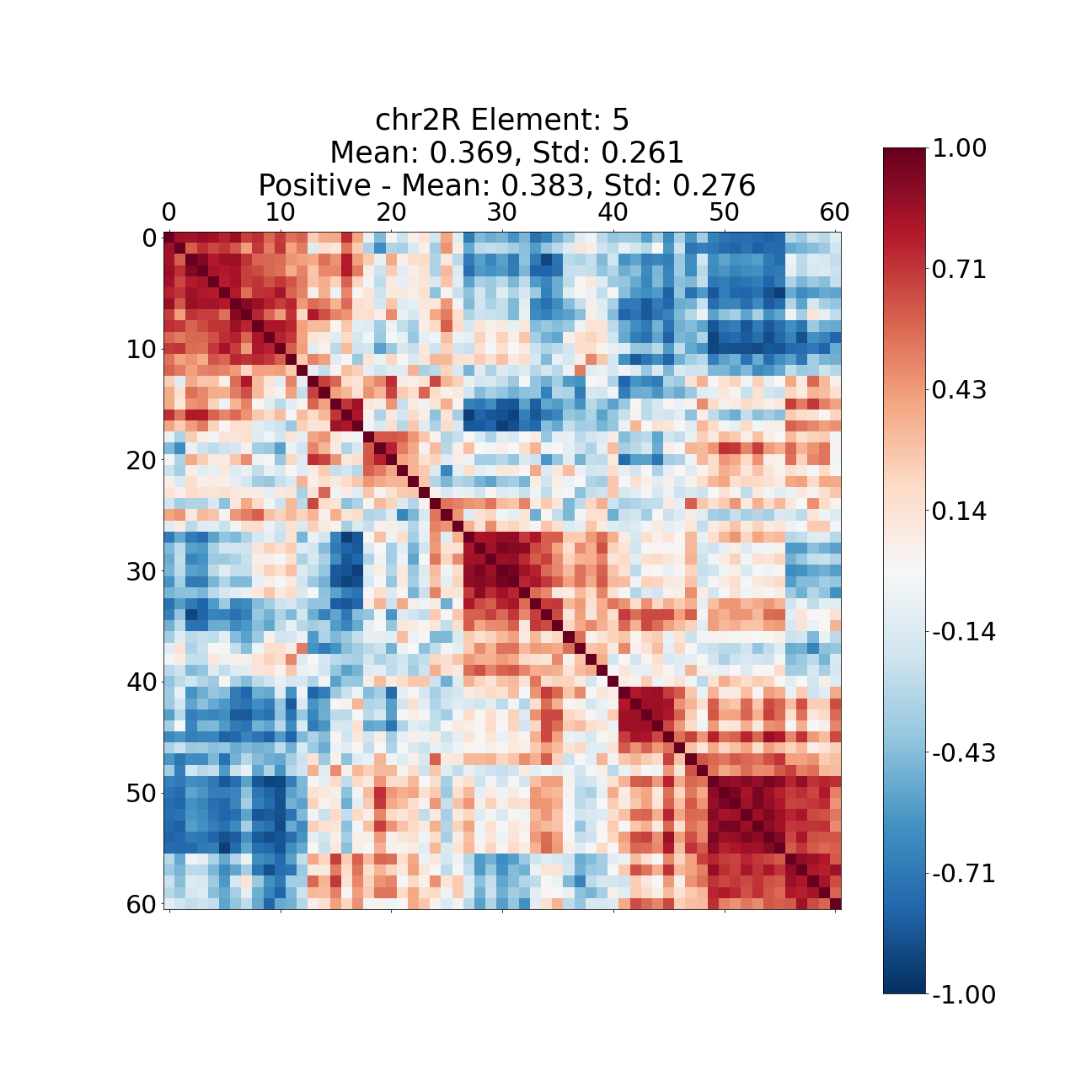}
     \end{subfigure}
     \hfill
     \begin{subfigure}[b]{0.32\textwidth}
         \centering
         \includegraphics[width=\textwidth]{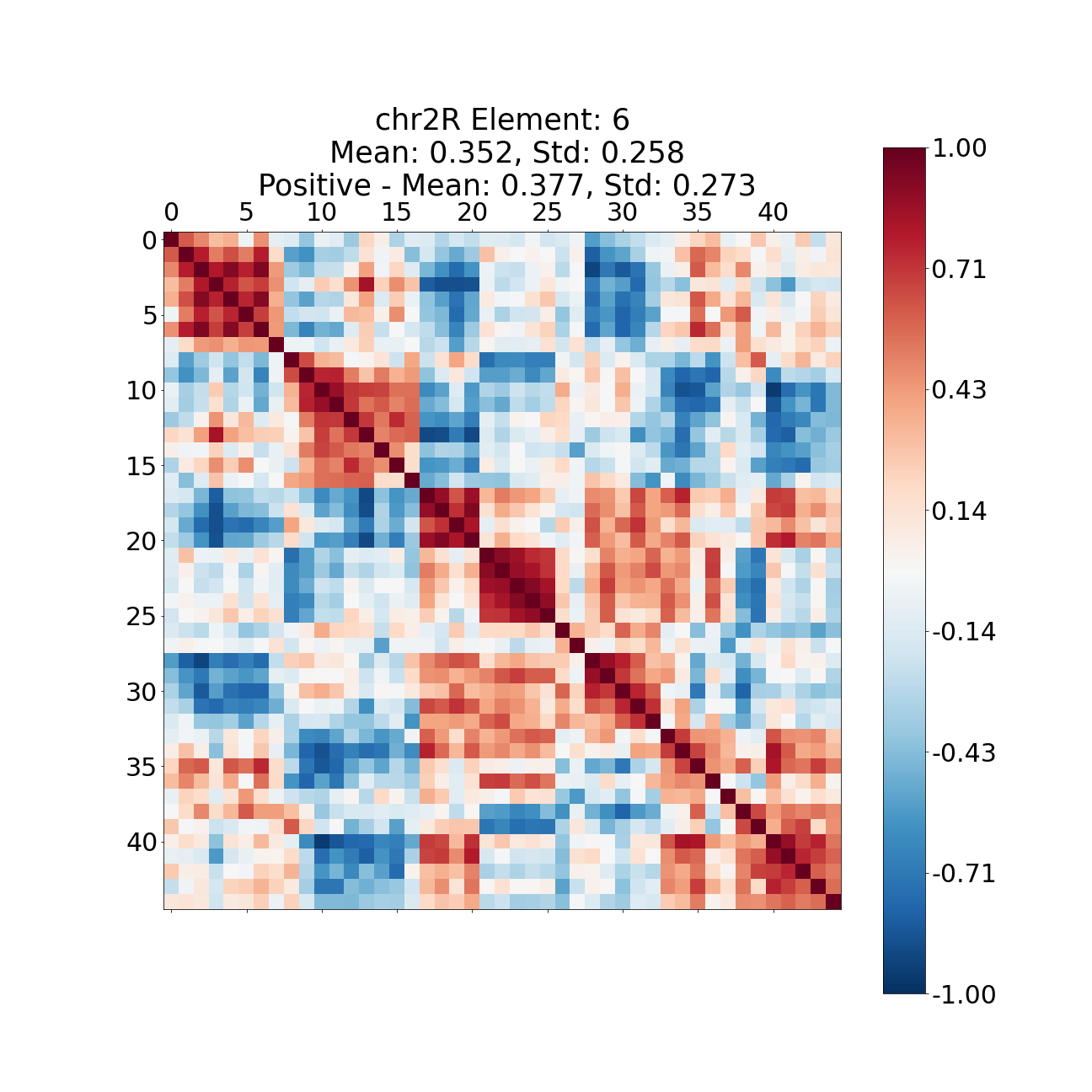}
     \end{subfigure}
      \hfill
     \begin{subfigure}[b]{0.32\textwidth}
         \centering
         \includegraphics[width=\textwidth]{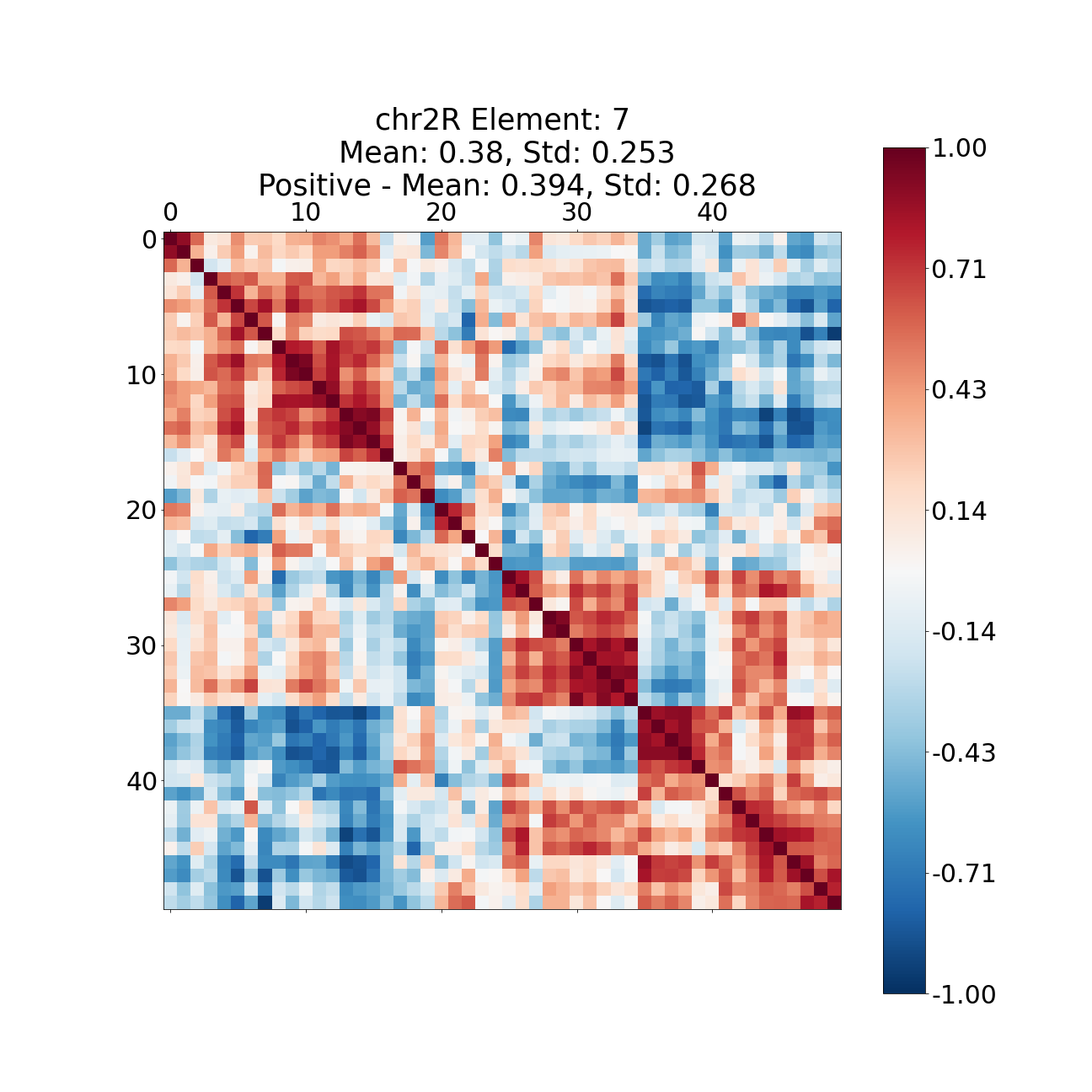}
     \end{subfigure}
     \hfill
     \begin{subfigure}[b]{0.32\textwidth}
         \centering
         \includegraphics[width=\textwidth]{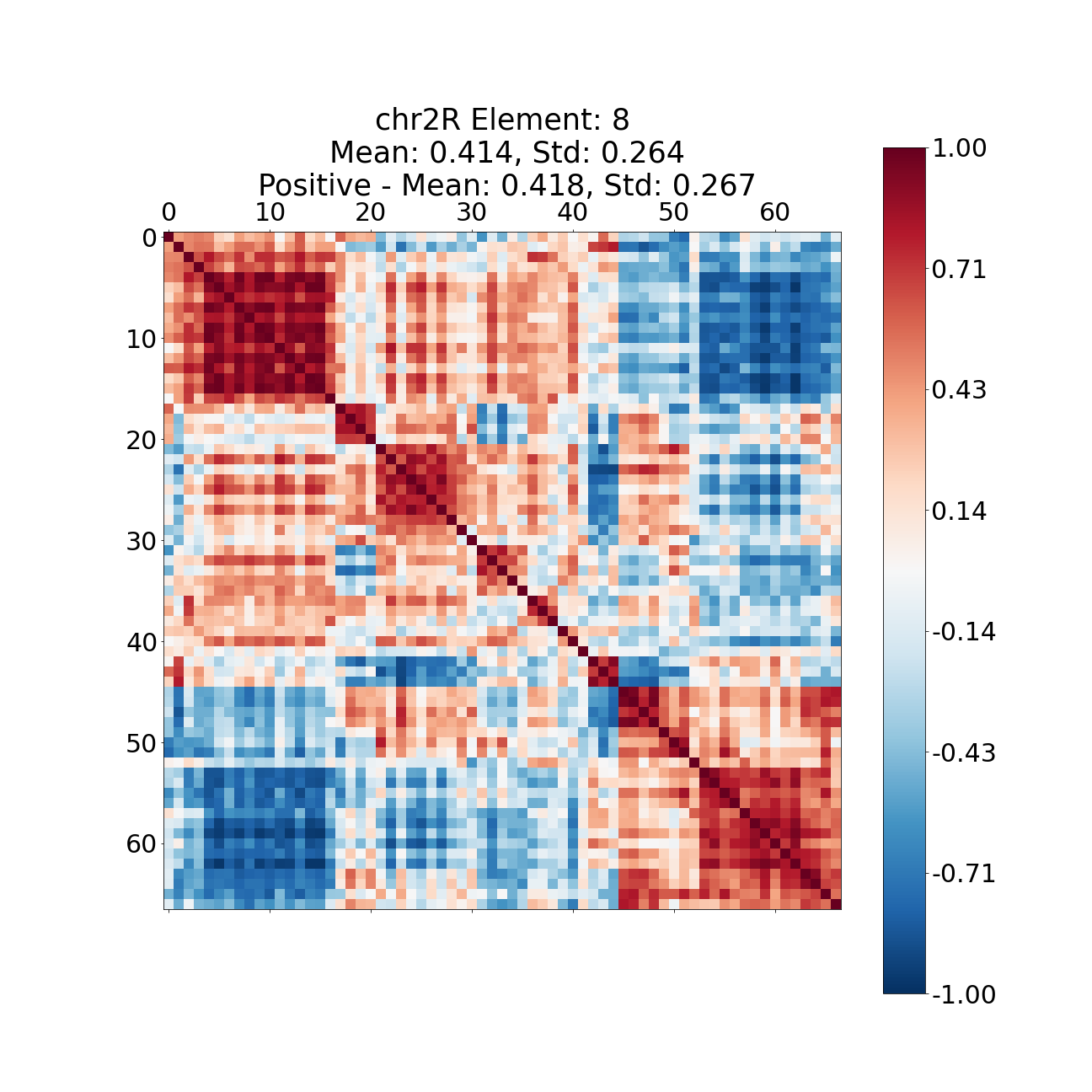}
     \end{subfigure}
     \hfill
     \begin{subfigure}[b]{0.32\textwidth}
         \centering
         \includegraphics[width=\textwidth]{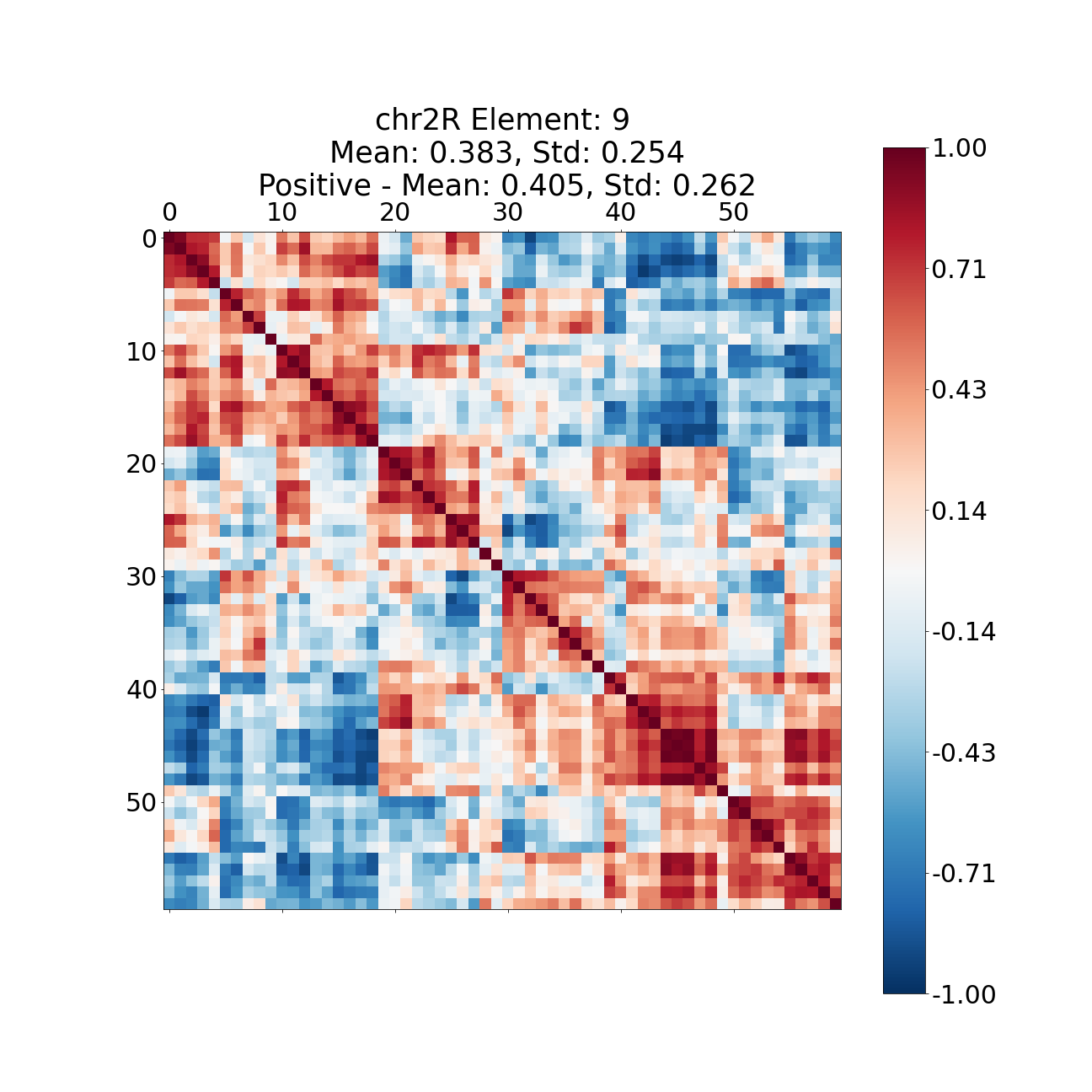}
     \end{subfigure}
     \hfill
     \begin{subfigure}[b]{0.32\textwidth}
         \centering
         \includegraphics[width=\textwidth]{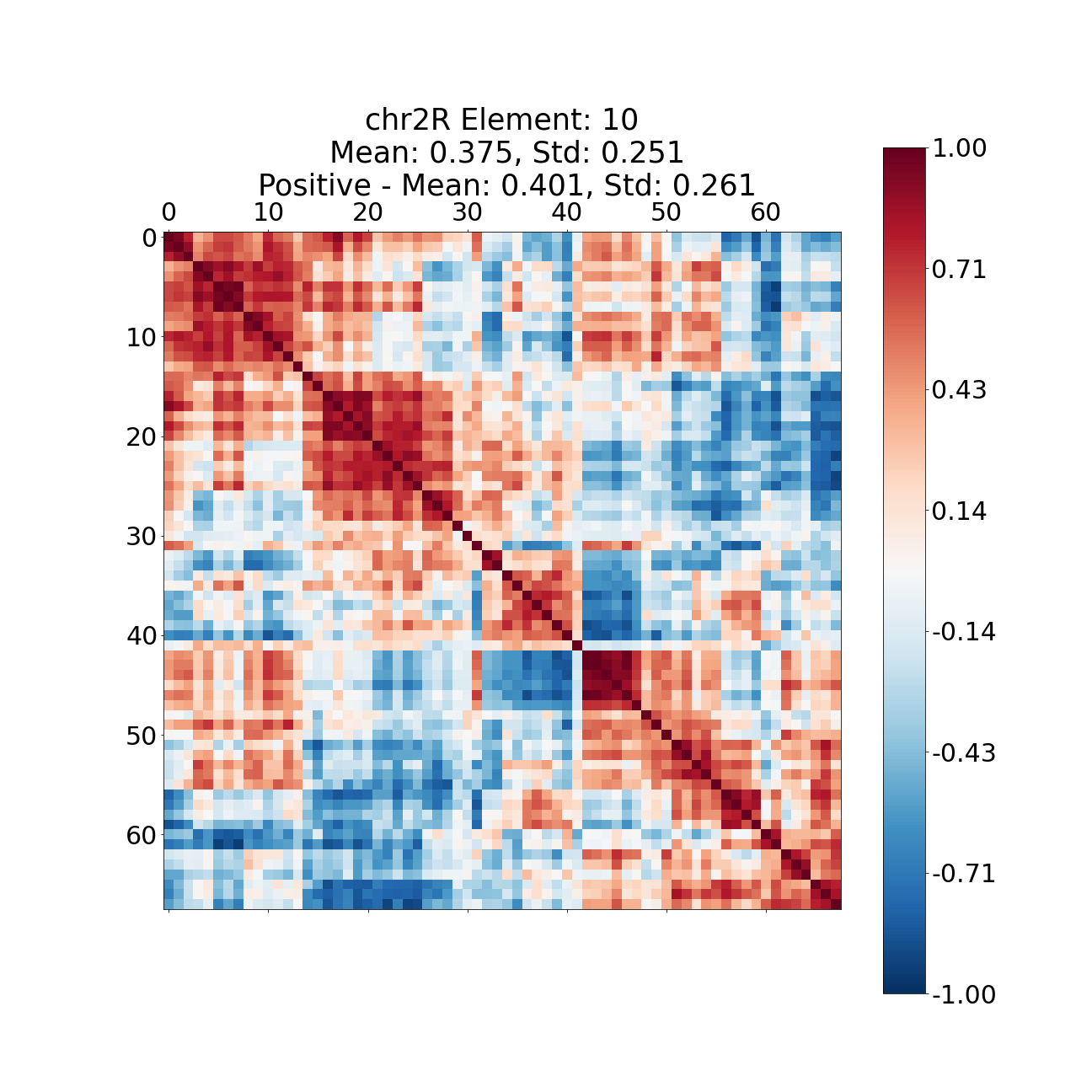}
     \end{subfigure}
      \hfill
     \begin{subfigure}[b]{0.32\textwidth}
         \centering
         \includegraphics[width=\textwidth]{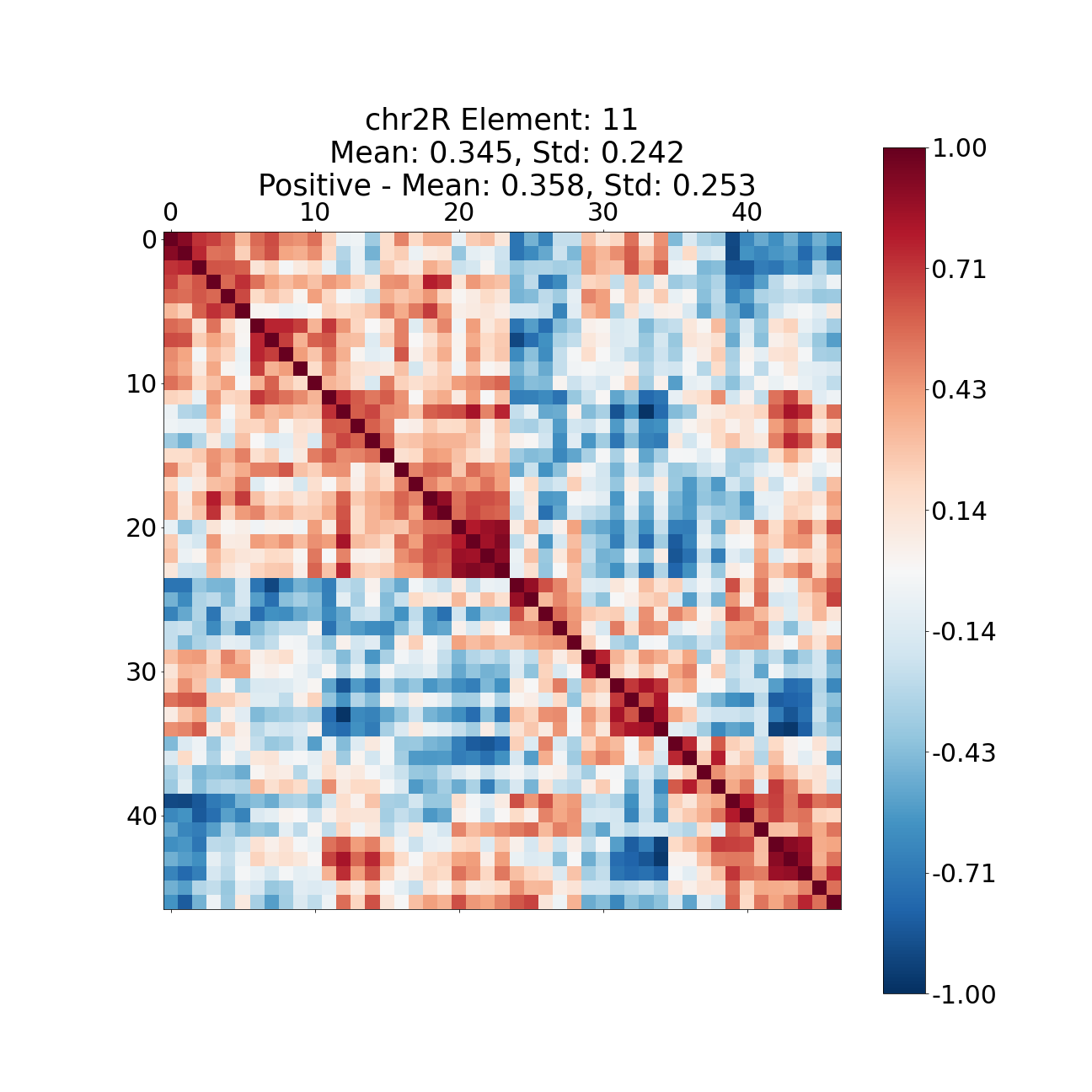}
     \end{subfigure}
        \caption{Pairwise coexpression of genes covered by various dictionary elements for chr 2R obtained through online cvxNDL. We calculated the mean and standard deviation of absolute pairwise coexpression values, along with the mean and standard deviation of coexpression values specifically for all positively correlated gene pairs.}
        \label{fig:dee2_pearson2R}
\end{figure}

\begin{figure}[h]
\ContinuedFloat
     \centering
     \begin{subfigure}[b]{0.32\textwidth}
         \centering
         \includegraphics[width=\textwidth]{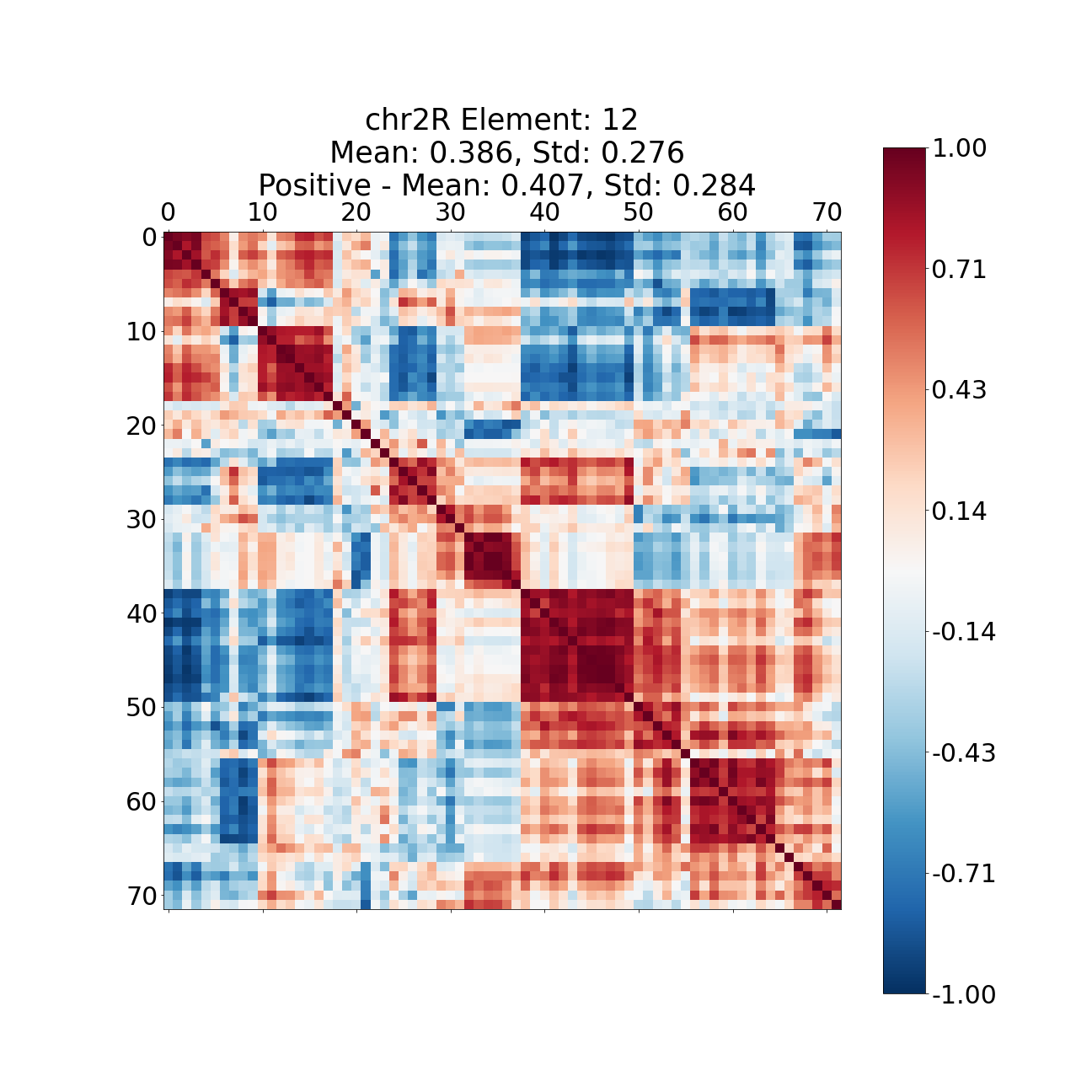}
     \end{subfigure}
     \hfill
     \begin{subfigure}[b]{0.32\textwidth}
         \centering
         \includegraphics[width=\textwidth]{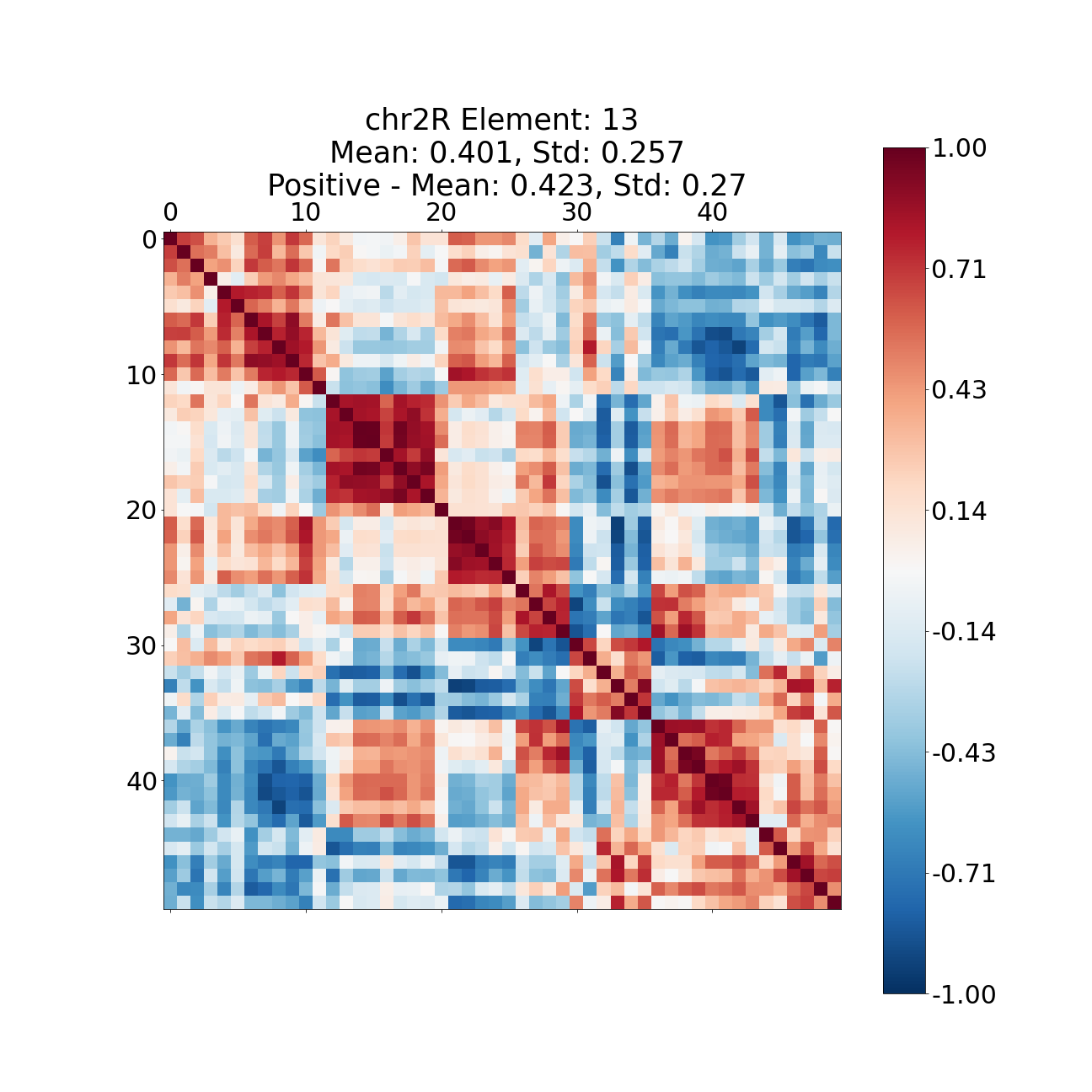}
     \end{subfigure}
     \hfill
     \begin{subfigure}[b]{0.32\textwidth}
         \centering
         \includegraphics[width=\textwidth]{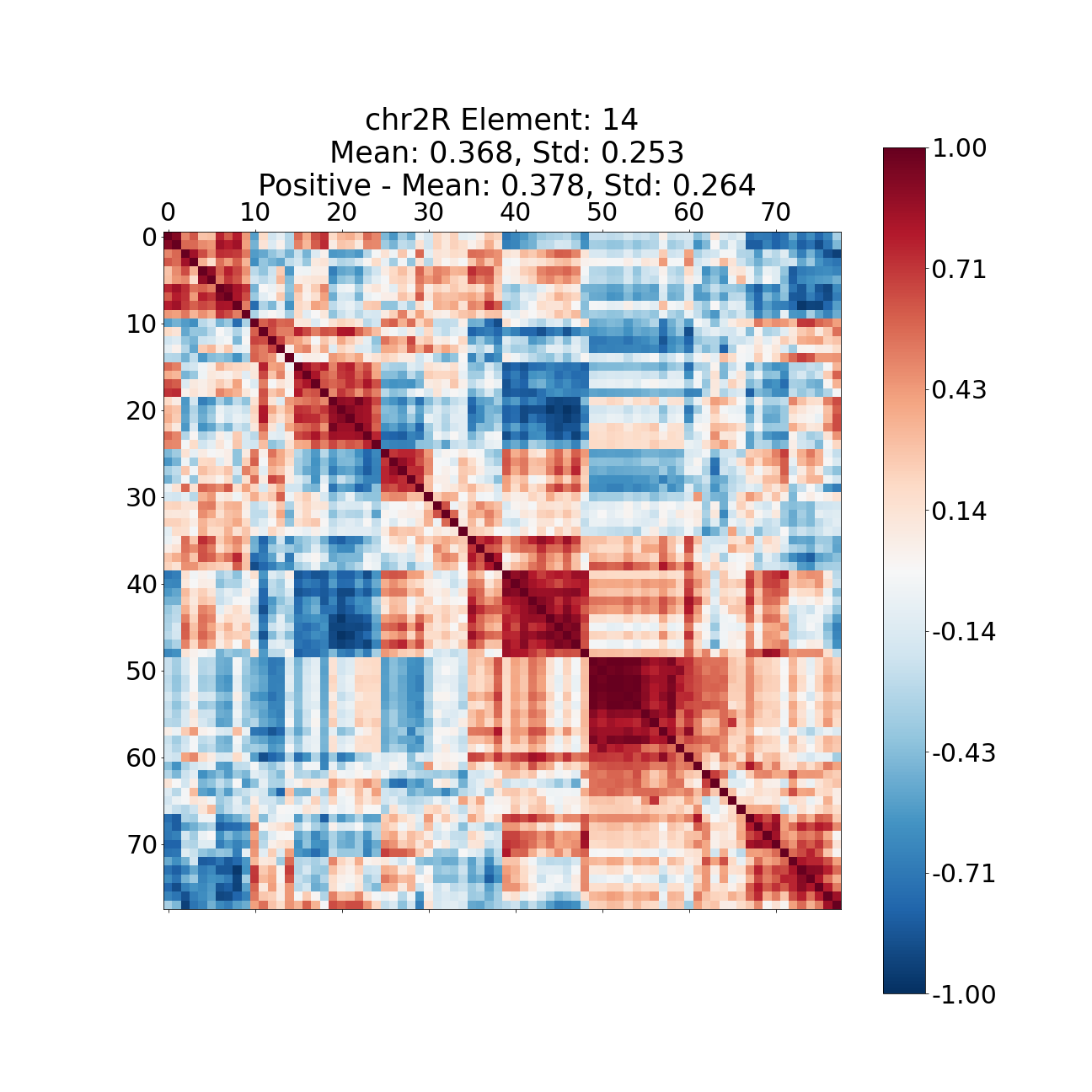}
     \end{subfigure}
      \hfill
     \begin{subfigure}[b]{0.32\textwidth}
         \centering
         \includegraphics[width=\textwidth]{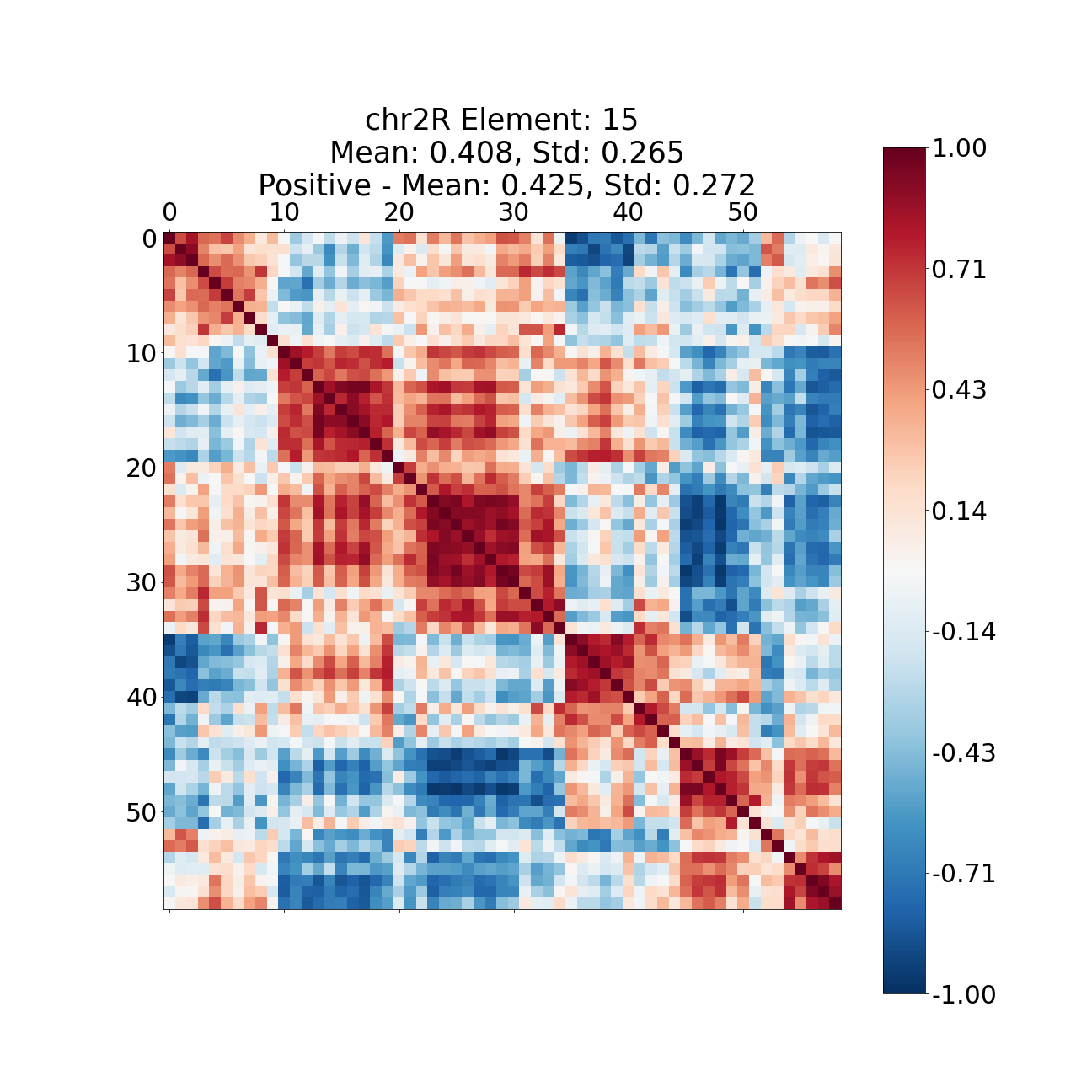}
     \end{subfigure}
     \hfill
     \begin{subfigure}[b]{0.32\textwidth}
         \centering
         \includegraphics[width=\textwidth]{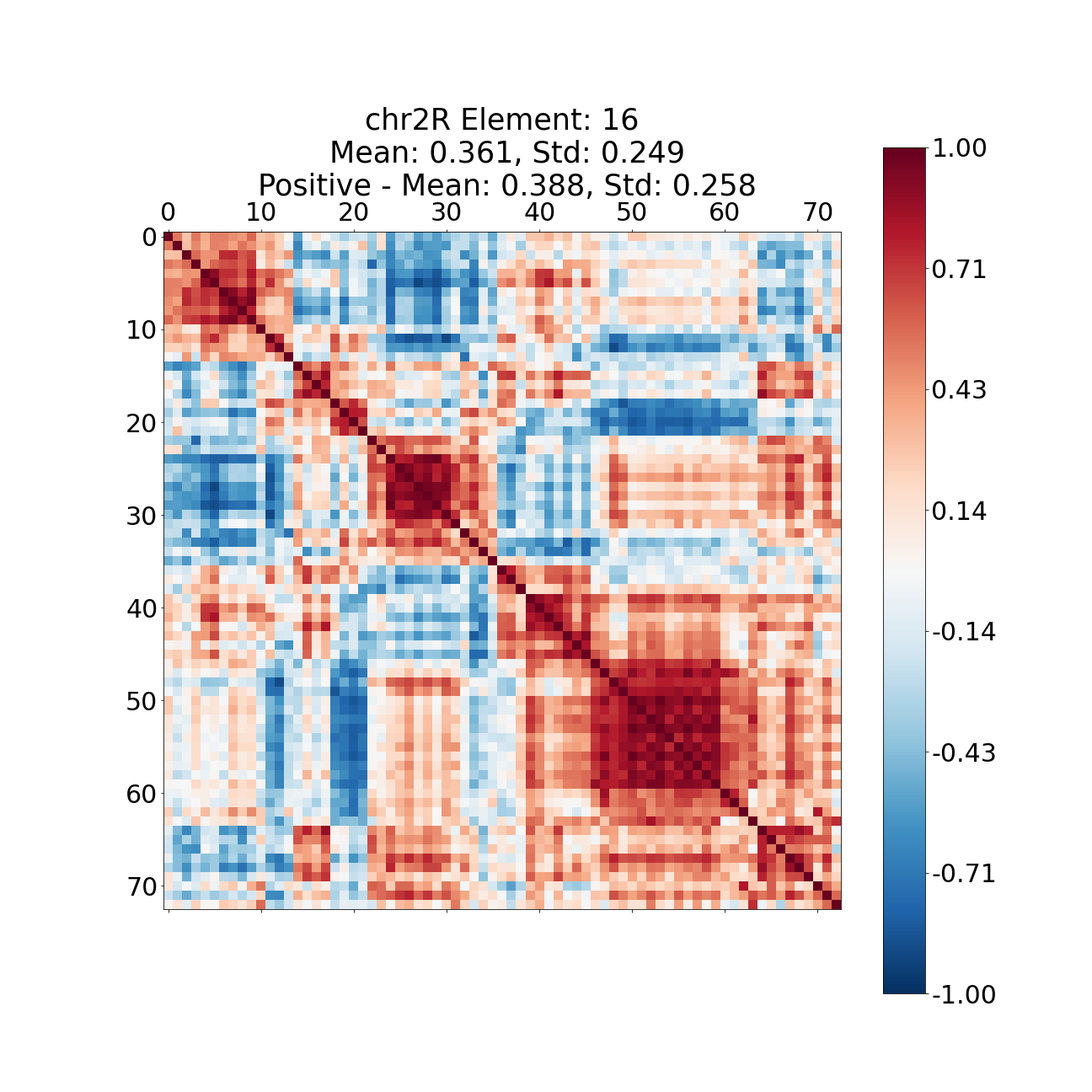}
     \end{subfigure}
     \hfill
     \begin{subfigure}[b]{0.32\textwidth}
         \centering
         \includegraphics[width=\textwidth]{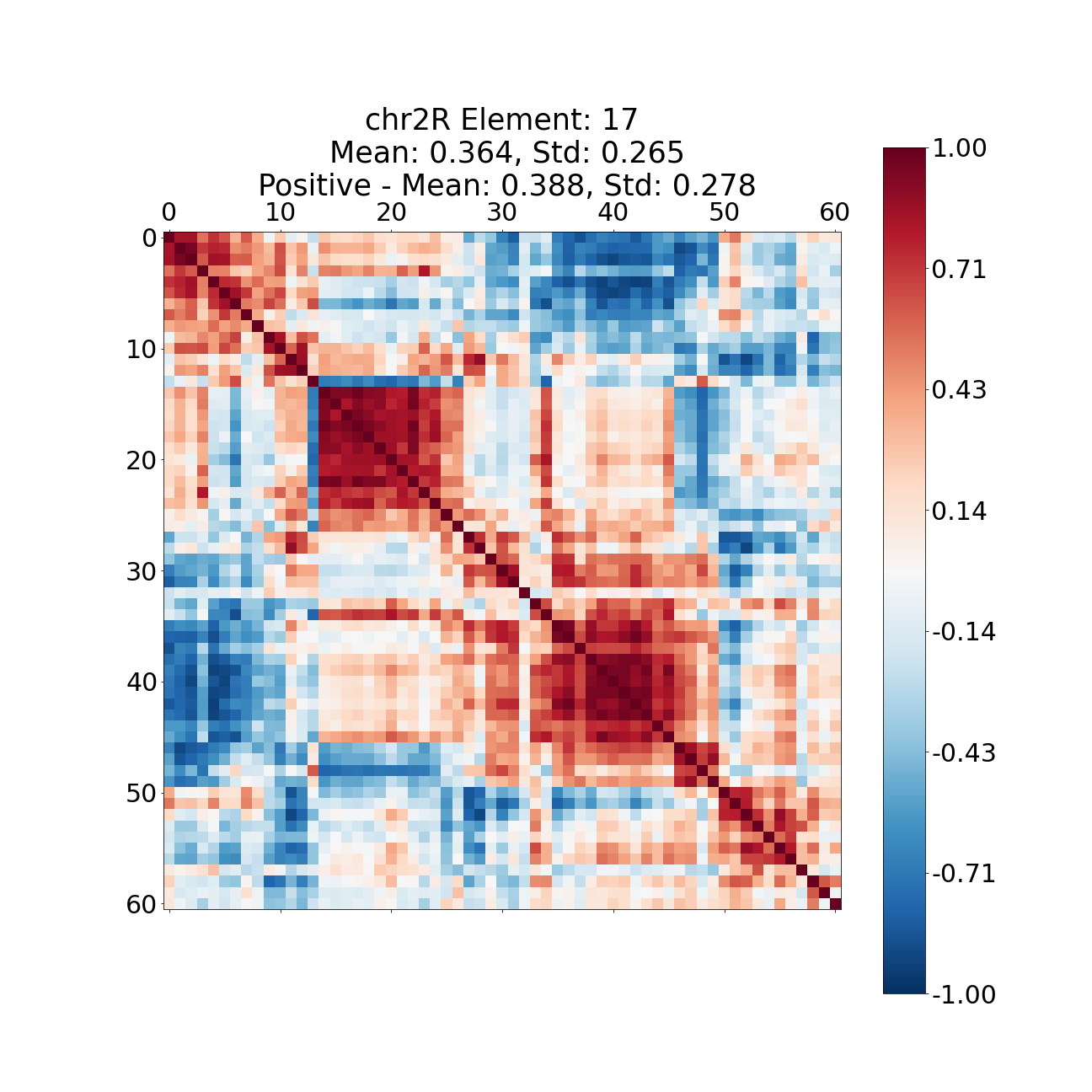}
     \end{subfigure}
     \hfill
     \begin{subfigure}[b]{0.32\textwidth}
         \centering
         \includegraphics[width=\textwidth]{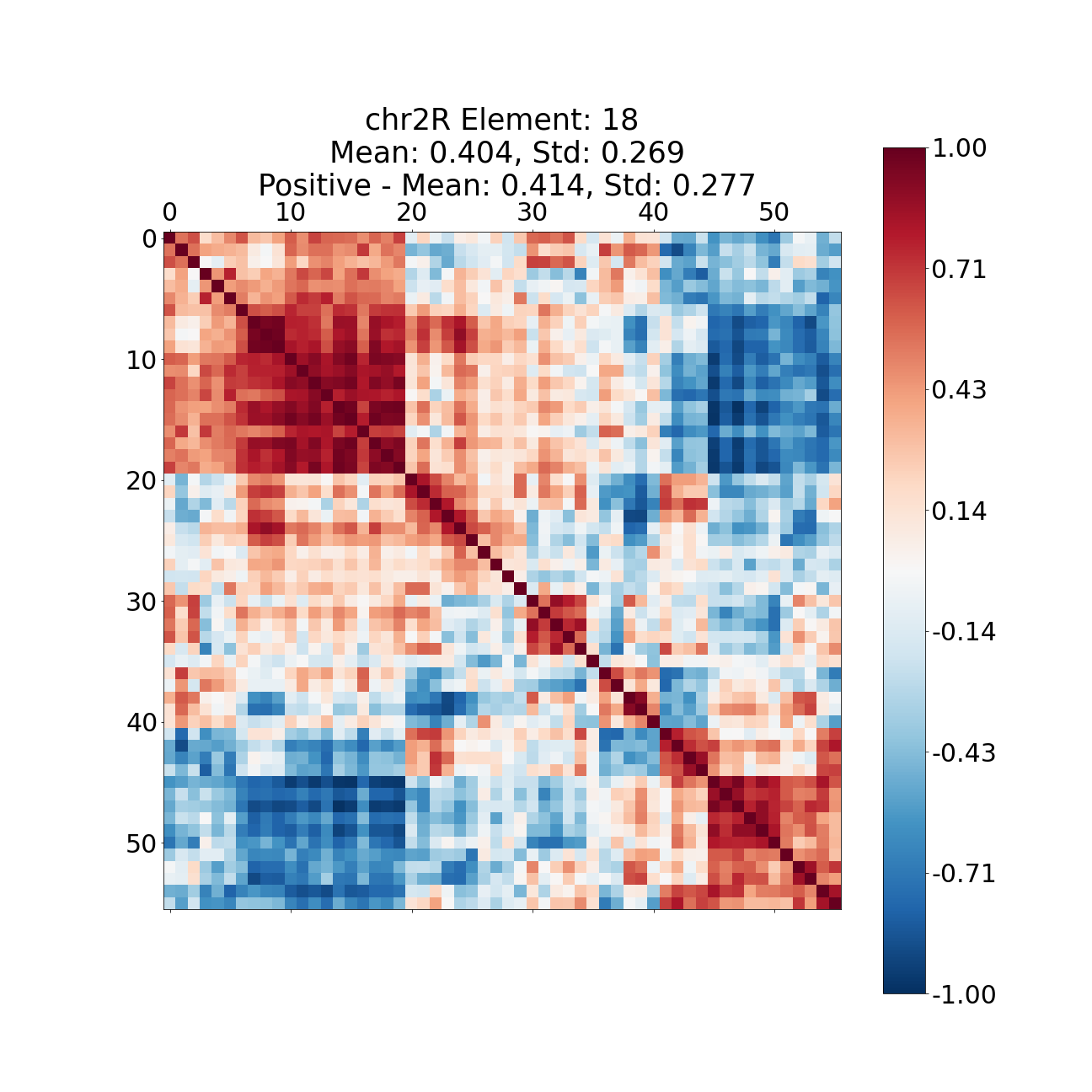}
     \end{subfigure}
      \hfill
     \begin{subfigure}[b]{0.32\textwidth}
         \centering
         \includegraphics[width=\textwidth]{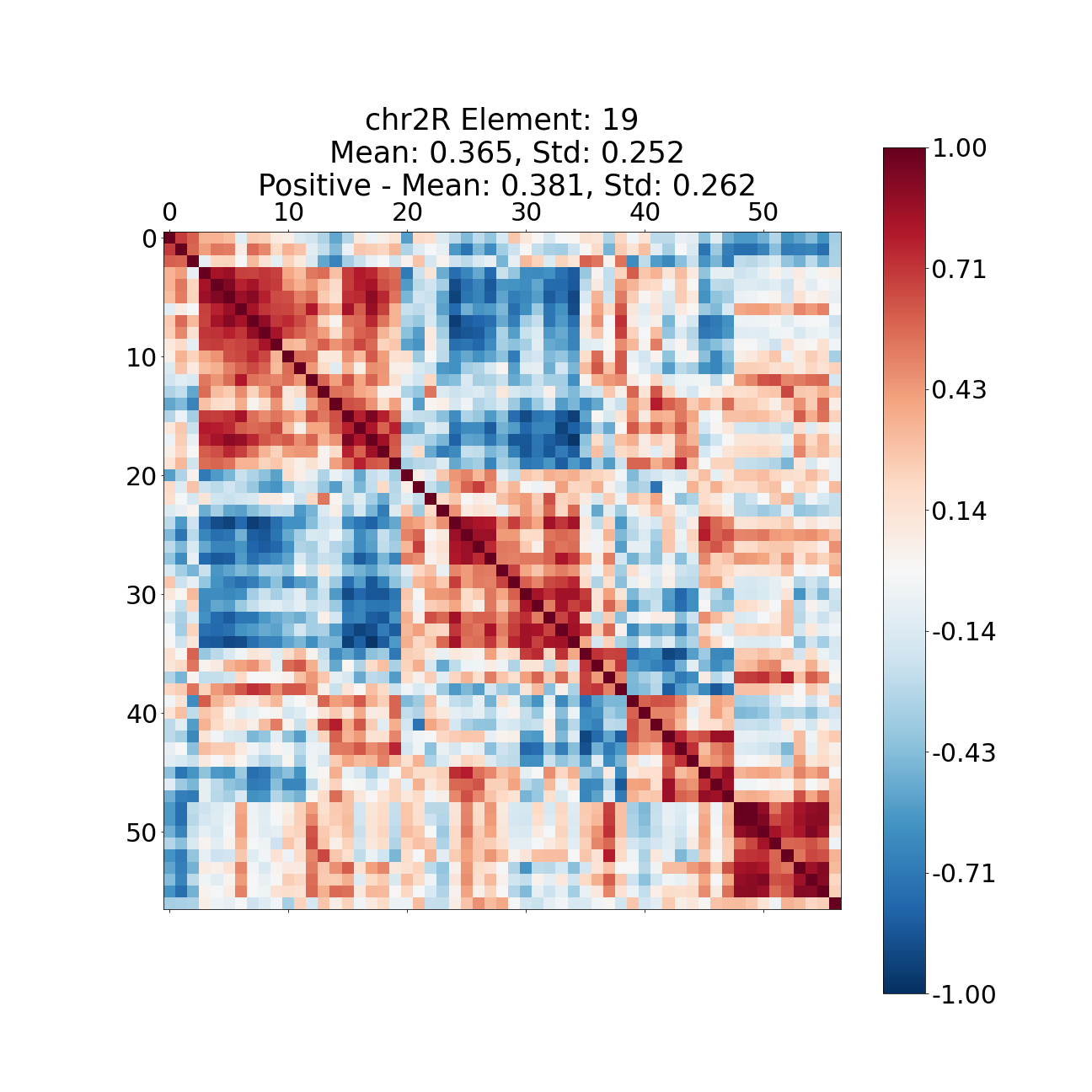}
     \end{subfigure}
     \hfill
     \begin{subfigure}[b]{0.32\textwidth}
         \centering
         \includegraphics[width=\textwidth]{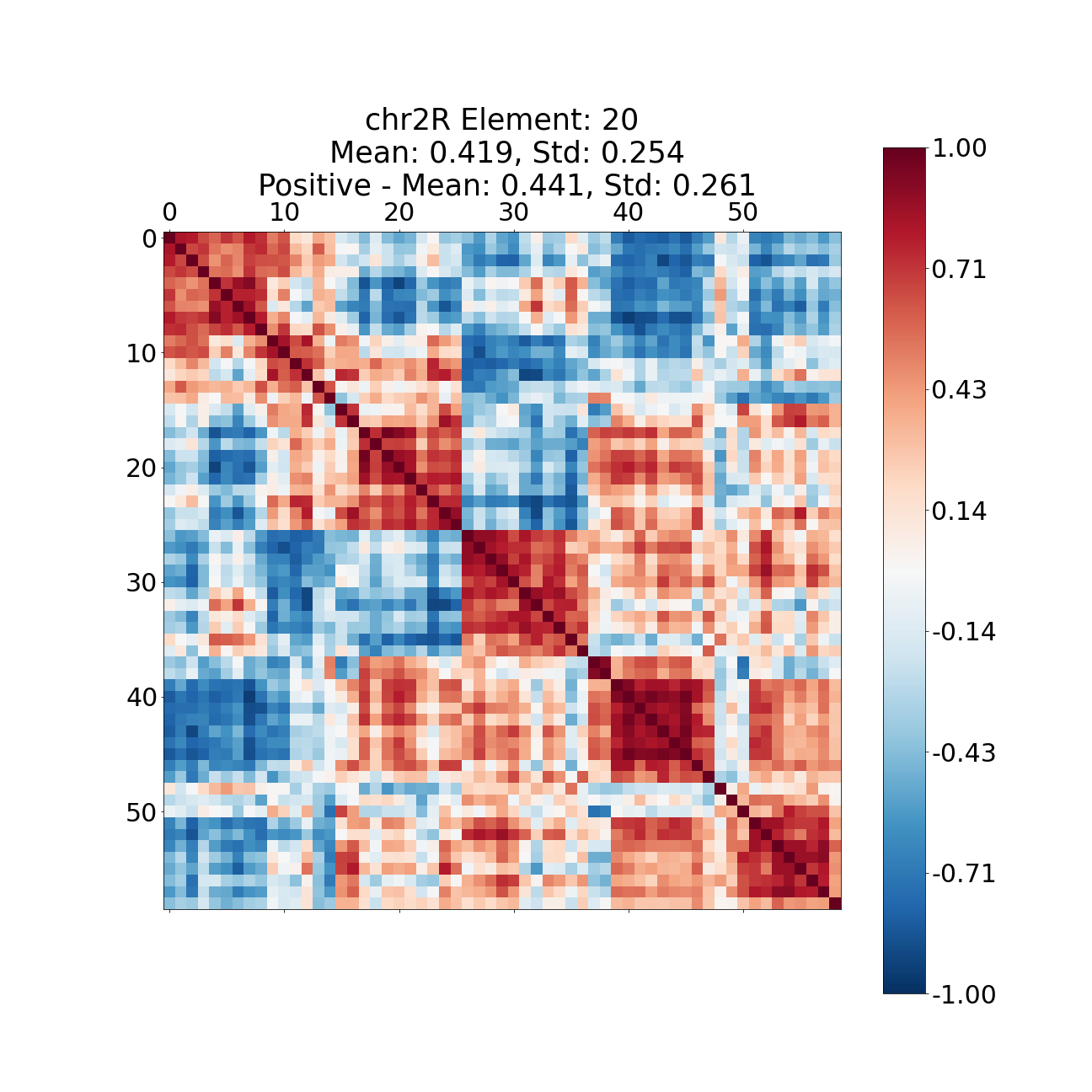}
     \end{subfigure}
     \hfill
     \begin{subfigure}[b]{0.32\textwidth}
         \centering
         \includegraphics[width=\textwidth]{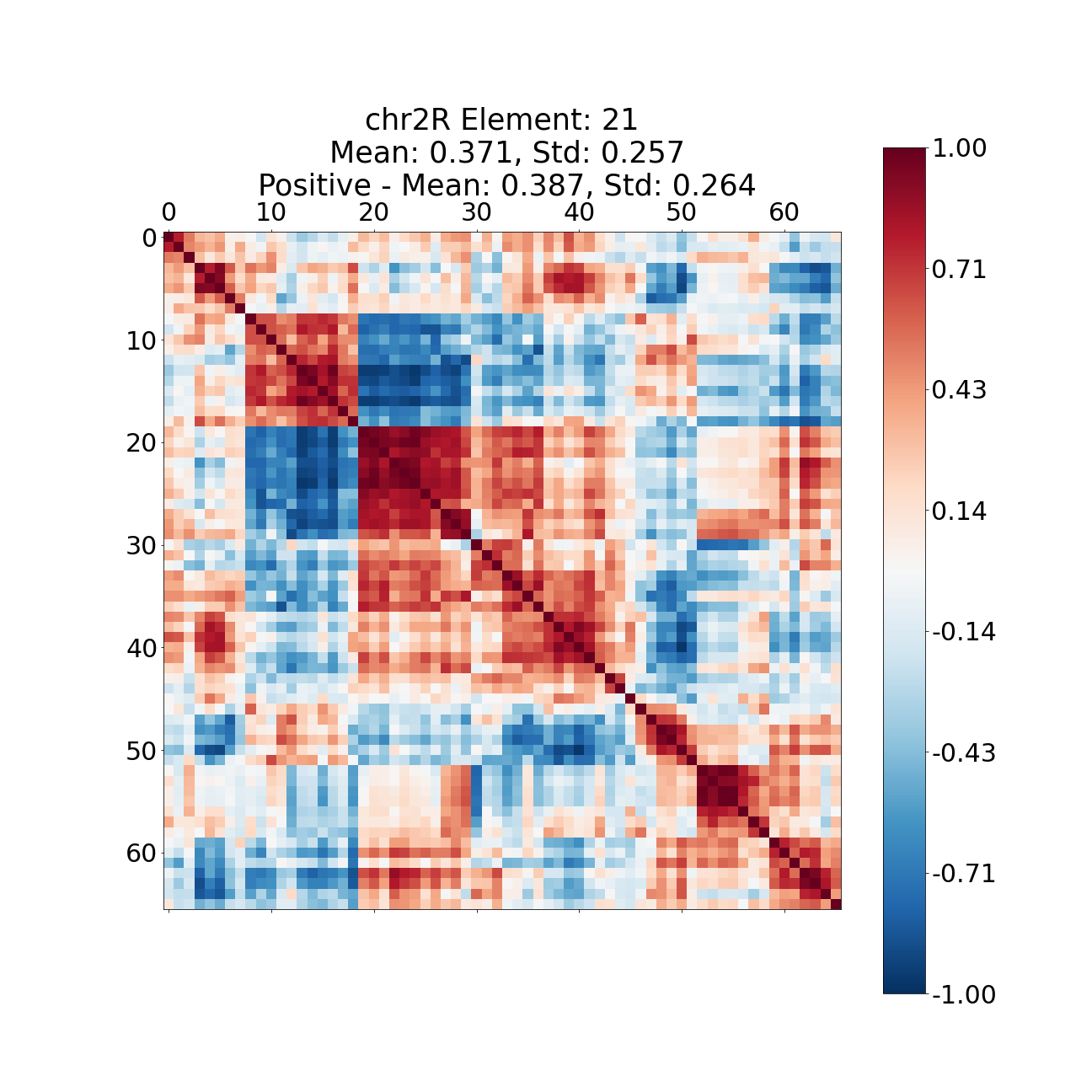}
     \end{subfigure}
     \hfill
     \begin{subfigure}[b]{0.32\textwidth}
         \centering
         \includegraphics[width=\textwidth]{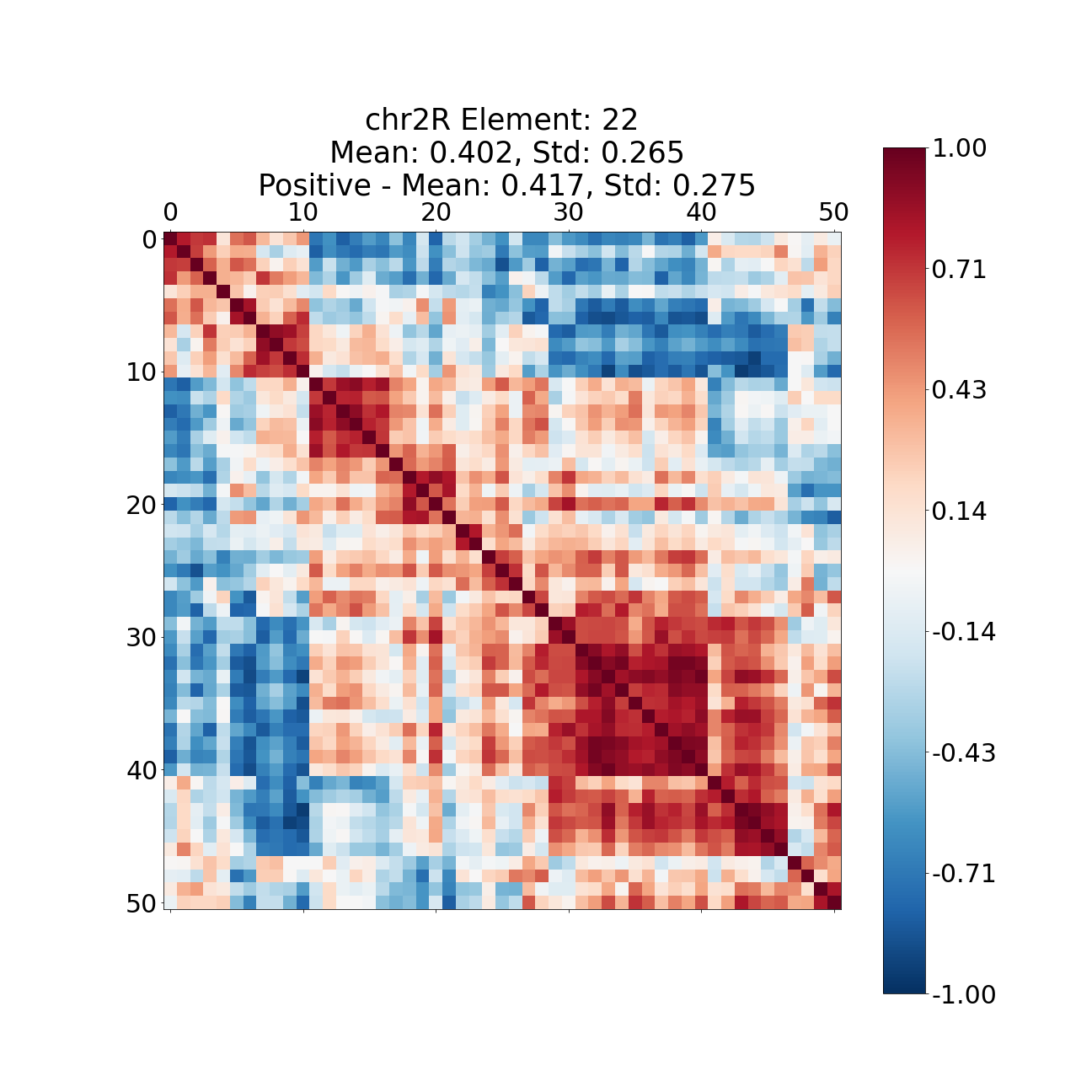}
     \end{subfigure}
      \hfill
     \begin{subfigure}[b]{0.32\textwidth}
         \centering
         \includegraphics[width=\textwidth]{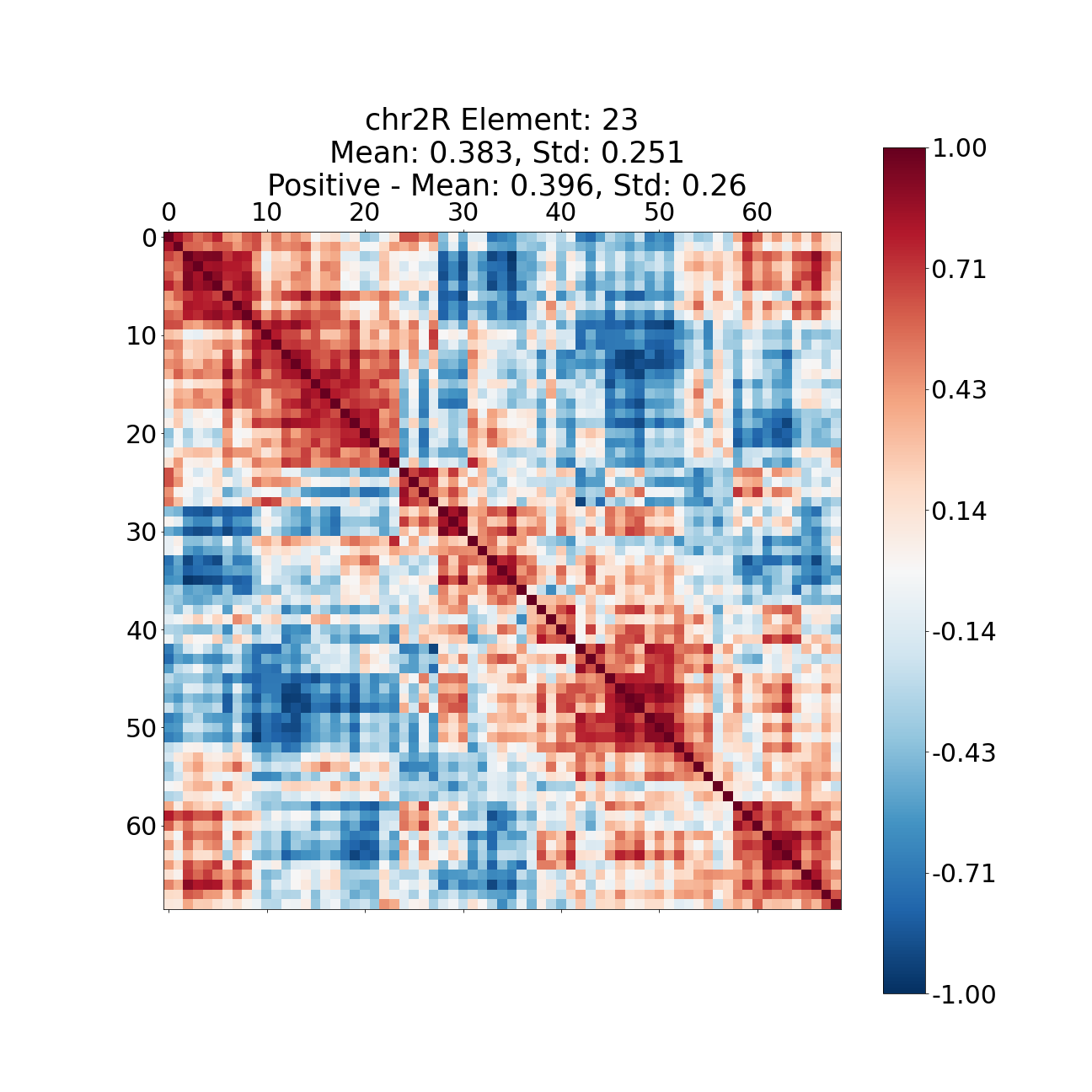}
     \end{subfigure}
        \caption{Pairwise coexpression of genes covered by various dictionary elements for chr 2R obtained through online cvxNDL. We calculated the mean and standard deviation of absolute pairwise coexpression values, along with the mean and standard deviation of coexpression values specifically for all positively correlated gene pairs.}
\end{figure}

\begin{figure}[h]
\ContinuedFloat
     \centering
     \begin{subfigure}[b]{0.32\textwidth}
         \centering
         \includegraphics[width=\textwidth]{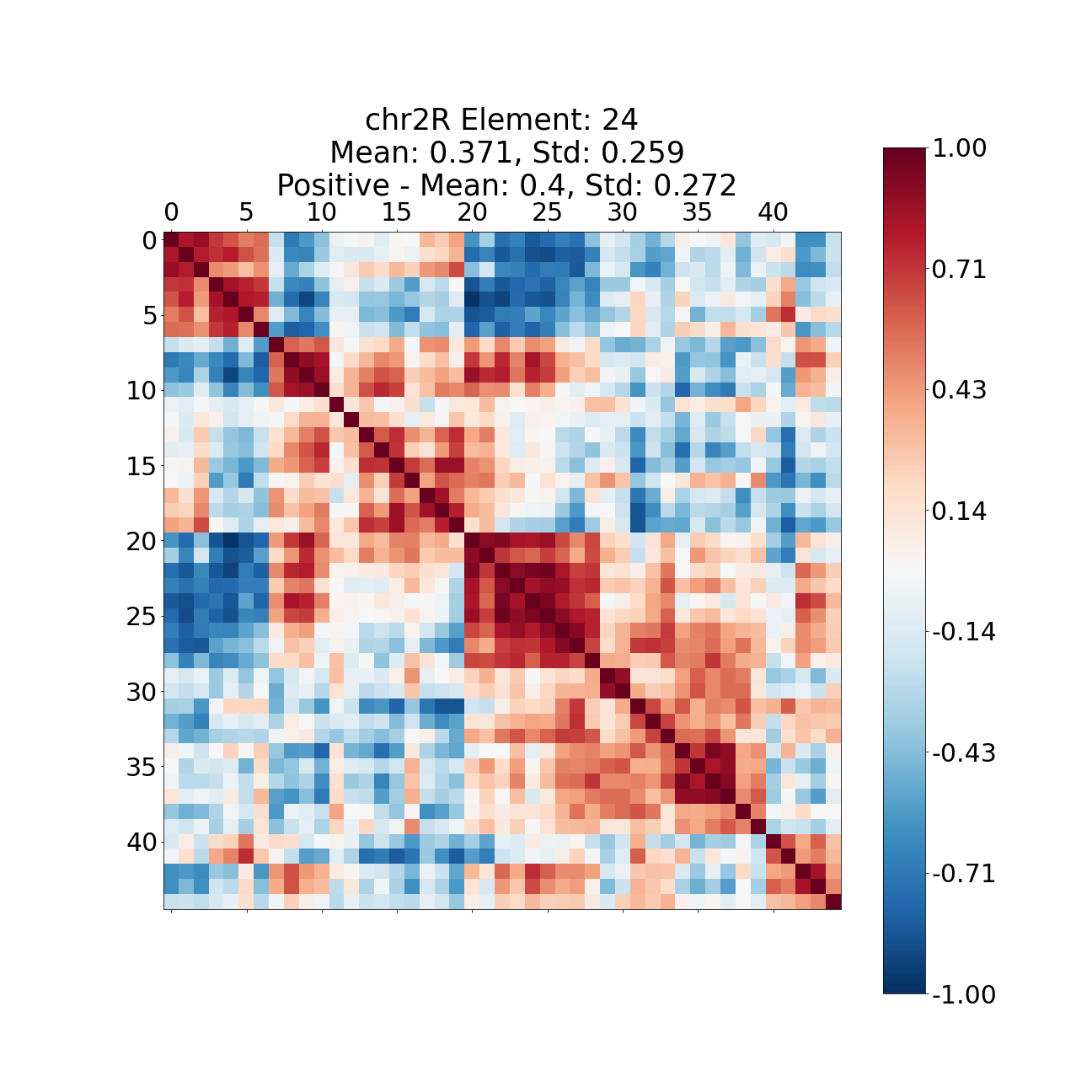}
     \end{subfigure}
        \caption{Pairwise coexpression of genes covered by various dictionary elements for chr 2R obtained through online cvxNDL. We calculated the mean and standard deviation of absolute pairwise coexpression values, along with the mean and standard deviation of coexpression values specifically for all positively correlated gene pairs.}
\end{figure}


\begin{figure}[h]
     \centering
     \begin{subfigure}[b]{0.32\textwidth}
         \centering
         \includegraphics[width=\textwidth]{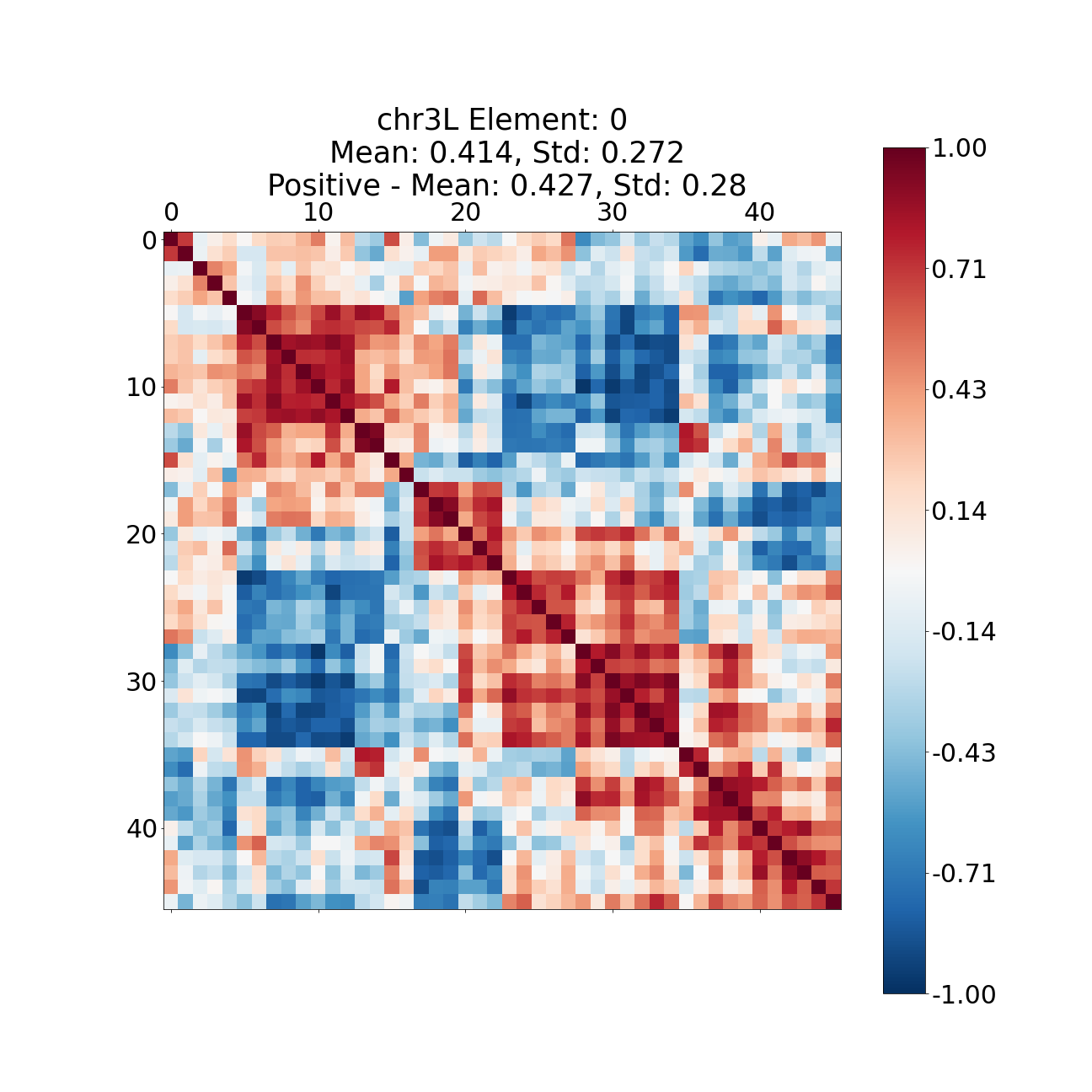}
     \end{subfigure}
     \hfill
     \begin{subfigure}[b]{0.32\textwidth}
         \centering
         \includegraphics[width=\textwidth]{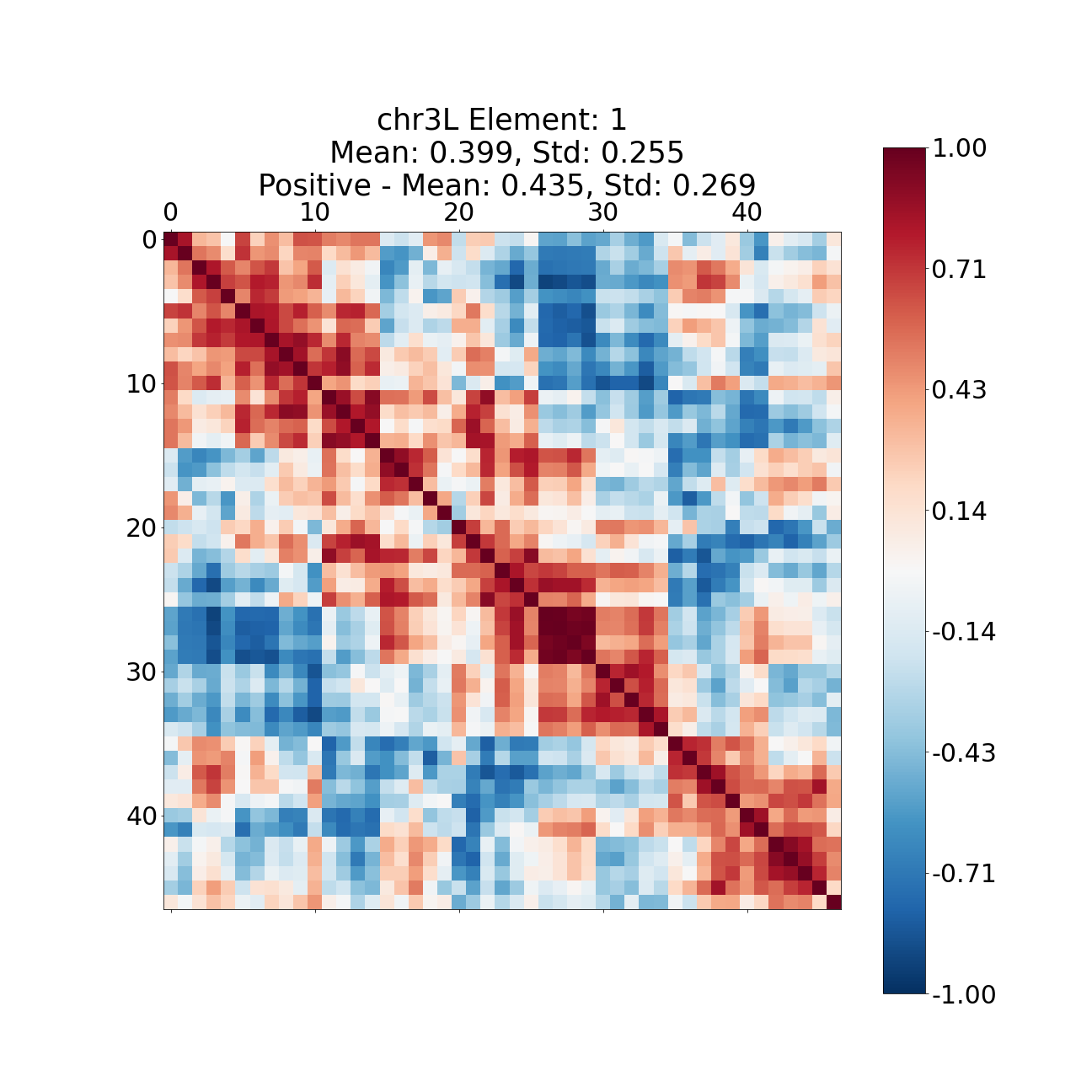}
     \end{subfigure}
     \hfill
     \begin{subfigure}[b]{0.32\textwidth}
         \centering
         \includegraphics[width=\textwidth]{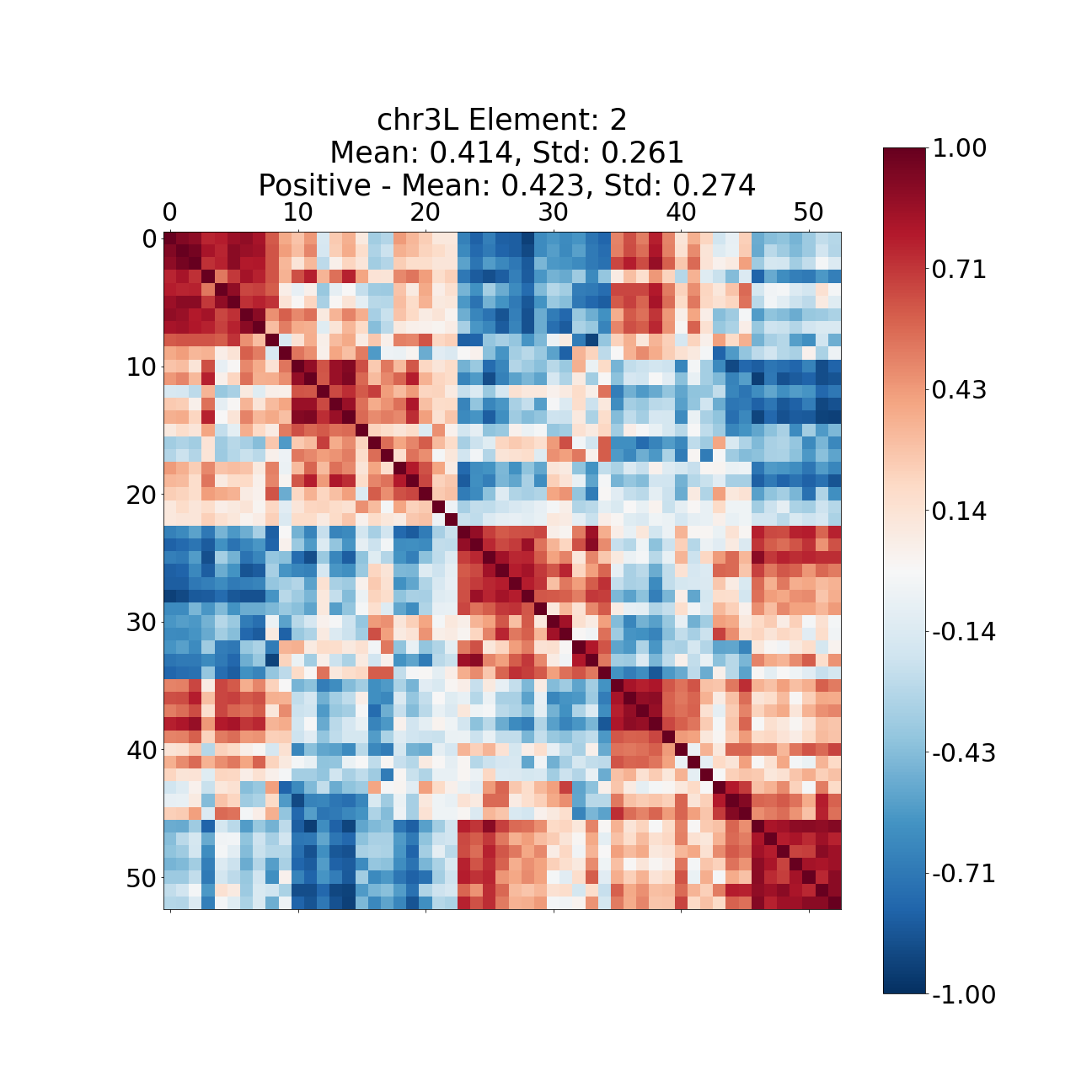}
     \end{subfigure}
      \hfill
     \begin{subfigure}[b]{0.32\textwidth}
         \centering
         \includegraphics[width=\textwidth]{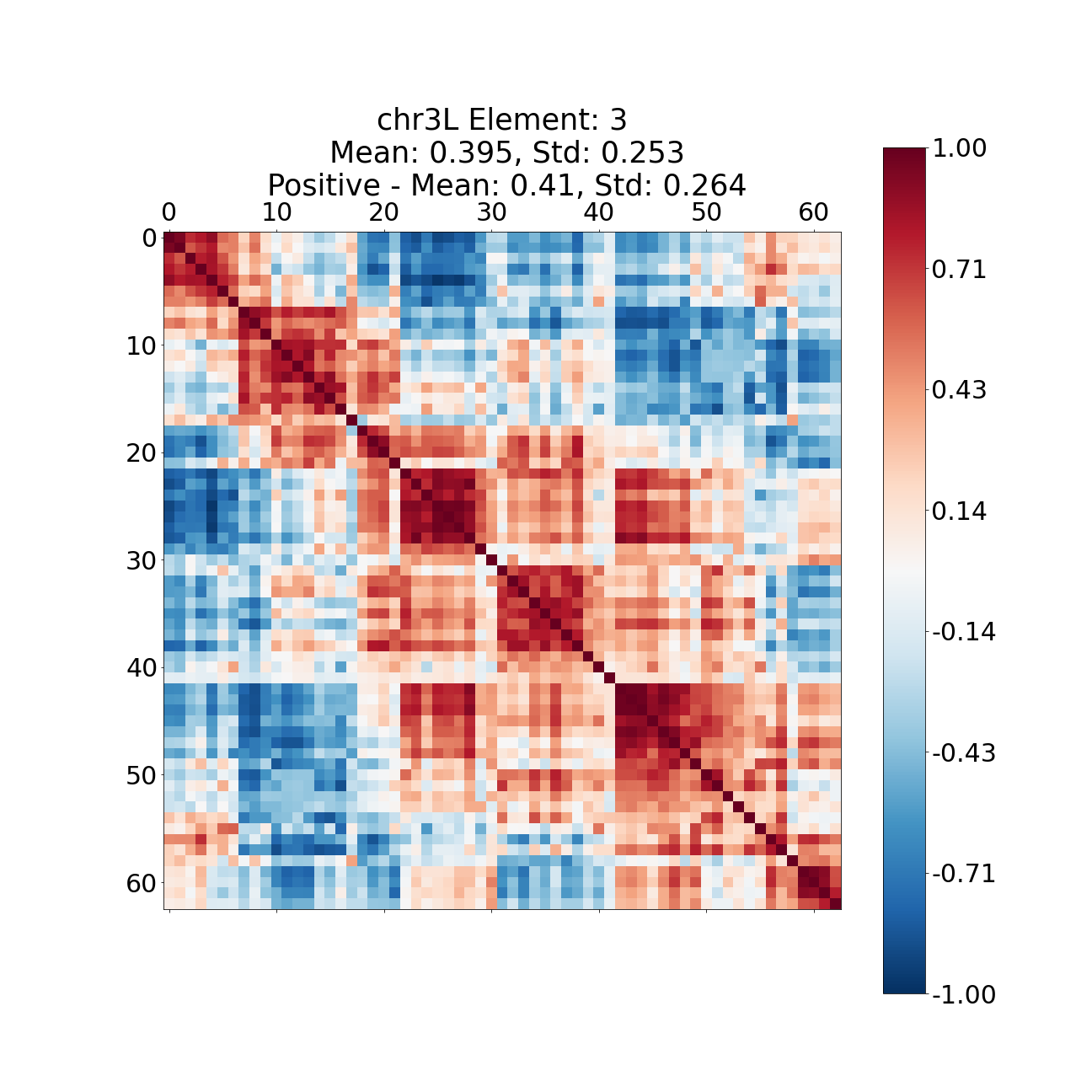}
     \end{subfigure}
     \hfill
     \begin{subfigure}[b]{0.32\textwidth}
         \centering
         \includegraphics[width=\textwidth]{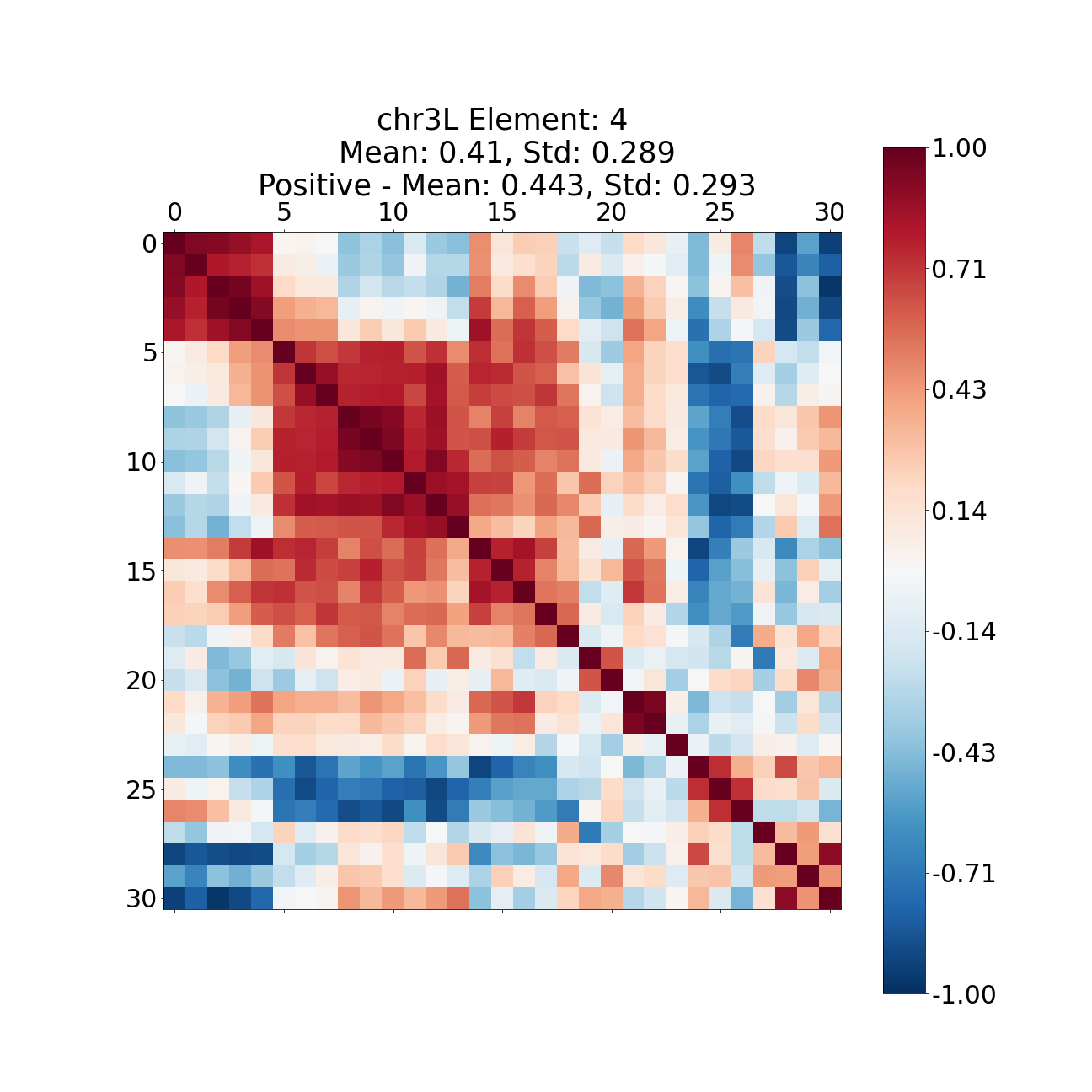}
     \end{subfigure}
     \hfill
     \begin{subfigure}[b]{0.32\textwidth}
         \centering
         \includegraphics[width=\textwidth]{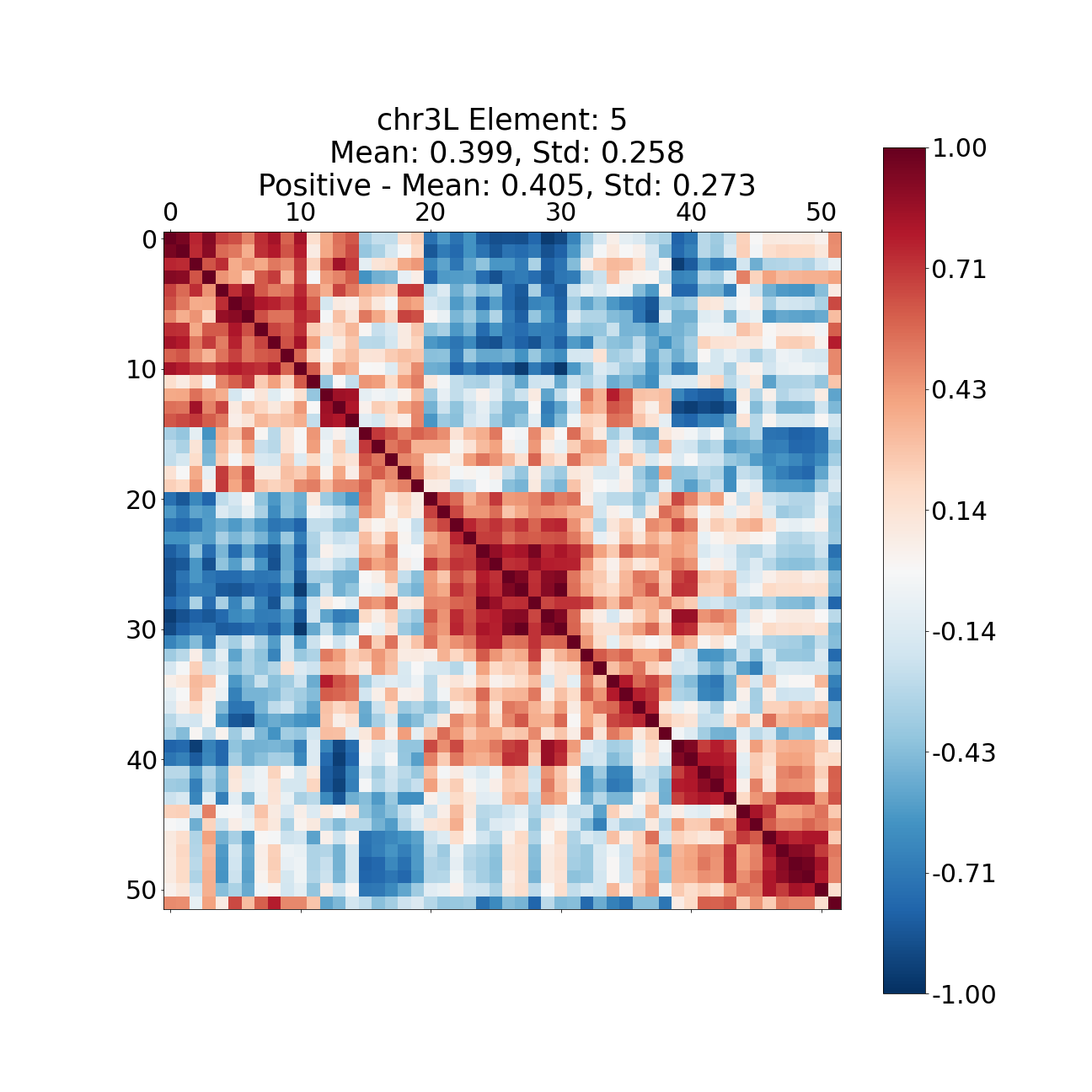}
     \end{subfigure}
     \hfill
     \begin{subfigure}[b]{0.32\textwidth}
         \centering
         \includegraphics[width=\textwidth]{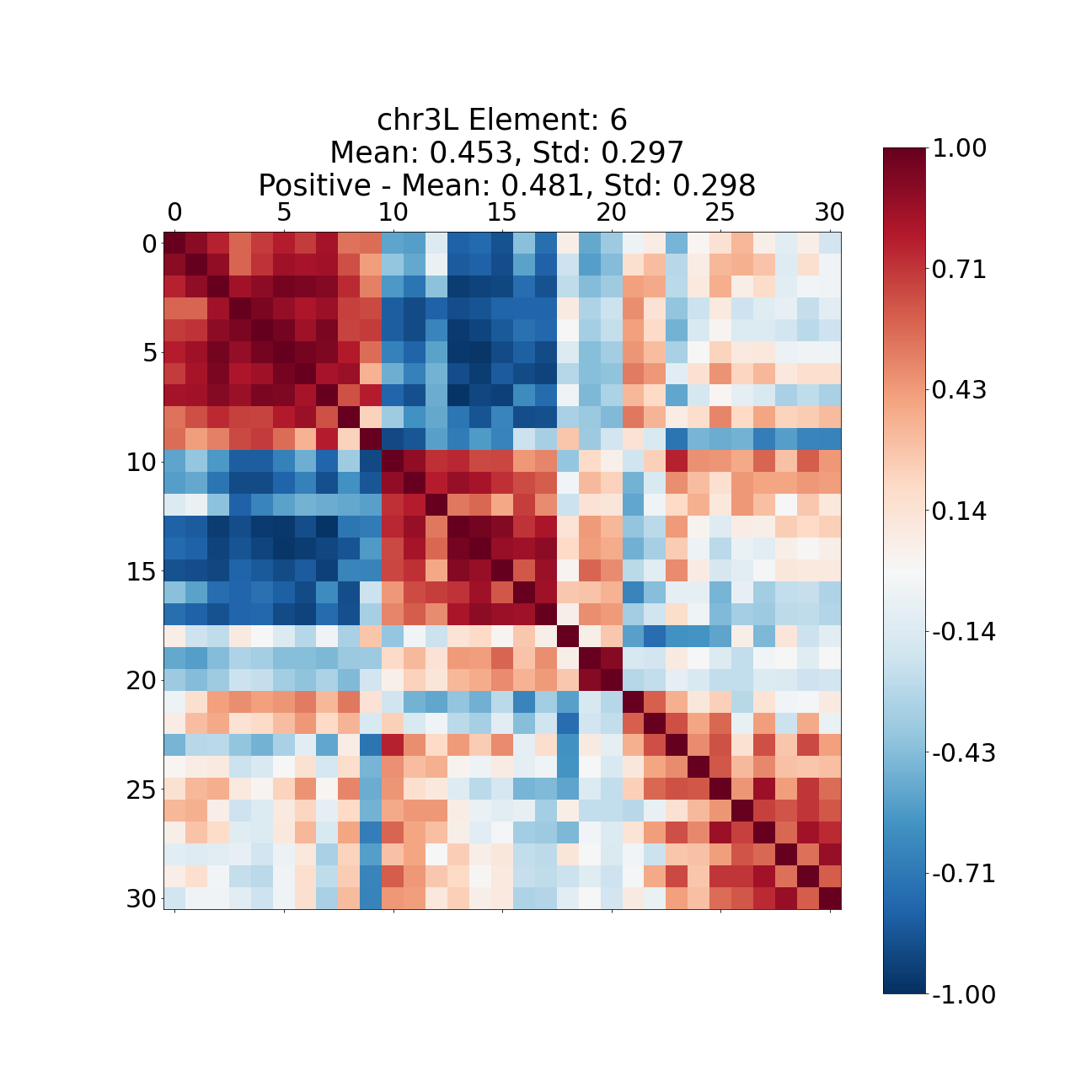}
     \end{subfigure}
      \hfill
     \begin{subfigure}[b]{0.32\textwidth}
         \centering
         \includegraphics[width=\textwidth]{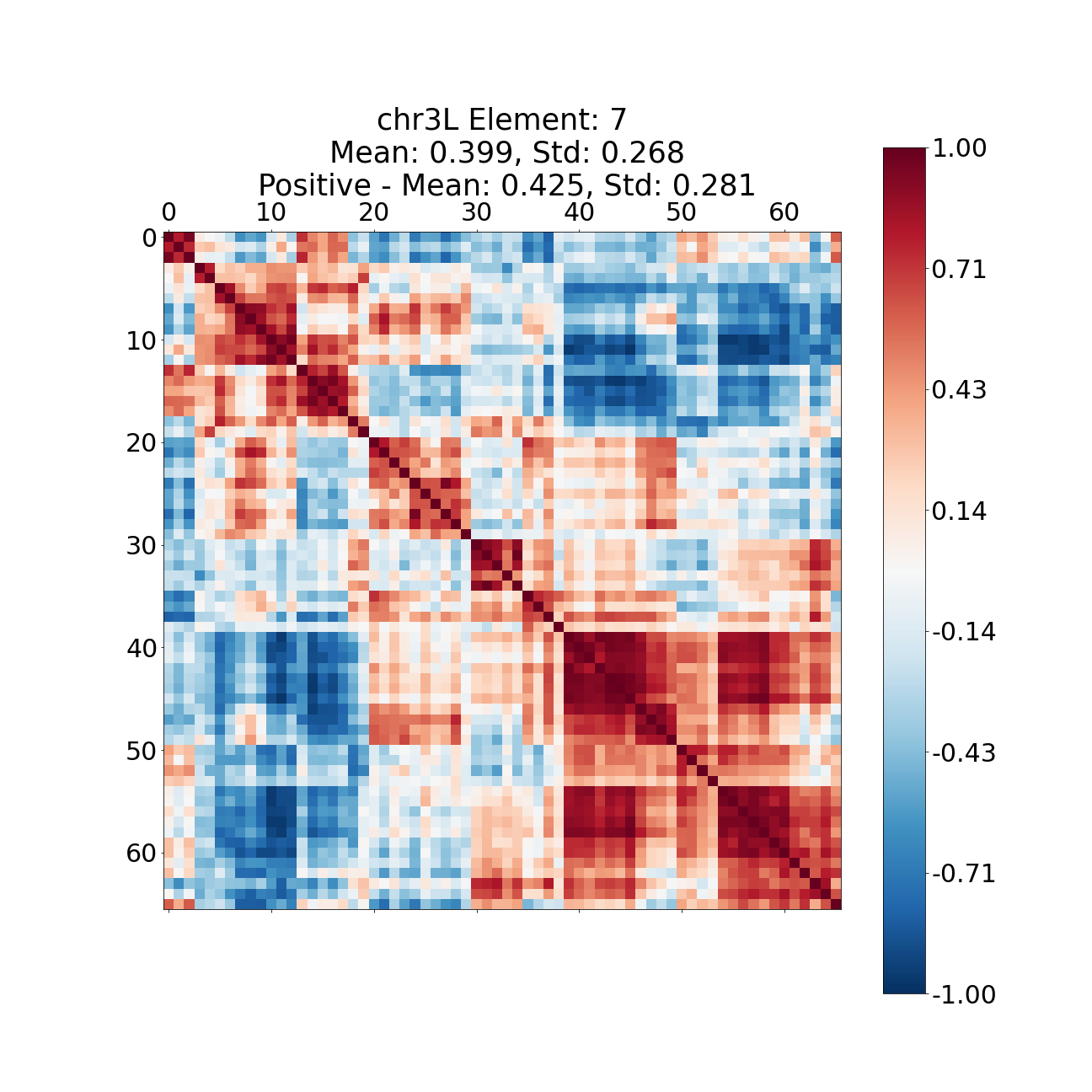}
     \end{subfigure}
     \hfill
     \begin{subfigure}[b]{0.32\textwidth}
         \centering
         \includegraphics[width=\textwidth]{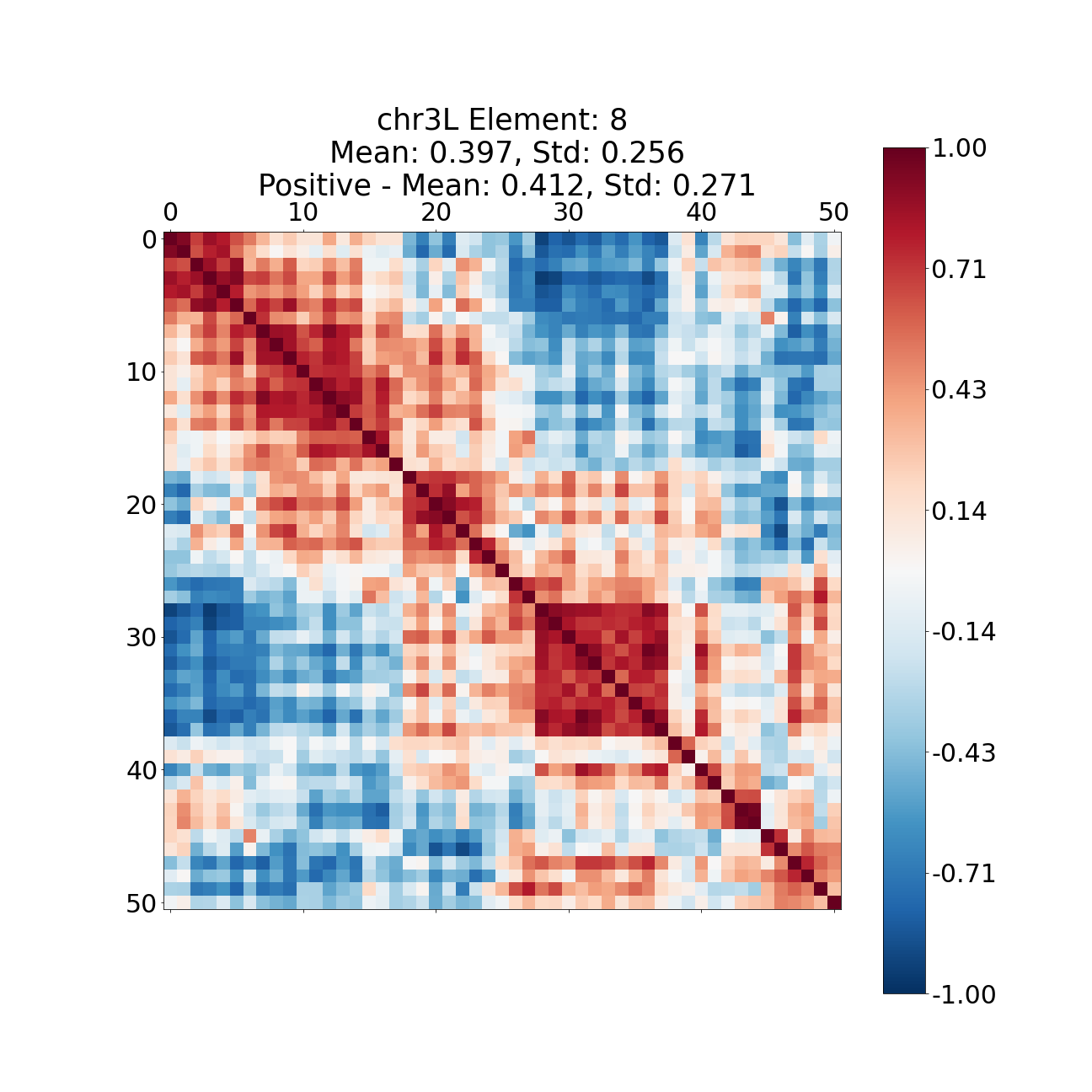}
     \end{subfigure}
     \hfill
     \begin{subfigure}[b]{0.32\textwidth}
         \centering
         \includegraphics[width=\textwidth]{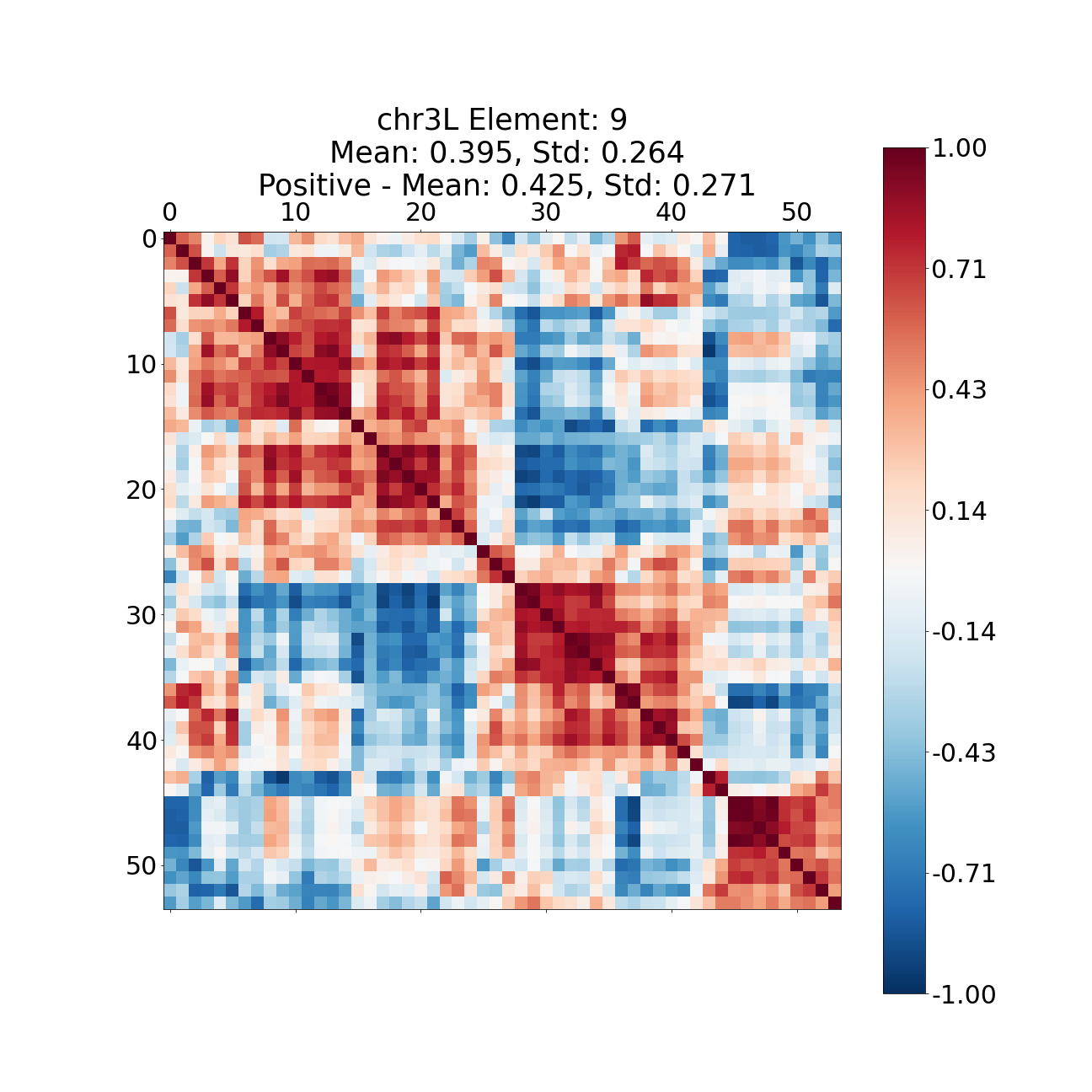}
     \end{subfigure}
     \hfill
     \begin{subfigure}[b]{0.32\textwidth}
         \centering
         \includegraphics[width=\textwidth]{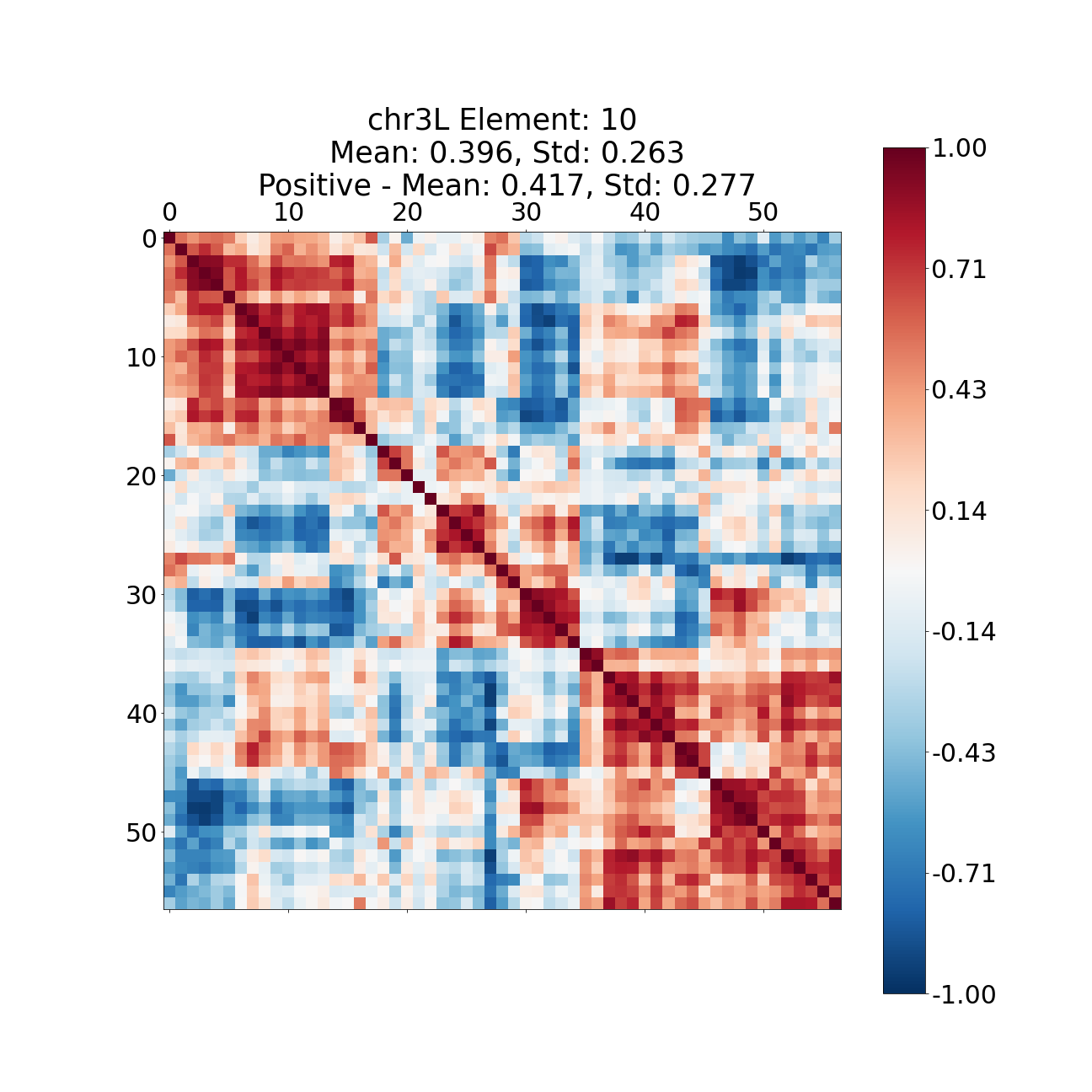}
     \end{subfigure}
      \hfill
     \begin{subfigure}[b]{0.32\textwidth}
         \centering
         \includegraphics[width=\textwidth]{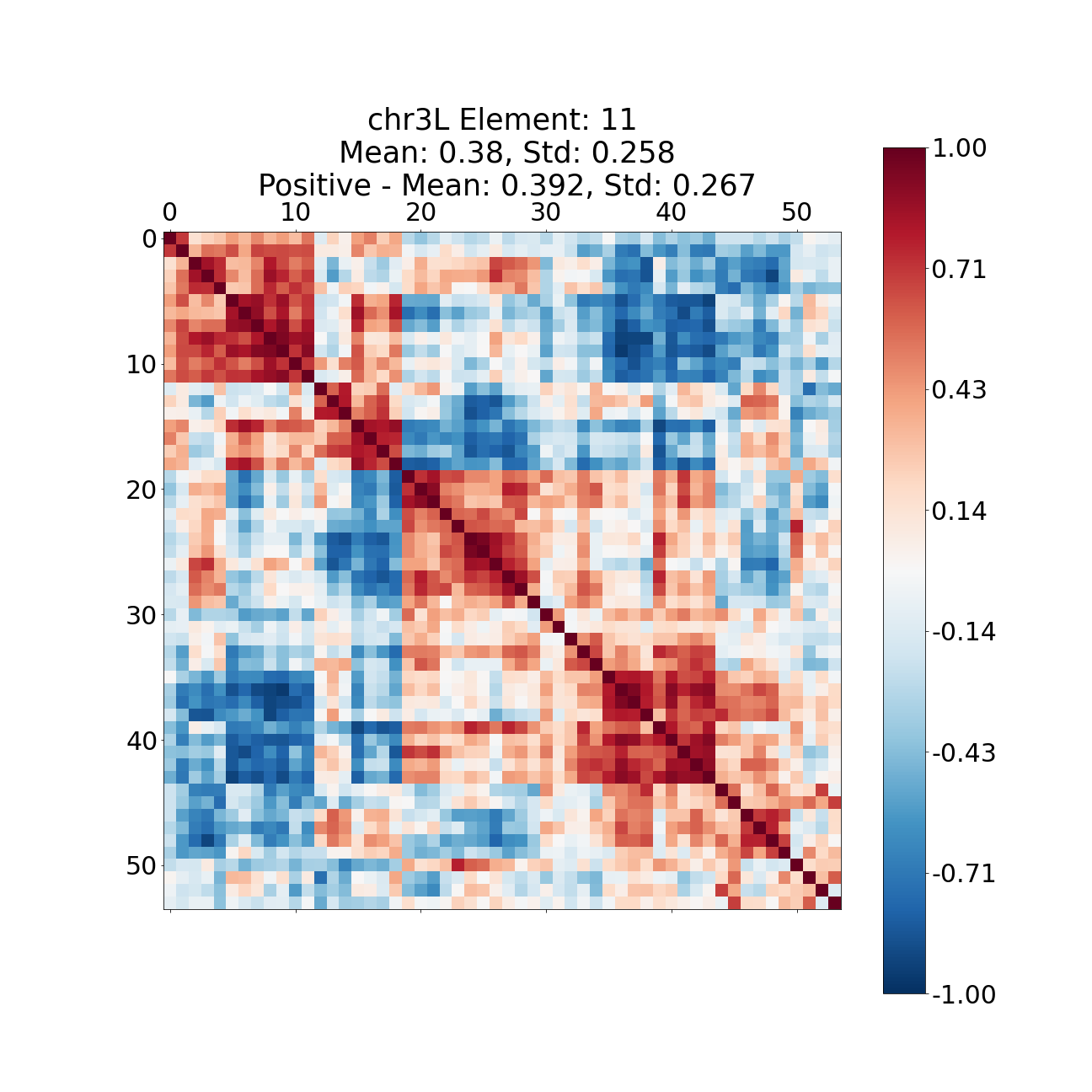}
     \end{subfigure}
        \caption{Pairwise coexpression of genes covered by various dictionary elements for chr 3L obtained through online cvxNDL. We calculated the mean and standard deviation of absolute pairwise coexpression values, along with the mean and standard deviation of coexpression values specifically for all positively correlated gene pairs.}
        \label{fig:dee2_pearson3L}
\end{figure}

\begin{figure}[h]
\ContinuedFloat
     \centering
     \begin{subfigure}[b]{0.32\textwidth}
         \centering
         \includegraphics[width=\textwidth]{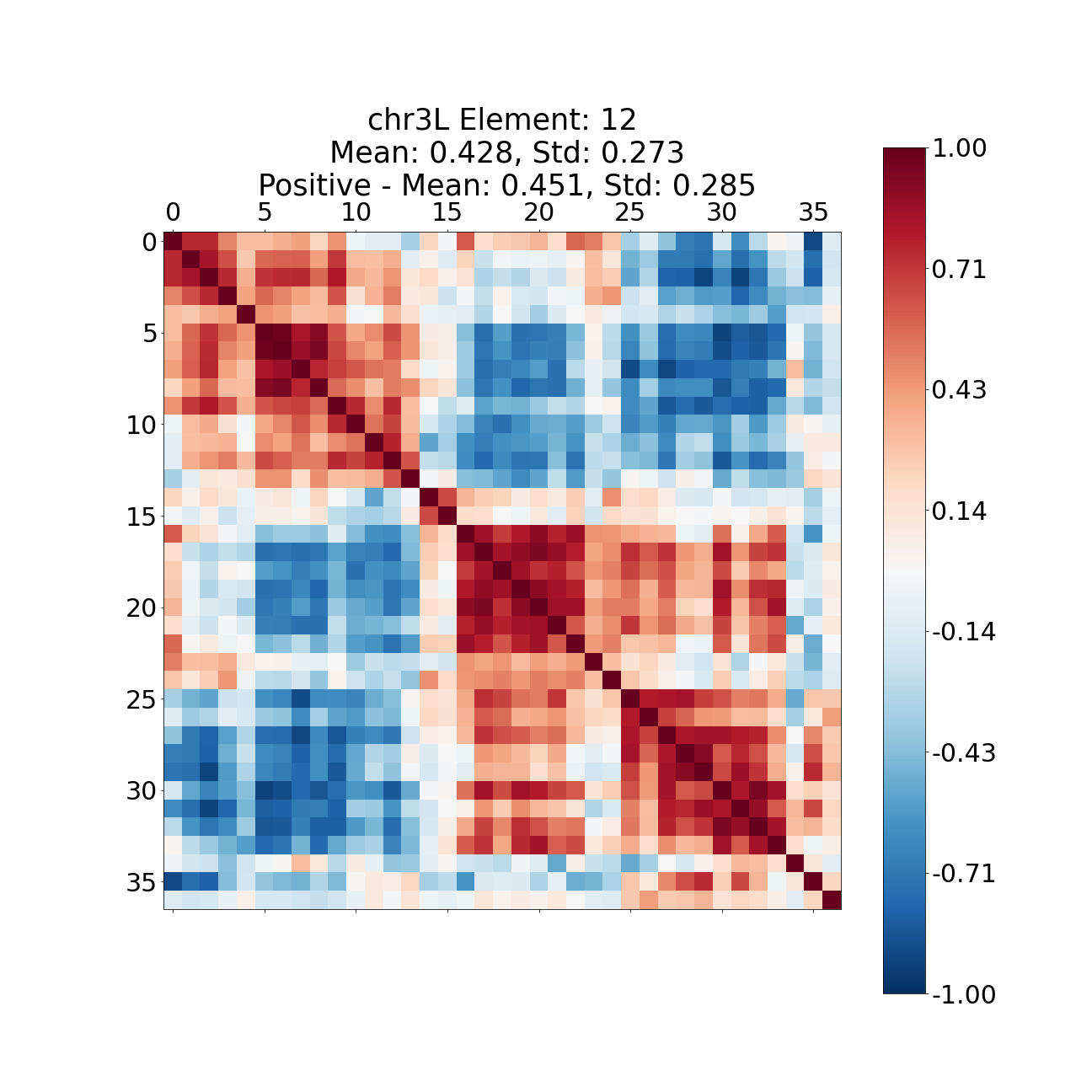}
     \end{subfigure}
     \hfill
     \begin{subfigure}[b]{0.32\textwidth}
         \centering
         \includegraphics[width=\textwidth]{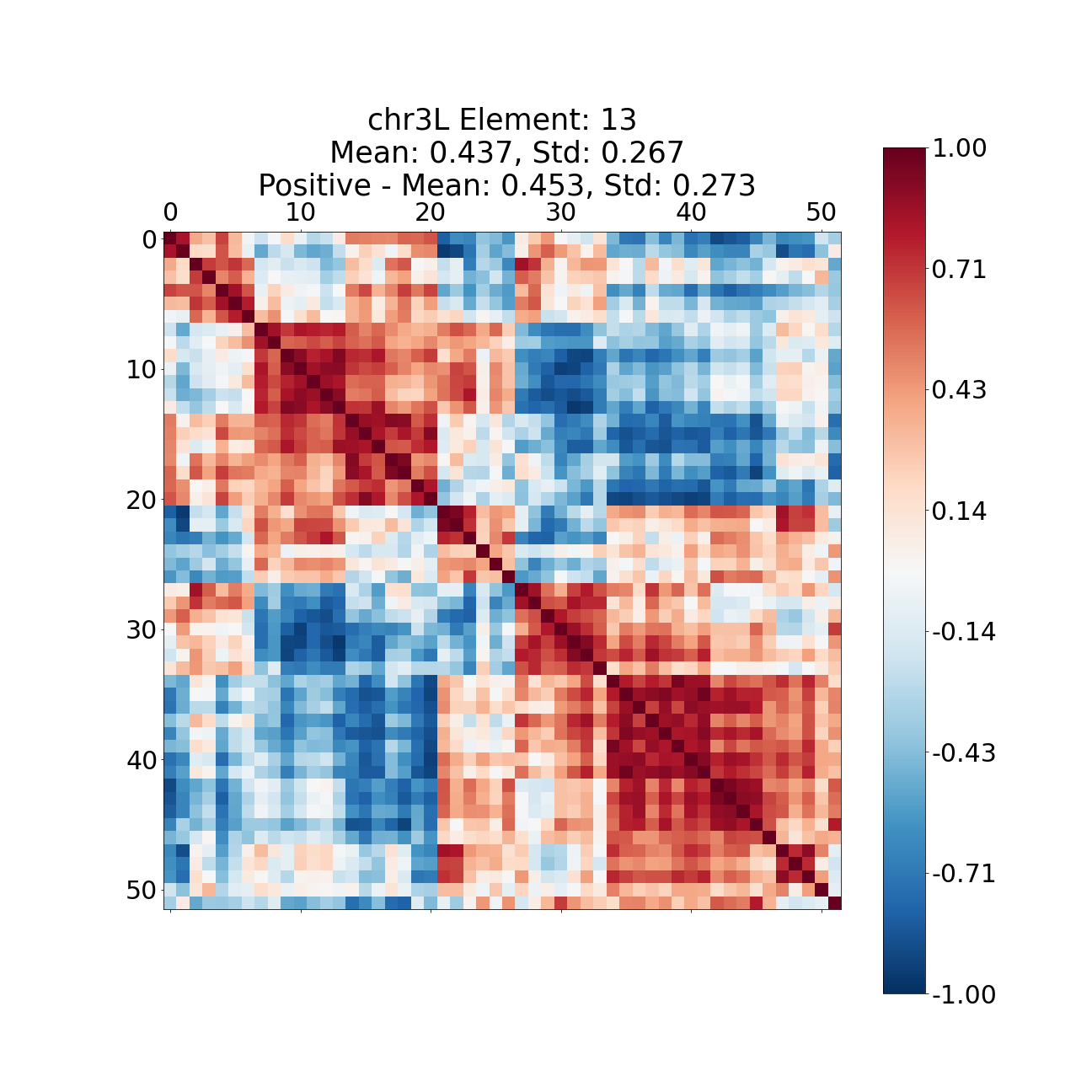}
     \end{subfigure}
     \hfill
     \begin{subfigure}[b]{0.32\textwidth}
         \centering
         \includegraphics[width=\textwidth]{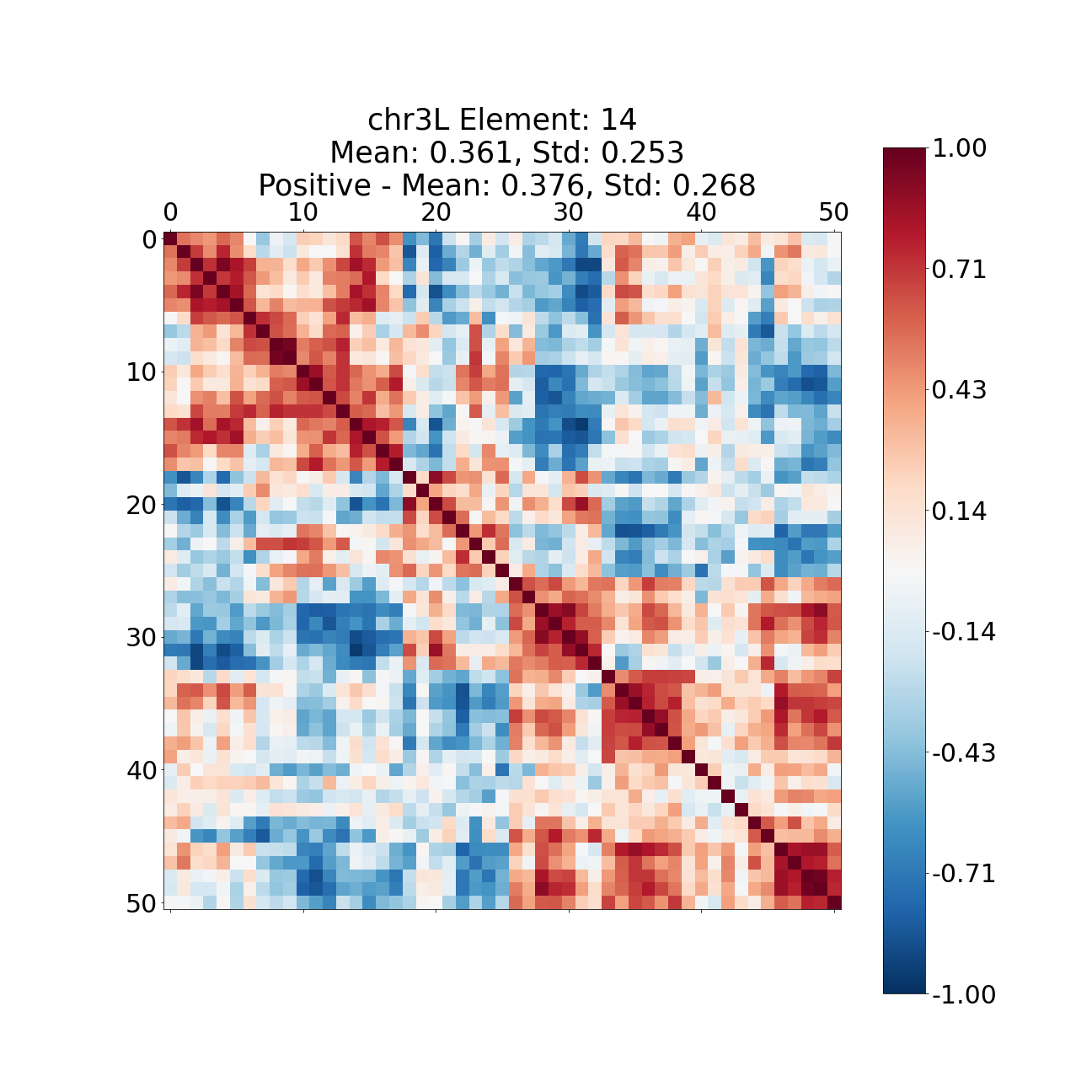}
     \end{subfigure}
      \hfill
     \begin{subfigure}[b]{0.32\textwidth}
         \centering
         \includegraphics[width=\textwidth]{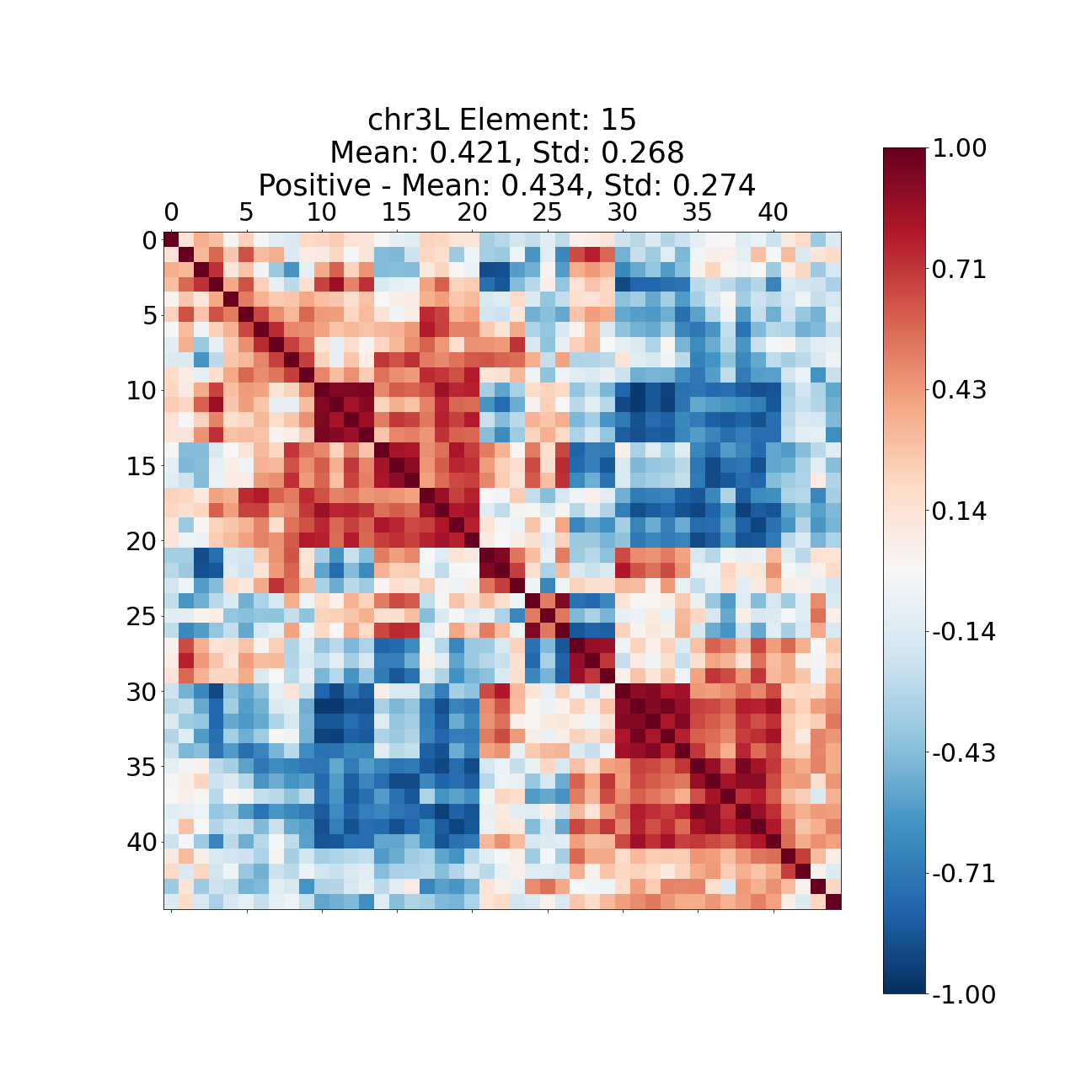}
     \end{subfigure}
     \hfill
     \begin{subfigure}[b]{0.32\textwidth}
         \centering
         \includegraphics[width=\textwidth]{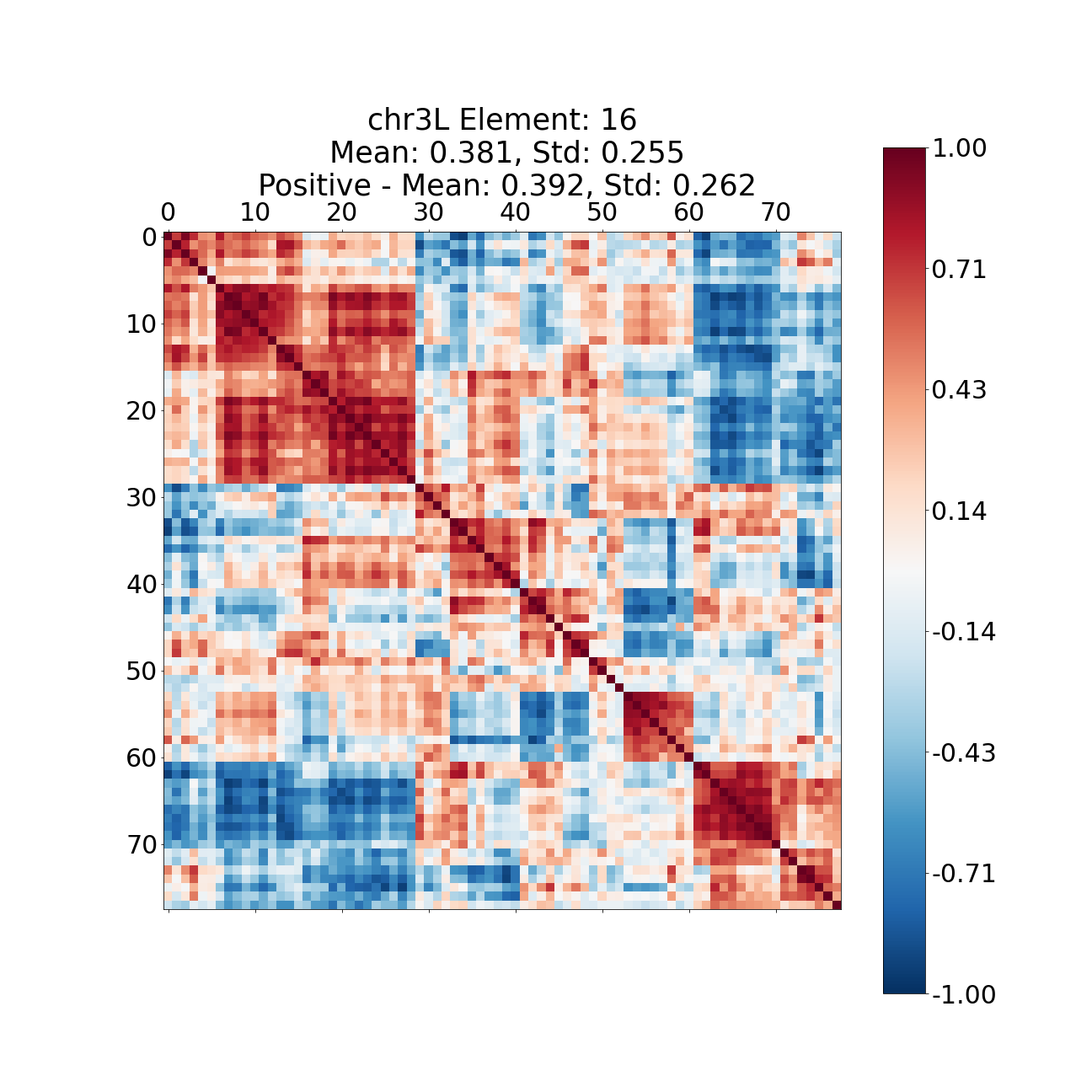}
     \end{subfigure}
     \hfill
     \begin{subfigure}[b]{0.32\textwidth}
         \centering
         \includegraphics[width=\textwidth]{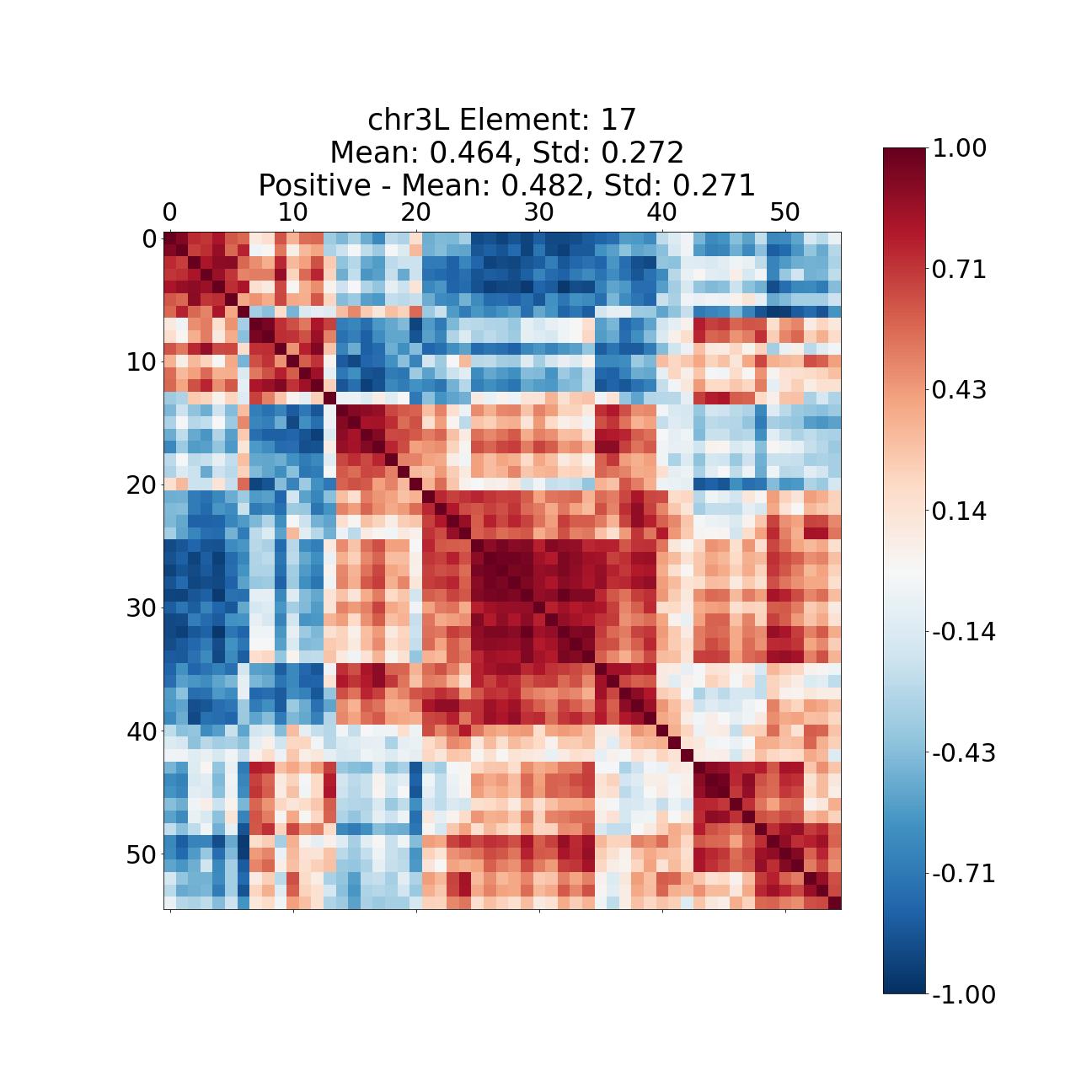}
     \end{subfigure}
     \hfill
     \begin{subfigure}[b]{0.32\textwidth}
         \centering
         \includegraphics[width=\textwidth]{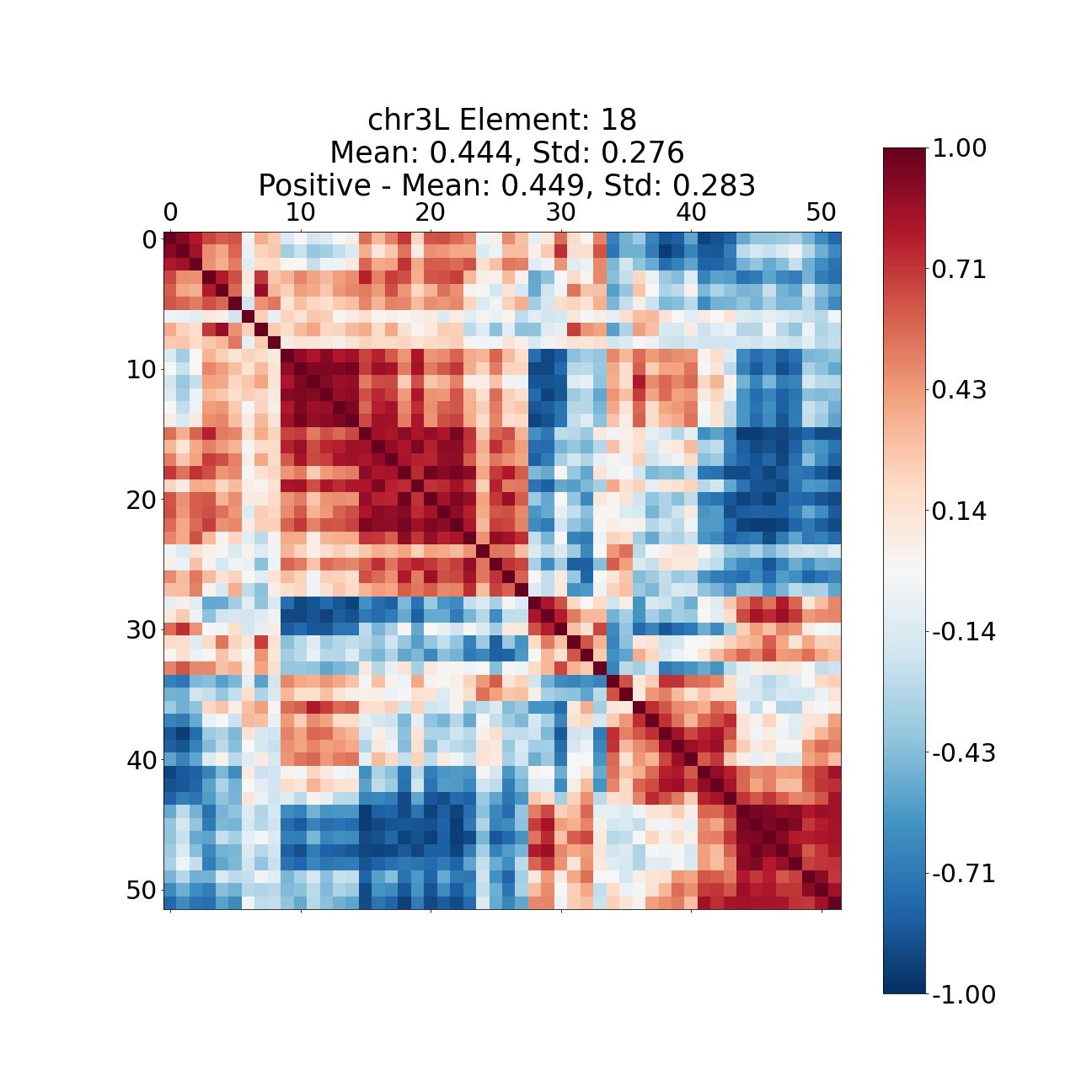}
     \end{subfigure}
      \hfill
     \begin{subfigure}[b]{0.32\textwidth}
         \centering
         \includegraphics[width=\textwidth]{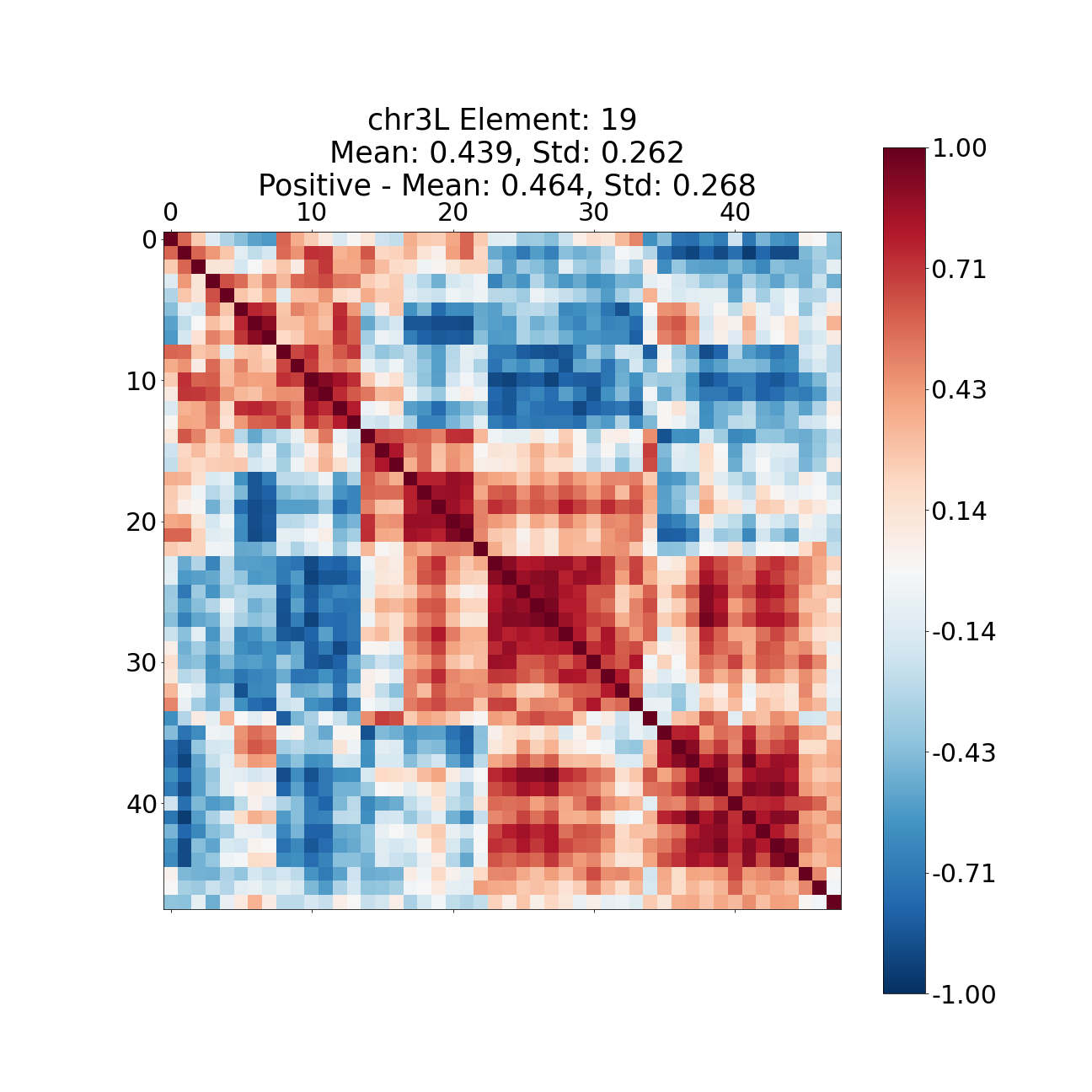}
     \end{subfigure}
     \hfill
     \begin{subfigure}[b]{0.32\textwidth}
         \centering
         \includegraphics[width=\textwidth]{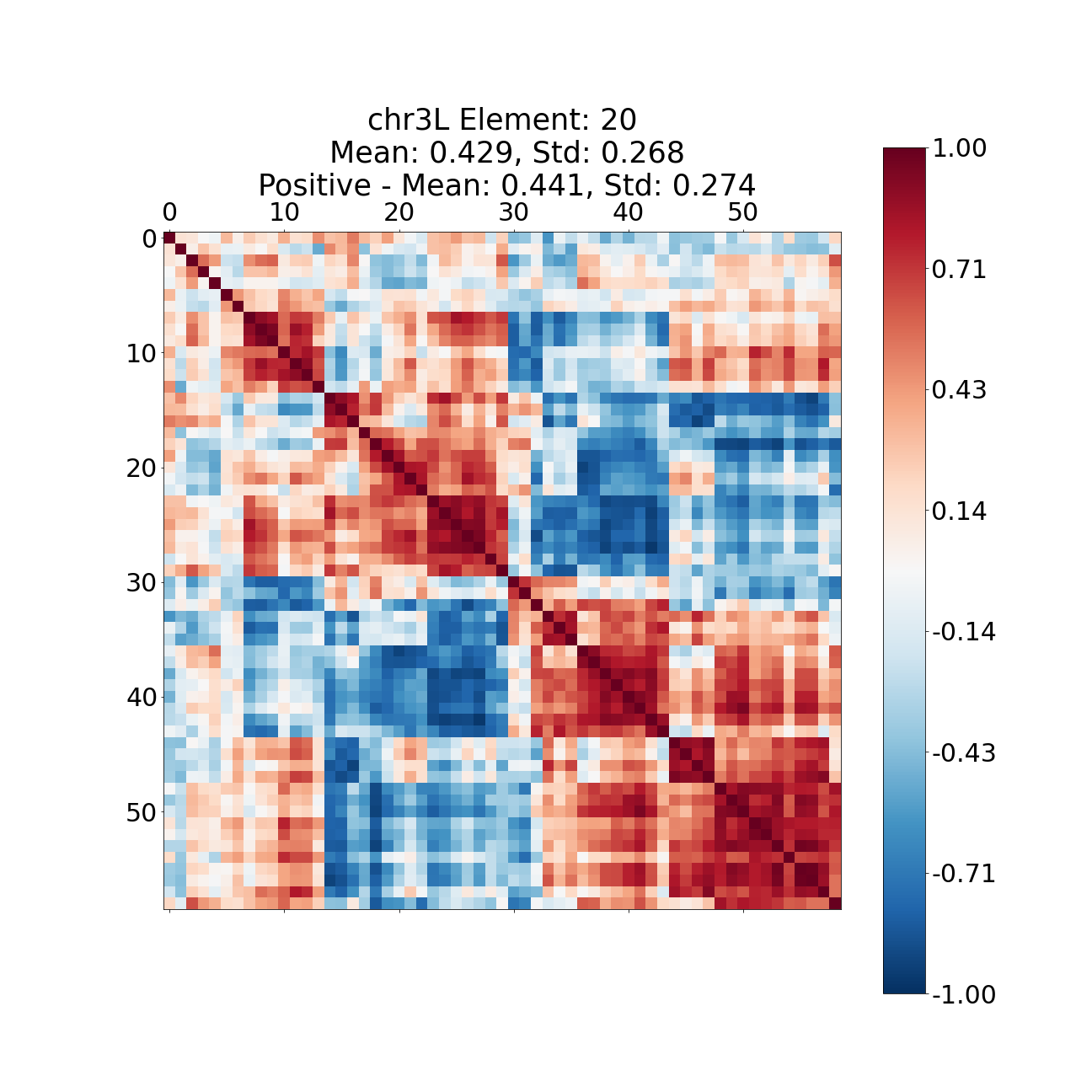}
     \end{subfigure}
     \hfill
     \begin{subfigure}[b]{0.32\textwidth}
         \centering
         \includegraphics[width=\textwidth]{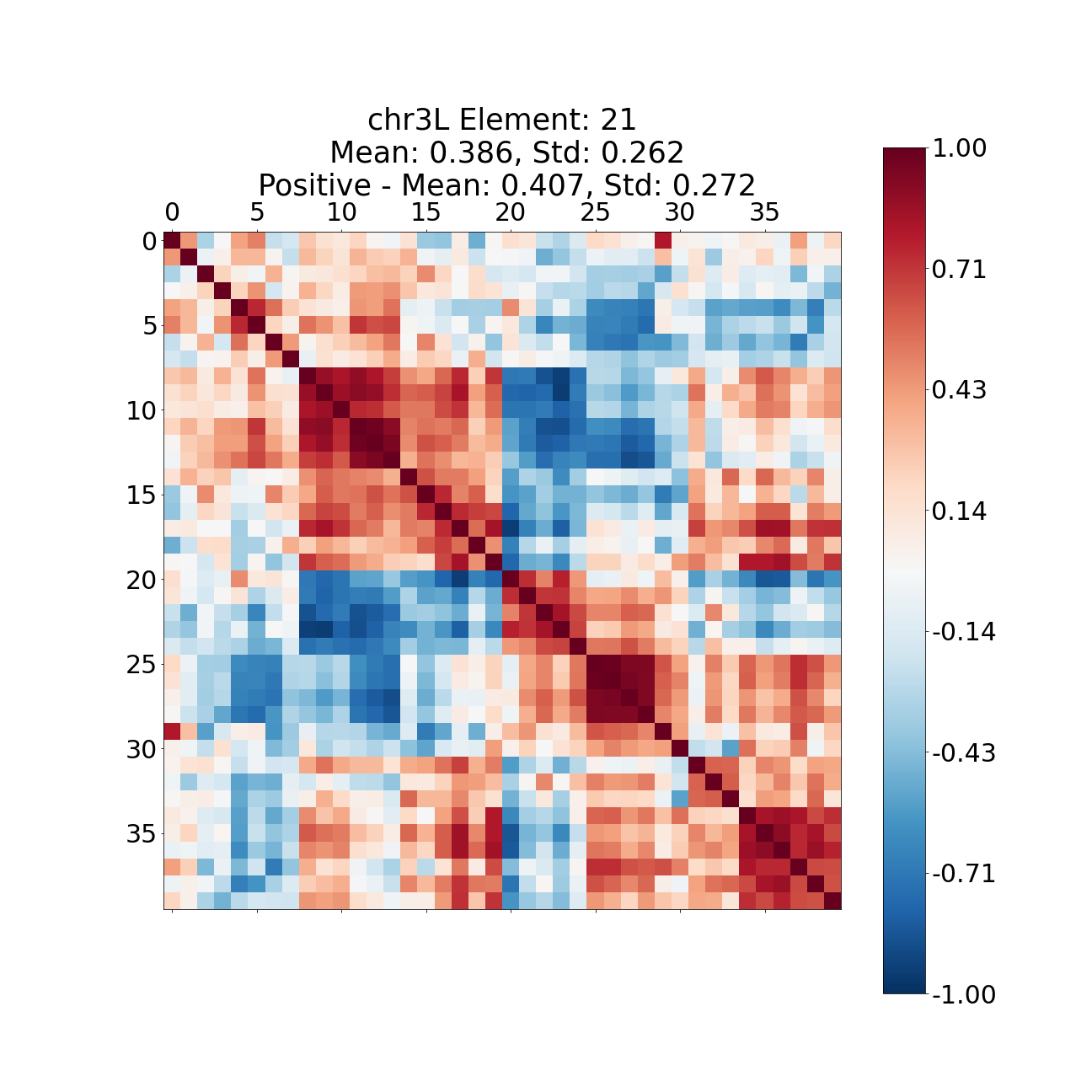}
     \end{subfigure}
     \hfill
     \begin{subfigure}[b]{0.32\textwidth}
         \centering
         \includegraphics[width=\textwidth]{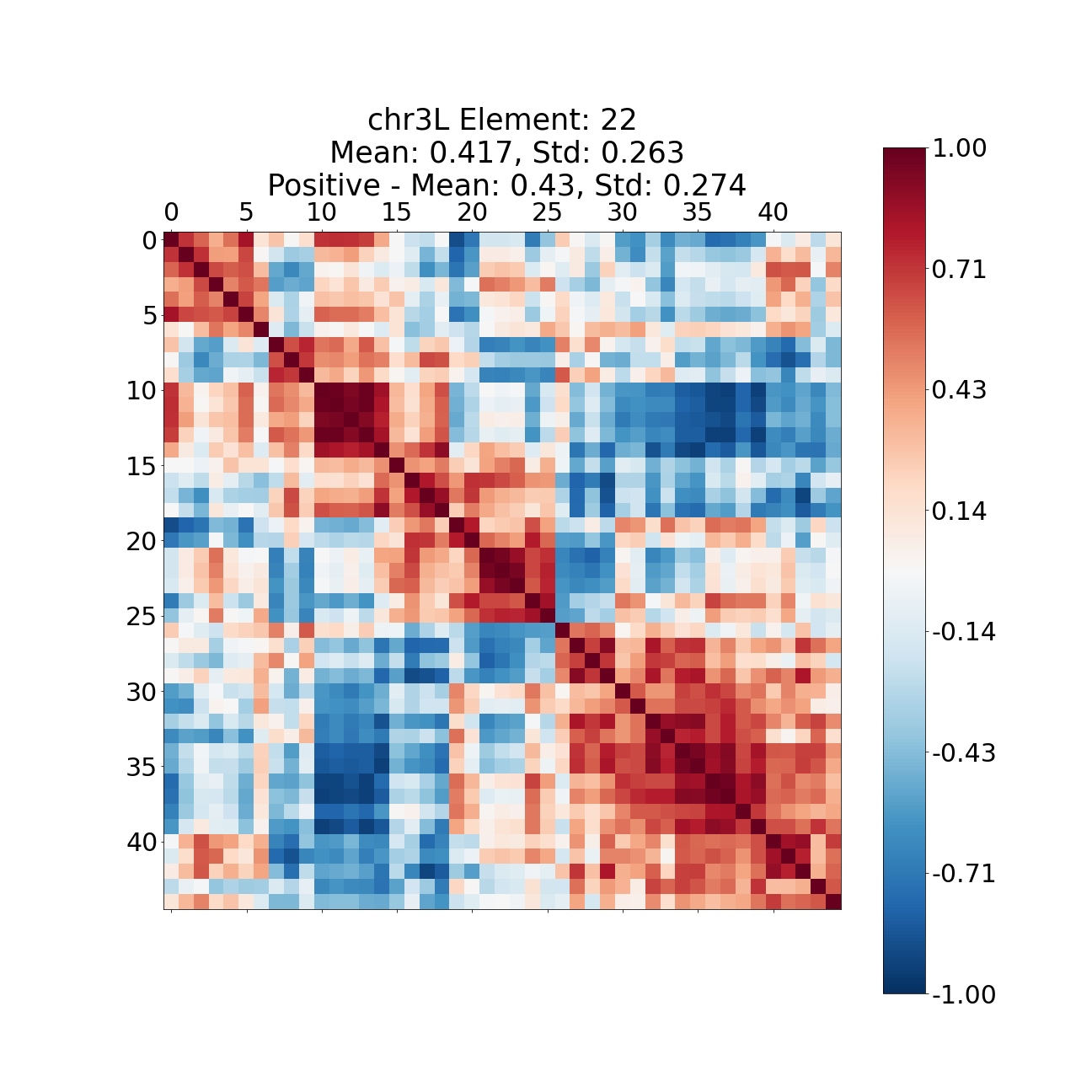}
     \end{subfigure}
      \hfill
     \begin{subfigure}[b]{0.32\textwidth}
         \centering
         \includegraphics[width=\textwidth]{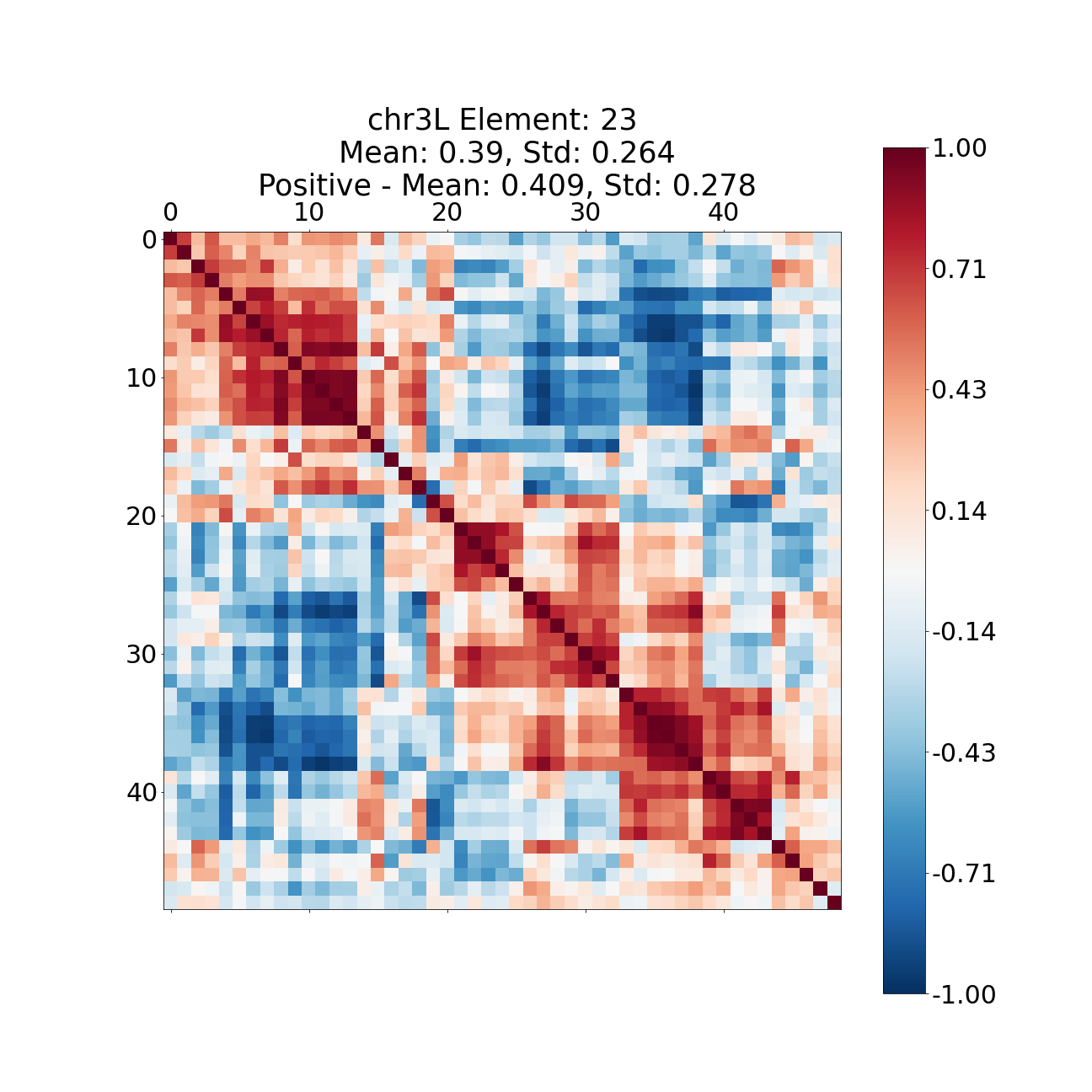}
     \end{subfigure}
        \caption{Pairwise coexpression of genes covered by various dictionary elements for chr 3L obtained through online cvxNDL. We calculated the mean and standard deviation of absolute pairwise coexpression values, along with the mean and standard deviation of coexpression values specifically for all positively correlated gene pairs.}
\end{figure}

\begin{figure}[h]
\ContinuedFloat
     \centering
     \begin{subfigure}[b]{0.32\textwidth}
         \centering
         \includegraphics[width=\textwidth]{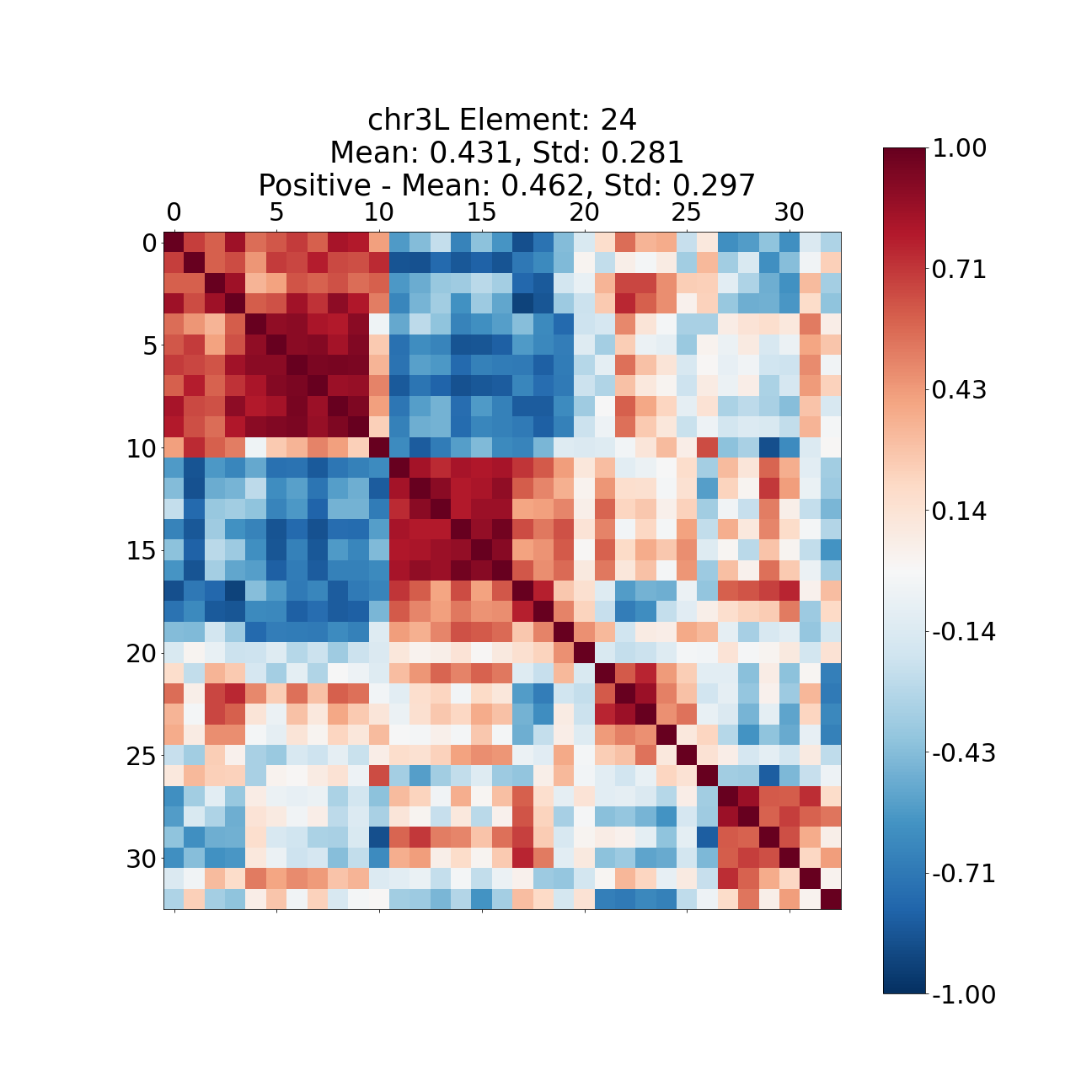}
     \end{subfigure}
        \caption{Pairwise coexpression of genes covered by various dictionary elements for chr 3L obtained through online cvxNDL. We calculated the mean and standard deviation of absolute pairwise coexpression values, along with the mean and standard deviation of coexpression values specifically for all positively correlated gene pairs.}
\end{figure}


\begin{figure}[h]
     \centering
     \begin{subfigure}[b]{0.32\textwidth}
         \centering
         \includegraphics[width=\textwidth]{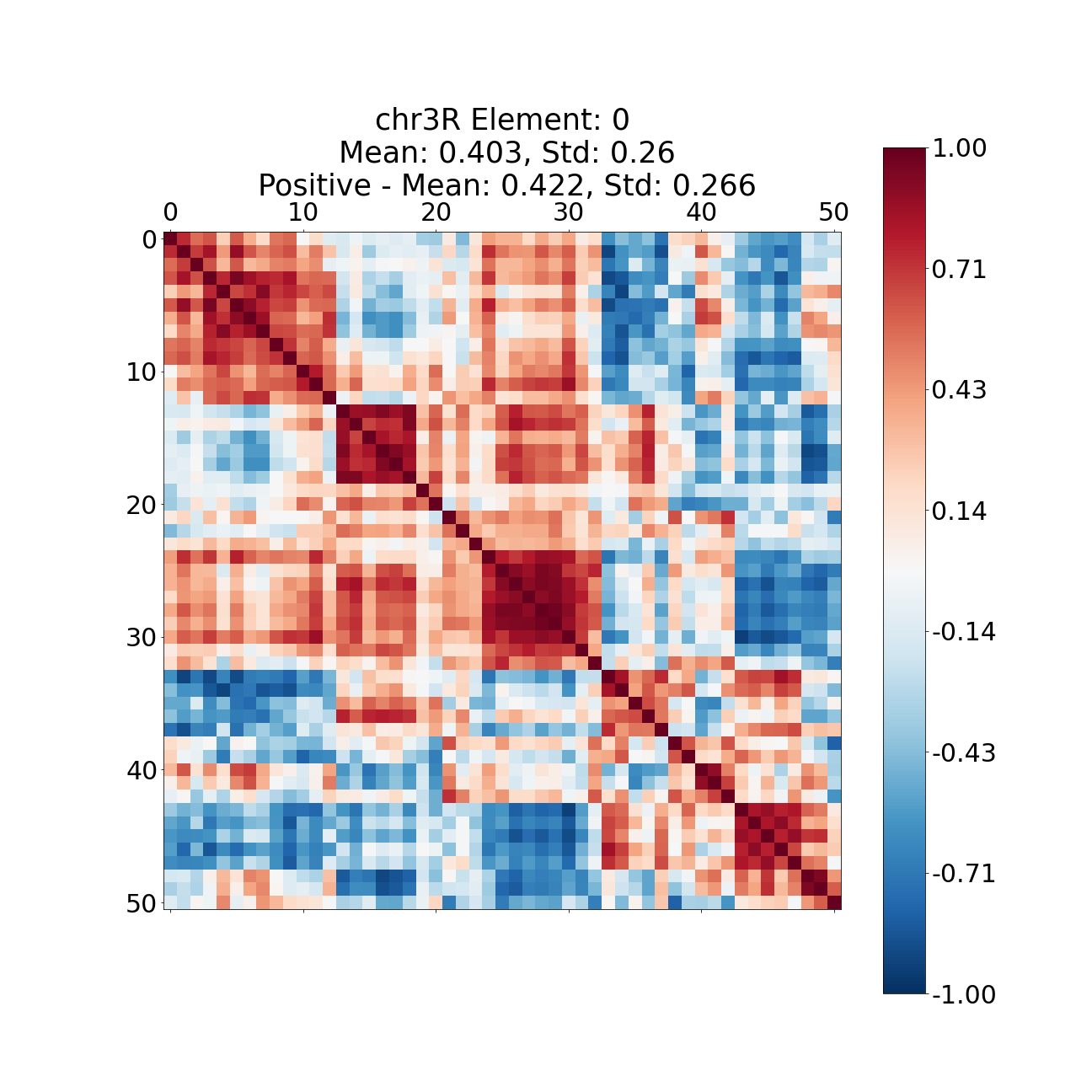}
     \end{subfigure}
     \hfill
     \begin{subfigure}[b]{0.32\textwidth}
         \centering
         \includegraphics[width=\textwidth]{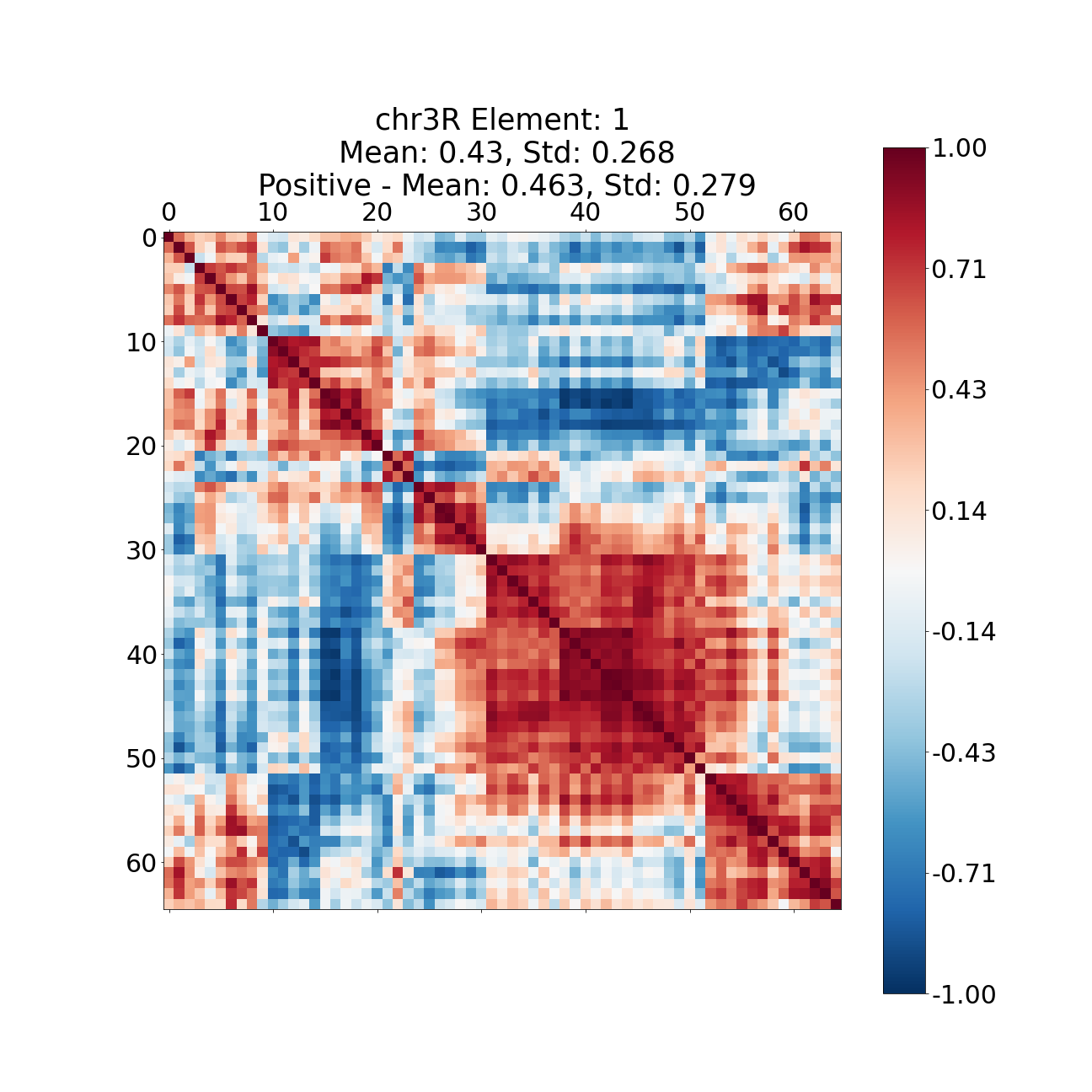}
     \end{subfigure}
     \hfill
     \begin{subfigure}[b]{0.32\textwidth}
         \centering
         \includegraphics[width=\textwidth]{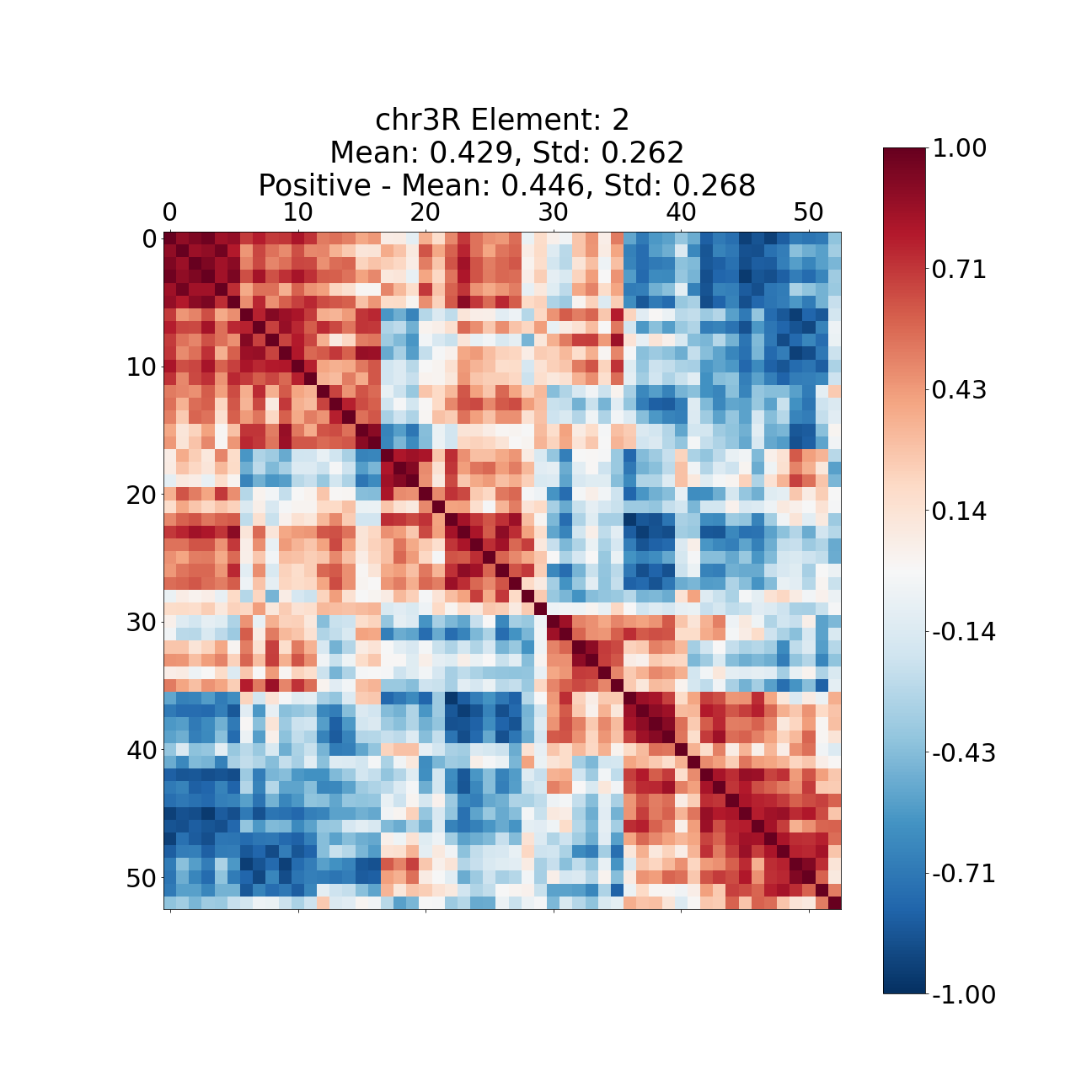}
     \end{subfigure}
      \hfill
     \begin{subfigure}[b]{0.32\textwidth}
         \centering
         \includegraphics[width=\textwidth]{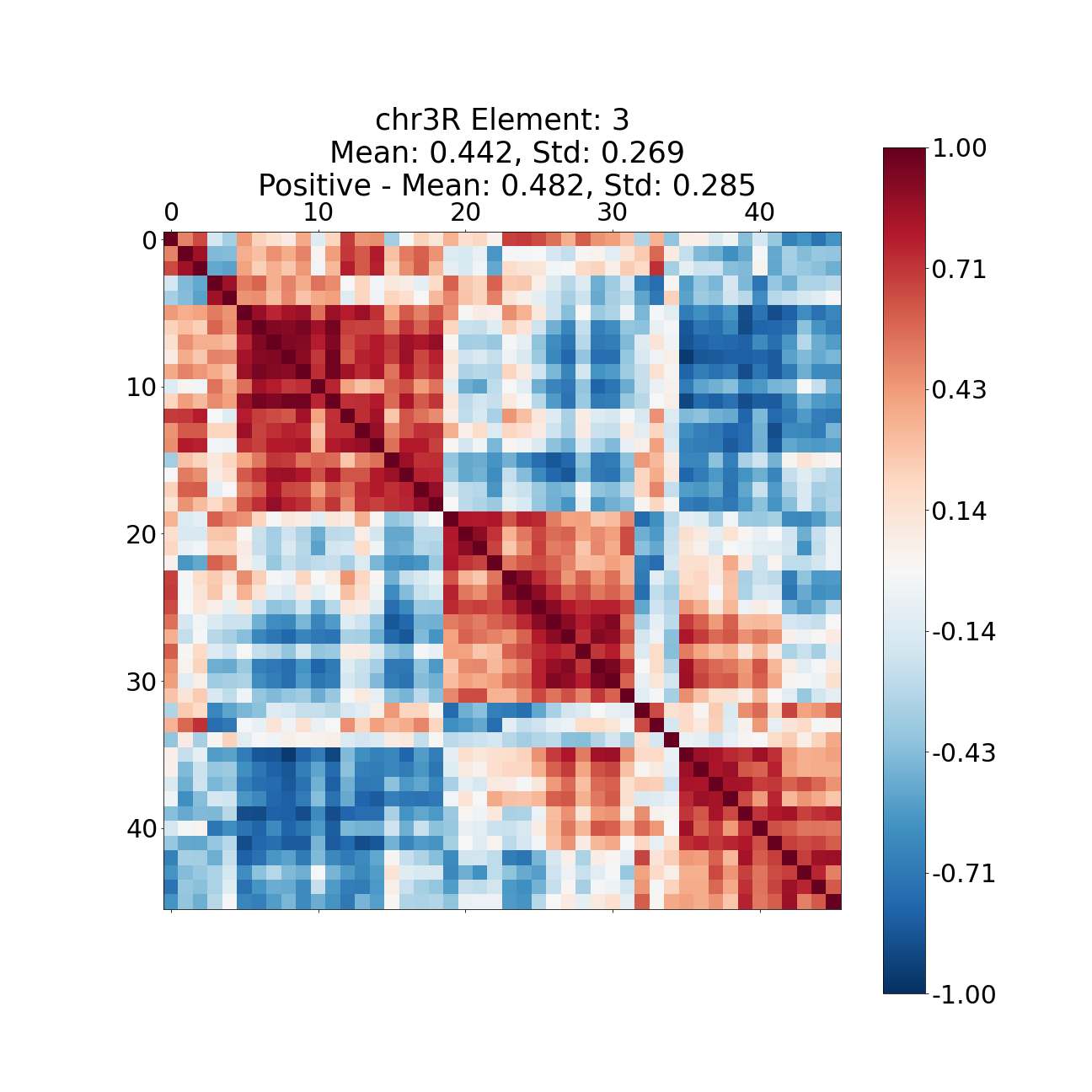}
     \end{subfigure}
     \hfill
     \begin{subfigure}[b]{0.32\textwidth}
         \centering
         \includegraphics[width=\textwidth]{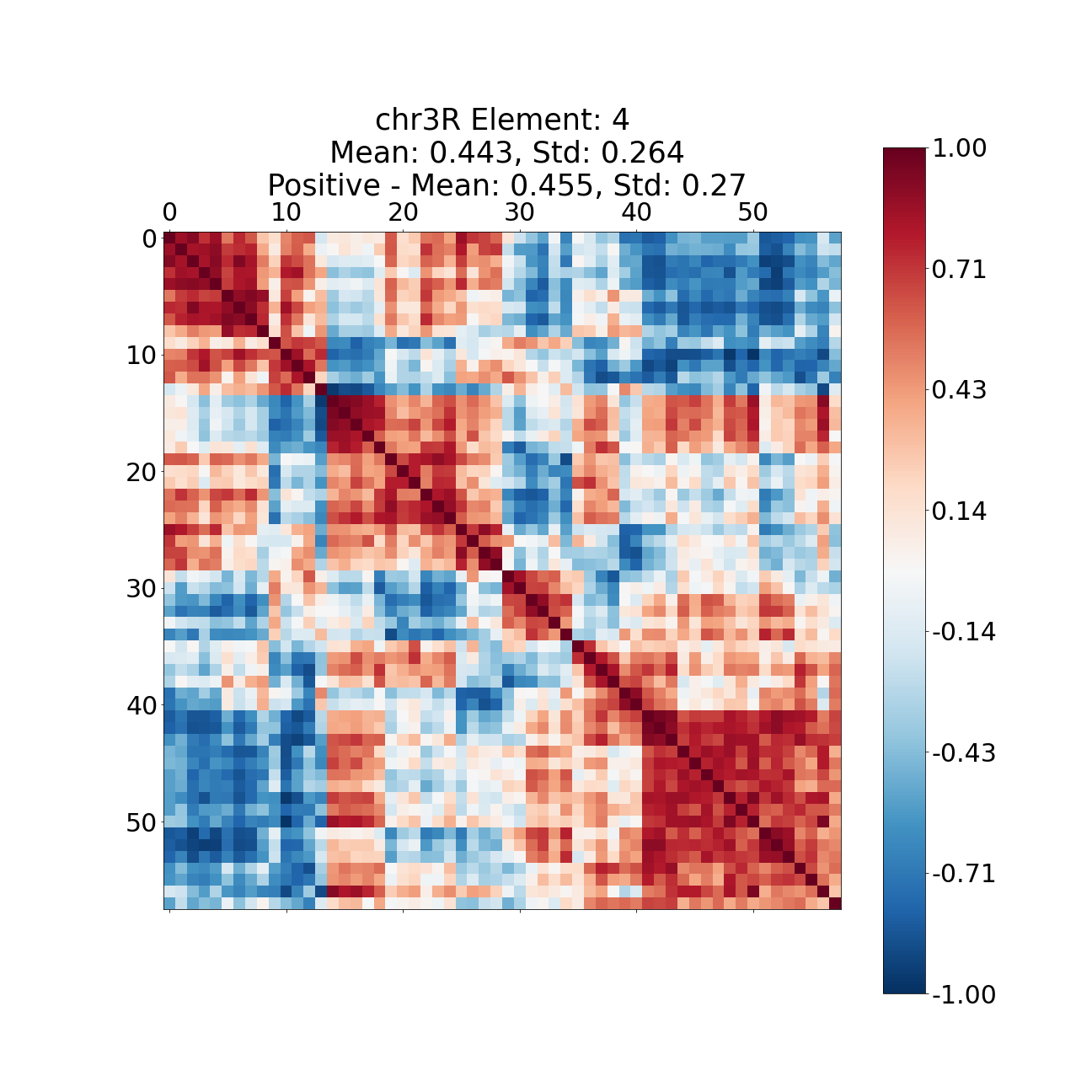}
     \end{subfigure}
     \hfill
     \begin{subfigure}[b]{0.32\textwidth}
         \centering
         \includegraphics[width=\textwidth]{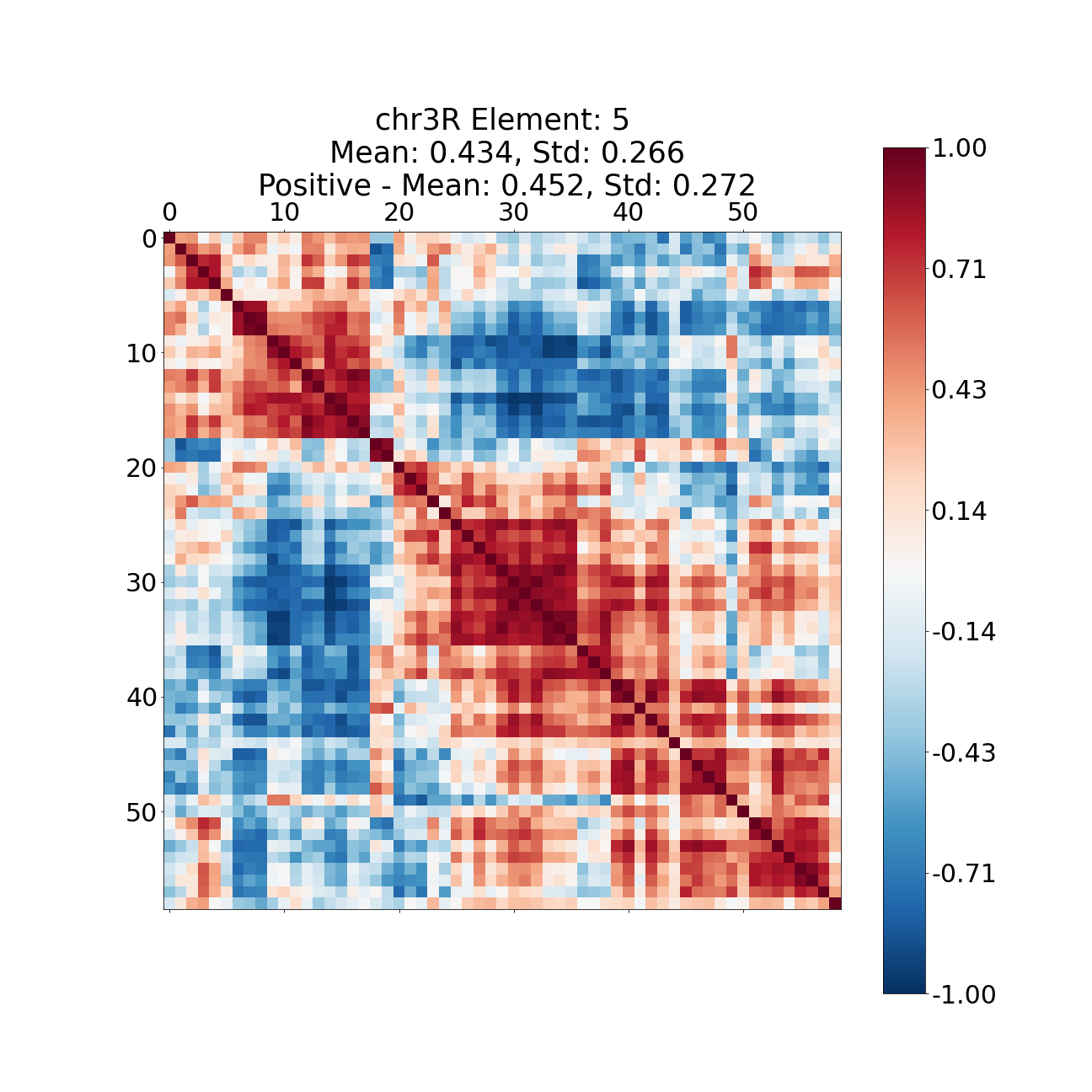}
     \end{subfigure}
     \hfill
     \begin{subfigure}[b]{0.32\textwidth}
         \centering
         \includegraphics[width=\textwidth]{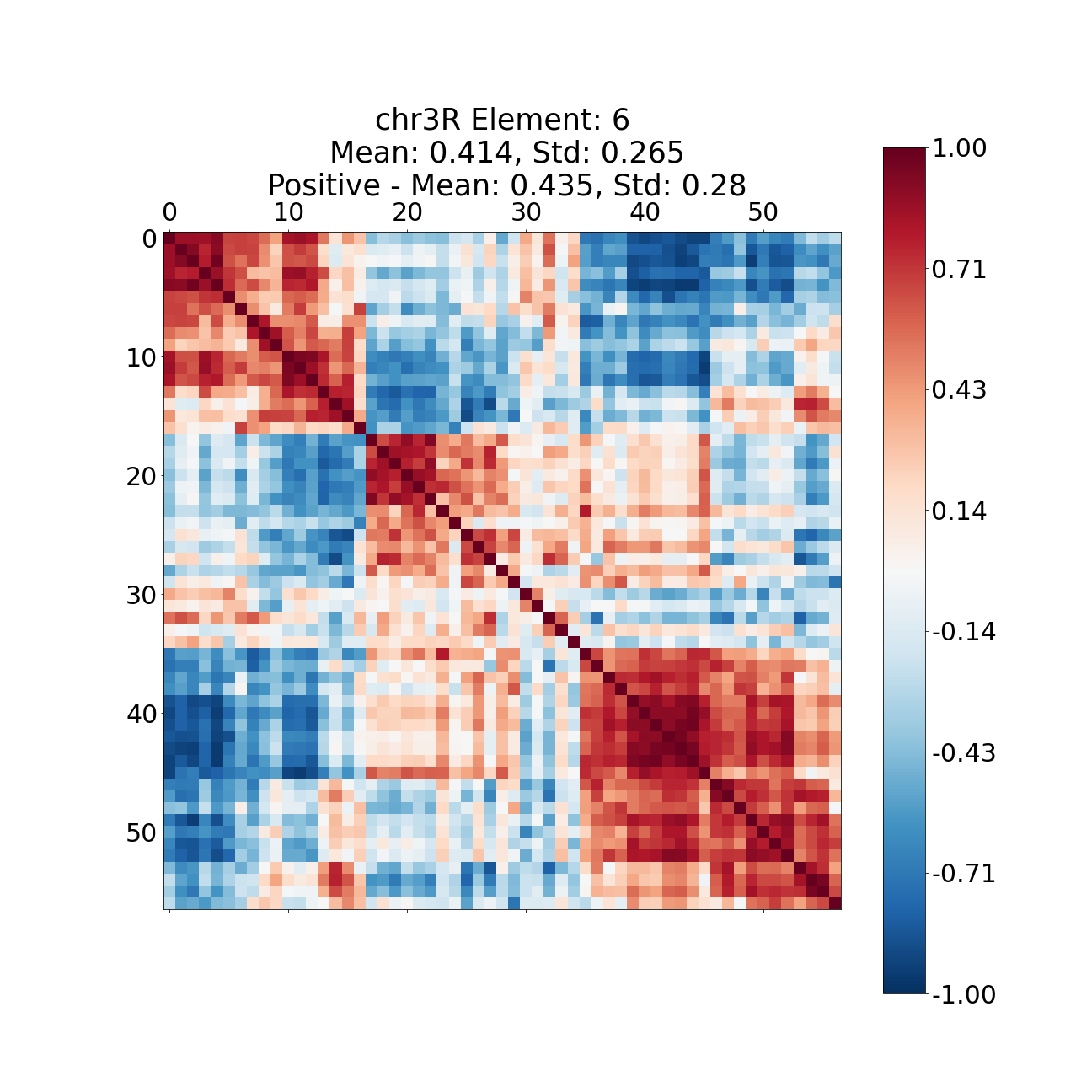}
     \end{subfigure}
      \hfill
     \begin{subfigure}[b]{0.32\textwidth}
         \centering
         \includegraphics[width=\textwidth]{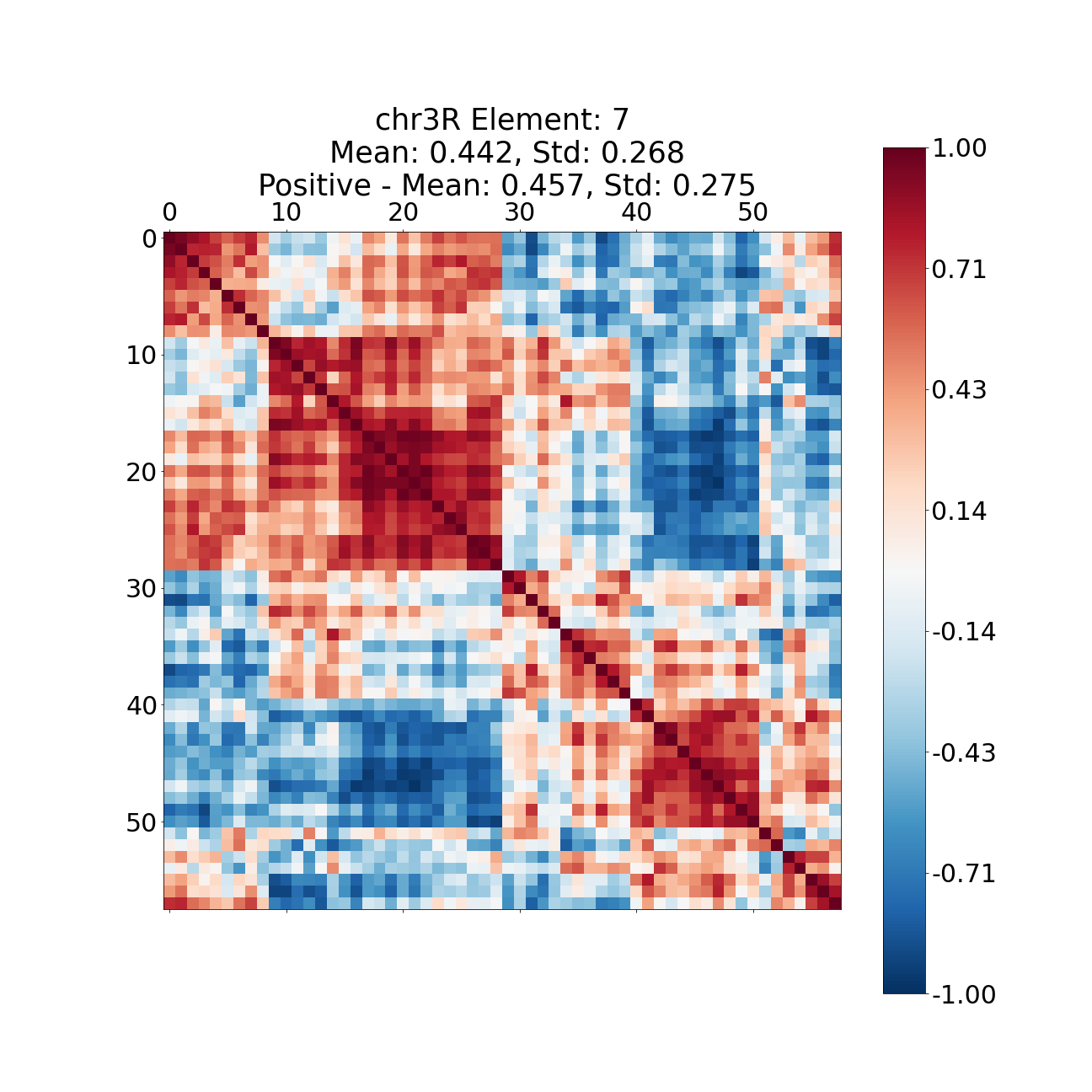}
     \end{subfigure}
     \hfill
     \begin{subfigure}[b]{0.32\textwidth}
         \centering
         \includegraphics[width=\textwidth]{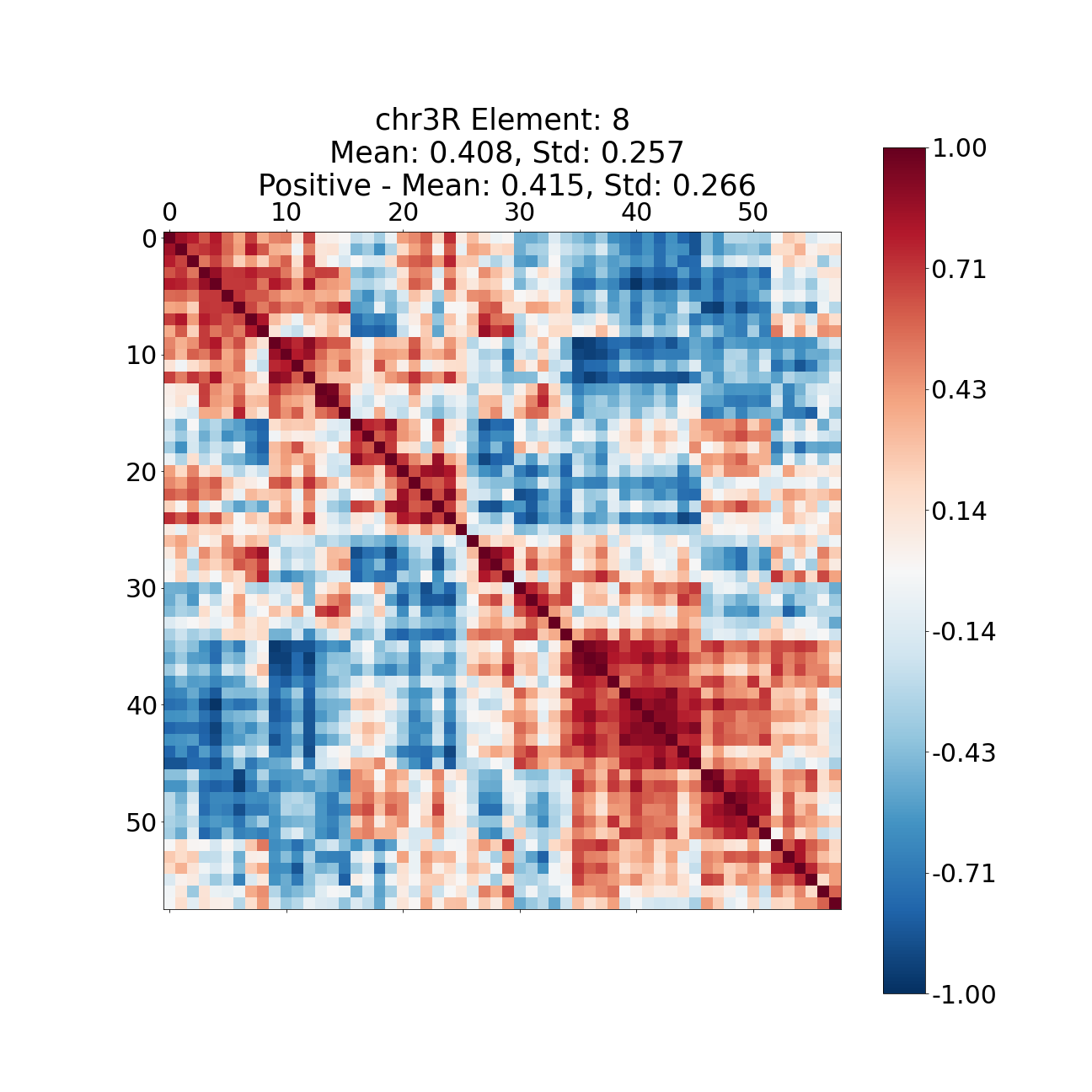}
     \end{subfigure}
     \hfill
     \begin{subfigure}[b]{0.32\textwidth}
         \centering
         \includegraphics[width=\textwidth]{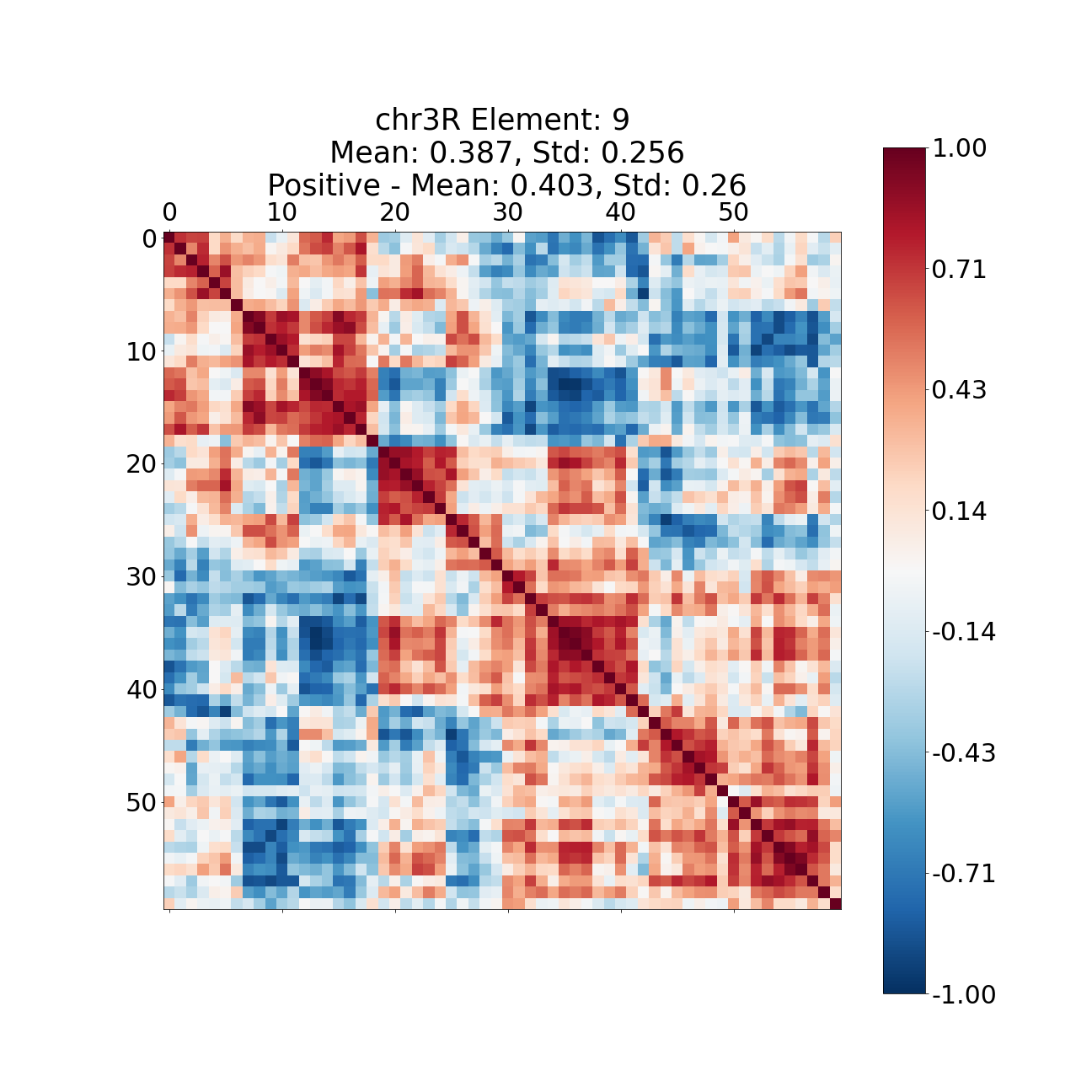}
     \end{subfigure}
     \hfill
     \begin{subfigure}[b]{0.32\textwidth}
         \centering
         \includegraphics[width=\textwidth]{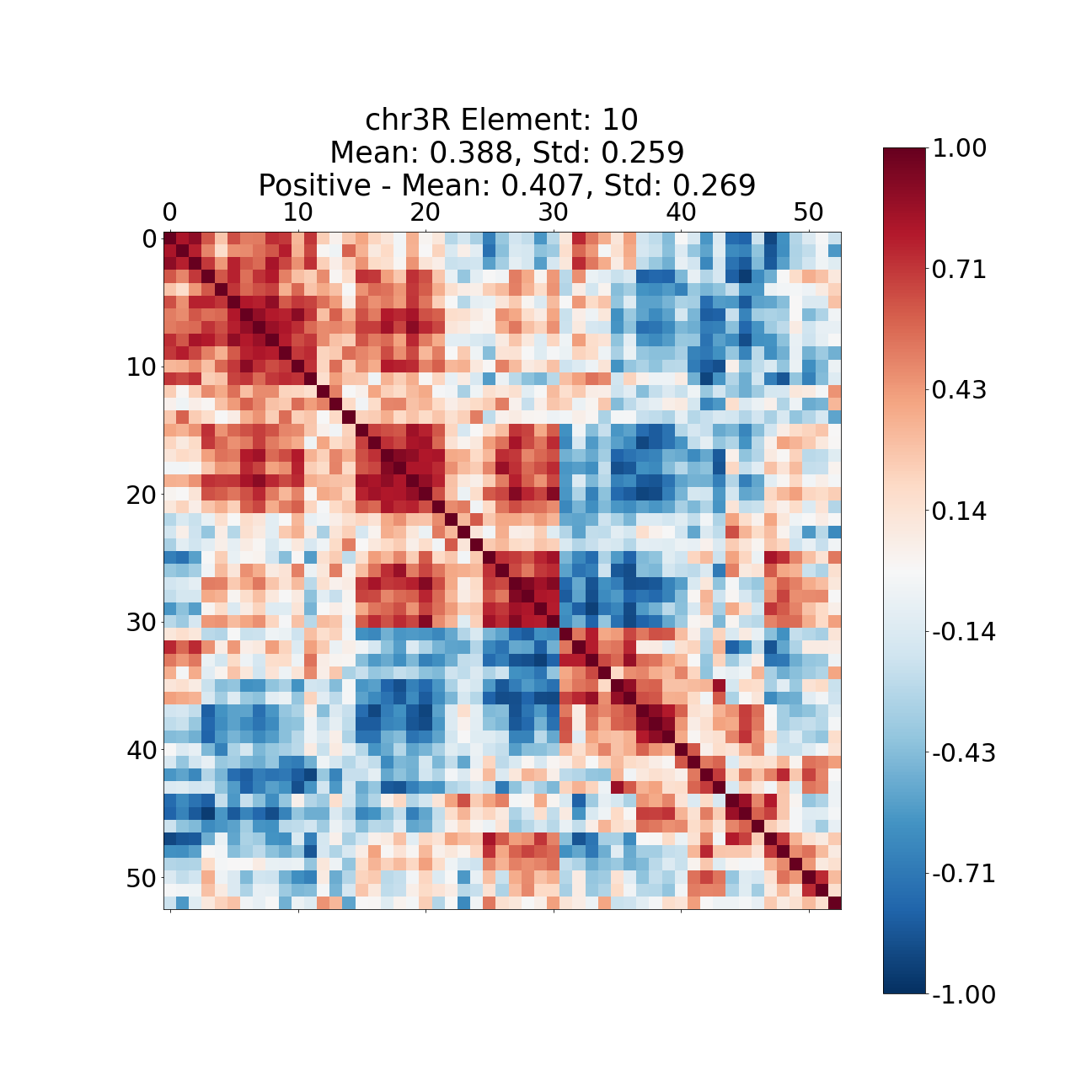}
     \end{subfigure}
      \hfill
     \begin{subfigure}[b]{0.32\textwidth}
         \centering
         \includegraphics[width=\textwidth]{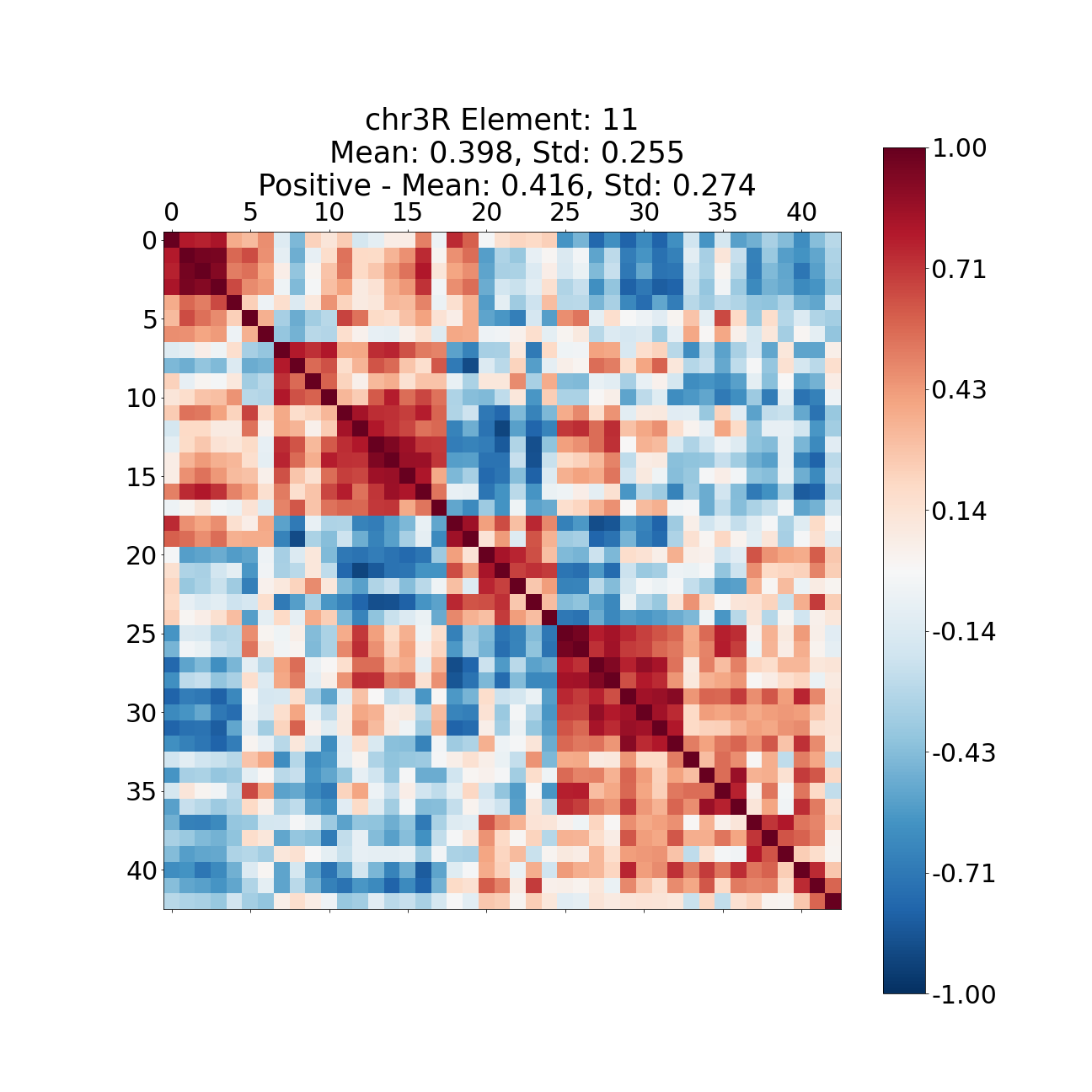}
     \end{subfigure}
        \caption{Pairwise coexpression of genes covered by various dictionary elements for chr 3R obtained through online cvxNDL. We calculated the mean and standard deviation of absolute pairwise coexpression values, along with the mean and standard deviation of coexpression values specifically for all positively correlated gene pairs.}
        \label{fig:dee2_pearson3R}
\end{figure}

\begin{figure}[h]
\ContinuedFloat
     \centering
     \begin{subfigure}[b]{0.32\textwidth}
         \centering
         \includegraphics[width=\textwidth]{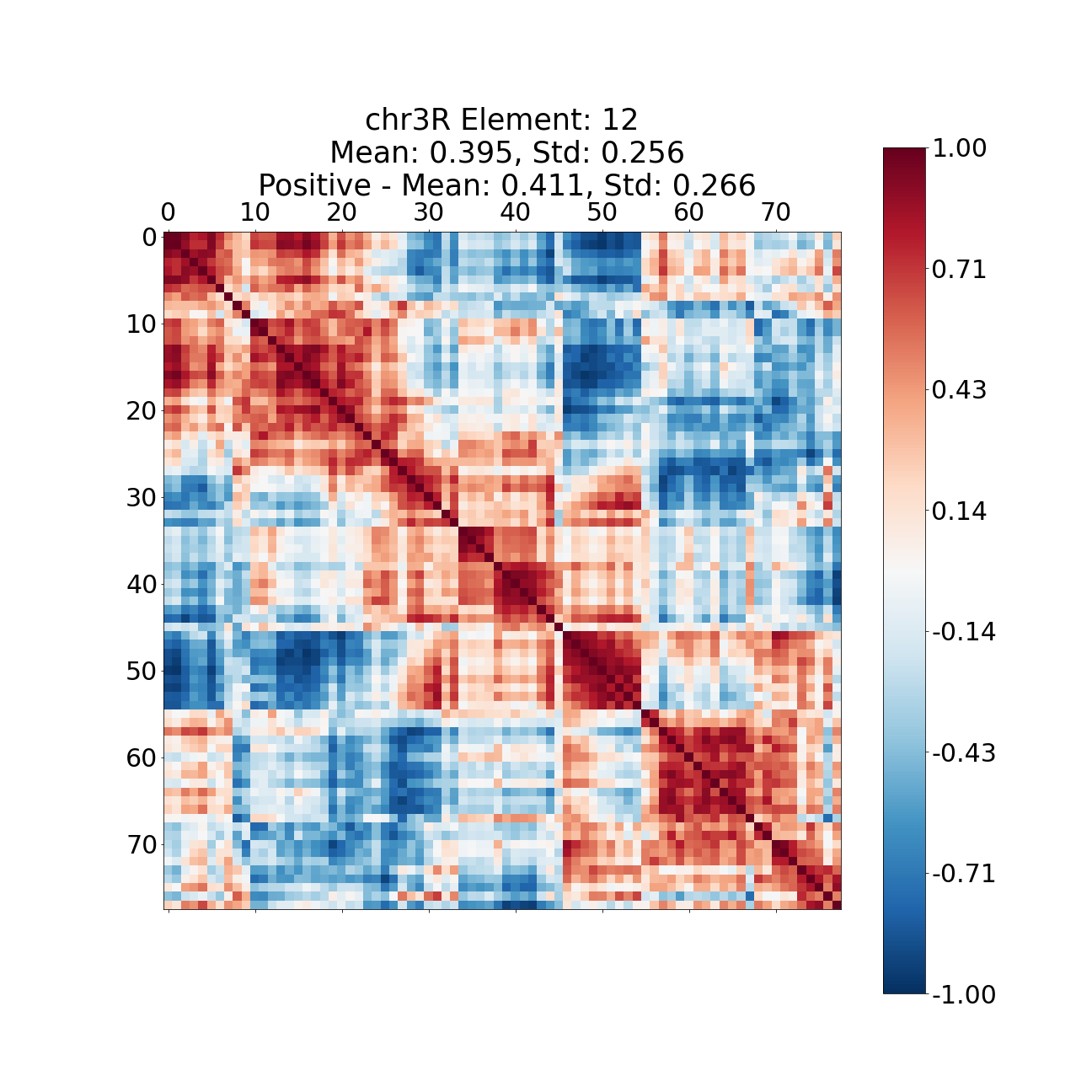}
     \end{subfigure}
     \hfill
     \begin{subfigure}[b]{0.32\textwidth}
         \centering
         \includegraphics[width=\textwidth]{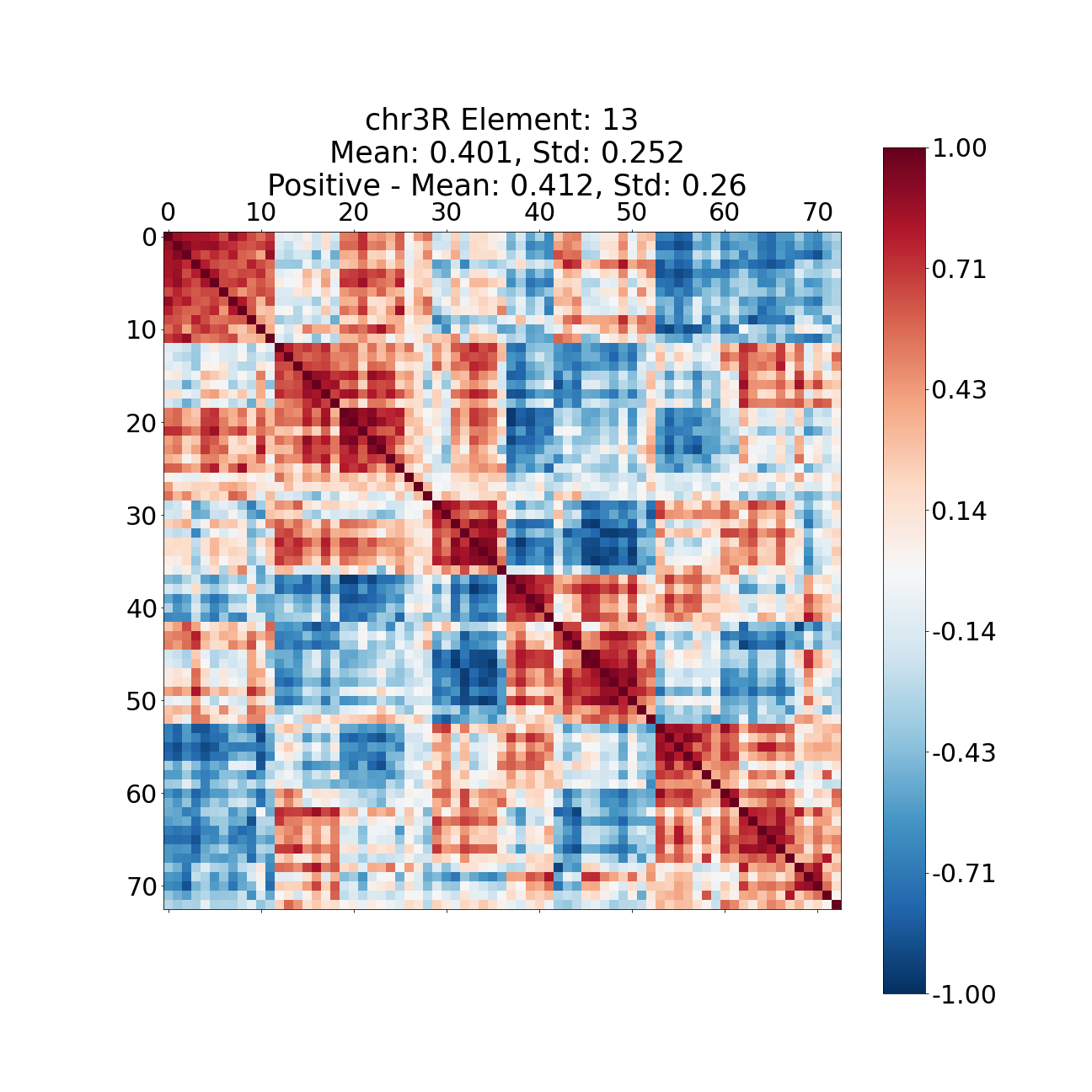}
     \end{subfigure}
     \hfill
     \begin{subfigure}[b]{0.32\textwidth}
         \centering
         \includegraphics[width=\textwidth]{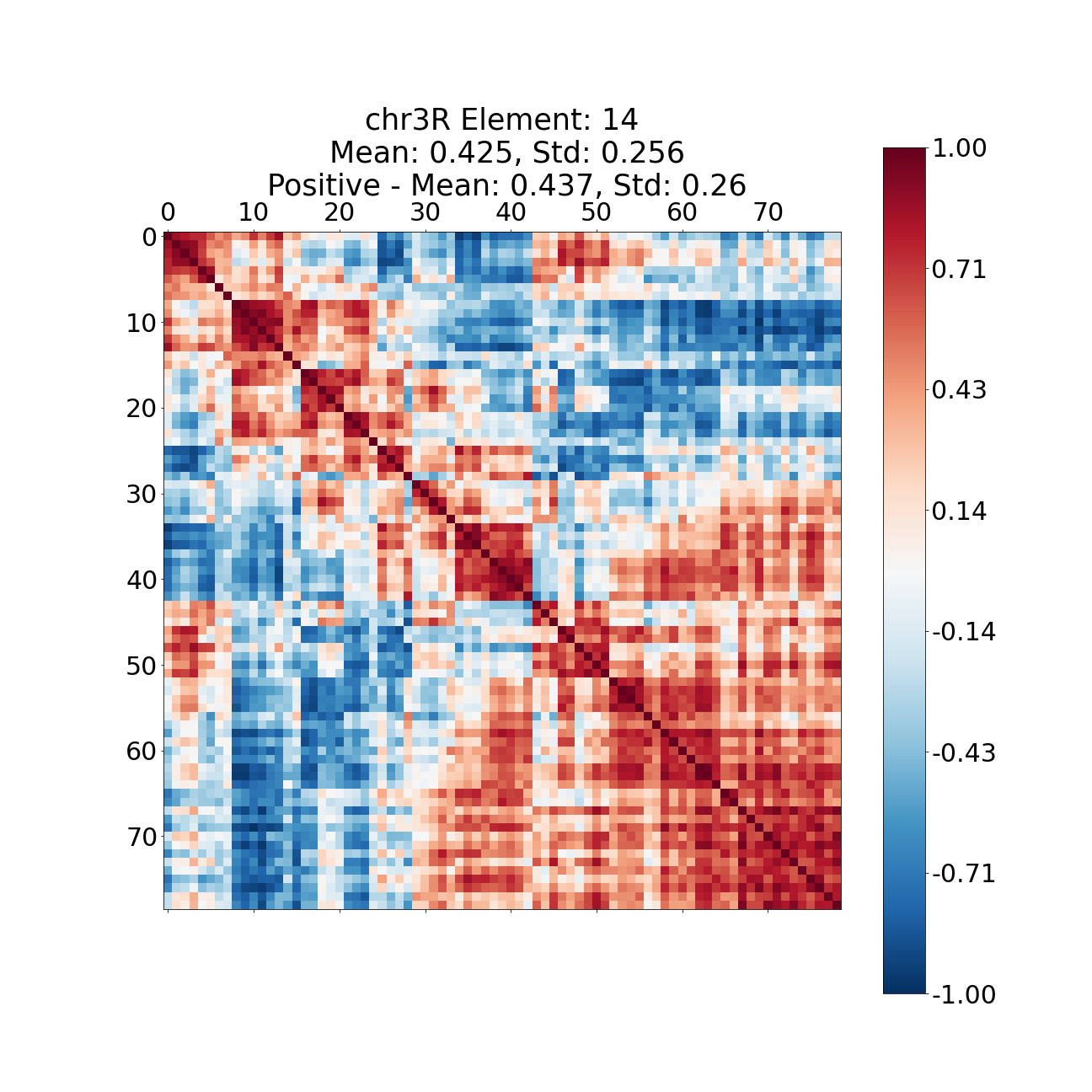}
     \end{subfigure}
      \hfill
     \begin{subfigure}[b]{0.32\textwidth}
         \centering
         \includegraphics[width=\textwidth]{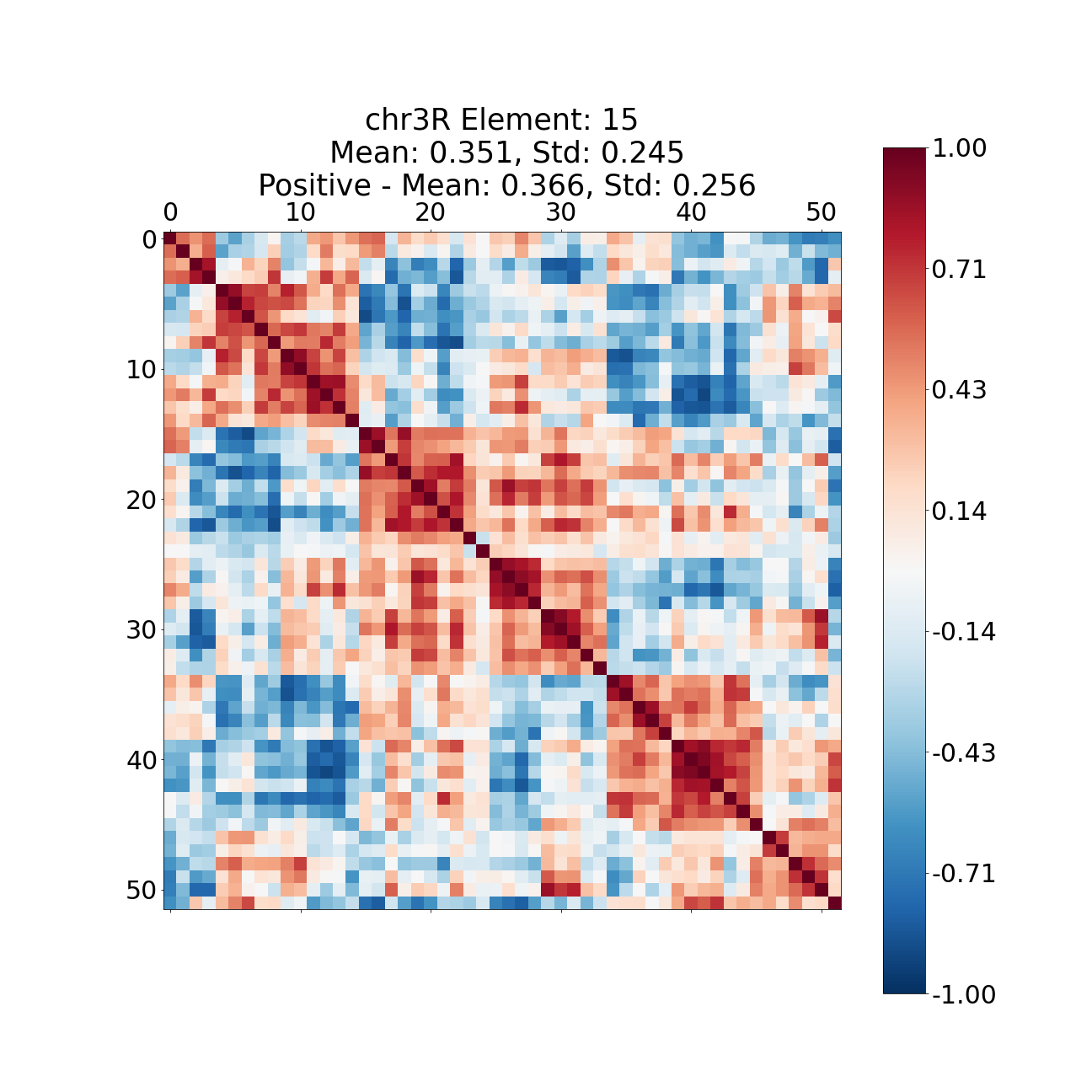}
     \end{subfigure}
     \hfill
     \begin{subfigure}[b]{0.32\textwidth}
         \centering
         \includegraphics[width=\textwidth]{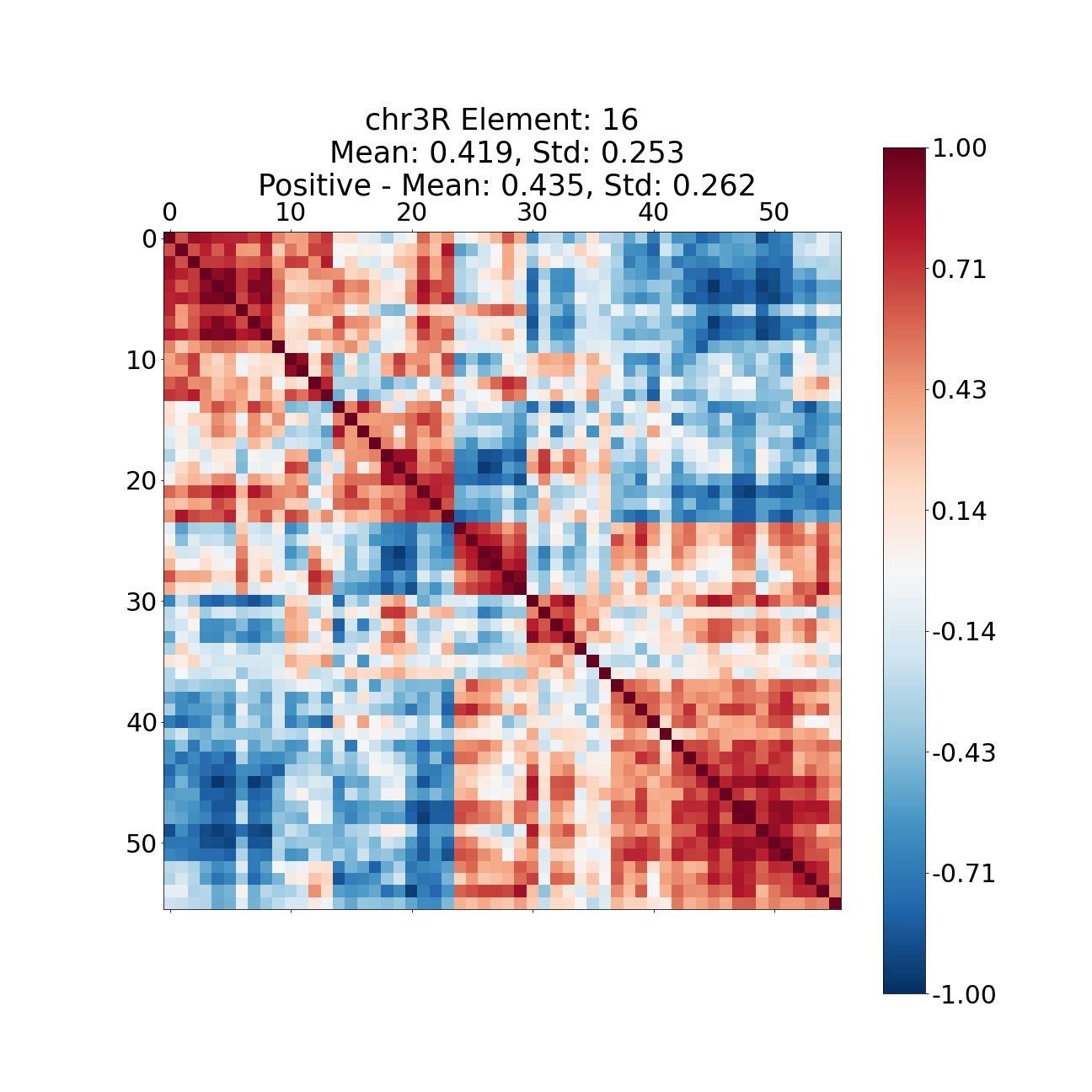}
     \end{subfigure}
     \hfill
     \begin{subfigure}[b]{0.32\textwidth}
         \centering
         \includegraphics[width=\textwidth]{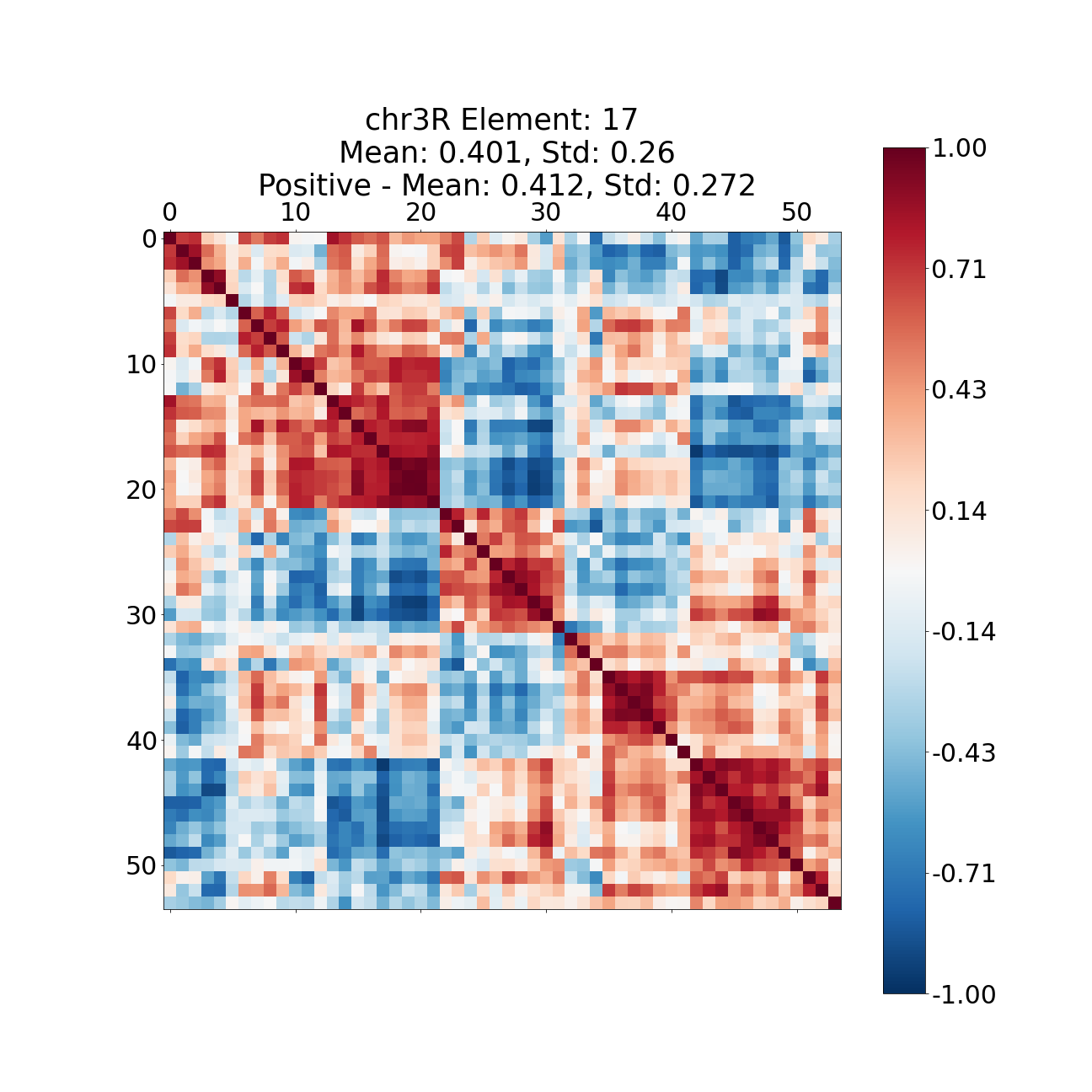}
     \end{subfigure}
     \hfill
     \begin{subfigure}[b]{0.32\textwidth}
         \centering
         \includegraphics[width=\textwidth]{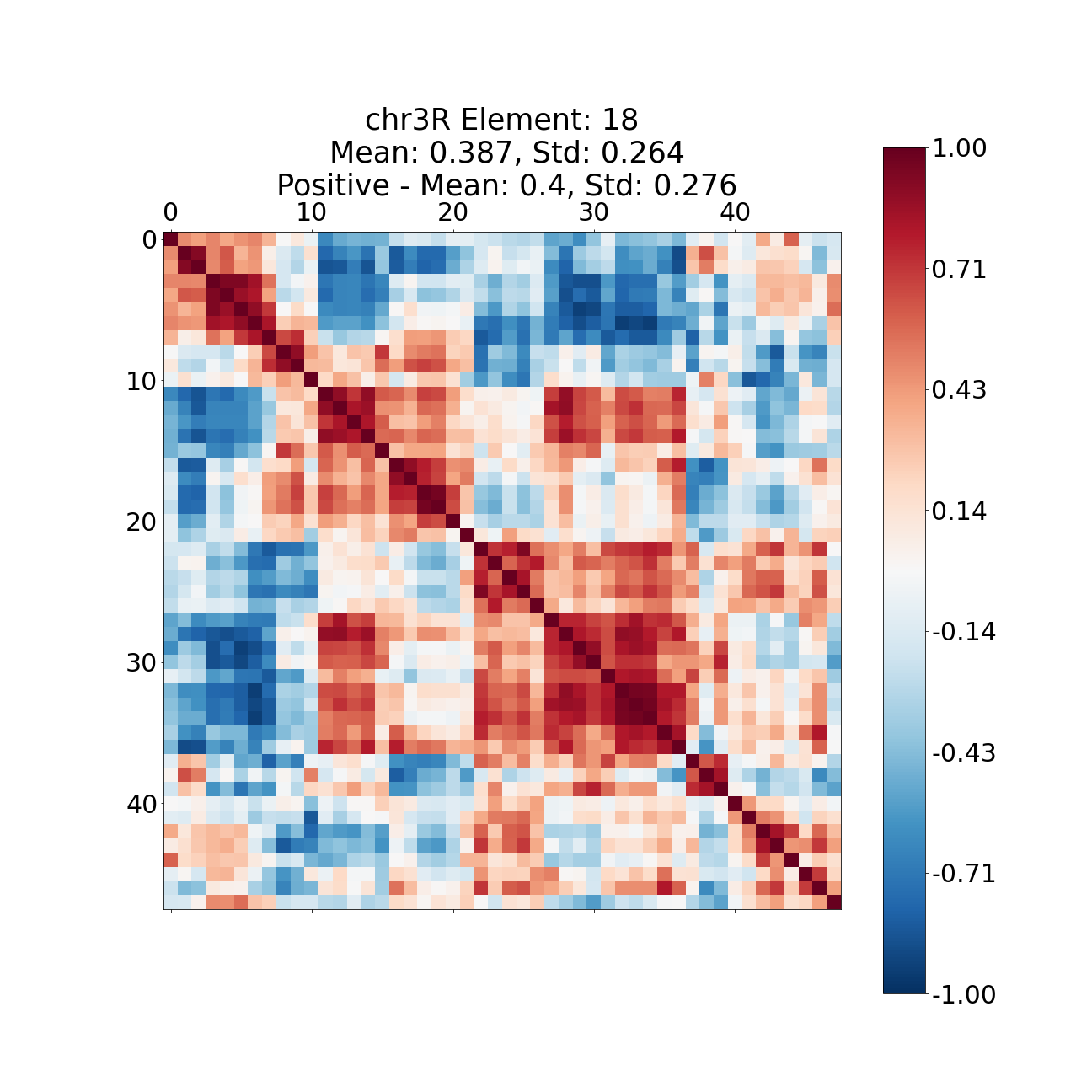}
     \end{subfigure}
      \hfill
     \begin{subfigure}[b]{0.32\textwidth}
         \centering
         \includegraphics[width=\textwidth]{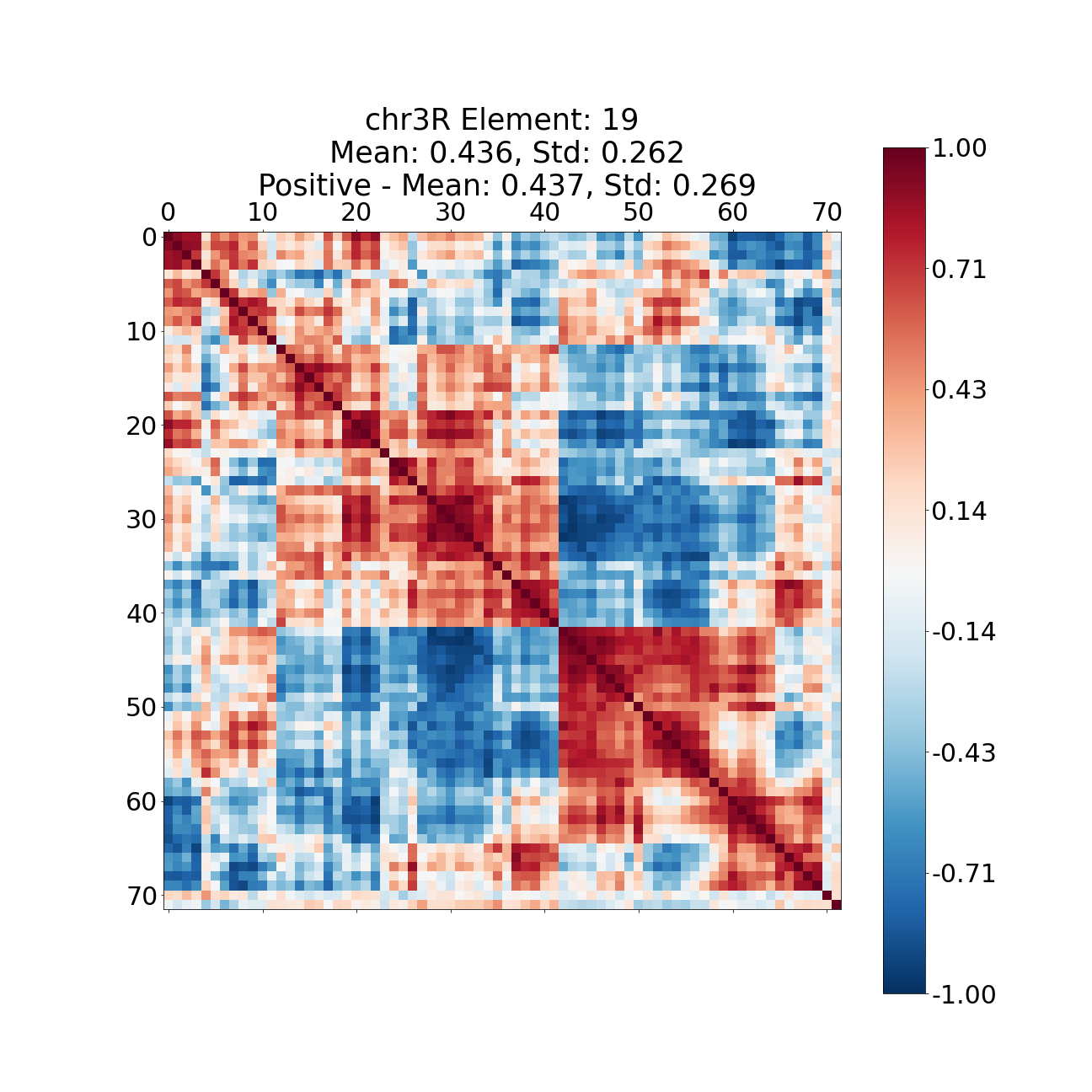}
     \end{subfigure}
     \hfill
     \begin{subfigure}[b]{0.32\textwidth}
         \centering
         \includegraphics[width=\textwidth]{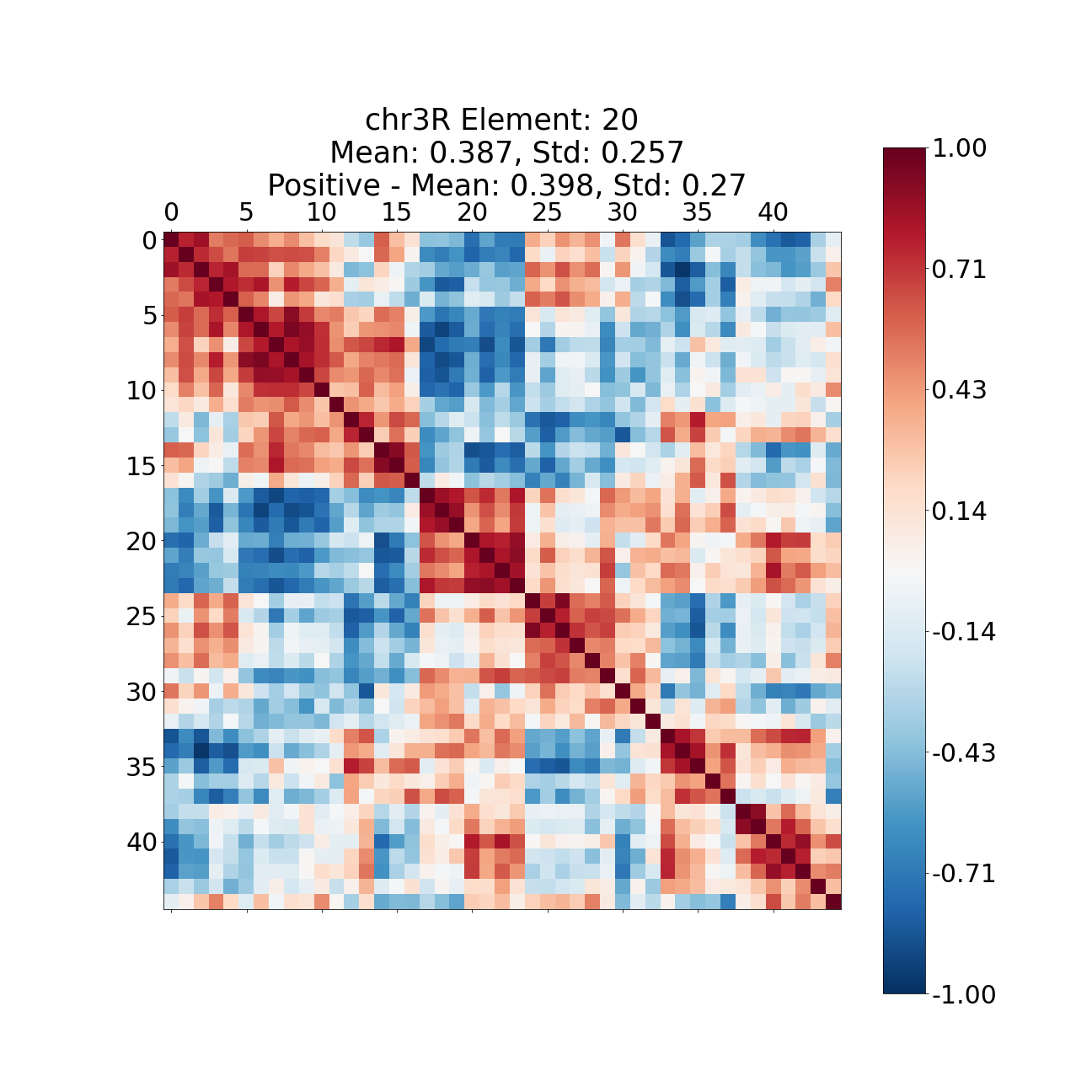}
     \end{subfigure}
     \hfill
     \begin{subfigure}[b]{0.32\textwidth}
         \centering
         \includegraphics[width=\textwidth]{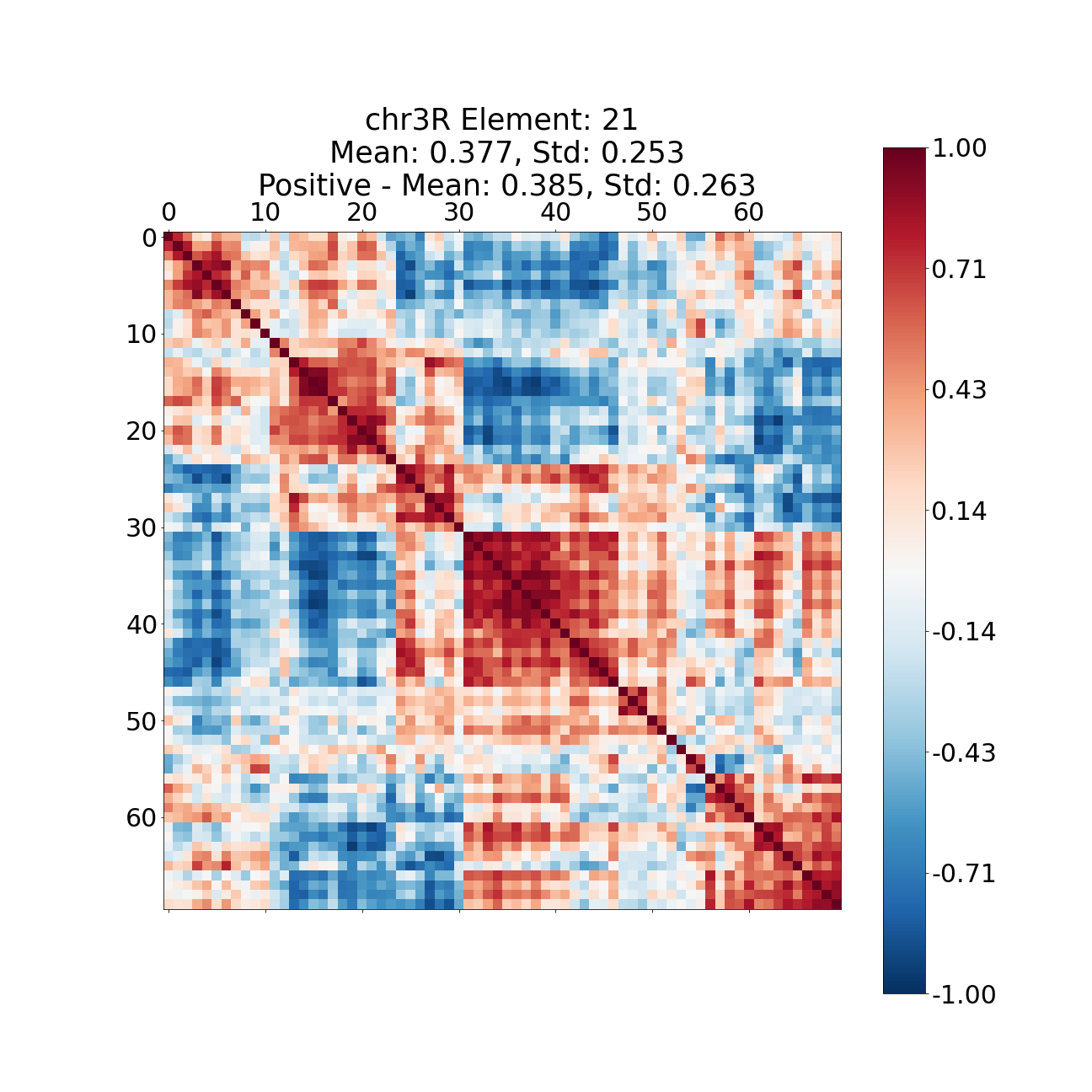}
     \end{subfigure}
     \hfill
     \begin{subfigure}[b]{0.32\textwidth}
         \centering
         \includegraphics[width=\textwidth]{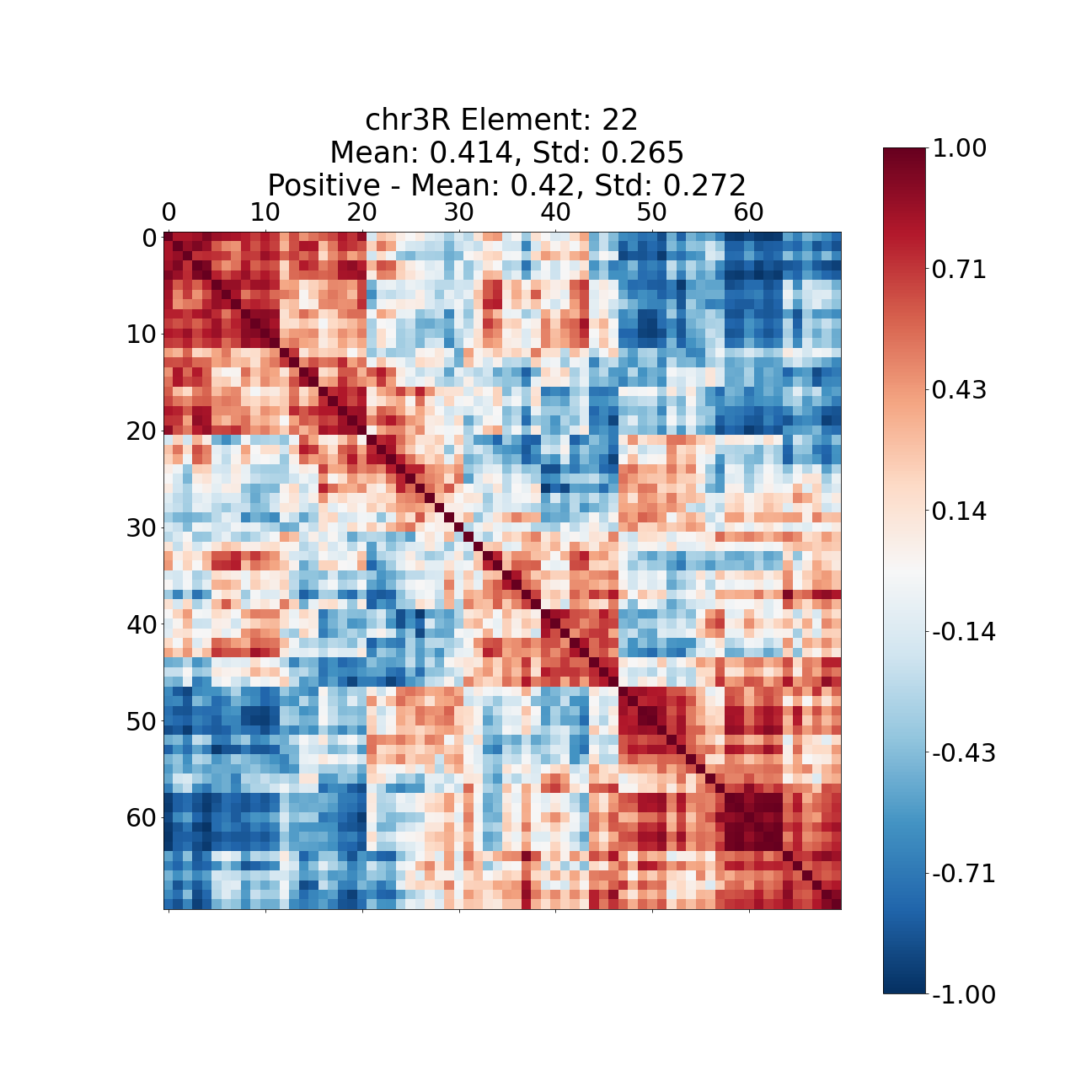}
     \end{subfigure}
      \hfill
     \begin{subfigure}[b]{0.32\textwidth}
         \centering
         \includegraphics[width=\textwidth]{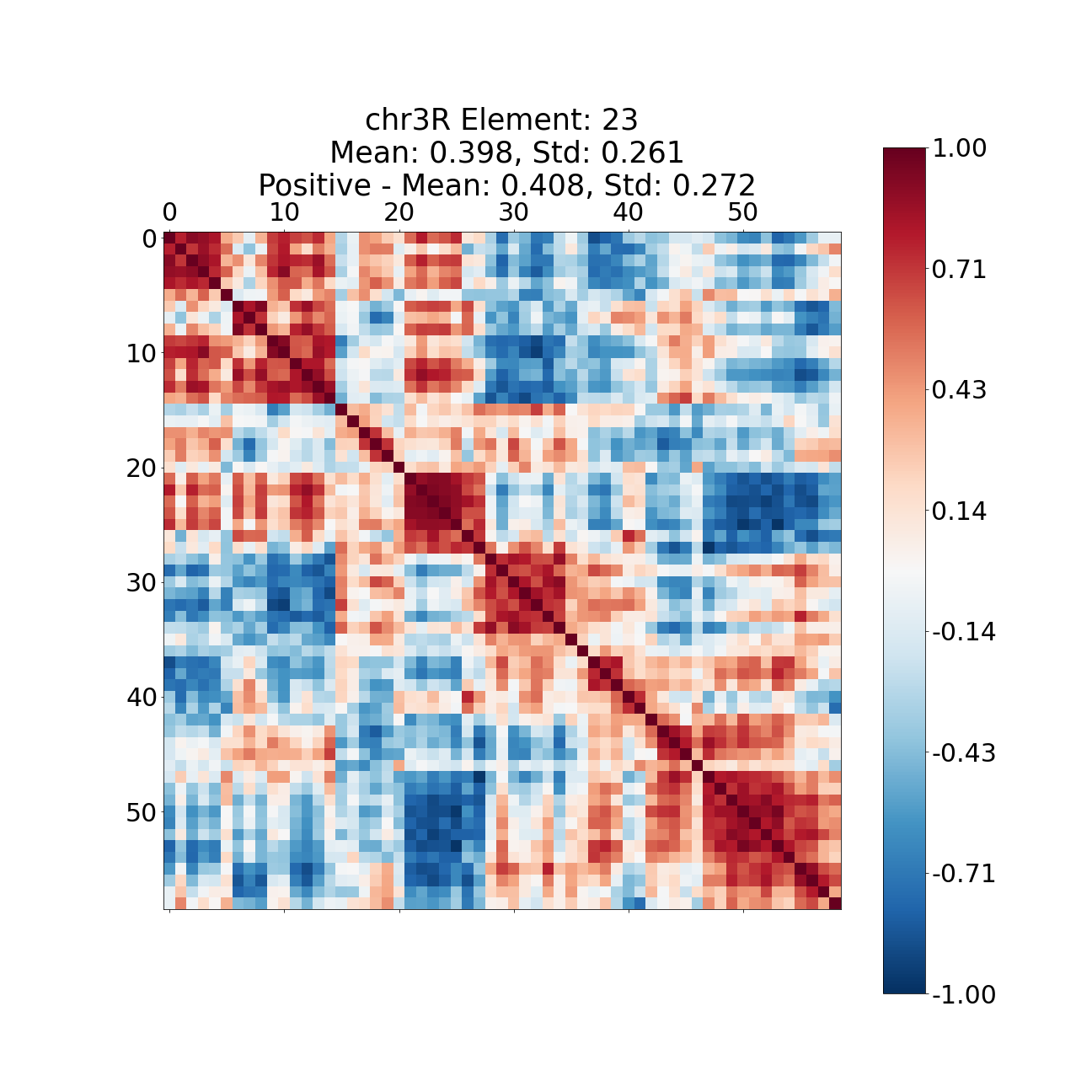}
     \end{subfigure}
        \caption{Pairwise coexpression of genes covered by various dictionary elements for chr 3R obtained through online cvxNDL. We calculated the mean and standard deviation of absolute pairwise coexpression values, along with the mean and standard deviation of coexpression values specifically for all positively correlated gene pairs.}
\end{figure}

\begin{figure}[h]
\ContinuedFloat
     \centering
     \begin{subfigure}[b]{0.32\textwidth}
         \centering
         \includegraphics[width=\textwidth]{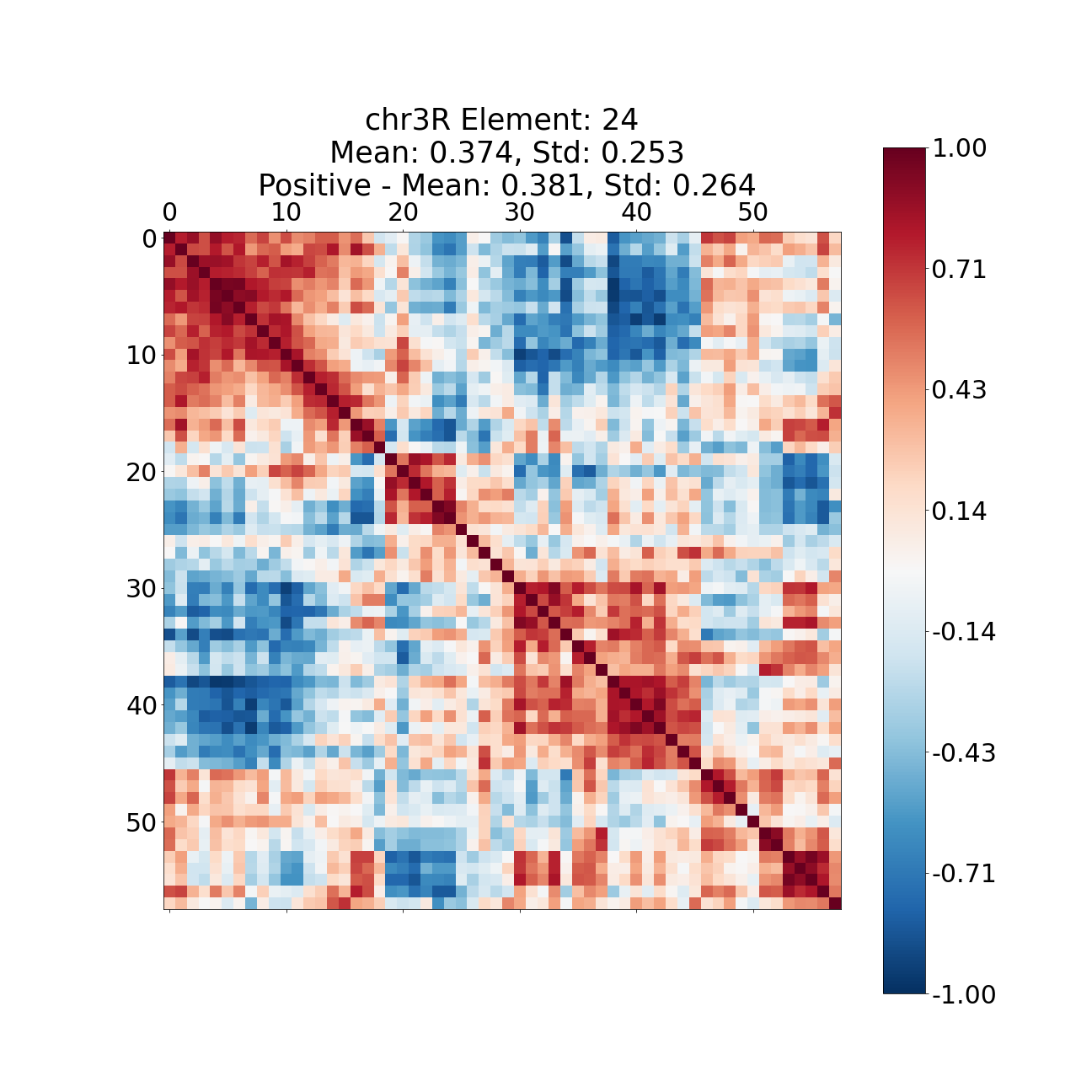}
     \end{subfigure}
        \caption{Pairwise coexpression of genes covered by various dictionary elements for chr 3R obtained through online cvxNDL. We calculated the mean and standard deviation of absolute pairwise coexpression values, along with the mean and standard deviation of coexpression values specifically for all positively correlated gene pairs.}
\end{figure}


\clearpage

\section{STRING interaction network and FlyMine}

\begin{figure}[h]
\begin{subfigure}{.5\textwidth}
  \centering
  \includegraphics[width=.8\linewidth]{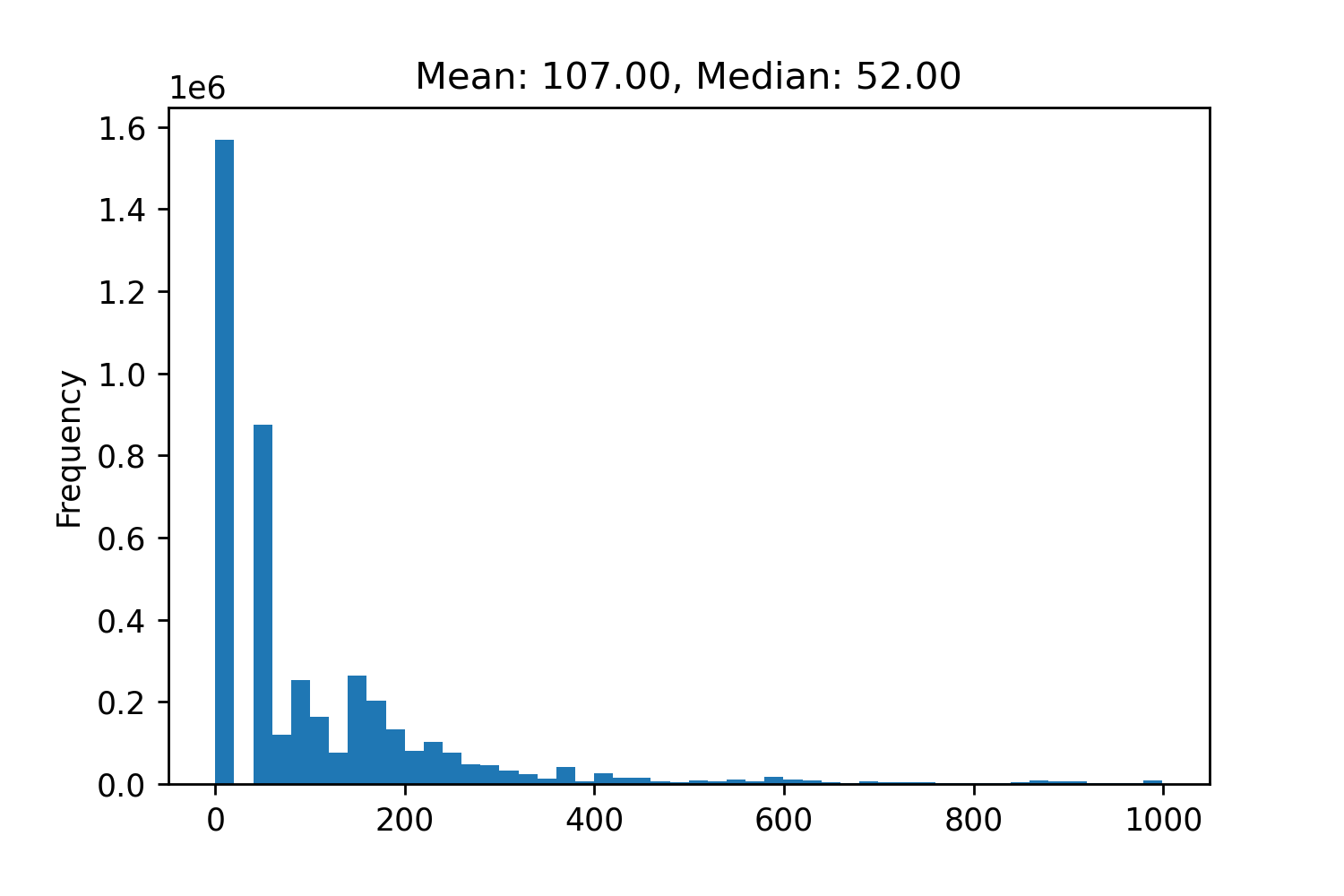}
  \caption{All confidence values.}
  \label{fig:string_confidence_a}
\end{subfigure}
\begin{subfigure}{.5\textwidth}
  \centering
  \includegraphics[width=.8\linewidth]{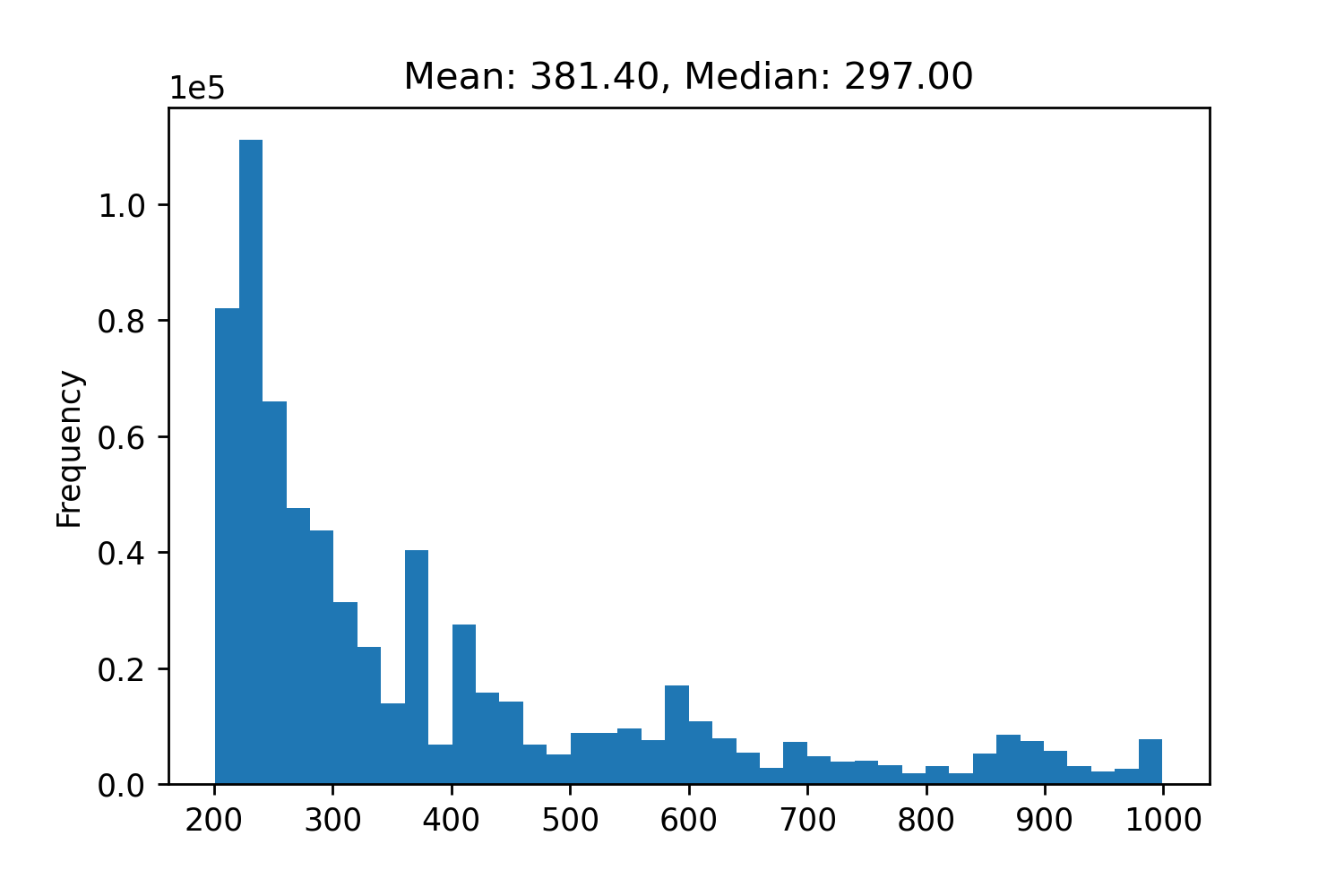}
  \caption{Confidence values filtered for $>200$.}
  \label{fig:string_confidence_b}
\end{subfigure}
\caption{Histogram of confidence values for pairwise interaction of proteins in the STRING interaction network for \emph{Drosophila Melanogaster}.}
\label{fig:string_confidence}
\end{figure}

The STRING interaction network~\cite{szklarczyk2019string} provides a confidence score indicating the interaction likelihood between a pair of proteins within an organism. This score reflects both direct interactions via physical protein binding and indirect interactions by virtue of the proteins participating in the same cellular pathways. The confidence level of interaction between a pair of proteins can vary from $0$, indicating very low confidence, to $1000$, indicating very high confidence. Figure \ref{fig:string_confidence_a} shows the distribution of confidence levels between all pairs of proteins in the STRING database for \emph{Drosophila Melanogaster}. A large majority of these interactions are very low confidence. To focus on more reliable interactions, we filtered the protein interactions to retain only those with a confidence score exceeding $200$, resulting in a refined dataset shown in Figure \ref{fig:string_confidence_b}. By mapping these proteins back to their corresponding genes, we derived an induced network representing gene-gene interactions.

\begin{figure}[h]
\begin{subfigure}{.5\textwidth}
  \centering
  \includegraphics[width=.8\linewidth]{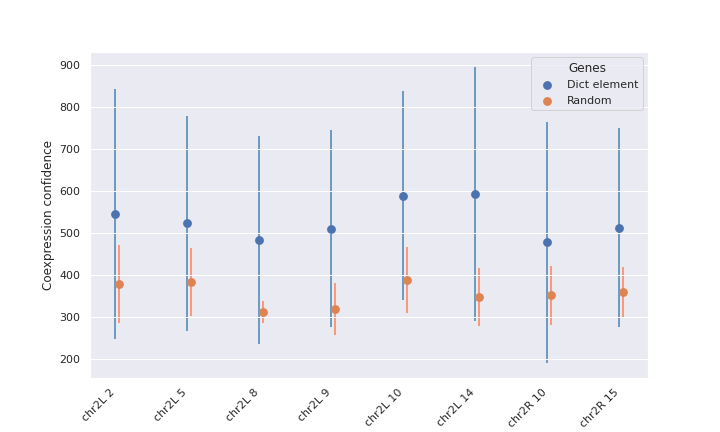}
  \caption{Mean confidence value for dictionary elements from chr2L and chr2R.}
  \label{fig:string_confidence_chr2}
\end{subfigure}%
\begin{subfigure}{.5\textwidth}
  \centering
  \includegraphics[width=.8\linewidth]{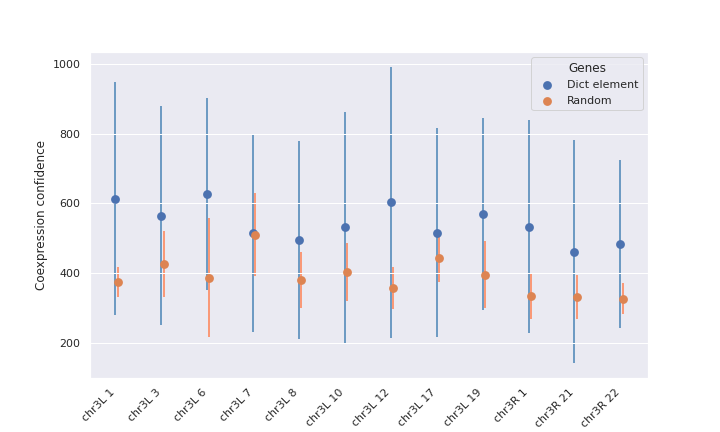}
  \caption{Mean confidence value for dictionary elements from chr3L and chr3R}
  \label{fig:string_confidence_chr3}
\end{subfigure}
\caption{Confidence levels for pairwise interaction of proteins for dictionary elements based on STRING interaction network.}
\label{fig:string_confidence_dicts}
\end{figure}

For the online cvxNDL dictionary, we calculated the mean confidence level for all pairs of proteins. We also repeated the same experiments with a randomly constructed dictionary as a control. Figure \ref{fig:string_confidence_dicts} shows the mean confidence level and confidence interval for a subset of dictionary elements. We performed a K-S test with the null hypothesis that the two sets of confidence scores for pairwise interactions belonging to online cvxNDL dictionaries and randomly constructed dictionaries are drawn from the same distribution. We rejected the null hypothesis with p-value $<0.05$.

Flymine~\cite{lyne2007flymine} is a large genomic and proteomic database for \emph{Drosophila Melanogaster}. We used FlyMine to retrieve a list of upregulated genes in S2 cell lines. We observe that the upregulated genes are overrepresented in our dictionary elements. To test our hypothesis, we performed the hypergeometric overrepresentation test. Our null hypothesis is that the proportion of upregulated genes in our dictionary elements is no higher than the overall proportion of upregulated genes in S2 cell lines. We rejected the null hypothesis (p-value $<0.05$) for all dictionary elements for all chromosomes except a small subset of $4$ dictionary elements ($1$ dictionary element from chr2R and $3$ dictionary elements from chr3L). The p-values for all dictionary elements are shown in Table~\ref{tab:flymine_hypergeometric}.  

\begin{table}[!h]
\centering
\scriptsize
\caption{Results for hypergeometric overrepresentation test for all dictionary elements. We report the p-values corresponding to the null hypothesis that the proportion of upregulated genes in our dictionary elements is no higher than the overall proportion of upregulated genes in S2 cell lines.}
\label{tab:flymine_hypergeometric}
\begin{tabular}{l||c|c|c|c}
\hline\hline
\begin{tabular}[c]{@{}l@{}}dictionary \\ element\end{tabular} & chr2L & chr2R & chr3L & chr3R \\ \hline
0 & 1.18E-03 & 5.90E-07 & 7.96E-05 & 3.24E-05 \\
1 & 1.93E-08 & 8.13E-06 & 5.38E-04 & 9.23E-09 \\
2 & 4.36E-08 & 4.44E-07 & 1.40E-03 & 1.36E-02 \\
3 & 8.13E-06 & 7.92E-05 & 1.65E-04 & 4.49E-08 \\
4 & 4.50E-06 & 1.83E-04 & 2.54E-03 & 4.88E-12 \\
5 & 1.23E-06 & 3.93E-04 & 3.53E-03 & 5.84E-05 \\
6 & 1.26E-03 & 2.88E-03 & 5.58E-03 & 6.07E-06 \\
7 & 1.60E-03 & 3.88E-06 & 1.76E-03 & 1.39E-05 \\
8 & 3.50E-05 & 9.15E-07 & 1.22E-04 & 3.03E-05 \\
9 & 2.17E-04 & 2.17E-06 & 2.73E-04 & 4.36E-07 \\
10 & 1.02E-05 & 3.57E-02 & 5.23E-06 & 2.37E-06 \\
11 & 1.82E-05 & 8.94E-04 & 8.92E-02 & 1.96E-04 \\
12 & 2.08E-06 & 8.90E-04 & 2.01E-01 & 3.23E-05 \\
13 & 8.12E-05 & 8.52E-03 & 3.40E-05 & 1.73E-04 \\
14 & 1.95E-05 & 1.41E-04 & 1.93E-03 & 1.84E-10 \\
15 & 6.95E-08 & 5.78E-05 & 1.20E-02 & 8.32E-05 \\
16 & 5.02E-03 & 7.60E-04 & 1.78E-03 & 4.82E-06 \\
17 & 3.24E-04 & 5.41E-02 & 9.17E-06 & 7.53E-04 \\
18 & 1.78E-03 & 6.04E-06 & 1.96E-02 & 3.89E-06 \\
19 & 3.89E-04 & 3.56E-05 & 8.10E-04 & 6.86E-08 \\
20 & 1.75E-08 & 2.90E-04 & 5.02E-03 & 1.50E-04 \\
21 & 6.41E-03 & 1.55E-02 & 3.72E-06 & 8.88E-10 \\
22 & 2.99E-03 & 1.40E-03 & 2.24E-05 & 9.23E-09 \\
23 & 1.65E-05 & 6.78E-03 & 5.98E-03 & 3.42E-07 \\
24 & 2.54E-06 & 1.03E-04 & 6.22E-02 & 7.19E-08 \\
\hline\hline
\end{tabular} 
\end{table}

\bibliographystyle{plos2015}
\bibliography{supplement.bib}